\documentclass[openany, 12pt]{book}
\usepackage[utf8]{inputenc}
\usepackage[left=1.5in,right=1in,top=1.3in,bottom=1.3in]{geometry}
\usepackage{amssymb}
\usepackage{amsmath}
\usepackage{feynmp}
\usepackage[force]{feynmp-auto}
\usepackage[numbers]{natbib}
\usepackage{cancel}
\usepackage[titletoc]{appendix}
\usepackage{esvect}
\usepackage{titlesec}
\usepackage{slashed}
\usepackage{adjustbox}
\usepackage{graphicx}
\usepackage[dvipsnames]{xcolor}
\usepackage{dsfont}
\usepackage{ulem}
\usepackage{hyperref}

\newcommand{\be}{\begin{eqnarray}}
\newcommand{\ee}{\end{eqnarray}}
\newcommand{\conclusion}[1]{%
  \chapter*{#1}%
  \markboth{#1}{#1}%
  \addcontentsline{toc}{chapter}{#1}}

\begin{document}
\begin{fmffile}{diagram}

\title{
\Huge{\textbf{Aspects of Self-Dual Yang-Mills and Self-Dual Gravity}}}
\author{Pratik Chattopadhyay}
\date{Thesis submitted to the University of Nottingham for the degree\\ of Doctor of Philosophy}
\maketitle

\chapter*{Abstract}
In this thesis, we study the all same helicity loop amplitudes in self-dual Yang-Mills and self-dual gravity. These amplitudes have long been conjectured to be interpreted as an anomaly and are recently linked to the UV divergence of two-loop quantum gravity. In the first part of the thesis, we study the loop amplitudes in self-dual Yang-Mills. We show that the four point one-loop amplitude can be reduced to a computation of shifts, which strongly suggests a case for an anomaly interpretation. We next propose a new formula for the one-loop amplitudes at all multiplicity, in terms of the Berends-Giele currents connected by an effective propagator. We prove the formula by observing that it readily implies the correct collinear properties. To demonstrate the validity of our formula, we do an explicit computation at 3, 4 and 5 points and reproduce the known results. The region momenta variables play an important role in our formula and thus it points to both the worldsheet and the momentum twistor interpretations.
In the second part of the thesis, we study the one loop behaviour of chiral Einstein-Cartan gravity and the one-loop amplitudes in self-dual gravity. We develop the ghost Lagrangian in chiral Einstein-Cartan gravity for a general Einstein background using the BRST formalism and compute the ghost contribution to the one-loop effective action. We next construct the one-loop graphs contributing to the four point same helicity amplitude. The double copy property is manifest in the diagrams. We also perform a shift computation of the self-energy bubble in gravity and show that the result is the square of Yang-Mills. The bubble is interpreted as an effective propagator, in complete analogy with Yang-Mills. However, the interpretation of the shift parameters in this case is not clear and thus the computation of the four point amplitude remains incomplete. We comment on a possible way to resolve this ambiguity. 

\addcontentsline{toc}{chapter}{Abstract}

\chapter*{Acknowledgement}
I would first and foremost like to thank my supervisor, Kirill Krasnov, without whose guidance and continuous support, this thesis would not have come into existence. He always encouraged me to be as independent and original as possible, but nevertheless  discussed, understood and helped me through my concerns whenever there was a need. It was an exciting as well as a challenging experience to work under his supervision.  
\\~\\
I am deeply grateful to Sven Gnutzmann and Silke Weinfurtner for their valuable suggestions and intensive discussions, which formed a crucial role in my overall development. Apart from learning how to work in a strategic way, this also led me to introspect myself and bring in a culture of healthy criticism, which is an integral part of any good research. 
\\~\\
I would also like to pay my sincere gratitude to K.P Yogendran for his continuous motivation and intense discussions regarding research and life in general. He taught me how to keep an eye on stuffs other than one's own research and to enlarge the research skills. 
\\~\\
Next I want to acknowledge all those friends and colleagues who made this journey enjoyable and without whom, it would have been much harder to sustain: Amit Bhunia (Amit Da), Arindam Chatterjee, Sanjib Kumar Das, Vikramaditya Mondal, Shreeya Rajadhyaksha, Akash Sharma, Gaurav Sharma, Neekar M. Mohammed etc. I would like to thank Guy Jehu for his insightful comments on the research which I have undertaken and Devashish Singh for interesting discussions. 
\\~\\
Finally, I would like to thank my parents for their immense contribution and sacrifice towards making me competent and boosting my confidence time and again in this journey. Without the support of them and my close relatives, this would not have been possible.

\addcontentsline{toc}{chapter}{Acknowledgement}

\newpage 
\tableofcontents
\newpage
\chaptermark{Introduction}

\chapter{Introduction}
\section{Motivation}
\subsection*{Quantum gravity and amplitudes}
It has been well established since a long time now that quantum field theory approach to quantum gravity fails. It gives results which are not consistent with perturbative unitarity and leads to loss of predictability. There are several ways to understand this. The perturbative expansion of the Einstein-Hilbert Lagrangian yields terms which are second order in derivatives. Thus, the coupling constant in gravity has negative mass dimensions. The tree amplitude for $2\rightarrow 2$ graviton scattering grows with the energy scale and indicates power counting non-renormalizability. Further, if one computes loop corrections, it is not hard to see that one would encounter more and more divergent integrals at each loop order because of increasing number of derivatives, which is an implication of negative mass dimension coupling.
Thus, one can say that perturbative quantum gravity does not make sense. However, this argument is too naive as it stands and it will be inappropriate to immediately arrive at this conclusion without explicitly computing physical quantities of interest. It may happen that there are cancellations between the divergent parts of the diagrams and the final result does make sense. Indeed, as it turns out, pure gravity with zero cosmological constant is finite at one-loop \cite{tHooft:1974toh}. This can be understood by analysing the arising counter-terms. The only possible counter terms that may arise at one-loop are of the form $R^2$, $R_{\mu\nu}R^{\mu\nu}$ and $R_{\mu\nu\lambda\rho}R^{\mu\nu\lambda\rho}$, where $R$ is the Ricci scalar, $R_{\mu\nu}$ is the Ricci tensor and $R_{\mu\nu\lambda\rho}$ is the Riemann curvature tensor. Thus, the counter-term Lagrangian takes a form $\mathcal{L}_{R^2}=a_1R^2+a_2R_{\mu\nu}R^{\mu\nu}+a_3 R_{\mu\nu\lambda\rho}R^{\mu\nu\lambda\rho}$, for arbitrary constants $a_i$. However, the first two terms can be eliminated by field re-definitions and the last term can be expressed in terms of the first two by adding the Gauss Bonnet curvature square. The Gauss Bonnet term is topological and therefore it integrates to zero in topologically trivial backgrounds. Thus, the one-loop counter-term completely vanishes. Four dimensional quantum gravity without cosmological constant is then finite at one-loop. If one assumes a non-zero cosmological constant, divergences do arise but they can be absorbed into the tree level action. Despite behaving nicely at one-loop, gravity does diverge at two-loops. This was first observed explicitly by Goroff and Sagnotti \cite{Goroff:1985th} and the non-zero two loop divergence is given by the term $\mathcal{L}_{R^3}=R^{\mu\nu}_{~~\sigma\delta}R^{\sigma\delta}_{~~\lambda\nu}R^{\lambda\nu}_{~~\mu\nu}$. This term can neither be eliminated by field re-definitions nor is a total derivative. Thus, quantum gravity at two-loops is said to be non-renormalizable. This fact led to abolishing perturbative treatments of quantum gravity and motivated interest in building new frameworks like string theory, loop quantum gravity and causal sets to name a few. However, each of these frameworks have their own technical limitations and this is why the problem of consistently quantizing gravity in four dimensions remains open to this day. 
\\~\\
In recent years, and in particular over the last two decades, there has been tremendous development in the field of scattering amplitudes \cite{Brandhuber:2007vm, Brandhuber:2006bf, Cachazo:2004kj, Cachazo:2004zb, Bern:2013uka, Mason:2009qx, Bern:1998sv, Witten:2003nn}. On one hand, new techniques such as recursion relations and on-shell methods are implemented to simplify computations in Yang-Mills and gravity. While on the other, conceptual developments are made in both these theories and their supersymmetric counterparts \cite{Cachazo:2004zb, Bern:2013uka, Mason:2009qx}. Several works in the past \cite{Bern:1993sx, Bern:1993qk, Mahlon:1993fe, Berends:1987me, Ossola:2008xq, Brandhuber:2006bf} have outlined the detailed computations of multi-photon/gluon/graviton scattering amplitudes. A corollary of these developments indicate that the UV behaviour of gravity is in some sense, better than Yang-Mills. Indeed, the BCFW recursion relations \cite{Britto:2004ap} imply that the fall-off behaviour of gravity tree amplitudes at large values of complex momenta goes like $1/z^2$, as opposed to Yang-Mills which goes like $1/z$, where $z$ is the complex momentum parameter. This is related to the fact that the group of diffeomorphisms is at play in gravity. However, the real surprise of the developments is that gravity is rather linked to Yang-Mills. This has been realized in the so called double copy relations where gravity amplitudes are a certain square of the Yang-Mills ones \cite{Bern:2010ue}.
\\~\\
Alongside these developments, there has been a resurgence of interest during the last few years in probing the ultraviolet structure of gravity, using on-shell techniques. In recent years, concrete results by Bern and collaborators \cite{Bern:2015xsa, Bern:2017puu} have shown that the UV behaviour of gravity is much more subtle and interesting than was thought earlier. In particular, the two-loop divergence was analyzed and it was observed that its coefficient is sensitive to off-shell degrees of freedom in the theory. In their work \cite{Bern:2015xsa}, they have added non-dynamical three-forms to gravity and found that the coefficient of the two-loop divergence changes. Also, when pseudo-scalar fields are replaced by their duals, i.e anti-symmetric two-forms, the divergence once again changes. The Gorof and Sagnotti computation of the two-loop divergence, although gives a direct result for the coefficient, falls short of giving any understanding of the particular number. Thus, it is not very surprising that the coefficient depends on off-shell degrees of freedom. However, in a subsequent paper \cite{Bern:2017puu}, what they found is that the coefficient of the renormalization scale dependence remains unchanged if one changes the off-shell contents of the theory. So, while the coefficient of the divergence does change, the renormalization scale dependence is not sensitive to the unphysical contents of the theory.  Thus, it comes as a rather surprise that there is no direct relation between the coefficients of the divergent part of the two-loop amplitude and that of the renormalization scale dependence in gravity, unlike in conventional quantum field theories. 
Using unitarity cuts, they have analyzed the divergent contribution and the associated renormalization scale dependence of the identical helicity four graviton scattering amplitude at two loops. It is quite clear from their computation that this coefficient gets a non-vanishing contribution from the two-particle cut, where the identical helicity one-loop four graviton amplitude appears on one side of the cut and a tree level four graviton amplitude appears on the other side. The three particle cuts do not contribute and vanish identically. Thus, overall the coefficient of the divergent part of the four graviton two-loop amplitude reduces to the same helicity one-loop amplitude. The relevant expression is given by  \cite{Bern:2017puu}
\be 
\mathcal{M}^{2-loop}_4(++++)=\Big(\frac{\kappa}{2}\Big)^6\frac{i}{(4\pi)^4}s_{12}s_{23}s_{13}\Bigg(\frac{[12][34]}{\langle 12\rangle\langle 34\rangle}\Bigg)^2\Bigg(\frac{209}{24\epsilon}-\frac{1}{4}\textrm{ln} \mu^2\Bigg)+\textrm{finite},\nonumber 
\ee 
where $s_{12}s_{23}s_{13}\Big(\frac{[12][34]}{\langle 12\rangle\langle 34\rangle}\Big)^2$ is the one loop amplitude. Thus, the divergence of quantum gravity at two-loops is related to the non-vanishing of the all plus four point amplitude at one-loop. It is therefore very important to understand this non-vanishing and its origin. The main aim of this thesis is to sharpen the understanding of this amplitude and thus the two loop divergence. However, such an amplitude is already non-vanishing in simpler theories like massless QED and self-dual Yang-Mills (SDYM). It then makes sense to simplify the question and understand such a non-vanishing first in these simpler theories. In the order of complexity, the theories are scalar QED, spinor QED, SDYM and self-dual gravity (SDGR). Thus, the study of the all same helicity (minus in our convention) amplitudes in these theories is the unifying theme of this thesis.  
\\~\\
An important feature of the identical helicity amplitudes in SDYM and SDGR is that
the integrand of this amplitude vanishes
if we take the loop momenta to be four-dimensional. This is easily verified in the case of gravity from unitarity principles as shown in \cite{Bern:2015xsa}. In SDYM, this behaviour is tied to an anomaly of the currents which make the theory integrable \cite{Bardeen:1995gk}. However, the existing computations of this amplitude in massless QED and SDYM \cite{Mahlon:1993fe, Bern:1993sx} do not reveal any anomaly type of behaviour and neither give any deeper understanding of the physics behind the non-vanishing. In this thesis, a new understanding of these amplitudes first at four points and then at all multiplicity is achieved in the case of Yang-Mills (and SDYM). At four points an analogous understanding is also achieved for the case of scalar QED and massless spinor QED. In particular, the four point one-loop amplitude of all these theories is reduced to a computation of shifts of the so-called quadratically divergent integrals. The computation is carried out in four dimensions and the result comes out without performing any integral explicitly, but rather from the shift technology that is well established in the case of chiral anomaly triangle diagram calculation. Although it does give some hint that the result may be related to an anomaly like behaviour, it directly does not give any interpretation of these amplitudes. The more prominent new understanding in the case of SDYM amplitudes at all multiplicity is achieved in terms of our formula with the Berends Giele currents of SDYM. It is now clear that the amplitudes are constructed from simpler building blocks, such as the currents, which are completely understood, and the self-energy diagram that is ambiguous (shift-dependent) but finite. The amplitude puts these blocks together in a way that removes the ambiguity and produces an expression that has all the right property such an amplitude should have. This can be understood as following from the fact that the loop momenta can be chosen so that the complete one-loop integrand vanishes. The special choice of loop momenta is related to the so-called region momenta (or dual momenta) variables which has interpretations in both the worldsheet formulation of Yang-Mills (YM) \cite{Chakrabarti:2005ny} and its twistor descriptions \cite{Mason:2009qx}. Thus, the interpretation in the case of YM reduces to the question of interpreting the bubbles, which is a much simpler and more sharply posed question than the original question of interpretation of the non-vanishing of the all same helicity amplitudes. The interpretation of the bubbles is not answered in this thesis and is left for future work. However, it is clear that considerable progress has been made to gain understanding of the same helicity amplitudes in YM.
\\~\\
We have also developed partial understanding of the said amplitudes in the case of gravity. It is now clear that the one-loop same helicity amplitudes in self-dual gravity (SDGR) can be constructed from the flat SDGR Feynman rules. The integrand of these amplitudes show the double copy structure. These can also be constructed from the full GR Feynman rules but we work in the simpler theory of SDGR. The amplitudes can be understood from the perspective of simpler building blocks like the currents and the self-energy bubble in an analogous way like SDYM. In particular we have performed a detailed computation of the self-energy bubble in this case and the result comes out to be the square of YM. More so, it is also clear that one can glue Berends Giele currents to this bubble and sum all possible contributions to represent the full amplitude. However, the interpretation of specific choices of loop momenta is not clear in this case and this is left for future work. Thus, for gravity we have detailed out the first steps of the understanding of these amplitudes in this thesis, which can now be phrased in terms of the 
understanding of the bubble. We expect to do the remaining part in a future work. 
\newpage 
\section{Thesis outline} 
Let us now sketch the outline of this thesis. It is divided into two parts. Part I deals with chiral formulation of Yang-Mills and amplitudes in self-dual Yang-Mills. Part II deals with chiral formulation of gravity and amplitudes in self-dual gravity. In the next section we review some basic concepts in scattering amplitudes. In chapter 2, we summarise the mathematical tools relevant for the computations in this thesis. In chapter 3, we give a brief introduction to chiral Yang-Mills.
In chapter 4, we expand on the chiral formulation of Yang-Mills, namely the Chalmers-Siegel action and compute the $\beta$-function in this theory using Feynman diagrams. We then explain the theory of self-dual Yang-Mills in its covariant form, describe the gauge fixing and write the Feynman rules. We next compute the BG current in this theory.\\
In chapter 5, we elaborate on the amplitude sector in self-dual Yang-Mills. We perform a new computation of the same helicity one-loop four point amplitude and show that this is related to a shift computation. We also emphasize that the same holds for scalar QED and massless QED. Next we elaborate on the sum of all possible geometries (four points) at one-loop and do an explicit calculation to show that the sum vanishes using appropriate choice of loop momentum variables. We use the self-energy bubbles to reproduce the one-loop four point amplitude. We then propose a new formula of the amplitude at all points in terms of the bubble and the Berends Giele currents. We give a general proof of our conjectured formula by showing that it satisfies the correct collinear limits. We interpret the self-energy bubble in SDYM as an effective propagator. We also sketch the explicit computation of the amplitude at 3, 4 and 5 points from our formula.
\\~\\
In chapter 6, we give a brief introduction to chiral formulation of gravity.
In chapter 7, we expand on the chiral Einstein-Cartan gravity, which was proposed recently \cite{Krasnov:2020bqr}. We study the BRST quantization of this theory on a general Einstein background and develop the ghost Lagrangian. We thereby compute the one-loop ghost contribution to the effective action using heat kernel methods. We then pass to the flat background case and write the corresponding Feynman rules. Our contribution in the flat background case is the development of the ghost Lagrangian and the ghost Feynman rules.
In chapter 8, we describe the recently proposed flat-SDGR action. We explain the gauge fixing and Feynman rules in this theory and sketch the computation of the Berends-Giele currents. 
\\~\\
In chapter 9, we construct the Feynman graphs for the one-loop four point same helicity amplitude in gravity. We demonstrate that the arising loop integrals are finite. We then construct the self energy bubble and compute it after projecting to two positive helicity states. We then expand on the gluing of currents to the self energy bubble but our computation stops because of the unavailability of the interpretation of bubble momenta. We thus leave this to future work. 

\newpage 
\section{Review}
\subsection{Scattering Amplitudes}    
The scattering amplitude in a field theory is a probability amplitude for one set of particles to scatter into another set of particles via some interaction. To compute this object in the conventional way, one considers the set of all possible Feynman diagrams with external propagators amputated, external legs put on-shell and finally these legs are projected to polarization states. The compact notation of such an amplitude in say, Yang-Mills theory is given as 
\be 
\mathcal{A}^{a_1a_2..a_n}(k_1,\epsilon_1,....,k_n,\epsilon_n),
\ee 
where $a_i$ are the colour indices which are left free, $k_i$ are the momenta of the particles scattered and $\epsilon_i$ are the polarization states of the particles. While the colour indices in such an amplitude are left free, the Lorentz indices are projected to polarization states. Thus the amplitude is a Lorentz scalar. Scattering in the classical limit is captured by the tree level amplitudes, while quantum corrections are perturbatively taken into account by loop level amplitudes. Amplitudes have few essential features which severely constraint their form. Let us point out and explain these. 
\begin{itemize}
    \item \textbf{Unitarity}\\
    The unitarity principle states that the sum of all probabilities for a process to happen will be unity. Thus, the complete S-matrix for a process must be a unitary operator 
    \be 
    SS^{\dagger}=\mathds{1}.
    \ee 
    The unitarity principle restricts the type of quantum field theories one can study to model reality. The fact that non-renormalizability is seen as a breakdown of the predictability of a theory stems from the fact that perturbative unitarity is lost. Beside constraining the theoretical structure, the principle of unitarity is very much applicable to compute actual processes. Based on it, the ideas of computing loop level amplitudes were developed many years ago and such techniques remain immensely fruitful to this day. 
\end{itemize}

\begin{itemize}
    \item \textbf{Locality}\\
    The interactions of a field theory are local in character, due to relativistic constraints. The amplitude contains the interactions in a given Feynman graph. Locality then manifests in the kinematic structure of an amplitude, particularly in its pole structure. For instance, the intermediate line in a Feynman graph corresponds to a propagator which is off-shell. The only poles which may occur are when some subset of these intermediate particles become on-shell. Thus, these poles determine the energy width in which some of the particles can appear to be observables.  
\end{itemize}

\begin{itemize}
    \item \textbf{Crossing symmetry}\\
In a scattering amplitude, we use a particular convention to apply momentum conservation. The convention can be that all the particles are taken to be incoming or all outgoing or some of them to be incoming and the rest could be outgoing. Whatever the convention, crossing symmetry states that the amplitude and thus the S-matrix is unchanged if we convert an incoming particle to an outgoing one and also change the corresponding signs of its momentum and helicity. Thus, consider a scattering process where $X(p_1)+Y(p_2)\rightarrow Z(p_3)+W(p_4)$. Then we can alter the positions of the particles across the arrow, e.g $X(p_1)\rightarrow \bar{Y}(-p_2)+Z(p_3)+W(p_4)$ where $\bar{Y}(-p_2)$ is the corresponding anti-particle with the sign of the momentum and helicity changed. Crossing symmetry says that these two processes are exactly identical and fetches the same amplitude. 
\end{itemize}

\subsection{Colour decomposition in Yang-Mills} 
Yang-Mills is an $SU(N)$ gauge theory. In such a theory, we have a gauge group where there are $(N^2-1)$ generators, labelled $T^a$. These are traceless Hermitian matrices of order $N$. The colour structure constants of the theory are related to the generators by $f^{ijk}=-\frac{i}{\sqrt{2}}(Tr(T^iT^jT^k)-Tr(T^iT^kT^j))$. In the chiral reformulation of Yang Mills, as we will see, there is just a single cubic vertex. The structure constant is a coefficient in the vertex. Any tree or loop level amplitude is constructed using this vertex and thus is a function of colour factors and kinematic factors (polarizations, momenta). The colour decomposition allows us to write any such amplitude as a sum over permutations of products of colour coefficients with purely kinematic factors. This leads us to define the so called colour ordered amplitude. Thus, the simplifcation allows us to just compute the colour stripped part of the amplitude, without bothering about the colour factors. These factors are put in at the end of the computation to write the complete amplitude. Let us write such a decomposition for the tree and loop level amplitudes. At tree level, there are terms with only single traces. It is given by 
\be 
\mathcal{A}^n_{tree}(\{a_i,k_i,\epsilon_i\})=g^{n-2}\sum_{\sigma\in S_n/Z_n}Tr(T^{a_{\sigma(1)}}...T^{a_{\sigma(n)}})A(k_{\sigma(1)},\epsilon_{\sigma(1)},...,k_{\sigma(n)}\epsilon_{\sigma(n)}),
\ee 
where $Tr(AB..)$ stands for the matrix trace, $\sigma(i)$ are the permutations over the labels of the external particle and $A(k_i,\epsilon_i)$ is the colour ordered/ colour-stripped amplitude, which is just a function of the momenta and the polarizations. In contrary to tree amplitudes, one-loop amplitudes have both single and double trace terms. However, in the large $N$ limit, i.e when $N\rightarrow\infty$ only the single trace terms contribute. Moreover, there is a relation between the colour ordered amplitudes in the single trace terms and those in the double trace ones. Thus, it is reasonable to deal with the single trace term amplitudes. These amplitudes consists of planar Feynman diagrams. The colour decomposition is similar to that in the tree level case and is given by 
\be 
\mathcal{A}^n_{1-loop}(\{a_i,k_i,\epsilon_i\})=g^{n}\sum_{\sigma\in S_n/Z_n}Tr(T^{a_{\sigma(1)}}...T^{a_{\sigma(n)}})A_{n,1}(k_{\sigma(1)},\epsilon_{\sigma(1)},...,k_{\sigma(n)}\epsilon_{\sigma(n)}).
\ee 
\subsection{Double copy}
The double copy is a conjectured relation between the amplitudes of gauge theories and of gravity \cite{Bern:2008qj}. This relation stems from the fact that there is a duality between colour and kinematics observed in Yang-Mills theory. The duality is proven to be true at tree level and is a conjecture at loop level. To understand it briefly, note that any tree level Yang-Mills amplitude is represented by a sum of all Feynman graphs. There are two kinds of vertices in Yang-Mills, cubic and quartic. However, the quartic vertex can be re-written as a sum of cubic vertices and the tree amplitudes can then be constructed from just cubic graphs. The form of such a generic amplitude is given by  
\be 
\mathcal{A}^n_{tree}=g^{n-2}\sum_{j\in\Gamma}\frac{1}{S_j}\frac{c_jn_j}{D_j}.
\ee 
Here the sum $j$ runs over all Feynman graphs with cubic vertices, $n_j$ are the kinematic numerators which store the information about momenta and polarizations of the particles scattered, $c_j$ are the colour factors which store the information about the gauge group of the theory, $S_j$ are the symmetry factors for the diagrams and $D_j$ are the appropriate propagator factors in the denominator for each diagram. The colour factors obey a Jacobi identity because they are built from the Lie algebra structure constants for some gauge group. The Jacobi identity is given by 
\be 
c_i+c_j+c_k=0.
\ee 
Here the labels $i,j, k$ stands for different cubic graphs. The important idea to realize is that the kinematic numerators $n_j$ are built from gauge dependent objects and are thus far from unique. Indeed, one can do gauge transformations on the fields and the kinematic numerators change appropriately. However, it has been noted that there exists a certain choice of a gauge such that the kinematic numerators can be obtained in such a way that they satisfy an analogous Jacobi like relation, in a similar way like the colour factors. 
\be
n_i+n_j+n_k=0.
\ee 
This kind of a relation of the kinematic numerators does not arise from Feynman rules. It is only in a specific representation that such a choice of numerators occurs. Although the choice is in no way unique, it has been proven at tree-level that such a choice always exists. This suggests that there might be some hidden symmetry at play in the Yang-Mills theory, manifesting itself in the form of a Jacobi like identity in the kinematic sector. So far this symmetry has resisted understanding except in the self-dual sector of Yang-Mills. In the self dual sector, this duality is manifest from the Feynman rules as was shown by Monteiro and O'Connell \cite{Boels:2013bi}. Let us now understand the double copy relation which follows from this duality. Whenever one finds a representation of the cubic graphs in a Yang-Mills amplitude such that the kinematic numerators obey the same Jacobi relation like the colour factors, it is straightforward to obtain the corresponding gravity amplitude. This is done by just replacing the colour factors by another copy of the kinematic factors. Note, the two kinematic factors can come from different gauge theories. Thus in the tree level case, with the simple replacement 
\be 
c_j\rightarrow \tilde{n}_j
\ee 
we obtain the corresponding gravity amplitude as 
\be 
\mathcal{M}^n_{tree}=\sum_{j\in \Gamma}\frac{1}{S_j}\frac{\tilde{n}_jn_j}{D_j}.
\ee 
where the $S_j$ and $D_j$ are the same symmetry factors and propagators which enter the Feynman graphs in gravity and the numerator now comprises of only two copies of kinematic factors. The two copies can come from two different gauge theories and thus amplitudes for a variety of different gravity theories can be obtained by considering the appropriate products of gauge theories. The loop level generalization is also straightforward. We will deal with only one loop amplitudes in this thesis and in this case, let us first write a generic one-loop amplitude in Yang-Mills (ignoring couplings)
\be 
\mathcal{A}^n_{one-loop}=\sum_{j\in \Gamma}\int \frac{d^4l}{(2\pi)^4}\frac{1}{S_j}\frac{c_jn_j}{D_j}.
\ee 
The procedure to obtain the corresponding gravity amplitude is to replace the colour factors in the integrand of the one-loop amplitude by another copy of the numerator factor. Thus, at the loop level, it is the numerator of the integrand of a colour-stripped Yang-Mills amplitude which gets squared to produce the corresponding gravity one-loop amplitude.
\be 
\mathcal{M}^n_{one-loop}=\sum_{j\in\Gamma}\int \frac{d^4l}{(2\pi)^4}\frac{1}{S_j}\frac{\tilde{n}_jn_j}{D_j}.
\ee 
\newpage
\subsection{Dual momentum variables}
In planar Feynman graphs with a canonical ordering, we can associate dual momentum coordinates by a re-labelling of all the usual momentum coordinates. Sometimes these are called region momenta because they belong to the regions separated by lines on the plane. It is important that the graphs must be planar and has to have a well defined ordering to consistently admit such a representation. The dual momenta are related to the ordinary momenta as 
\be 
k_i=p_i-p_{i-1},
\ee 
where $k_i$ is the $i$-th ordinary momentum and $p_i$ is the i-th dual momentum variable. 
Thus on a given graph with $n$ ordinary momenta, we have the following set of relations
\be 
\begin{split} 
k_0&=p_0-p_{n-1},\\ 
k_1&=p_1-p_{0},\\
&.\\
&.\\
&.\\
k_{n-1}&=p_{n-1}-p_{n-2}.
\end{split} 
\ee 
We can then see that momentum conservation is automatically satisfied in dual momentum variables. Indeed, we have 
\be 
k_0+k_1+......k_{n-1}=(p_0-p_{n-1})+(p_1-p_0)+....+(p_{n-1}-p_{n-2})=0.
\ee 
We can interpret this graphically. Consider a Feynman graph with six external momenta. The convention is that all momenta are incoming.\\~\\

\be 
\begin{gathered}
~~~\begin{fmfgraph*}(150,90)
     \fmfleft{i,i4,i3}
     \fmfright{o1,o4,o7}
     \fmfblob{.15w}{o}
     \fmf{vanilla,label=$p_1$}{o,v2}
     \fmf{vanilla}{v2,i4}
     \fmf{vanilla,label=$p_2$}{v6,o}
     \fmf{vanilla}{v6,i3}
     \fmf{vanilla,label=$p_0$}{v1,o}
  \fmf{vanilla}{v1,i}
     \fmf{vanilla,label=$p_4$}{v3,o}
     \fmf{vanilla}{v3,o4}
     \fmf{vanilla,label=$p_5$}{o,v5}
     \fmf{vanilla}{v5,o1}
     \fmf{vanilla,label=$p_3$}{o,v4}
     \fmf{vanilla}{v4,o7}
     \fmflabel{$k_1$}{i}
     \fmflabel{$k_6$}{o1}
      \fmflabel{$k_5$}{o4}
     \fmflabel{$k_4$}{o7}
     \fmflabel{$k_3$}{i3}
      \fmflabel{$k_2$}{i4}
     \end{fmfgraph*}
\end{gathered}~~~~~~~\Longleftrightarrow~~~~~~
\begin{gathered}
~~~\begin{fmfgraph*}(120,100)
     \fmfleft{i,i4,i3}
     \fmfright{o1,o4,o7}
     \fmf{vanilla,label=$k_1$}{i,i4}
     \fmf{vanilla,label=$k_2$}{i3,i4}
     \fmf{vanilla,label=$k_3$}{i3,o7}
     \fmf{vanilla,label=$k_4$}{o7,o4}
     \fmf{vanilla,label=$k_5$}{o1,o4}
     \fmf{vanilla,label=$k_6$}{o1,i}
     \fmf{dotted}{o1}
     \fmflabel{$p_0$}{i}
     \fmflabel{$p_1$}{i4}
     \fmflabel{$p_2$}{i3}
     \fmflabel{$p_3$}{o7}
     \fmflabel{$p_4$}{o4}
     \fmflabel{$p_5$}{o1}
     \end{fmfgraph*}
     \end{gathered}\nonumber 
     \ee 
     \\~\\
The figure on the left hand side depicts a scattering process of six external massless particles, where the usual momenta are labelled by the particle number at the end of each line. The region momenta are labelled in the region between any two lines. The right hand side is a dual diagram where the region/dual momenta are placed at each vertices and the edges represent the usual line momenta. The edges are ordered in that the vertices are joined head to head in an oriented way. The edges are null for an on-shell scattering process, i.e when all the momenta are on-shell. The null polygon in dual space takes momentum conservation into account automatically in that the vector sum of all edges add up to zero. This is the main motivation behind introducing the dual momentum space. 
\\~\\
In any loop diagram, we also ascribe a dual momentum for the region enclosed by the loop. The original loop momentum can be expressed as a difference of the dual loop momentum and one of the adjacent dual momenta outside the loop. Note, all tree and one-loop diagrams are planar and they admit dual representations. However, one also needs cyclic ordering in a graph to ensure that momentum conservation is trivialized in the dual space. Thus for a graph with cyclic ordering, the identification $p_{n}=p_0$ ensures momentum conservation automatically. In Yang-Mills we already have the colour ordering which allows to consistently define dual momenta. In gravity there is no such ordering and it is more subtle to deal with it. We will elaborate on this further in the second part of the thesis when we construct one loop amplitudes in gravity. Although dual variables trivialize momentum conservation, the on-shell condition ($k^2=0$) is an additional constraint. It is convenient and simpler to work with amplitudes in a set-up where both these conditions are trivialized. This motivated Hodges to introduce the notion of momentum twistors \cite{Hodges:2009hk}. We will not use momentum twistors in this thesis, but let us briefly review this idea for completeness.
\subsection{Momentum twistors}
The idea of twistors dates back to Penrose \cite{Penrose:1967wn}, where he established the connection between null rays in spacetime and complex points in some auxiliary space, called the twistor space. Thus, the space of twistors consists of points, with complex coordinates. Let us call the twistor coordinates $Z=(v^A,\lambda_{A'})$ where we use 2-component spinor notations, which we describe in the next chapter. Then the incidence relation relates the spacetime point $x^{AB'}$ to the twistor $v^A=x^{AB'}\lambda_{B'}$. Points in spacetime thus gets mapped to $\mathbb{CP}^1$ lines in twistor space. On the other hand, if two lines in twistor space intersect, the corresponding points are null separated. So the null-condition is trivialized in twistor space by the fact that one needs to associate a pair of twistors for each null-separated point in spacetime and then the lines formed by these two pairs automatically intersect. This motivated Hodges to introduce the notion of the so called momentum twistors. 
The pivotal point to note is that the square of the difference of two consecutive dual momentum coordinates is null 
\be 
(p_i-p_{i-1})^2=0.
\ee 
Thus, instead of spacetime points, one can associate the dual momentum coordinates to a new auxiliary space by the mapping~$p_i\leftrightarrow (Z_i,Z_{i+1})$. Then for an $n$-point scattering with all external particles on-shell, the $n$ dual momentum coordinates can all be mapped to a set of $n$ momentum twistors $\{Z_1,Z_2,...Z_n\}$. Then it is guaranteed that the on-shell/null condition is automatically satisfied in the momentum twistor space.

\newpage 
\chapter{Mathematical Preliminaries}
In this chapter, we describe the technology of 2-spinors, which will be heavily used in the rest of the portions of the thesis. This part is mostly based on \cite{Krasnov:2020lku} and \cite{Krasnov:2020bqr}, but we describe it in our notations. The formalism of 2-spinors is unavoidable in any amplitude computation. However, in addition to that, it is also very natural to use spinors in the development of chiral theories, for e.g self-dual Yang Mills. Thus, the role played by the spinor technology is two-fold. The first is to describe the Lagrangian and the related Feynman rules of chiral theories in a convenient way and the second is to compute scattering amplitudes of interest from those. Let us then start building the relevant techniques and explain our notations. 
\section{Two Component Spinor Techniques}
There exists a local isomorphism between the Lorentz group $SO(4,\mathbb{C})$ and the group $SL(2,\mathbb{C})\times SL(2,\mathbb{C})$ of unimodular transformations. This is the underlying idea that leads to the development of spinor calculus in a flat 4-dimensional manifold. The isomorphism between groups induces an isomorphism on the Lie algebras $so(4,\mathbb{C})\equiv sl(2,\mathbb{C})\times sl(2,\mathbb{C})$. Therefore, any vector valued in $so(4,\mathbb{C})$ index on the complex manifold $\mathbb{M}_C$ can be represented by a pair of spinor indices taking values in $sl(2,\mathbb{C})$. 
\\~\\
We define at each point of our complex 4-dimensional space-time, a complex 2-dimensional linear space, the spinor space. The elements of the spinor space are 2-component complex quantities $\phi^A$. We label these spinors using unprimed indices. The elements of the complex conjugate spinor space are labelled with primed indices. 
The 2-component spinors are subject to the group of unimodular spin transformation, the group $SL(2,\mathbb{C})$ of $2\times 2$ complex matrices
\be
 D=\begin{bmatrix} 
a & b \\
c & d 
\end{bmatrix}  
\ee
with unit determinant,
\be
 |D|=ad-bc=1.
\ee
Therefore, if $\phi^A$ is a spinor, it undergoes a transformation
\be
    \phi^{'A}=\phi^BS^A_{~B}.
\ee
The elements of the complex conjugate space are subjected to analogous transformations with matrices replaced by $S^{A'}_{~B'}$. Given the general notion of spinors in spaces of any signature, let us now stick to the Lorentzian signature case. In this case, the four coordinates of spacetime can be represented in the form of a matrix 
\be 
\label{x}
x=i\begin{bmatrix}
x_0+x_3 & x_1-ix_2\\
x_1+ix_2 & x_0-x_3
\end{bmatrix}
\ee 
such that the norm of a vector is given by the determinant of the matrix 
\be 
det(x)=-x_0^2+x_1^2+x_2^2+x_3^2.
\ee 
The group $SL(2,\mathbb{C})$ acts on the space of the such matrices and the action is given by 
\be 
x\rightarrow hxh^{\dagger}, ~~~~h\in SL(2,\mathbb{C}).
\ee 
As already mentioned, the spinors come in two different types. They are the two irreducible representations of the group $SL(2,\mathbb{C})$. We classify them as unprimed and primed spinors. By convention, the unprimed spinors are represented as two component columns and are labelled by an unprimed index. They undergo the following transformation under the $SL(2,\mathbb{C})$ group action
\be 
\mu_{M}\rightarrow h^{~~N}_{M}\mu_{N}.
\ee 
The space of such spinors are denoted by $S_+$. We then have the primed spinors which are represented by two component columns but this time with complex entries. We denote the space of such spinors to be $S_-$.  complex conjugated group elements act on such spinors as 
\be 
\mu_{M'}\rightarrow (h^*)^{~~N'}_{M'}\mu_{N'}.
\ee 
Let us now discuss about raising and lowering of spinor indices. In the flat Minkowski space, the invariant metric tensor is the object which raises and lowers tensor indices. We thus need to define an analogous metric through the inner product in the space of spinors. The spinors metrics in $S_+$ and $S_-$ will be complex conjugates of each other. For the space $S_+$ we define it using the bilinear form (inner product)
\be 
\label{spin1}
\langle \mu\lambda\rangle:=(\epsilon\mu)^T\lambda=-\mu^T\epsilon\lambda,
\ee 
where the object $\epsilon$ is a $2\times2$ matrix 
\be 
\epsilon=\begin{bmatrix}
0 & 1\\
-1 & 0
\end{bmatrix}.
\ee 
The entries of the matrix $\epsilon$ are all real and thus the complex conjugation results in the same matrix. The definition of the inner product in (\ref{spin1}) helps us identify the spinor contraction and raising of indices. The object $(\epsilon\mu)^T$ is interpreted to be the action of $\epsilon$ on $\mu$ such that it raises the index of $\mu$. This is reminiscent of what we do for 4-vectors. In that case we write the inner product of two vectors $x_{\mu}, y_{\nu}$ as $\eta^{\mu\nu}x_{\mu}y_{\nu}$ where $\eta^{\mu\nu}x_{\mu}=x^{\nu}$. Thus in a similar way, our spinor contraction takes the form $\langle \mu\lambda\rangle=\mu^B\lambda_B$. However, the difference with 4-vectors is that our spinor contraction is anti-symmetric with respect to the two spinors. Indeed, in index free notations, we have 
\be 
\langle \mu\lambda\rangle=-\langle \lambda\mu\rangle.
\ee 
We can understand the anti-symmetry from the anti-symmetric structure of $\epsilon$. In particular, let us show how it arises using indexed notations. We have 
\be 
\mu^B\lambda_B=\epsilon^{BA}\mu_A\lambda_B=-\epsilon^{AB}\lambda_B\mu_A=-\lambda^A\mu_A.
\ee 
We see that the unprimed index in the upper left gets contracted to the one on the lower right. The crucial step is interchanging the indices of $\epsilon$ and this picks up a minus sign. The order of spinors are irrelevant and thus it is now the spinor $\lambda_B$ on which $\epsilon^{AB}$ acts from the left to raise its index. This gives the required identity. Next, we define the lowering of indices. To do so, we use the fact that consecutive operations of lowering and raising the index of a spinorial object in any order returns the same spinor. We then define the lowering operation as 
\be 
\mu^M\epsilon_{MN}=\mu_N.
\ee 
With this definition, it is easy to see that if we simultaneously raise and lower a spinor index, we end up with the spinor with its indices in the original position. In particular, we have  $\epsilon^{MN}\mu_{N}\epsilon_{MK}=\mu_K$. This immediately gives us an identity between the two $\epsilon$. 
\be 
\epsilon^{MN}\epsilon_{MK}=\mathds{1}^{~N}_K.
\ee 
Also we note that the same raising and lowering operation can be applied in the above identity to conclude that the identity matrix in the spinor space is just the matrix $\epsilon$ with the first index lowered and the second index raised. In other words, we can represent the above equation as $
\epsilon^{MN}\epsilon_{MK}=\epsilon^{~N}_K$. This leads us to the equality of two $\epsilon$, one with all upper indices and another with all lower indices. 
\be
    \epsilon_{AB}=\epsilon^{AB}=\begin{bmatrix}
    0 & 1\\
    -1 & 0
    \end{bmatrix}.
\ee
A similar set of relations hold for the primed spinor space. We have the primed $\epsilon$ which is the same as the unprimed one. The primed spinor contraction is labelled by a square bracket. It follows the anti-symmetry property. However, the difference in this case is that the indices are contracted from the below left to the upper right direction. 
\be 
\mu_{B'}\lambda^{B'}=[\mu\lambda]=-[\lambda\mu].
\ee 
The completely antisymmetric tensor in 4-dimensional spacetime is represented with spinor objects in the following way
\be
    e^{\mu\nu}_{~\eta\chi}=i(\epsilon^C_A\epsilon^D_B\epsilon^{D'}_{A'}\epsilon^{C'}_{B'}-\epsilon^D_A\epsilon^C_B\epsilon^{C'}_{A'}\epsilon^{D'}_{B'}).
\ee
Clearly, if we exchange a pair of spinors say $AA'$ with another pair say $BB'$, then right hand side picks up a minus sign. This justifies the spinorial representation of the completely anti-symmetric tensor. It is important to note that the matrix in (\ref{x}) can be interpreted as a bi-spinor. Indeed, we can decompose such a matrix in a basis of soldering forms. Let us elaborate on this a bit further. 
\subsection{Soldering form}
To define a map from the space of four vectors to the space of rank two spinors, we introduce an object known as the soldering form. We use a Hermitian soldering form defined by
\be 
(e_{\nu}^{BB'})^*=e_{\nu}^{BB'}.
\ee 
The metric is obtained as a square of the soldering form:
\be 
\eta_{\mu\nu}=-e_{\nu A}^{A'}e_{\nu B}^{B'}\epsilon^{AB}\epsilon_{A'B'},
\ee 
where a minus sign appears because we wish to work with a Hermitian soldering form while at the same time have signature $(-,+,+,+)$. The sign can be traced to the convention we use in defining contractions of primed and unprimed spinors. The primed spinors are contracted as $\mu_{B'}\lambda^{B'}$ while the unprimed ones are contracted like $\mu^B\lambda_B$. We can also rewrite the above formula as
\be 
\label{theta}
\eta_{\mu\nu}=-e_{\mu~B'}^B e_{\nu~B}^{B'}.
\ee 
The contraction which appears in this formula, i.e  primed spinors contracting bottom left to up right and primed spinors contracting oppositely, will be referred to as the natural contraction. Let us then express the bi-spinor in terms of the soldering forms and the 4-vector 
\be 
\label{x2!}
x^{BB'}=i\sqrt{2}e_{\mu}^{BB'}x^{\mu}.
\ee 
It is then customary to identify (\ref{x}) with (\ref{x2!}) and write an explicit representation of the soldering forms. We have 
\be 
e_0=\frac{1}{\sqrt{2}}\begin{bmatrix}
    1 & 0\\
    0 & 1
    \end{bmatrix},
    ~~~
e_1=\frac{1}{\sqrt{2}}\begin{bmatrix}
    0 & -1\\
    -1 & 0
    \end{bmatrix}
\nonumber\\ 
e_2=\frac{1}{\sqrt{2}}\begin{bmatrix}
    0 & i\\
    -i & 0
    \end{bmatrix},
    ~~~
e_3=\frac{1}{\sqrt{2}}\begin{bmatrix}
    -1 & 0\\
    0 & 1
    \end{bmatrix},
\ee 
where $e_{\mu}$ are the components of the particular form  
\subsection{Momentum spinor}
We deal with massless fields in our work. The momentum of a massless particle is null. This is is used in amplitude computations where we have external states and the momentum of such states are constrained to be null. To incorporate this feature in the language of spinors, let us quickly introduce the notion of null vectors and their decomposition into the so called momentum spinors.
\\~\\
Any null vector has the property that the square of its 4-momentum vanishes. In spinorial notation, this can be written as $k^{AA'}k_{AA'}=0$. But as we know, the vector inner product can be expressed as the determinant of its equivalent matrix representation, so we write, for any null vector,
\be 
|k^{AA'}|=0.
\ee 
A vanishing determinant implies that the rank of such a matrix is less than the dimension of the space, i.e, 2 in this case. One can always write such a matrix as a tensor product of a row vector and a column vector. Therefore, any such null vector $k^{AA'}$ can be decomposed as 
\be 
k^{AA'}=k^A k^{A'}.
\ee 
Also, the inner product of two null vectors, $k_{1AA'}$ and $k_{2BB'}$ is given by 
\be 
k^{AA'}_1k_{2AA'}=\langle k_1k_2\rangle [k_1k_2].
\ee 
\textbf{Contraction in index free notation:}
Let us now explain some further notations which will be used in in subsequent parts of this work. We represent any one-loop amplitude as an integral over loop momenta which is non-null and project it to appropriate polarization states. The integrand will be a scalar, built from contractions of different kinds of spinors. We give some examples of these contractions and introduce a convenient notation.
The loop momentum $l^{\mu}$ becomes a spinorial object $l_M^{~~M'}$. This can get contracted by a primed and an unprimed spinor. In index free notations: 
\be 
\mu^{M} l_{M}^{~~M'}\lambda_{M'}\equiv \langle \mu|l|\lambda].
\ee 
\textbf{Labelling of external momenta:}\\
For external momenta $k_1,k_2,..$ we refer them just by their number, dropping the letter $k$.
Our standard notation:~$k_{1\mu}\equiv 1_{\mu}$. In index free notation
\be 
k_{1M}^{~~M'}\equiv 1_{M}^{~~M'}=|1\rangle [1| .
\ee 
Let us illustrate this with an example. We will use a string of contraction of spinors (both primed and unprimed) and write it in index free notations as follows 
\be 
\label{not}
\mu^M2^{~~M'}_M l_{M'}^{~~N}\lambda_{N}\equiv \langle \mu|2\circ l|\lambda\rangle . 
\ee
Let us start from the left. The unprimed spinor index on $\mu$ gets contracted to the unprimed index of $2$ which we can represent as $\langle \mu|2$. Next, the primed spinor index on 2 gets contracted to the primed one on $l$. This contraction of indices between two mixed spinors is denoted by $\circ$. So far we thus have the object $\langle \mu|2\circ l|$. This can next get contracted to an unprimed spinor and we recover (\ref{not}).
\subsection{Self-dual and anti self-dual forms}
We want to introduce the basis of self-dual $2$-forms which are constructed from the soldering forms $e_{\mu}^{MM'}$. This is going to be used in order to explain the development of the chiral formalism in gravity, where the notion of self-duality plays an important role. The precise definition for the 2-form is 
\be 
\label{sd11}
\Sigma^{MN}_{\mu\nu}:=e^{M}_{[\mu M'}e^{NM'}_{\nu]}.
\ee 
As we can see, the right hand side is a product of two soldering forms wedged with respect to their vector indices $\mu, \nu$. We can instead strip of these indices and write the self-dual 2-form as 
\be 
\label{form1}
\Sigma^{MN}=\frac{1}{2}e^{M}_{~M'}e^{NM'}.
\ee 
However, it is convenient if we write the 2-form in completely spinorial notations. This helps us to do further computations which will be heavily based on these notations. Another important aspect is about the definition of the particular self-dual 2-form as in (\ref{sd11}). Clearly, this definition is far from unique since the soldering forms can admit many different representations. However, if we completely convert it to spinor indices, it gives us a simple expression for the particular 2-form as we will see. This is the main motivation behind the choice. We convert the spacetime indices $\mu\nu$ using the soldering form. We have 
\be 
\Sigma^{MN}_{PP'QQ'}:=\Sigma^{MN}_{\mu\nu}e^{\mu}_{PP'}e^{\nu}_{QQ'}=\epsilon_P^{~(M}\epsilon^{~N)}_{Q}\epsilon_{P'Q'}.
\ee 
We now compute the 2-forms explicitly in their matrix representation. Note that we do not have to compute all of them separately because some are related to the others due to self-duality. In particular, we have a triplet of independent self-dual 2-forms and we compute the components $\Sigma^{MN}_{0j}$. We have 
\be 
\Sigma_{01}=\frac{1}{2}(e_0\epsilon e_1^T-e_1\epsilon e_4^T)=\frac{1}{\sqrt{2}}\begin{bmatrix}
    -1 & 0\\
    0 & 1
    \end{bmatrix},\nonumber\\
\Sigma_{02}=\frac{1}{2}(e_0\epsilon e_2^T-e_2\epsilon e_4^T)=\frac{1}{\sqrt{2}}\begin{bmatrix}
    i & 0\\
    0 & i
    \end{bmatrix},
\nonumber\\ 
\Sigma_{03}=\frac{1}{2}(e_0\epsilon e_3^T-e_3\epsilon e_4^T)=\frac{1}{\sqrt{2}}\begin{bmatrix}
    0 & 1\\
    1 & 0
    \end{bmatrix}.
\ee 
We can write the other components using self-duality. For instance, the component $\Sigma^{MN}_{23}=-i\Sigma^{MN}_{41}$. Let us now define the anti self-dual 2-form basis. A convenient definition is to identify this basis with the complex conjugation of the self-dual 2-forms and put a minus sign. Thus we have for the anti self-dual basis 2-forms 
\be 
\label{sd12}
\bar{\Sigma}^{M'N'}_{\mu\nu}:=e^{MM'}_{[\mu}e^{~~~N'}_{\nu]M}.
\ee 
Once again, we can strip off the indices $\mu,\nu$ and write the anti self-dual 2-form basis in the compact form 
\be 
\label{form2}
\bar{\Sigma}^{MN}=\frac{1}{2}e^{M}_{~M'}e^{NM'}.
\ee 
Next we want to write down some useful spinor identities. These will be later used in computations and thus it is instructive to collect most of these here. There are identities involving the $\epsilon$, momentum spinors and the basis of self-dual/anti self-dual 2-forms. We also mention which of these identities can be translated to any number of dimensions.

\subsection{Spinor identities}
Let us start with the identities involving the soldering form and the self-dual/anti self-dual 2-forms. The product of two soldering forms when contracted in one of the spinor indices gives 
\be 
\label{iden1}
e^M_{\mu~M'}e^{NM'}_{\nu}=-\frac{1}{2}\eta_{\mu\nu}\epsilon^{MN}+\Sigma^{MN}_{\mu\nu}.\nonumber\\
e^{MM'}_{\mu}e^{N'}_{\nu M}=-\frac{1}{2}\eta_{\mu\nu}\epsilon^{M'N'}+\bar{\Sigma}^{M'N'}_{\mu\nu}. 
\ee 
In the first one, the primed spinor is contracted after taking the wedge product of two soldering forms. The identity can be verified to be true as follows. We can contract the unprimed indices with the $\epsilon$ on both sides and this gives us the coefficient of the first term on the right hand side using the identity (\ref{theta}). The second term then follows from the identity (\ref{sd11}). A similar reasoning holds for the second identity. Let us write the wedge product for two soldering forms. 
\be 
e^{MM'}e^{NN'}=\epsilon^{M'N'}\Sigma^{MN}-\epsilon^{MN}\bar{\Sigma}^{M'N'}.
\ee 
It is easy to check that the above identity holds by contraction of both sides with $\epsilon^{MN}, \epsilon^{M'N'}$ and then using the identity of the self-dual/anti self-dual 2-forms in (\ref{form1}) and (\ref{form2}). To complete the list of identities with the 2-forms, we write down yet another one which can be derived from the identities in (\ref{iden1}).
\be 
\bar{\Sigma}^{M'N'~\alpha}_{~~\beta}\bar{\Sigma}^{P'Q'~\gamma}_{~~\alpha}=\frac{1}{2}\eta^{\gamma}_{\beta}\epsilon^{M'(P'}\epsilon^{Q')N'}\nonumber\\
+
\frac{1}{2}\Bigg(\bar{\Sigma}^{M'(P'~\gamma}_{~~\beta}\epsilon^{Q')N'}+\bar{\Sigma}^{N'(P'~\gamma}_{~~\beta}\epsilon^{Q')M'}\Bigg).
\ee 
Let us now state some of the identities which we will use extensively in amplitude computations. These are the ones associated with momentum spinors and the spinor metric. First we set the notation for the symmetrized addition of a pair of spinor metrics in the following way 
\be
\epsilon^A_{~C}\epsilon^B_{~D}+\epsilon^A_{~D}\epsilon^B_{~C}=\epsilon^{AB}_{~~(CD)}.
\ee
Note that in the right hand side, the symbol $\epsilon$ stands for just notation purposes. The round bracket on the indices in the subscript imply symmetrization of these. We could have equally written the expression with round brackets on the superscript. The following identities with the metric spinors can be seen to hold.
\be
\begin{split} 
    \epsilon^A_{~B}\epsilon^D_{~A}&= \epsilon^{~D}_{B},\\
    \epsilon^D_{~B}\times\epsilon^{AB}_{~~(CD)} &=\epsilon^{AB}_{~~(BC)} =3\epsilon^A_{~C}.
\end{split} 
\ee
For any arbitrary momentum vector $k$, we have the following identities
\be
    k^{T'}_{~M}k_{T'J}=\frac{1}{2}\epsilon_{MJ}k^2,
\ee
\be
    k_{MJ'}k_{M'J}-k_{MM'}k_{JJ'}=\frac{1}{2}\epsilon_{M'J'}\epsilon_{JM}k^2.
\ee
\subsection{Spinor basis}
In each of the spinor spaces denoted by $V^{(1/2,0)}$ and its complex conjugate $V^{(0,1/2)}$ there are two independent basis spinors
\be
o_A,\tau_A\in V^{(1/2,0)},\\
\nonumber
o_{A'},\tau_{A'}\in V^{(0,1/2)},
\nonumber
\\
\tau_{A'}=(\tau_A)^*,
\nonumber
\\
\o_{A'}=(o_A)^*.
\ee
Their normalization is
\be
\tau^Ao_A=1, \tau^{A'}o_{A'}=1,
\\
\nonumber
\epsilon_{AB}=o_A\tau_B-\tau_Ao_B,
\\
\nonumber
\epsilon_{A'B'}=o_{A'}\tau_{B'}-\tau_{A'}o_{B'}.
\ee
\subsection{The soldering form in spinor basis}
The soldering form $e_{\mu}^{~BB'}$can be explicitly expressed in terms of the basis one forms $t_{\mu}$ and $x_{\mu}$,$y_{\mu}$,$z_{\mu}$ as well as the spinor basis vectors $o^A$,$o^{A'}$,$\tau^A$,$\tau^{A'}$ as follows:
\be 
e_{\mu}^{BB'}=\frac{t_{\mu}}{\sqrt{2}}(o^Bo^{B'}+\tau^B\tau^{B'})+\frac{x_{\mu}}{\sqrt{2}}(o^B\tau^{B'}+\tau^Bo^{B'})\nonumber\\+\frac{iy_{\mu}}{\sqrt{2}}(o^B\tau^{B'}-\tau^Bo^{B'})+\frac{z_{\mu}}{\sqrt{2}}(o^Bo^{B'}-\tau^B\tau^{B'}).\nonumber
\ee 
Note that the expression above is explicitly Hermitian .
\subsection{Parametrization of Momentum Spinors}
Let us consider a massless particle with 3-momentum vector $\vec{k}$. The 4-vector $k_{\mu}$ in this case is null. It can be expressed as a product of two spinors $k^Bk^{B'}=e_{\mu}^{BB'}k_{\mu}$. The spinors $k^B k^{B'}$ are complex conjugates of each other modulo a sign. The sign determines whether the momentum vector is future or past directed in the light-cone. As the unit vector $n=\vec{k}/|k|$ varies over the sphere $S^2$, there is no continuous choice of the spinor $k^B$.
We make the following choice.
\be 
\label{formula}
k^A(\vec{k}):=2^{1/4}\sqrt{w_k}(\sin{\theta/2}e^{-i\phi/2}\tau^A+\cos(\theta/2)e^{i\phi/2}o^A),
\ee
where $o^B,\tau^B$ is a basis in the space of unprimed spinors, $w_k=|k|$. $\theta,\phi$ are the usual coordinates on $S^2$ so that the momentum vector in the positive z-axis corresponds to $\theta=\phi=0$. The expression of the soldering form can be used to check the formula \ref{formula}.
\\
Let us now see the effects of the change of momentum direction. Consider the case where the momentum direction gets reversed. This corresponds to $\theta\rightarrow \theta+\pi$. We then get
\be 
\label{formula2}
k^A(-\vec{k}):=i2^{1/4}\sqrt{w_k}(-\cos{\theta/2}e^{-i\phi/2}\tau^A+\sin(\theta/2)e^{i\phi/2}o^A).
\ee
\\
Let us now consider the scattering of four mass-less particles where we take the convention of all momenta incoming. We take the particles 1 and 2 moving in the z-axes, positive and negative respectively. The particles 3, 4 we take to be scattered ones moving at an angle $\theta$ to the z-axis. This gives using (\ref{formula})
\be
k_1^B=2^{\frac{1}{4}}\sqrt{w_k}o^B,k_1^{B'}=2^{\frac{1}{4}}\sqrt{w_k}o^{B'},
\\
\nonumber
k_2^B=i2^{\frac{1}{4}}\sqrt{w_k}\tau^B,k_2^{B'}=-i2^{\frac{1}{4}}\sqrt{w_k}\tau^{B'},
\\
\nonumber
k_3^B=2^{\frac{1}{4}}\sqrt{w_k}(\sin{\frac{\theta}{2}}\tau^B+\cos{\frac{\theta}{2}}o^B),k_3^{B'}=-2^{\frac{1}{4}}\sqrt{w_k}(\sin{\frac{\theta}{2}}\tau^{B'}+\cos{\frac{\theta}{2}}o^{B'}),
\\
\nonumber
k_4^B=i2^{\frac{1}{4}}\sqrt{w_k}(\sin{\frac{\theta}{2}}o^B-\cos{\frac{\theta}{2}}\tau^B),k_4^{B'}=i2^{\frac{1}{4}}\sqrt{w_k}(\sin{\frac{\theta}{2}}o^{B'}-\cos{\frac{\theta}{2}}\tau^{B'}).
\ee
Note the extra minus sign appearing in the primed spinor parameters for particles 3 and 4. This tells that these particles are past directed(so that 4-momentum conservation is obeyed).
We also introduce the three Mandelstam variables in terms of the parameters
\be
s=(k_1+k_2)^2=(k_3+k_4)^2=2k_1.k_2=2k_3.k_4=-4w_k^2,
\\
\nonumber
t=(k_1+k_3)^2=(k_2-k_4)^2=2k_1.k_3=2k_2.k_4=4w_k^2\sin^2{\frac{\theta}{2}},
\\
\nonumber
u=(k_1+k_4)^2=(k_2-k_3)^2=2k_1.k_4=2k_2.k_3=4w_k^2\cos^2{\frac{\theta}{2}}.
\ee
\newpage 
\section{Spinor Helicity formalism}
This part is mostly taken from an expository account on spinor helicity and amplitudes in general \cite{Dixon:2013uaa}. The spinor-helicity formalism is natural for amplitude computations because one can exploit it to parametrize the polarization vectors (or tensors) in a canonical way. Particularly, when all the helicities are same, one can choose a single auxiliary spinor\footnote{An auxiliary spinor $q^A$ encodes the gauge freedom of the theory. Gauge transformations are understood as shifts on the auxiliary spinor: $q^A\rightarrow q^A+\eta k^A$, $k^A$ being some arbitrary momentum spinor and $\eta$ is the gauge parameter} for all the polarization vectors, making the calculation extremely simple. This would not be possible in the usual four vector notation.
Thus it is efficient to use a better choice of variables, in this case spinors, which form the smallest fundamental representation of the Lorentz group. The notational simplicity which we already presented will be introduced from the perspective of Weyl spinors. One can then use this notation not just for massless fermions, but for any particle with spin degree of freedom. To this end, we consider trading the Lorentz vectors
$k^{\mu}_i$
for a pair of spinors
\be 
k^{\mu}_i \implies u_+(k_i)\equiv|i^+\rangle\equiv k^{A}_i,~u_-(k_i)\equiv|i^-\rangle\equiv k^{A'}_i.
\ee 
Here
$u_+(k_i)=\frac{1}{2}(1+\gamma_5)u(k_i)$ is a right-handed spinor written in four-component Dirac
notation, and
$k_i^{A}$
is its two-component version,
 with $A 
= 1,2$. Similarly,
$u_-(k_i)=\frac{1}{2}(1-\gamma_5)u(k_i)$  a left-handed spinor in Dirac notation, and
$k_i^{A'}$ is the two-component version. The massless Dirac equation is satisfied
by these spinors,
\be 
\slashed{k_i}u_{\pm}(k_i)=\slashed{k_i}|i^{\pm}\rangle=0.
\ee
There are also negative-energy solutions
$v_{\pm}(k_i)$
$k_i^2= 0$ they are not distinct
from the earlier one. The unprimed and primed spinor indices correspond to two different spinor
representations of the Lorentz group.
We can build some Lorentz-invariant quantities out of the spinors, which we already illustrated in the previous section. We define the spinor products,
\be 
\langle ij\rangle\equiv \epsilon^{AB}(k_i)_A(k_j)_B=\bar{u}_-(k_i)u_+(k_j),
\nonumber
\\
\big[ij\big]\equiv \epsilon^{A'B'}(k_i)_{A'}(k_j)_{B'}=\bar{u}_+(k_i)u_-(k_j).
\ee
We have the positive energy projector for $m=0$
\be 
\bar{u}_+(k_i)u_+(k_i)=|i^+\rangle\langle i^+|=\frac{1}{2}(1+\gamma_5)\slashed{k_i}(1-\gamma_5).
\ee 
In two-component notation, this relation becomes, using the explicit form of the Pauli
matrices,
\be 
k_{iA}k_{iA'}=k^{\mu}_i(\sigma_{\mu})_{AA'}:=k_{iAA'}.
\ee
We note that the determinant of this $2\times 2$ matrix vanishes, which is
consistent with its factorization into a column vector $k_{iA}$ times a row vector $k_{iA'}$.
Also note that if the momentum vector
$k^{\mu}_i$ 
is real, then complex conjugation is equivalent to transposing the matrix
$\slashed{k_i}$
, which corresponds to exchanging the left-
and right-handed spinors, $k_{iA'}\leftrightarrow k_{iA}$. In other words, for real momenta, a chirality flip
of all spinors (which could be induced by a parity transformation) is the same as complex
conjugating the spinor products,
\be 
[ij]=\langle ij\rangle^*.
\ee 
The Mandelstam invariants in any scattering process can be defined using the spinorial notations. Upto factors, they are various combinations of products of two momenta. We denote them as $s_{ij}=\langle ij\rangle[ij]$.
Now we provide two useful spinor product identities
\be
\begin{split} 
    \textrm{Momentum conservation}&:~~ \Sigma_{i=1}^n \langle ji\rangle[ik]=0,
    \\
    \textrm{Schouten identity}&:~~\langle ij\rangle\langle kl\rangle-\langle ik\rangle\langle jl\rangle=\langle il\rangle\langle kj\rangle.
    \end{split} 
    \ee 
    
The main advantage of the helicity formalism is that the helicity states can be conveniently written with a canonical choice of auxiliary spinors. Both the spin-1 and spin-2 particles come with two physical polarization states. The states can be combined in the complex plane in two different ways which then gives rise to circular polarizations. These are the states which form a helicity basis. Let us write down the circular polarization states for each of the spin-1 and spin-2 massless particles in spinorial notation. For the spin-1 particle, we have 
\be 
\label{sp-1}
e^{AA'}_+(q,k)=\frac{q^{A'}k^A}{[qk]},~~
e^{AA'}_-(q,k)=\frac{q^{A}k^{A'}}{\langle qk\rangle}.
\ee 
where $q^{A'}$ and $q^{A}$ are the auxiliary spinors for each of the states respectively and $k^A$, $k^{A'}$ are the momentum spinors which identify the states with a particular momentum $k^{AA'}$. Note, the polarization states are dimensionless quantities. We will later show that these are the two solutions for the linearised Yang-Mills. Next, we give the states for the spin-2 massless particle (graviton) which we will use for computations. It is important to realize that for gravity, there can arise many different representations for the polarization states in terms of spinors. We stick to one of the representations which will appear in the chiral perturbation theory. 
\be 
\label{sp-2}
e^{AA'BB'}_+(q,k)=\frac{q^{A'}q^{B'}k^Ak^B}{[qk]^2},~~
e^{AA'BB'}_-(q,k)=\frac{q^{A}q^Bk^{A'}k^{B'}}{\langle qk\rangle^2}.
\ee 
The structure of the states in (\ref{sp-2}) appears in the form such that taking two copies of the ones in (\ref{sp-1}) produces these. The representation of such states in gravity inherits the double copy property of amplitudes, which we will elaborate on later.

\part{Aspects of self-dual Yang-Mills}
\chapter{Introduction}
Non abelian gauge theories are central to high energy scattering processes. However, when it comes to theoretical computations, they pose many difficulties. The usual route to compute any amplitude is carried out via Feynman diagrams in a perturbative expansion of the theory. However, the complexity of diagrammatic computations increase drastically if we start going to higher loop processes or increase the number of external legs. On top of it, it is evident that individual Feynman diagrams are not gauge invariant quantities. The amplitudes are however gauge invariant. The gauge dependency gets cancelled when we take the appropriate sum of all Feynman diagrams. However, this process to extract gauge independent quantities from some inherently gauge dependent objects turns out to be very complicated. While in many cases the final result happens to be quite simple. It thus motivates us to understand if there are simpler ways to compute amplitudes which may also fetch deeper insights. Here we would like to state particularly two such directions.\\~\\
The first one has to do with avoiding the Lagrangian formulation altogether and develop on-shell methods to determine amplitudes. After Witten's seminal paper on twistor strings \cite{Witten:2003nn}, it was realised that there exists recursion relations for amplitudes in Yang-Mills and gravity. These are the so called BCFW relations \cite{Britto:2004ap}. The idea is that the simplest non-vanishing amplitude, namely the 3-point tree level amplitude is first continued to the complex plane and is thereby fixed using scaling behaviour of the helicity spinors. Next, a generic tree level amplitude is analytically continued to the complex plane and it is observed that the amplitude vanishes as the complex parameter goes to infinity, due to gauge (or diffeomorphism in case of gravity) symmetries. The amplitude consists of simple poles, where one of the propagators go on-shell. Thus, the complete amplitude can be constructed from its residues. However, the residues themselves are sub-amplitudes in the complex plane, which are non-zero. Thus one can obtain all higher point tree-level amplitudes from lower order pieces using such a recursive technology. 
This further generalizes to loop level. Using the well developed one-shell technique, namely generalized unitarity, it is possible to construct all higher loop amplitudes using the tree level ones \cite{Brandhuber:2005jw}, which are themselves determined by the recursion relations. Thus, without resorting to the Lagrangian formulation, it is possible to determine amplitudes at each order in perturbation theory in both Yang-Mills and gravity. 
\\~\\
Even though the computation of amplitudes becomes simpler using on-shell techniques, it motivates us to understand what are the underlying reasons behind this simplicity. To gain deeper understanding, it is essential to develop better theoretical formulations which are expected to have close connections with the simpler observables which we compute, like the amplitudes. The chiral formulation is such an attempt. We work with the chiral formulation of Yang-Mills, which was introduced by Chalmers and Siegel \cite{Chalmers:1996rq}. \\~\\
The basis of chiral formulations is to consider the dimensionality of spacetime to be four and to use mathematical ingredients which are specific to this many dimensions. For instance, the Hodge dual operator, in a four-dimensional spacetime, maps any $p$-form to a $(4-p)$-form.
\be 
\star: \Omega^p\rightarrow \Omega^{4-p}.
\ee 
Particularly, any $2$-form gets mapped to another $2$-form thus preserving self-duality. 
The $2$-forms are eigenvectors of such an operator with eigen values $\pm 1$ in the Euclidean signature or $\pm i$ in Lorentzian signature. We work with the former because it is in this signature that momentum and other observables become real. The Hodge dual operator then allows any $2$-form to decompose into its self-dual and anti-self dual parts.
These two sectors are the two chiral halves of the particular $2$-form. The main motivation of the chiral formulation is that it makes full Yang-Mills as an extension of self-dual Yang-Mills, which is a nice theory. The theory is formulated by considering just one chiral half of the $2$-form curvature/field strength. Thus such theories inevitably treat the two helicities of the gluon on a different footing. The chiral decomposition is best understood in the spinor notations. In these notations, the curvature $2$-form decomposes as
\be 
F_{MM'NN'}^i = \frac{1}{2} F_{MN}^i \epsilon_{M'N'} + \frac{1}{2} F_{M'N'}^i \epsilon_{NM},
\ee 
where $F_{MM'NN'}^i$ is the 2-form field strength,  $F_{MN}^i$ and $F_{M'N'}^i$ are the self-dual and anti self-dual parts of the field strength respectively. Clearly, the type of spinor index (unprimed for self-dual and primed for anti self-dual) distinguishes the two sectors. The chiral action for Yang-Mills is then given by 
\be 
\label{YM-sd1}
\mathcal{L}_{YM}=-\frac{1}{4g^2}(F^{+\, i}_{MN})^2.
\ee
which takes into account just one of the chiral halves of the curvature. We will elaborate on this further in the second chapter of this part. The important point to emphasize is that the Lagrangian in (\ref{YM-sd1}) can be written in a first order form by introducing a non-propagating self-dual field $B^+$. In the first order form, one just has a single cubic interaction term along with a kinetic term. The form of the interaction term leads to very simple Feynman rules, when expressed in 2-spinor notations. Moreover, in the absence of the quartic vertex, unlike in the usual Yang-Mills, perturbative computations becomes much simpler. As we will see, a further benefit of such a formalism lies in its very simple gauge fixing procedure, which we outline in the next chapters. Thus the advantage to work with the chiral formulation of Yang-Mills is twofold. First it gives rise to very simple Feynman rules which reduces the number of diagrams at each loop order significantly. Second, the use of 2-spinors is naturally embedded in chiral theories and this leads to efficient computation of scattering amplitudes.
\\~\\
In the chiral setup, it is natural to pass from full Yang-Mills to self-dual Yang-Mills. One just needs to remove the quadratic term in the auxiliary field, by taking the gauge coupling parameter $g\rightarrow 0$. Thus, one is left with just one propagator and the cubic vertex. In SDYM, the connection carries one of the polarizations of the gluon while the auxiliary field carries the other polarization. The equation of motion for the connection is basically the self-duality condition of the curvature 2-form
\be 
F^+=0.
\ee 
As is elaborated in \cite{Krasnov:2016emc}, the theory is quantum finite since the arising divergences are proportional to the Pontryagin number and are thus topological in character. Therefore they do not contribute to the S-matrix. Then, it is easy to see that all tree level amplitudes vanish. The theory is one-loop exact and the one loop amplitudes are non-vanishing. As we will see, there is a striking similarity between the Feynman rules of SDYM and massless quantum electrodynamics in the spinorial notations. In particular, both the theories have one propagator and just the cubic interaction. Thus, the arising one-loop diagrams are very similar. It is worth mentioning to note that supersymmetric Ward identities relate the one-loop diagrams for different spin particles circulating in the loop. In particular it is well established that one can embed massless QED and pure Yang-Mills in a supersymmetric theory and then, the one-loop (all same helicity) diagrams where a fermion circulates in the loop is proportional to that where a gluon circulates in the loop. However, this is most easily seen using the spinorial version of Feynman rules in SDYM and massless quantum electrodynamics as we explain later. 
\\~\\
SDYM is an integrable theory, integrability being prominent in its twistor description, see \cite{Krasnov:2016emc}. Integrability can be seen to be the fundamental reason as to why tree level amplitudes are trivial in the covariant formalism of SDYM and no higher loops except one loop diagrams exist. The one loop amplitudes are non-trivial and cut-free, except for two particle poles. Some understanding as to why this is so comes from unitarity principles, where for any one-loop amplitude, the tree level sub-amplitudes on the cut vanish. However, from the Lagrangian point of view, there is no understanding as to why the properties of one-loop amplitudes has a striking similarity to that of tree amplitudes. Such a structure of the one-loop amplitudes led Bardeen to conjecture that the same helicity (and all but one same helicity) four point one-loop amplitudes may be related to some anomaly in the currents responsible for integrability of the self-dual sector \cite{Bardeen:1995gk}. So far, this has not been realized in the literature. In this thesis, we focus on the all same helicity amplitude, which is also the correct one captured by SDYM theory. The previous calculations which compute this amplitude does not shed light on the anomaly interpretation. Moreover, the computations use dimensional regularization and thus it is difficult to see the origin of such a simple result, which should be more transparent in a four dimensional calculation. In our work, we use the language of 2-spinors and it is well known that dimensional regularisation has its problems when dealing with such objects which are inherently described in four dimensional spacetime. Also, the amplitudes in question are finite and thus it is not necessary to introduce a regularization to compute them. We first propose a four-dimensional computation of the one loop same helicity four point amplitude which mimicks the chiral anomaly computation. It gives some idea about the origin of the simple result which comes from a complicated loop integral. Although it does not give rise to an interpretation, it does share some key features on the lines of an anomaly. 
\\~\\
Alternatively, there are well known worldsheet formulations of Yang-Mills, developed by Thorn and co, see \cite{Thorn:2002fj}. Bardakci and Thorn first proposed a way of expressing the planar Feynman diagrams,
selected by ’t Hooft’s $N_c\rightarrow \infty$ limit, of a quantum matrix field theory in the language of lightcone interacting string diagrams. Subsequently, they formulated the planar sector of Yang-Mills on the string worldsheet in the lightcone gauge \cite{Thorn:2002fj}. In \cite{Chakrabarti:2005ny}, they give the lightcone gauge calculation of (++++) gluon scattering amplitudes, using a novel choice of regulator. They realized an interesting fact that the complete one loop integrand for the on-shell Green function
associated with the (++++) amplitude vanishes, with a particular routing of momentum in each diagram contributing to this amplitude. In a subsequent paper \cite{Brandhuber:2007vm}, Brandhuber and collaborators showed that a two-point one-loop counterterm is the generating function for the infinite sequence of the one-loop all same helicity amplitudes in pure Yang-Mills. Their computation was carried out in the light-cone formalism and has close connections with the well developed MHV diagram formulation of pure Yang-Mills. The computation relies on a specific choice of regularization similar to the one by Thorn and collaborators. However, any such analogue of these results is missing in the covariant formalism of Yang-Mills.
\\~\\ 
In this part of the thesis, we show as to how the two point function can be interpreted as an effective propagator which can connect Berends Giele (BG) currents and thereby generates the series of all same helicity amplitudes in SDYM in its covariant formulation. The main observation is that the two-point one-loop amplitude is sensitive to shifts of the loop momentum variable. Thus, it is a very specific choice of shifts which we use in order to first show that the complete one loop integrand vanishes in SDYM theory. However, the diagrams contributing to the S-matrix are the box and triangles. There are no tadpole diagrams in SDYM and the bubbles do not contribute. Therefore, what we achieve is that the four point one loop (same helicity) amplitude, which is the sum of the box and the triangle diagrams can be expressed entirely in terms of the bubbles. The important aspect of the bubbles in the covariant formalism is that they can be treated as an operator to which all possible combinations of BG currents can be glued. Thus, it is this active participation of the tree level BG currents, generating the one-loop amplitudes which naturally gives an understanding as to why these amplitudes have striking similarities with the tree level amplitudes, in that they are cut-free and only possess two particle poles.
\chapter{Chiral formulation of Yang-Mills}
The same helicity sector of Yang-Mills has many simple properties. The tree amplitudes in this sector altogether vanish while the one-loop amplitudes do not possess any branch cut and resemble the features of tree amplitudes, in that they only possess soft and collinear singularities. The usual process to derive these amplitudes from the Yang-Mills Lagrangian is complicated. First, because the number of Feynman diagrams needed to compute is quite large which can be traced to the fact that such a Lagrangian contain both cubic and quartic vertices. The usual Yang-Mills admits a first order formulation by integrating in a generic 2-form field. This gives rise to a non-vanishing propagator of the 2-form field with itself, in addition to the usual gauge field propagator and this makes perturbation theory complicated. However, if a chiral version of the Yang-Mills is used, one needs to integrate in a self-dual 2-form field and the propagator of this field with itself vanishes. This makes perturbation theory simpler and thus motivates such a formulation. In a chiral theory, the two helicities of the gluon are described differently and chiral projections are used. The covariant formulation of the chiral Yang-Mills theory was first proposed by Chalmers and Siegel in \cite{Chalmers:1996rq}. The resulting chiral theory is first order in the fields and the action contains two terms. There is a self-dual Lagrange multiplier which enters the action whose purpose is to restrict the field strength to be anti-self dual. It is then easy to pass from the full Yang-Mills to self-dual Yang-Mills by just truncating the action and taking appropriate limits of the coupling constant. We will elaborate on this further in the remaining part of this chapter.
\\~\\
Such a formulation led to many other developments, most notably in the twistor community. For a long time, only very special cases of the Yang-Mills theory could be studied in the twistor space, for instance the self-dual or the anti self-dual configurations. These configurations describe the instanton sector of the theory and is classically integrable. Integrability can be understood in a convenient way in the twistor space in contrast to ordinary spacetime. Also, the amplitudes in these sectors admit elegant representations when they are uplifted to the twistor space. However, most of the studies were confined to one of these helicity sectors and the full theory could not be studied because of a lack of a chiral action. The Chalmers-Seigel formulation thus led to a perturbative study of the full Yang-Mills in the twistor space around the self-dual (instanton) sector and opened up many further developments. Studies of scattering amplitudes in twistor space in perturbative Yang-Mills led to on-shell recursion relations (BCFW), MHV rules and the construction of all-loop integrands for the planar $\mathcal{N}=4$ super Yang-Mills.
\\~\\
The main purpose of this chapter is to develop the chiral Yang-Mills formulation in the spinorial notations which is suitable for all pertubative calculations, derive the Feynman rules and compute the one-loop $\beta$-function. The simplicity of the Feynman rules with just a cubic vertex makes this formulation very useful for computations. This will also establish the notations that we use in the next chapter. We will then sketch the self-dual Yang-Mills theory from the chiral action and describe some of the essential features of this theory. Let us then start with the Lagrangian description of usual Yang-Mills theory and appropriately recast it into the well known chiral formulation. 
\section{From second order to first order}
The Lagrangian of the full Yang-Mills is given by
\be
    \mathcal{L}_{YM}=-\frac{1}{4g^2}F^i_{\mu\nu}F^{\mu\nu}_i,
\ee
where
\be
    F^i_{\mu\nu} = \partial_\mu A_\nu^i - \partial_\nu A_\mu^i + f^{ijk} A_\mu^j A_\nu^k
\ee
is the field strength and $f^{ijk}$ are Lie algebra structure constants. We work in Lorentzian signature to start with, and the convention is mostly plus, to simplify the analytic continuation necessary for evaluation of the loop integrals. 
\\~\\
It is convenient to pass to the spinor notations immediately. This converts each spacetime index into a pair of spinor indices $\mu \to MM'$. Whatever the signature, the field strength two-form can be split into its self- and anti-self-dual parts
\be
F_{\mu\nu}^i = F_{\mu\nu}^{+ \, i} + F_{\mu\nu}^{-\, i}.
\ee
The spinor formalism equivalent of this formula is
\be
F_{MM'NN'}^i = \frac{1}{2} F_{MN}^i \epsilon_{M'N'} + \frac{1}{2} F_{M'N'}^i \epsilon_{NM}.
\ee
Note the order of indices in the second $\epsilon$ is different from that in the first. This is in order to produce the following expressions for the self- and anti-self-dual parts of the field strength $F_{MN} = F_{MM' N}{}^{M'}, F_{M'N'}=F^M{}_{M'MN'}$. In particular, we have $F^i_{MN} = 2\partial_{MM'} A^i_{N}{}^{M'} + f^{ijk} A^j_{MM'} A^k_N{}^{M'}$.

Using the fact that the self-dual part of the field strength squared equals to the anti-self-dual part squared modulo a surface term, we can rewrite the YM Lagrangian as
\be
    \mathcal{L}_{YM}=-\frac{1}{2g^2}(F^{+\, i}_{\mu\nu})^2, 
\ee
which in spinor notations becomes
\be\label{YM-sd}
    \mathcal{L}_{YM}=-\frac{1}{4g^2}(F^{+\, i}_{MN})^2.
\ee
In Lorentzian signature this Lagrangian is complex, with the imaginary part being a total derivative. 
\subsection{Chiral YM Lagrangian}
The second order Lagrangian with just one-half of the field strength is not very convenient for perturbative computations. Indeed, we want a Lagrangian where the quartic vertex can be eliminated which in turn could simplify the algebraic complexity in the computations. Such a version is possible by switching to a first order formulation. To do so, we introduce an auxiliary self-dual Lie algebra valued field $B^i_{MN}$ and write the full YM Lagrangian 
\be\label{L-YM}
    \mathcal{L}_{YM}= B^{i MN} F^i_{MN} + g^2 (B^i_{MN})^2.
    \ee
Indeed, integrating out the $B^i_{MN}$ field we get back  (\ref{YM-sd}), and setting $g=0$ we get back the SDYM Lagrangian. We now linearize the theory on background $B^i_{\mu\nu}=0, A^i_\mu=0$. We denote the perturbations of $B, A$ by $b,a$. The linearisation of (\ref{L-YM}) reads
\be\label{L-lin}
 \mathcal{L}_{YM}= 2 b^{i\, MN} \partial_{MM'} a_N^i{}^{M'} + b^{i\, MN} f^{ijk} a^j_{MM'} a^k_{N}{}^{M'} + g^2 (b^i_{MN})^2.
 \ee
 \subsubsection{Gauge fixing}
 Let us now discuss the gauge-fixing. We take the BRST gauge-fixing fermion to be
 \be
 \label{gfermion}
 \Psi = \bar{c}^i \epsilon^{NM} (2  \partial_{MM'} a_N^i{}^{M'}) + g^2 \bar{c}^i \epsilon^{NM} h_c \epsilon_{NM}.
 \ee
 The BRST variation of this is
 \be
 s\Psi = h_c^i \epsilon^{NM} (2  \partial_{MM'} a_N^i{}^{M'}) + g^2 h_c^i \epsilon^{NM} h_c \epsilon_{NM} \\ \nonumber 
 + 2 \bar{c}^i \epsilon^{NM}   \partial_{MM'} (\partial_N{}^{M'} c^i + f^{ijk} a^j_N{}^{M'} c^k).
 \ee
This is what needs to be added to the Lagrangian (\ref{L-lin}). One then notices that the terms in the first line here can be combined with the terms already present in (\ref{L-lin}) by introducing
\be
\label{bh}
\bar{b}^{i\, MN} := b^{i\, MN} + h_c \epsilon^{NM}.
\ee
The gauge-fixed Lagrangian is then
\be
 \mathcal{L}_{YM} + s\Psi = 2 \bar{b}^{i\, MN} \partial_{MM'} a_N^i{}^{M'} + \bar{b}^{i\, MN} f^{ijk} a^j_{MM'} a^k_{N}{}^{M'} + g^2 (\bar{b}^i_{MN})^2 \\ \nonumber
 +2 \bar{c}^i    \partial^M{}_{M'} (\partial_M{}^{M'} c^i + f^{ijk} a^j_M{}^{M'} c^k),
 \ee
 where to write the second term in the first line in terms of $\bar{b}$ we used the fact that the combination $f^{ijk} a^j_{MM'} a^k_{N}{}^{M'}$ is automatically $MN$ symmetric, and so extending the symmetric object $b^{i\, MN}$ in front of it to the object $\bar{b}^{i\, MN}$ without any symmetry does not change this term.

\subsection{Feynman rules in 2-spinor notation}

We start deriving the Feynman rules for the chiral formulation of YM. To derive the propagators, we use the generating functional method. Let us write down the path integral for the kinetic part of the YM Lagrangian.\\
\be
    Z_0(J)=\int \mathcal{D}\bar{b}^{iMN}\mathcal{D}a_{MM'}^ie^{i\int d^4x[\mathcal{L_0}+J_{1MN}\bar{b}^{iMN}+J^N_{2M'}a_N^i{}^{M'}]}
\ee
where 
\be
   \mathcal{L_0}=2 \bar{b}^{i\, MN} \partial_{MM'} a_N^i{}^{M'}+ g^2 (\bar{b}^i_{MN})^2 
\ee
is the lagrangian density.\\~\\
We introduce Fourier transform of the fields
\be
\bar{b}^{MN}(x)=\int \frac{d^4k}{(2\pi)^4}e^{ikx}\bar{b}^{MN}(k),~~a_N^i{}^{M'}(x)=\int \frac{d^4k}{(2\pi)^4}e^{ikx}a_N^i{}^{M'}(k)
\ee
It is feasible to work in the momentum space because the derivative has a simpler expression here. This is why we first express our position space field variables in terms of the momentum space ones and then solve for the fields. To do so, let us write the action in momentum space. We now omit colour indices.
\be
\label{action22} 
S_0=\int \frac{d^4k}{(2\pi)^4}\Bigg[2i
\bar{b}^{MN}(k)k_{MM'}a_N^{~M'}(-k)+g^2\bar{b}^{MN}(k)\bar{b}_{MN}(-k)+\\
\nonumber
J_{1MN}(-k)\bar{b}^{MN}(k)+J^N_{2~M'}(k)a_N^{~M'}(-k)\Bigg].
\ee
Let us now solve for the fields using their equations of motion. The two equations of motion are respectively the Euler-Lagrange equations for the auxiliary and gauge fields. We have 
\be
\label{eqn11}
2ik_{MM'}a_N^{~M'}(-k)+2g^2\bar{b}_{MN}(-k)+J_{1MN}(-k)=0
\ee
and 
\be
\label{eqn22}
2i\bar{b}^{MN}(k)k_{MM'}+J_{2~M'}^N(k)=0.
\ee
To solve these equations simultaneously, we first consider the equation (\ref{eqn22}), take the current to the other side and multiply both side with $k^{AA'}$. Then taking the contraction on both sides, we get for the auxiliary field 
\be
\bar{b}^{MN}(k)=\frac{iJ^N_{2~M'}(k)k^{MM'}}{k^2}.
\ee
Next, we put the solution of the auxiliary field in (\ref{eqn11}). Again, keeping everything except the term with the gauge field on the right hand side, we multiply by another copy of momentum and take the contraction, giving us the solution for the gauge field in terms of the currents 
\be
a_N^{~M'}(k)=\frac{iJ_{1MN}(k)k^{MM'}}{k^2}-\frac{g^2J_{2N}^{~M'
}(k)}{k^2},
\ee
where we have used the spinor identity
\be
k^{MM'} k_{MN'} = -(1/2) \epsilon^{M'}{}_{N'} k^2.
\ee
Let us now integrate out the fields from the action in  (\ref{action22}). To do so, we plug their equations of motion into the action, which then expresses the action solely in terms of the currents. We next write the generating functional for this action 
\be
 Z_0[J]=exp\Bigg(i\int \frac{d^4k}{(2\pi)^4}\Bigg[\frac{iJ^N_{2M'}(k)k^{MM'}J_{1MN}(-k)}{k^2}-\frac{J^N_{2M'}(k)g^2J^{M'}_{2N}(-k)}{k^2}\Bigg]\Bigg).
 \ee
\subsubsection{Propagators}
We can now define the propagators in this theory by taking second order derivatives of the generating functional with respect to the currents. Clearly, there are two kinds of propagators. The $\langle aa\rangle$ propagator is given by 
\vspace{1em}
\vspace{1em}
\be 
\begin{gathered} 
\begin{fmfgraph*}(120,0)
\fmfleft{i1}
\fmfright{o1}
\fmf{photon}{i1,v1,o1}
\fmflabel{}{i1}
\fmflabel{}{o1}
\end{fmfgraph*}
\end{gathered}\nonumber\\~\nonumber\\
\langle a^i_{MM'}(k)a^j_{NN'}(-k)\rangle=ig^2\delta^{ij}\frac{\epsilon_{MN}\epsilon_{M'N'}}{k^2}.
\ee 
\\
where we use the spinor translation for the Minkowski metric as a product of two spinor metrics. The $\delta^{ij}$ factor appears as a result of the contraction of colour factors. There is no copy of momentum in the numerator whereas there is a factor of $k^2$ in the denominator and this makes the gauge field propagator trivial. Let us now write the other propagator in this theory, which connects the gauge field to the auxiliary self-dual field. 

\be 
\begin{gathered} 
\begin{fmfgraph*}(140,0)
\fmfleft{i1,i2}
\fmfright{o1,o2}
\fmf{vanilla}{i1,v1}
\fmf{photon}{v1,o1}
\fmflabel{}{i1}
\fmflabel{}{o1}
\end{fmfgraph*}
\end{gathered}\nonumber\\~\nonumber\\
\langle b^i_{AB}(k)a^j_{CC'}(-k)\rangle=\delta^{ij}\frac{\epsilon_{AC}k_{BC'}}{k^2}.
\ee 
In the $\langle ba\rangle$ propagator, we find that there is both a factor of the spinor metric and a single copy of momentum sitting in the numerator. The spinor metric connects two unprimed indices while the momentum connects the unprimed index of the auxiliary field to the primed index of the gauge field. The $k^2$ factors appears as usual in the denominator.\\
\vspace{1em}
\be 
\label{ghprop1}
\begin{gathered} 
\begin{fmfgraph*}(140,0)
\fmfleft{i1}
\fmfright{o1}
\fmf{ghost}{i1,o1}
\end{fmfgraph*}
\end{gathered}\nonumber\\~\nonumber\\
\langle c^i(k)c^j(-k)\rangle=\delta^{ij}\frac{i}{k^2}.
\ee 
The ghost propagator is quite similar to the ordinary scalar propagator with the exception that it is dressed with a colour indexed kronecker delta. The propagator is symbolically expressed as a directed line connecting the ghost field to the anti-ghost field.
\subsubsection{Vertices}
\vspace{1em}
\be 
\label{ghver1}
\begin{gathered} 
\begin{fmfgraph*}(120,90)
\fmftop{i1,i2}
\fmfbottom{o1}
\fmf{photon}{i1,v1}
\fmf{photon}{i2,v1}
\fmf{vanilla}{v1,o1}
\fmflabel{}{i1}
\fmflabel{}{i2}
\fmflabel{}{o1}
\end{fmfgraph*}
\end{gathered}\nonumber\\~\nonumber\\
\langle b^{iMN}a^j_{BB'}a^k_{CC'}\rangle=if^{ijk}\epsilon_{B'C'}\epsilon^{(M}_{~B}\epsilon^{N)}_{~C}.
\ee 
The vertex factor is symmetric in the two unprimed spinor indices of the auxiliary field. This is expressed as a symmetrised sum of a product of two spinor metrics where we indicate the symmetrization by putting a round bracket before and after the indices $M$ and $N$ respectively. In other words, the round bracket notation explicitly means 
\be 
\epsilon^{(M}_{~B}\epsilon^{N)}_{~C}=\epsilon^{M}_{~B}\epsilon^{N}_{~C}+\epsilon^{N}_{~B}\epsilon^{M}_{~C}.\nonumber 
\ee 
\vspace{1em}
\be 
\begin{gathered} 
\begin{fmfgraph*}(120,90)
\fmftop{i1,i2}
\fmfbottom{o2}
\fmf{ghost}{i1,v1}
\fmf{ghost}{v1,i2}
\fmf{photon}{v1,o2}
\fmflabel{}{o2}
\end{fmfgraph*} 
\end{gathered}\nonumber\\~\nonumber\\
\langle a^{i\,}_{~MM'}c^jc^k\rangle=if^{ijk}k_{MM'}.
\ee 
\subsection{One loop Feynman diagrams}
The aim of this section is to compute the one-loop Feynman diagrams in the chiral formalism and use them to calculate the beta function in the theory. The presence of the two kinds of propagators in this formalism increases the number of diagrams than in conventional Yang-Mills theory. However, the only vertex we have is the cubic vertex and this simplifies the computation process in that there are no diagrams with quartic vertices, which we have in the conventional case.
\\~\\
We use Feynman parametrization technique to evaluate the diagrams. The details of this technique is summarised in the appendix. We use dimensional regularization to regulate the UV divergences in each individual diagram. It is to be noted that there are well known problems to reconcile dimensional regularization with two component spinors. However, some of the spinor identities do hold in dimensional regularization while the others do not. To keep things unambiguous, we give a brief review of dimensional regularization and Feynman parameter integral and mention the spinor identities which we use in this scheme.
\subsubsection{Dimensional regularization} 
Integrals which diverge in four dimensions may not diverge in $d$ dimensions, where $d$ is some arbitrary spacetime dimension. This fact is exploited in the regularization of Feynman integrals. We take the spacetime dimension in this case to be $d=4-\epsilon$ where $\epsilon\neq 0$ and carry out the integration. The result obtained can then be expanded as a Laurent expansion in $\epsilon$, where we drop all the terms which are of the order $\epsilon$ or higher and keep only those which have orders $\epsilon^{n}$, $n<=0$. In the one-loop case, we find all the diagrams to be order $1/\epsilon$ and therefore the divergences are manifested as poles in $\epsilon$. \\~\\ 
To carry out the integral in $d$-dimensions, we first write the denominator of the integrand in a particular way. The denominator is a product of terms of the form $\Pi_{i=1}^n(l+a)^{n_i}$ where $n_i=2$ for the relevant diagrams in the computation. It is then straightforward to write it in a Feynman parametrized way. The Feynman parametrization method is used to write an algebraic fraction consisting of a product of terms in the denominator as an integral over some parameters of a sum of terms weighted by those parameters. Explicitly, we have a Feynman parametrized integral as 
\be
\frac{1}{B_1B_2...B_n}=
(n-1)!\int_0^1dx_1\int_0^1dx_2...\int_0^1dx_n \frac{\delta(1-x_1-x_2-..x_n)}{(B_1x_1+B_2x_2+...B_nx_n)^n}.
\ee
For a generic one loop process, it looks like 
\be 
\int d^dl\frac{1}{l^2(l+a)^2}=\int_0^1dx_1 \int d^dl\frac{1}{(x_1l^2+(1-x_1)(l+a)^2)^2}.
\ee 
where we have used the delta function and performed the integral over $x_2$.
It is then useful to write the denominator of the integrand such that we only have a quadratic power of the loop variable. We then make a shift 
\be 
l\rightarrow l-(1-x_1)a
\ee 
and this puts our integral into a form 
\be 
\int \frac{d^dq}{(2\pi)^d}\frac{1}{(q^2+D)^2}.
\ee 
where $D$ is a function of the Feynman parameters and $q$ is the shifted loop momentum variable. In dimensional regularization, we use $d=4-\epsilon$ and then there are standard results on integrals like the one we explained, with which we compute the result. We review the list of standard integrals in Appendix A.
\newpage 
\subsection{One-loop computation}
We start computing the one loop divergent integrals which arise in the chiral formalism. There are two kinds of propagators and only interaction vertex in the theory. Amongst them, the $\langle aa\rangle$ propagator has a factor of the gauge coupling in it. So there will be a correction to this propagator at one-loop and this will be the relevant diagram for the renormalization of the coupling constant. We also need to add to it, the ghost diagram where the ghost runs in the loop. For the renormalization of the fields, we need several diagrams. There will be a one-loop correction to the $\langle ba\rangle$ propagator and to the $\langle baa\rangle$ vertex. The correction to the vertex is given by two inequivalent triangle diagrams. Note, the vertex correction does not have any factors of the gauge coupling, in contrast to conventional Yang-Mills. There will also be a $b-b$ self energy diagram and this, along with others is needed to extract the renormalization constant for the fields. The $a-a$ self energy diagram does not contribute to the renormalization. It gives the usual transverse term, consistent with Lorentz covariance with the addition of the ghost diagram.
\subsubsection{Self energy for auxiliary field}
\be 
\begin{fmfgraph*}(150,75)
     \fmfleft{i}
     \fmfright{o}
     \fmf{vanilla,tension=3}{i,v1}
     \fmf{photon,left=1}{v1,v2}
     \fmf{photon,left=1}{v2,v1}
     \fmf{vanilla,tension=3}{v2,o}
     \fmflabel{AG}{i}
     \fmflabel{PQ}{o}
\end{fmfgraph*}\nonumber
\ee 
This diagram contributes the following
 \begin{equation}
 iM=f^{jki}f^{jkm}\epsilon^{~(B}_{A}\epsilon^{~C)}_{G}\epsilon_{(BC}\epsilon_{P)Q}\int \frac{d^{4-\epsilon}l}{(2\pi)^4}\frac{g^4}{l^2(l+k)^2}.
 \end{equation}
\\
To compute the integral, we use the conventional method of changing the loop momentum variable by introducing a new loop momentum $q=l+xk$ and put the integral in a Feynman parametrized form. Once we have done that, we calculate the integral in dimensional regularization. A complete list of standard Feynman parametrized integrals is reviewed in Appendix A. In particular, we use the result for the integral in (\ref{1}) and compute. The result is 
 \be
 iM=iT(A)\delta^{im}\Big(\epsilon_{((AP}\epsilon_{QG))}\Big)\Big(\frac{g^4}{16\pi^2\epsilon}\Big),   
\ee
where $\epsilon_{((AP}\epsilon_{QG))}:= \epsilon_{AP}\epsilon_{QG}+\epsilon_{AQ}\epsilon_{PG}$
and the structure constants in the contributing diagram is written in terms of the scalars of the gauge group, i.e $f^{jki}f^{jkm}=T(A)\delta^{im}$. The result is symmetric in the indices $A,G$ and $P,Q$. This is reminiscent of the fact that the auxiliary field is the self-dual part of the 2-form, which is symmetric in two of its spinor indices. 
\subsubsection{One loop correction to propagators}
The next diagram is the gauge field-auxiliary field diagram with the two kinds of propagators running in the loop. This diagram is interpreted as a one-loop correction to the $\langle ba\rangle$ propagator.
\be 
\begin{fmfgraph*}(150,75)
     \fmfleft{i}
     \fmfright{o}
     \fmfbottom{v}
     \fmftop{u}
     \fmf{photon,tension=3}{i,v1}
     \fmf{photon}{v1,u}
     \fmf{photon}{u,v2}
     \fmf{photon}{v2,v}
     \fmf{vanilla}{v,v1}
     \fmf{vanilla,tension=3}{v2,o}
     \fmflabel{CC'}{i}
     \fmflabel{PQ}{o}
\end{fmfgraph*}
\ee
The contribution of this diagram is the following:
\be
\label{bfpp}
 iN=if^{ijp}f^{jpn}\epsilon_{C'A'}\epsilon^{B}_{(A}\epsilon^{C}_{~G)}\epsilon^{P}_{(B}\epsilon^{A}_{~Q)}\int \frac{d^{4-\epsilon}l}{(2\pi)^4}\frac{g^2l^{A'}_{~G}}{(k+l)^2l^2},
 \ee
 where the indices $A,A',B,G$ lie inside the loop. We do not explicitly write these in the diagram. 
 Changing variables appropriately, we get the Feynman parametrized form as in (\ref{1}). We then obtain the result for the correction to the $\langle ba\rangle$ propagator 
\be
    iN=T(A)\delta^{in}\frac{g^2}{16\pi^2\epsilon}k^{C'}_{(P}\epsilon^{C}_{Q)}.
\ee
Note, the result is symmetric in the indices $P$ and $Q$. Indeed, the unprimed indices $P,Q$ are on auxiliary field line and the auxiliary field is a symmetric object in its two spinor indices. This justifies the symmetrization of these unprimed spinor indices. The structure constants in (\ref{bfpp}) are written as usual in terms of the scalars of the gauge group. 
\be
\begin{fmfgraph*}(150,75)
     \fmfleft{i}
     \fmfright{o}
     \fmfbottom{v}
     \fmftop{u}
     \fmf{photon,tension=3}{i,v1}
     \fmf{photon}{v1,u}
     \fmf{vanilla}{u,v2}
     \fmf{photon}{v2,v}
     \fmf{vanilla}{v,v1}
     \fmf{photon,tension=3}{v2,o}
     \fmflabel{CC'}{i}
     \fmflabel{PP'}{o}
\end{fmfgraph*}
\ee 
We now compute the a-a self energy diagram with $\langle ba\rangle$ propagators inside. This diagram corresponds to the correction of the $\langle aa\rangle$ propagator. The contributing term for this diagram is 
\be
iK=f^{ijp}f^{pmj}\epsilon_{C'B'}\epsilon_{P'M'}\epsilon^{(B}_{A}\epsilon^{C)}_{G}\epsilon^{(M}_{N}\epsilon^{P)}_{T}\int \frac{d^4l}{(2\pi)^4}\frac{l^G_{~M'}(l+k)^T_{~B'}\epsilon^A_{~M}\epsilon^N_{~B}}{l^2(l+k)^2}.
\ee
We change variables to put the above integral in Feynman parametrized form. We obtain a tensor integral, one in which the numerator of the integrand contains free indices. There are well known methods to reduce general tensor integrals into scalar integrals. In this case, one can use the Passarino-Veltman reduction and the quadratic terms in the shifted loop momentum inside the integral can be simplified using the following spinor identity which holds in $d\neq 4$ dimensions
\be 
\int \frac{d^{4-\epsilon}q}{(2\pi)^4} q_{AA'}q_{BB'}=\frac{1}{4-\epsilon}\epsilon_{AB}\epsilon_{A'B'}\int \frac{d^{4-\epsilon}q}{(2\pi)^4}q^2.
\ee 
We next use the integral formula in (\ref{2}) to compute the integration on the right hand side above. For the part of the integral where there is no $q$ dependence in the numerator, we use the formula in (\ref{1}) to evaluate it.
The linear terms in $q$ all vanish upon integration and we get 
\be
iK=-iT(A)\delta^{im}\frac{1}{8\pi^2\epsilon}\Big(\frac{5}{6}k^2\epsilon_{P'C'}\epsilon_{CP}-\frac{3}{6}k_{PP'}k_{CC'}\Big).
\ee
\subsubsection{Ghost contribution} 
\be 
\begin{fmfgraph*}(150,70)
     \fmfleft{i}
     \fmfright{o}
     \fmf{photon,tension=3}{i,v1}
     \fmf{ghost,left=1}{v1,v2,v1}
     \fmf{photon,tension=3}{v2,o}
     \fmflabel{PP'}{i}
     \fmflabel{CC'}{o}
\end{fmfgraph*}\nonumber 
\ee 
We now compute the ghost diagram, where the external lines are composed of the gauge field and the ghost propagates in the loop.
The contribution of this diagram is 
\be
(-1)f^{ijp}f^{pmj}\int \frac{d^4l}{(2\pi)^4}\frac{16(k+l)^{PP'}l^{CC'}}{(k+l)^2l^2},
\ee
where the factor of $(-1)$ in front of the integral is the usual contribution from a fermion loop. We write this in a Feynman parametrized form as before and  extract the index structure which is independent of the loop momentum. We use the integral formula (\ref{1}) and compute it. The result is
\be
    -i\frac{iT(A)\delta^{im}}{8\pi^2\epsilon}\Big(\frac{4}{3}k^2\epsilon^{PC}\epsilon^{P'C'}+\frac{8}{3}k^{PP'}k^{CC'}\Big).
\ee
\subsubsection{Vertex correction} 
Let us now compute the one-loop correction to the vertex. We need these diagrams in the computation of the $\beta$-function. This is different from the usual YM theory where there is no triangle contribution in the computation of the $\beta$-function. In the chiral formalism, there is a single cubic vertex in the theory. So the relevant diagram at one-loop is the triangle graph, with one $\langle aa\rangle$ propagator and two $\langle ba\rangle$ propagators. Note, we cannot have any other combination of propagators which can consistently produce a triangle diagram. Further, the $\langle aa\rangle$ propagator can lie either adjacent to the two external gauge field lines or adjacent to the gauge field and auxiliary field lines. But for the latter case, the two possible ways are equivalent to each other and this is why there will be a total of two triangle graphs. Let us first compute the graph where the $\langle aa\rangle$ propagator is adjacent to the two external gauge fields.
\\~\\
\be 
\begin{fmfgraph*}(150,75)
\fmfbottom{i1,i2}
\fmftop{o1}
\fmf{vanilla,tension=3}{i1,v1}
\fmf{photon,tension=3}{i2,v2}
\fmf{photon}{v1,v5}
\fmf{vanilla}{v5,v2}
\fmf{photon}{v1,v6}
\fmf{vanilla}{v6,v3}
\fmf{photon}{v2,v3}
\fmf{photon,tension=3}{v3,o1}
\fmflabel{CG}{i1}
\fmflabel{BB'}{i2}
\fmflabel{AA'}{o1}
\end{fmfgraph*}\nonumber 
\ee 
This diagram contributes the following
 \be
    -ig^2f^{acj}f^{eia}f^{emc}\epsilon_{A'B'}\times Y\times\frac{1}{2}\times \int \frac{d^{4-\epsilon}l}{(2\pi)^4}\frac{l^Q_{~M'}(l+k_1)^{M'}_{~S}}{l^2(l+k_1)^2(l-k_2)^2},
\ee
where $Y$ is an unprimed spinorial object, composed by stacking many unprimed $\epsilon$ spinors. In explicit form, it is given by 
\be 
Y:=\epsilon^{A}_{~(T}\epsilon^{D}_{~S)}\epsilon^{N}_{~(C}\epsilon^M_{~G)}\epsilon^{D}_{~(P}\epsilon^B_{~Q)}\epsilon^P_{~M}\epsilon^N_{~T}.
\ee 
The factor of half in front of the integral is the symmetry factor for this diagram. As usual, we change variables and put the above integral in Feynman parametrized form. However, the difference in this case is that there are three Feynman parameters. This is because in the triangle diagram, there are three propagators. The only way to combine the propagator factors in the denominator is to introduce three such parameters, with the constraint that the sum of all these three adds up to unity. This is achieved by introducing a delta function in the integral. We can then integrate out one of the parameters using the delta function and this changes the limits of integration for the other parameters. Overall, the Feynman parametrized form of the integral is given by
 \be
 \label{fey}
     \int_0^1dx_1\int_0^{1-x_1}dx_2\int \frac{d^{4-\epsilon}q}{(2\pi)^4}\frac{(q-x_2k_1+x_1k_2)^Q_{~M'}(q-k
     _1(1-x_2)-x_1k_2)^{~M'}_{S}}{(q^2+D_1)^3},
 \ee
where 
\be 
D_1=x_1(1-x_1)k_2^2+x_2(1-x_2)k_1^2+2x_1x_2k_1.k_2.
\ee 
Note that this is a new type of Feynman parameter integral where we have a cubic factor in the denominator . Then we can use the standard integrals in (\ref{3}) and (\ref{4}) to evaluate it. We first simplify the index structure in (\ref{fey}) by using the spinor identity $k^{A}_{~B'}k^{~B'}_{C}=\frac{1}{2}\epsilon_{AC}k^2$.
After simplifying the algebra and noting that the divergent part will only contain terms of $q^2$ in the numerator, we use the integral formula (\ref{3}). The result for this diagram is
\be
    T(A)f^{jim}g^2    \epsilon^{B'}_{~A'}\epsilon^{A}_{~(C}\epsilon^{B}_{~G)}.
\times\frac{1}{16\pi^2\epsilon} 
\ee
There will be two other triangle diagrams where the $\langle aa\rangle$ propagator is adjacent to the external auxiliary field and one of the external gauge fields. However, both these diagrams are equal due to permutation symmetry of the external gauge field lines. We then consider this particular diagram 
\\~\\
\be 
\begin{fmfgraph*}(150,75)
\fmfbottom{i1,i2}
\fmftop{o1}
\fmf{vanilla,tension=3}{i1,v1}
\fmf{photon,tension=3}{i2,v2}
\fmf{photon}{v1,v5}
\fmf{vanilla}{v5,v3}
\fmf{photon}{v1,v2}
\fmf{vanilla}{v2,v4}
\fmf{photon}{v4,v3}
\fmf{photon,tension=3}{v3,o1}
\fmflabel{CG}{i1}
\fmflabel{BB'}{i2}
\fmflabel{AA'}{o1}
\end{fmfgraph*}\nonumber 
\ee 
The contribution of this diagram reads
 \be
    iK=-if^{aej}f^{ica}f^{emc}\epsilon_{A'M'}\epsilon^{AM}_{TS}\epsilon_{N'D'}\epsilon^{ND}_{CG}\epsilon_{D'B'}\epsilon^{DB}_{PQ}\times\int \frac{d^4l}{(2\pi)^4}\frac{g^2(l-k_2)_Q^{M'}(l+k_1)^{N'}_S\epsilon_P^M\epsilon^N_T}{l^2(l+k_1)^2(l-k_2)^2}.
\ee
Using the same technique as we described in the previous diagrams, we obtain the result for this particular diagram. It reads
\be
    3T(A)f^{jim}g^2\epsilon^{B'}_{~A'}\epsilon^{A}_{~(C}\epsilon^{B}_{~G)} \times\frac{1}{16\pi^2\epsilon}.
    \ee 
\section{Renormalization in chiral formalism}
In the previous section, we have regularized the one loop diagrams using dimensional regularization and expressed the divergences as the poles in the parameter $\epsilon$. We now describe the procedure to renormalize the fields and coupling constant in this formalism such that the physical quantities, like the scattering amplitudes yield finite results at each order of perturbation theory. Here we do renormalization at one-loop to extract the beta function. At higher loops, the process can be iterated following the BPHZ procedure. We first define the relation between the bare fields and the renormalized fields. The renormalization constants are the factors which relate the two. There will be four such factors, three for the fields and one for the gauge coupling constant.\\~\\ 
Once the bare fields are written in terms of the renormalized fields, we can split the Lagrangian into two parts. It is important to state that the renormalization constants are expanded in a series in $1/\epsilon$, where the first term is unity. Higher order terms in $\epsilon$ corresponds to higher loop orders. 
Once the Lagrangian is split into two parts, the factors appearing as coefficients in the second part are responsible to cancel the divergences which appear in the one-loop diagrams. We use the minimal subtraction scheme to renormalize the fields and the coupling.
\\~\\ 
Let us illustrate all this. First, we define the relation between the  bare fields and the bare coupling constant with the renormalized fields and the renormalized coupling constant 
\be
\begin{split} 
\label{relation!}
b^{AB}_0&=Z_b b^{AB}, \\
a^{MM'}_{0}&=Z_a a^{MM'},\\
g_0&= Z_g g \mu^{\frac{\epsilon}{2}},\\
c_0&= Z_c c.
\end{split} 
\ee
where the bare fields are denoted by a $'0'$ subscript and $Z_b,Z_a,Z_g,Z_c$ are the renormalization constants. It is particularly important to focus on the third relation. We have introduced a parameter $\mu$ in order to define the relation of the bare coupling to the renormalized one. The reason is, the bare coupling is dimensionless in four spacetime dimensions. However, in a general $d$-dimensional spacetime, the renormalized coupling will have a mass dimension proportional to $\mu^{-\epsilon/2}$. This is why we make the replacement $g\rightarrow g\mu^{\epsilon/2}$ to keep this coupling dimensionless in any general $d$-dimensional spacetime. The parameter $\mu$ is not physical and physical observables should not depend on it. For instance, all the bare fields and the bare coupling is independent of this parameter. Therefore, one should get differential equations relating the parameter $\mu$ and the renormalized coupling so that the bare quantities remain invariant under the changes in $\mu$. \\~\\
We also note that only the symmetrised part of the auxiliary field enters the $\langle baa\rangle$ interaction term. So we split this field into its symmetrised and anti-symmetrised part and redefine the renormalized fields accordingly
\be
\begin{split} 
b^{AB}&= b^{(AB)}+b^{[AB]},\\
  b^{(AB)}_{0}&=Z_{b_{sym}} b^{(AB)}, \\
  b^{[AB]}_{0}&=Z_{b_{as}} b^{[AB]}. \\
   \end{split}
\ee
where $b^{(AB)}$ is the symmetric part, often denoted by a round bracket and $b^{[AB]}$ is asymmetric part, denoted by a square bracket.
It is the symmetric part of the auxiliary field and therefore $Z_{b_{sym}}$ which we use in the renormalization. It is then convenient, where confusions do not arise, to drop the notation for the symmetric part and just write the renormalization constant for the auxiliary field as $Z_b$. Let us then write the Lagrangian in terms of the bare fields and parameters. 
\be
\label{L1}
 \mathcal{L} = 2 \bar{b}^{i\, MN}_0 \partial_{MM'} a_{0N}^i{}^{M'} + \bar{b}^{i\, MN}_0 f^{ijk} a^j_{0MM'} a^k_{0N}{}^{M'} + g_0^2 (\bar{b}^i_{0MN})^2 \\ \nonumber
 +2 \bar{c}^i_0    \partial^M{}_{M'} (\partial_M{}^{M'} c^i_0 + f^{ijk} a^j_{0M}{}^{M'} c^k_0).
 \ee
We now use the relations in (\ref{relation}) to write the same Lagrangian in terms of the renormalized fields.
\be 
\label{L2}
\mathcal{L} = 2Z_bZ_a \bar{b}^{i\, MN} \partial_{MM'} a_{N}^i{}^{M'} + Z_bZ_a^2\bar{b}^{i\, MN} f^{ijk} a^j_{MM'} a^k_{N}{}^{M'} + g^2Z_g^2Z_b^2\mu^{\epsilon} (\bar{b}^i_{0MN})^2 \\ \nonumber
 +2Z_c^2 \bar{c}^i_0    \partial^M{}_{M'}\partial_M{}^{M'} c^i_0 + f^{ijk} Z_aZ_c^2 \bar{c}^i_0    \partial^M{}_{M'}a^j_{0M}{}^{M'} c^k_0.
 \ee
The first order perturbations of the Lagrangian in (\ref{L1}) generates the loop diagrams. Then the idea is to cancel the divergent parts of these one-loop contributions by the relevant tree level counter-terms obtained from the Lagrangian in (\ref{L2}). It is important to point out that the renormalization constants has the role to cancel the divergences, which manifests as poles in $\frac{1}{\epsilon}$. Also, at tree level, the couplings and the bare fields must match the renormalized ones. Then, it is reasonable to expand the renormalization constants in a power series in $\frac{1}{\epsilon}$, where the first term of the series is unity. Thus, 
\be 
\begin{split} 
Z_b&=1+\sum_{j=1}^{\infty}\frac{p_n}{\epsilon},\\
Z_a&=1+\sum_{j=1}^{\infty}\frac{q_n}{\epsilon},\\
Z_g&=1+\sum_{j=1}^{\infty}\frac{g_n}{\epsilon}.\\
\end{split} 
\ee 
Now we provide the values of the renormalization factors by equating them to the divergent parts of the loop diagrams. We carefully equate the counter-terms by taking into account the factors which appear in places of the traces accordingly. We get
\be
\label{first} 
    Z_bZ_a^2=1+\frac{7g^2T(A)}{16\pi^2\epsilon}+....,
\ee

\be
\label{second} 
    Z_bZ_a+g^2T(a)\eta=1+\frac{g^2T(A)}{16\pi^2\epsilon}+....,
\ee

\be
\label{third} 
    Z_g^2Z_b^2=1+\frac{g^4T(A)}{16\pi^2\epsilon}+....
\ee
The term $(\partial a)^2$ has a coefficient $\eta$. We get the value of this coefficient from the $\langle aa\rangle$ self energy diagram. Thus we set 
\be
\label{fourth} 
    \eta=-\frac{13}{48\pi^2\epsilon}.
\ee
We now solve the simultaneous equations for the renormalization constants. Let us first take the equations (\ref{first}) and (\ref{second}). In equation (\ref{second}), we move the $\eta$-term to the right hand side and noting that it comes with a factor $g^2/\epsilon$, we combine it with the other term having the same factor. Then we divide both these equations. This eliminates the $Z_b$ factor and we get for $Z_a$ upto first order in $1/\epsilon$
\be
Z_a=1+\frac{5g^2T(A)}{48\pi^2\epsilon}+....
\ee
We next plug the value of $Z_a$ into equation (\ref{first}) and we recover the value of $Z_b$
\be
    Z_b=1+\frac{11g^2T(A)}{48\pi^2\epsilon}+...
\ee
The remaining thing to compute is the value of the renormalization constant for the coupling. This can be easily obtained by plugging the value of $Z_b$ in (\ref{third}). Notice that the terms in (\ref{third}) start at order $g^4$. However, we divide this term by a factor $Z_b^2$. This then gives the correct order of the coupling in $Z_g$. Overall, simplifying the algebra, we recover
\be
\label{Z_g}
    Z_g=1-\frac{11g^2T(A)}{48\pi^2\epsilon}+ O(\epsilon^{-2}).
\ee
This is the expected result for the renormalization constant of the gauge coupling in pure Yang-Mills. In the chiral formalism, we thus see that the renormalization constant for the auxiliary field plays a role to determine the renormalization constant for the coupling constant. Also, the renormalization constant for the gauge field is consistent with the usual Yang Mills. We can now calculate the flow of the coupling as a function of the energy scale. In the third equation in (\ref{relation!}), we related the bare coupling to the renormalized coupling and we explained that a factor of $\mu^{\epsilon/2}$ appears to match the dimension on both sides. We also said that the bare fields and the coupling constant in the original Lagrangian should not depend on this arbitrary parameter and therefore a differential equation governing the dependence of the renormalized coupling on this parameter must arise. In the following section, we derive this equation and compute the well known beta function of the theory. 
\subsection{$\beta$-function}
The $\beta$-function provides the information of how the couplings evolve with the energy scale of the process. One can imagine that there is a parameter space of the theory in which one of the axes denotes the energy scale and the other one denotes the renormalized parameters as a function of this energy scale. There will be differential equations governing the flow of the parameters as we keep shifting the energy scales. We understand this by observing how the trajectories in the parameter space evolve as we go high up in the UV or down below in the IR. Note, the bare couplings are completely unchanged and so are the scattering amplitudes. Thus, it is the renormalized coupling which gives us the information about how the theory becomes either strongly or weakly coupled as we move up or down in the energy ladder. To make things precise, we start by relating the bare coupling and renormalized coupling as follows
\be
    g_0=Z_gg\mu^{\frac{\epsilon}{2}}.
\ee
Let us introduce two other parameters related to the couplings
\be
    \alpha=\frac{g^2}{4\pi},~~\alpha_0=\frac{g_0^2}{4\pi},
\ee
So that
\be
    \alpha_0=\alpha Z_g^2\mu^{\epsilon}.
\ee
The next idea is to take logarithm to both sides and expand the logarithm of the renormalization constant in a series in $1/\epsilon$. This helps us define the $\beta$-function.
\be
\label{r}
    \ln{\alpha_0}=\ln{Z_g^2}+\ln{\alpha} +\epsilon \ln{\mu}.
\ee
Then we expand $\ln{Z_g^2}$ in powers of $1/\epsilon$
\be
\ln{Z_g^2}=\sum\frac{G_n(\alpha)}{\epsilon^n},
\ee
where $\ln{Z_g^2}$ consists of terms in powers of $\alpha$.
To calculate the $\beta$-function, we take $G_1(\alpha)$ and compute its derivative with respect to $\ln{\mu}$. From (\ref{r}), we see that if we take the derivative on both sides with respect to $\ln{\mu}$, the left must be zero as we already mentioned. However, then the terms on the right side can rearranged to have a relation of the form
\be 
\frac{d\alpha}{d\ln{\mu}}= f(\alpha).
\ee 
The function appearing on the right hand side in the above equation is defined to be the beta function of the theory. It tells how the couplings of the theory vary with respect to the logarithm of the energy scale. We can then equate this function to the one we obtain by taking the derivative of $G_1(\alpha)$. We have
\be
    \beta(\alpha)=\alpha^2G_1^{'}(\alpha).
\ee
Note that here the $\beta$-function is defined in terms of the variable $\alpha$ instead of the couplings themselves. This is why we introduced this variable as a square of the coupling. We next see that this relation helps us re-define the $\beta$-function directly in terms of the coupling.
Using the value of $Z_g$ from (\ref{Z_g}) we compute the $\beta$-function to be
\be
\label{beta1} 
    \beta(\alpha)=-\frac{11\alpha^2T(A)}{6\pi}+O(\alpha^3).
\ee
We want to express the $\beta$-function as a function of the coupling. So we use 
\be 
\begin{split} 
\alpha&=\frac{g^2}{4\pi},\\ \frac{d\alpha}{d(ln\mu)}&=\frac{1}{2\pi}g\frac{dg}{d(ln\mu)}
\end{split} 
\ee 
to get
\be
    \beta(g)=\frac{g^4}{16\pi^2}\frac{dG_1}{dg}\Big(\frac{dg}{d\alpha}\Big)^2.
\ee
Plugging the relevant values on the right hand side of the above equation, we get the expected result for the $\beta$-function in pure Yang-Mills
\be
    \beta(g)=-\frac{11g^3T(A)}{48\pi^2}+O(g^5).
\ee
The negative sign in front of the factor in right hand side indicates that the coupling becomes weak as we move to UV scales. This phenomenon as known is asymptotic freedom. However, this also means that pure Yang-Mills becomes strongly coupled in the IR. This is the reason why free quarks or gluons are never observed experimentally and perturbation theory breaks down at low energies.

\newpage
\section{Self-Dual Yang-Mills}
Let us once again consider the Yang-Mills action functional 
\be 
\label{YMM}
\mathcal{S}_{YM}=\int_M Tr(F\wedge\star F),
\ee 
where $M$ is the spacetime manifold over which we integrate. The Hodge dual operator, denoted by $\star$ maps any $p$-form to an $(n-p)$-form, in an $n$-dimensional manifold. In particular, in a $4$-dimensional manifold, the $\star$ maps any $2$-form to another $2$-form. Thus in these many dimensions, a $2$-form field does does not change its degree under the action of this operator. This fact is important and is the basis of self-duality, as we discuss below. The trace stands for the trace over the Lie algebra of some compact gauge group $G$ and $F$ is the usual curvature or the field strength, which in terms of the gauge field reads 
\be 
F=dA+A\wedge A.
\ee 
The classical Yang-Mills equations of motion are obtained by varying this action with the gauge potential. These are the Euler Lagrange equations of motion. They are given by 
\be 
dF=0, ~~~ d\star F=0.
\ee 
One can now start finding solutions to this equations of motion. In general, it is non-trivial to solve these simultaneous sets of equations. However, the existence of the Hodge operator immediately tells us that the 2-form field strength can be split as follows 
\be 
F=\frac{1}{2}(F+\star F) + \frac{1}{2}(F-\star F).
\ee 
These two sectors are called the self-dual and anti self-dual parts of the field strength. We identify the part $\frac{1}{2}(F+\star F)=F^+$ and $\frac{1}{2}(F-\star F)=F^-$ respectively.
When only one of these two sectors vanish, we obtain what is called the self-dual (or anti self-dual) solution to the equations of motion. The equation of motion for this particular case is given by 
\be 
\label{sd22}
F^{+}= 0.
\ee 
where the positive sign in our convention denotes that the solution is anti self-dual.
This tells us that given any solution to the equation $dF=0$, it automatically solves the other equation of motion. Is there any covariant action functional which gives (\ref{sd22}) as its equation of motion? The answer is yes and is known as the Chalmers-Siegel action. We now describe this action and give some details of it. We work in either Euclidean or split signature because it is in this signature that the fields are real. Also, it is this action which captures the sector of all plus (or minus) helicity amplitudes, which we will study in the next chapter. 
The action for the so-called self-dual YM theory is given by
\be 
\label{asdym}
\mathcal{S}_{SDYM}(A,B)=\int_M Tr(B^+\wedge F),
\ee 
where $B^+$ is a self-dual auxiliary 2-form field and $F$ is the usual curvature 2-form. In Euclidean signature the self-dual condition of the 2-form is given by 
\be 
B^+_{\alpha\beta}=\frac{1}{2}\epsilon^{~~\delta\sigma}_{\alpha\beta}B^+_{\delta\sigma}.
\ee 
There will be two equations of motion. First, varying the action with respect to the $B^+$, we obtain the equation of motion 
\be
\label{asd}
F^+=0.
\ee 
This says that the self dual part of the field-strength vanishes. Then the corresponding theory is called anti self-dual YM, but this is merely a convention. It is important to realize that it is the self-dual 2-form $B^+$ which imposes the anti self-duality condition on the field strength. However, the 2-form is an auxiliary variable and does not propagate itself. It is important to point out that the SDYM Lagrangian can be seen to arise as a truncation of the full YM Lagrangian in (\ref{YMM}), by letting $g=0$. Solutions obeying (\ref{asd}) are called instantons. This is why the self-dual YM theory captures the instanton sector of the full YM. This will play an important role later, as we will see. One of the polarizations of the gluon (negative in our convention) is described by the field configurations which satisfy the above equation. Let us now write the other equation of motion. Varying the action with respect to connection gives 
\be 
d_A B^+=0.
\ee 
which says that the covariant derivative of the 2-form with respect to the connection vanishes. Solutions to this equation gives us the other polarization of the gluon. 
\subsubsection{Linearization} 
Let us linearize our theory of self-dual YM around a self-dual gauge field configuration. We choose our background field for the 2-form to be trivial, i.e $B^+=0$, while for the connection we choose our background such that it satisfies the equation of motion in (\ref{asd}). Let us call this background $A$. We then linearize around these backgrounds. We denote the perturbations as $b=\delta B^+$ and $a=\delta A$. The linearized Lagrangian reads 
\be 
\mathcal{L}_{linear}=Tr(bd_Aa) + Tr(baa),
\ee 
where $d_A$ is as before, the covariant derivative with respect to the background connection $A$. The first term is the kinetic one while the second describes interaction. There is only one interaction vertex in the self-dual theory, similar to the chiral formalism of YM. However, in this case there is only one kinetic term which restricts the number of propagators to one. It is convenient to explicitly write the Lagrangian and pass to the spinor notations. We then give the Feynman rules and discuss gauge-fixing.

\subsection{Gauge fixing and Feynman rules} 
In spinor notations, the Lagrangian in SDYM reads 
\be
\label{sdym2}
\mathcal{L}_{SDYM}=b^{iNM}\partial^{~M'}_Ma^{i}_{NM'}+2f^{ijl}b^{iMN}a^{j~M'}_Ma^{l}_{NM'},
\ee
where $i,j,l$ are Lie algebra indices, $b^{iNM},  a^{i}_{NM'}$ are perturbations of the self-dual auxiliary field and connection. The gauge-fixing procedure is similar to the one we described in full YM. Let us describe it briefly for brevity. One considers the gauge fixing fermion as is outlined previously. The variation of the gauge fixing fermion is then added to the Lagrangian in (\ref{sdym2}). One then notices that the first term in the variation is similar to the kinetic term of the original Lagrangian. Then it is possible to combine the self-dual auxiliary field perturbation with the BRST auxiliary field as is described in (\ref{bh}). The new field is no longer symmetric in its two unprimed indices since the presence of the antisymmetric spinor $\epsilon^{MN}$ kills the symmetry. The gauge-fixed Lagrangian is given by 
\be 
\label{gauge}
\mathcal{L}_{SDYM} + s\Psi = 2 \bar{b}^{i\, MN} \partial_{MM'} a_N^i{}^{M'} + \bar{b}^{i\, MN} f^{ijk} a^j_{MM'} a^k_{N}{}^{M'} + \\ \nonumber
 +2 \bar{c}^i    \partial^M{}_{M'} (\partial_M{}^{M'} c^i + f^{ijk} a^j_M{}^{M'} c^k).
 \ee
The Feynman rules can be read off from this. The $\langle ba\rangle$ propagator and the cubic vertex are given by \\~\\
\textbf{Propagator}:
 \be 
 \langle b^{i~M}_A(-k)a^j_{BM'}(k)\rangle=\frac{2}{k^2}\delta^{ij}\epsilon_{AB}k^{~M}_{M'}.
 \ee\\ 
  \textbf{Vertex}:
 \be 
 \langle b^{iMN}a^j_{AM'}a^l_{BN'}\rangle=2if^{ijl}\epsilon_{M'N'}\epsilon^{~M}_A\epsilon^{~N}_B.
 \ee 
The ghost propagator and vertex are similar to full YM and we refer the reader to (\ref{ghprop1}) and (\ref{ghver1}) for details.\\~\\
\subsection{Berends-Giele Current} 
Let us now discuss about the currents in this theory. This part is based on \cite{Krasnov:2016emc}. A current is defined as the sum of all tree level Feynman graphs with all but one leg on-shell. Thus one can have different states inserted into the on-shell legs and the off-shell leg is taken with the propagator on that leg. However, the most interesting current arises if we take all the on-shell legs to be of same helicity. By convention, we take the negative helicity state inserted into these legs. 
\\~\\
The one-current is the polarization state itself. We can write it as 
\be 
J^{NN'}(1)= \frac{q^Nq_{E}1^{EN'}}{\langle q1\rangle\langle 1q\rangle}. 
\ee 
\\~\\
The two-current is obtained by taking the cubic vertex, projecting the legs of the gauge field on to helicity states and then applying the propagator on the final leg. It is given by
\\~\\
\be 
\begin{gathered}
\begin{fmfgraph*}(180,50)
     \fmftop{i1}
     \fmfbottom{o1}
     \fmfright{o2}
     \fmf{vanilla}{i1,v2}
     \fmf{vanilla}{v2,o1}
     \fmf{vanilla}{v1,v2}
     \fmf{vanilla}{v2,v1}
     \fmf{vanilla}{v1,o2}
     \fmflabel{$2^-$}{i1}
      \fmflabel{$1^-$}{o1}
      \fmflabel{}{o2}
      \end{fmfgraph*}
\end{gathered}
\ee 
\be 
J^{NN'}(1,2)= \frac{q^Nq_{G}(1+2)^{GN'}}{\langle q1\rangle\langle 12\rangle\langle 2q\rangle}. 
\ee 
The three current can be obtained by taking the two current $J_{NN'}(1,2)$ and attach the third gluon to it and then sum over the cyclic permutations. Two such diagrams arise, one is the attachment of the third gluon to the current $J_{NN'}(1,2)$ and another is the attachment of the first gluon to the current $J_{NN'}(2,3)$. Let us then add the scalar parts of these contributions. 
\\~\\
\\~\\
The first diagram is 
\be 
\begin{gathered} 
\begin{fmfgraph*}(120,50)
     \fmftop{i1}
     \fmfbottom{o1}
     \fmfright{o2}
     \fmf{vanilla}{i1,v2}
     \fmf{vanilla}{v2,o1}
     \fmf{vanilla}{v1,v2}
     \fmf{vanilla}{v2,v1}
     \fmf{vanilla}{v1,o2}
     \fmflabel{$2^-$}{i1}
      \fmflabel{$1^-$}{o1}
      \fmflabel{}{o2}
      \end{fmfgraph*}
      \end{gathered}
      \begin{gathered} 
      \begin{fmfgraph*}(120,50)
     \fmftop{i1}
     \fmfbottom{o1}
     \fmfleft{o2}
     \fmf{vanilla}{i1,v2}
     \fmf{vanilla}{v2,o1}
     \fmf{vanilla}{v1,v2}
     \fmf{vanilla}{v2,v1}
     \fmf{vanilla}{v1,o2}
     \fmflabel{$3^-$}{i1}
      \fmflabel{$4$}{o1}
      \fmflabel{}{o2}
      \end{fmfgraph*}
      \end{gathered} 
      \ee 
      \\~\\
\be 
=\frac{J(1,2)J(3)}{(1+2+3)^2(1+2)^2}\langle q|1+2\circ 3|q\rangle\langle q|\langle q|(1+2+3)|.  \nonumber 
\ee 
\\~\\
For the second contribution we glue the current $J_{NN'}(2,3)$ with the current $J_{MM'}(1)$. We get \\~\\
\be 
\begin{gathered} 
\begin{fmfgraph*}(120,50)
     \fmftop{i1}
     \fmfbottom{o1}
     \fmfright{o2}
     \fmf{vanilla}{i1,v2}
     \fmf{vanilla}{v2,o1}
     \fmf{vanilla}{v1,v2}
     \fmf{vanilla}{v2,v1}
     \fmf{vanilla}{v1,o2}
     \fmflabel{$2^+$}{i1}
      \fmflabel{$3^+$}{o1}
      \fmflabel{}{o2}
      \end{fmfgraph*}
      \end{gathered}
      \begin{gathered} 
      \begin{fmfgraph*}(120,50)
     \fmftop{i1}
     \fmfbottom{o1}
     \fmfleft{o2}
     \fmf{vanilla}{i1,v2}
     \fmf{vanilla}{v2,o1}
     \fmf{vanilla}{v1,v2}
     \fmf{vanilla}{v2,v1}
     \fmf{vanilla}{v1,o2}
     \fmflabel{$1^+$}{i1}
      \fmflabel{$4$}{o1}
      \fmflabel{}{o2}
      \end{fmfgraph*}
      \end{gathered} 
      \ee 
      \\~\\
\be 
=\frac{J(2,3)J(1)}{(1+2+3)^2(2+3)^2}\langle q|1\circ (2+3)|q\rangle\langle q|\langle q|(1+2+3)|.  \nonumber 
\ee 
Adding the scalar part, we have 
\be 
\begin{split} 
J(1,2,3)&=\frac{1}{(1+2+3)^2\langle q1\rangle\langle q2\rangle\langle q3\rangle}\Bigg(\frac{\langle q1\rangle[13]+\langle q2\rangle[23]}{\langle 12\rangle}+ \frac{\langle q2\rangle[12]+\langle q3\rangle[13]}{\langle 23\rangle}\Bigg)\\
&= \frac{1}{(1+2+3)^2}\Bigg(\frac{\langle q1\rangle[13]\langle 23\rangle+\langle q2\rangle [23]\langle 23\rangle+\langle q2\rangle [21]\langle 21\rangle+\langle q3\rangle[31]\langle 21\rangle}{\langle 12\rangle\langle 23\rangle\langle 1q\rangle\langle 2q\rangle\langle 3q\rangle}\Bigg).
\end{split} 
\ee 
Now we use the Schouten identity 
\be 
\langle q3\rangle\langle 12\rangle+\langle q1\rangle\langle 23\rangle=\langle q2\rangle \langle 13\rangle
\ee 
and write 
\be
\begin{split} 
J(1,2,3)&= \frac{1}{(1+2+3)^2\langle q1\rangle\langle q2\rangle\langle q3\rangle}\Bigg(\frac{\langle q2\rangle([23]\langle 23\rangle+[13]\langle 13\rangle+[12]\langle 12\rangle}{\langle 12\rangle\langle 23\rangle}\Bigg)\nonumber\\
&=\frac{1}{\langle q1\rangle \langle 12\rangle\langle 23\rangle\langle 3q\rangle}. 
\end{split} 
\ee 
Thus the three current can be written as 
\be 
J_{NN'}(1,2,3)= J(1,2,3)q_Nq^{E}(1+2+3)_{EN'}.
\ee 
The pattern is becoming clear. The general $n$th-order current can be written as 
\be 
\label{curr}
J_{NN'}(1,2,...,n)= J(1,2,...,n)q_Nq^{E}(1+2+...+n)_{EN'}.
\ee 
where the scalar part is given by 
\be 
\label{sc}
J(1,2,...,n)=\frac{1}{\langle q1\rangle \langle 12\rangle\langle 23\rangle....\langle (n-1)n\rangle\langle nq\rangle}. 
\ee 
The structure of the scalar part of the current shows much resemblance with the well known Park-Taylor formula. We recall that the Park-Taylor formula gives the tree level MHV amplitude in YM and is given by 
\be 
\label{Park}
A(1^-,2^-,....,j^+,l^+,...,n^-)=\frac{\langle jl\rangle^4}{\langle 12\rangle\langle 23\rangle\langle 34\rangle....\langle n1\rangle}. \ee 
It is interesting to note that even though the BG current in self dual YM theory is non-trivial, the tree level amplitudes are trivial. One can argue this by noticing the form of the current in (\ref{curr}). Indeed, if we want to compute the tree level amplitude with more than three external gluons, we can as well compute the current with all but one leg on-shell and finally project the off-shell leg to some helicity states. The helicity states can either be taken positive or negative. However, to do so we need to multiply the current by the propagator which we included in the final leg. It is evident from the structure in (\ref{curr}) that the current does not possess any pole and this is why one cannot cancel the arising propagator in the numerator which vanishes on-shell. Therefore, all such amplitudes vanish. We cannot construct amplitudes with more than one positive leg in this theory. Thus, all tree level amplitudes with more than three external gluons vanish on the trivial background, i.e $B^+=0, A=0$ in the self dual YM theory.
\newpage 
\subsection{Yang-Mills Instantons}
Let us give a brief sketch of instantons here. For a detailed overview of the subject, we refer the reader to a more expository account in \cite{Tong:2005un}. A Yang-Mills instanton is a classical solution to the Euclidean equation of motion which minimize the action functional. They are topologically non-trivial solutions of the Yang-Mills equations and are non-perturbative in nature. In particular, an instanton solution is a gauge field connection which approaches to pure gauge at spatial infinity. 
\be 
A_{\nu}\rightarrow g^{-1}\partial_{\nu}g.
\ee 
Such gauge field configurations imply that the field strength vanishes at infinity and the Euclidean action functional remains finite at spatial infinity $S_{E}[A_{\nu}(x)]<\infty$. As we now explain, such solutions to the Yang-Mills equations are self-dual and anti-self-dual connections, belonging to two different topologically distinct classes. Consider the Yang Mills action functional 
\be 
\mathcal{S}=\int_M\textrm{Tr} F\wedge \star F.
\ee 
Using the decomposition of the curvature two-form into its self dual and anti-self-dual parts $F=F^++F^-$, we can write the action as 
\be 
\label{s1}
\int_M d^4x \Big(|F^+|^2+|F^-|^2\Big).
\ee 
This term is the sum of squares of the self dual and the anti self-dual parts of the curvature. Let us now consider an integral which is the difference of the squares of these two. This is a topological term because it can be expressed as a total derivative of a Chern-Simons form and can only have non-vanishing contribution on the boundary of the manifold. For a boundary with trivial topology, such a term vanishes by Stokes' theorem. Thus, consider the term 
\be 
\label{topo}
k=\int_M \textrm{Tr} F\wedge F=\int_M d^4x \Big(|F^+|^2-|F^-|^2\Big). 
\ee 
From (\ref{s1}) and (\ref{topo}), it is easy to see that the Yang-Mills action functional is bounded from below
\be 
S\geq |k|.
\ee 
The number $k$ is an integer and is known as the second Chern number $\int ch_2(F)$, where the Chern character is defined as $ch(F)=\sum ch_n(F)=exp(\frac{iF}{2\pi})$. The minimization of the action occurs when either the self-dual or the anti-self-dual part of the curvature vanishes. Such configurations correspond to the equation $F^{\pm}=0$, which is also the equation of motion for SDYM. Thus, we see that connections which correspond to instanton solutions of the Euclidean field equations are necessarily self-dual or anti-self-dual depending on the signature of the topological term (\ref{topo}). It is important to emphasize that the instanton condition in YM is first order which implies that one can find such solutions by solving a first order partial differential equation. The simplest such solution is the BPST instanton. The instantons have wide ranging applications starting from the vacuum structure of Yang-Mills to the classification of four manifolds. It is also worth mentioning that there exists a powerful method called the ADHM construction \cite{Atiyah:1978ri} which solves the self-dual Yang-Mills equations and construct instanton solutions. However, in this thesis we do not give any details of it. 
\chapter{Amplitudes in self-dual Yang-Mills}
The study of amplitudes in self-dual Yang-Mills is central in this thesis because they pose many interesting features. The vanishing of the same helicity tree amplitudes is understood from the fact that this theory is classically integrable and it is the currents (Berends-Giele) whose structure determine such a vanishing. Moreover, the one-loop amplitudes have recently been tied to UV divergences of quantum gravity and supergravity, see \cite{Bern:2015xsa}, \cite{Bern:2017puu}, \cite{Bern:2013uka}. This raises further interest because the whole problem of divergences in quantum gravity (and supergravity) boils down to the understanding of these amplitudes. In addition to all these, recent studies on higher spin theories of gravity have shown that the one-loop amplitudes in such theories bear close resemblance to that in self-dual Yang-Mills \cite{Skvortsov:2020gpn}. This is a rather interesting result and it hints that there might be close relations between self-dual Yang-Mills and chiral high spin gravity theories. So far, these avenues have not been explored in their full details and glory. \\~\\
In this chapter, we aim to understand the features of one-loop same helicity amplitudes in SDYM. However, it is customary to sketch a brief introduction on tree amplitudes and build on the subject of loop amplitudes from there. It is convenient to analyze the theory of SDYM in split signature because only in this signature the spinorial objects become real. To this end, let us start to work in a self dual background configuration. The simplest self-dual background is the zero connection. When the action is expanded around this, we get a kinetic term and a non-derivative cubic interaction. Thus let us deduce the physical polarization states in this theory. To do so, we first write the linearised field equations. There are two such, one for the gauge field and another for the auxiliary field. Using the Lagrangian in (\ref{sdym2}) we get
\be 
\label{dirac}
\partial_{A}^{~A'}a_{BA'}=0,~~~\partial^A_{~B'}b_{BA}=0.
\ee 
The operator which acts on these equations is the Dirac operator. This operator eats one primed/unprimed index and thus maps $S_+\times S_- \rightarrow S_+\times S_+$ in the first equation whereas $S_+\times S_+\rightarrow S_-\times S_+$ in the second. We then apply it once more and go to momentum space to deduce that (using Dirac squared = Laplacian) the momentum vector is null.
\be 
k^2=0.
\ee 
It is then easy to construct helicity states by solving (\ref{dirac}). The negative helicity state is given by 
\be 
\epsilon^-_{AA'}(k)=\frac{q_Ak_{A'}}{\langle qk\rangle}.
\ee 
where $k_{AA'}=k_Ak_{A'}$. The helicity state comprises of two components. One is the auxiliary spinor $q^A$ which takes care of the gauge freedom of the theory. Shifting $q^A$ to some $q^A+\xi \eta^A$ for some arbitrary vector $\eta^A$ keeps any physical quantity such as amplitudes invariant. The other is the momentum spinor $k_{A'}$ which results from the null condition. Then the general solution for the linearised field equation for the gauge field are plane waves weighted with such polarization states. The solution to the second equation in (\ref{dirac}) results in the positive polarization helicity state and is given by 
\be 
\epsilon^+_{AB}(k)=k_Ak_B.
\ee 
Note that the dimensions of the two helicity states are different. Whereas the negative helicity state is dimensionless as should be appropriate for a field of mass dimension one, the positive helicity has mass dimension one in accordance with the auxiliary field which has mass dimension two. Then the general solution to the equation for the auxiliary field are plane waves weighted with positive helicity states such as the one sketched above. It is important to realize that in SDYM, the gauge field propagates only one of the helicities. This is because the field configuration is such that only of the self-dual/anti-self dual part of the field strength is non-zero. In the full Yang-Mills, the connection (gauge field) carries both the helicities and on top of that the auxiliary field continues to carry the positive helicity. This leads to many more amplitude configurations in the full Yang-Mills than in self-dual Yang-Mills. 
\\~\\
Most of the tree amplitudes in self-dual Yang-Mills are vanishing. The argument relies on the structure of the Berends-Giele current. Indeed, the current consists of a sum of all Feynman graphs with all but one leg on-shell. The on-shell legs are projected to negative helicity states. Then, to recover the all minus or all but one minus amplitude, we must multiply the propagator with this current (in the convention that the off-shell propagator is included in the current) and then project the final leg to a negative helicity state. However, in the on-shell limit, the propagator is zero ($k^2=0$) and there is no pole in the current to cancel this. Thus all such amplitudes vanish. The only non-vanishing amplitudes are the so called MHV ones in which two of the helicities are of the same kind and the rest are of the opposite kind. Thus, the simplest non-vanishing amplitude is the colour-ordered 3 point one which is given by 
\be 
A^{++-}=\frac{\langle 3q\rangle^2[12]}{\langle 1q\rangle\langle 2q\rangle},
\ee 
where the angle bracket stands for unprimed index contraction and square brackets for primed index contraction. It is straightforward to eliminate the $q$-dependence by using the momentum conservation for 3-particle kinematics and write the final suggestive form for this amplitude 
\be 
A^{++-}=\frac{[12]^3}{[13][32]}.
\ee 
It is to be noted that this is the only non-zero tree amplitude in the theory. Let us now consider the loop amplitudes in this theory. It is straightforward to see from the Feynman rules that the theory of SDYM is one-loop exact. These one-loop amplitudes are special in that they do not contain any branch cuts and only possess collinear or soft singularities, very much resembling the tree amplitudes. The reason as to why such a structure emerges has resisted understanding till now from the point of view of Feynman diagrams. There is some understanding of the cut-free nature of these amplitudes from the principles of generalized unitarity, where the cuts vanish due to the vanishing of the same helicity tree amplitudes. The importance of such amplitudes is twofold. One is because it remains obscure till date whether the non-zero value of these amplitudes result from an anomaly of the self-dual currents and another because the understanding of the divergences in 2-loop quantum gravity (and 4-loop $\mathcal{N}=4$ supergravity) boils down to the understanding of these amplitudes, as we elaborate on it later. Let us then review the existing expressions of these amplitudes and thereafter we present our new computation of such an amplitude at four points which shares some similarity with anomaly like features.
\section{Review: Literature expression for one-loop amplitudes}
The series of all same helicity one loop amplitudes $A_n(1^+,2^+,...n^+)$ in YM are finite rational functions of the momenta involved and has cyclical symmetry in the arguments, see \cite{Bern:1993sx} and \cite{Bern:1993qk}. These amplitudes are singular in the region where two adjacent momenta become collinear or a momentum become soft. Let us write the first two in the series explicitly and then compactly write the ansatz at $n$ points.  
\be
\begin{split} 
A_4(1^+,2^+,3^+,4^+)&=\frac{s_{12}s_{23}}{\langle 12\rangle\langle 23\rangle\langle34\rangle\langle 41\rangle},\nonumber\\
A_5(1^+,2^+,3^+,4^+,5^+)&=\frac{s_{12}s_{23}+s_{23}s_{34}+s_{34}s_{45}+s_{45}s_{51}+s_{51}s_{12}+\epsilon(1,2,3,4)}{\langle 12\rangle\langle 23\rangle\langle34\rangle\langle 45\rangle\langle 51\rangle}.\nonumber\\
\end{split} 
\ee 
where $\epsilon(i,j,m,n)=[ij]\langle jm\rangle[mn]\langle ni\rangle-\langle ij\rangle[jm]\langle mn\rangle[ni]$.
As we can see, both the expression at four and five points are purely rational functions of the momenta involved. The denominator of these expressions share a particular pattern where a string of angle brackets starting with the contraction between the first and second momenta spinor and ends with the contraction between the last and the first ones. This structure of the denominator exhibits two particle poles, similar to tree amplitudes. The numerator of these expressions carry products of Mandelstam invariants in a cyclic fashion. There are no logarithms or polylogarithms involved in these expressions, unlike other loop amplitudes and this is why they are cut-free. 
The ansatz for the general $n$-point amplitude can be written as
\be 
\label{n}
A_n(1^+,2^+,....n^+)=\frac{E_n+O_n}{\langle 12\rangle\langle 23\rangle....\langle n1\rangle}
\ee 
where 
\be 
O_n=\sum_{1\leq i_1<i_2<i_3<i_4\leq n-1}\epsilon(i_1,i_2,i_3,i_4)=-\sum_{1\leq i_1<i_2<i_3<i_4\leq n}tr[i_1i_2i_3i_4\gamma_5]
\ee 
and 
\be 
E_n=-\sum_{1\leq i_1<i_2<i_3<i_4\leq n}tr[i_1i_2i_3i_4]
\ee
Everywhere, the trace stands for matrix trace. The difference between the expressions $E_n$ and $O_n$ is the factor of $\gamma_5$. Clearly, we can see that the denominator of the general $n$-point amplitude follows the same pattern like the lower points. Thus, the only singularities which appear are when one of the gluon becomes soft or two adjacent gluons become collinear. Such expressions have been deduced from the consideration of soft and collinear singularity arguments. On another side, the expressions for these amplitudes are derived from string theory considerations where appropriate field theory limits are taken. One can merge the expressions for $E_n$ and $O_n$ and write the amplitude in a compact way as 
\be 
A_n(1^+,2^+,....n^+)=\sum_{1\leq i_1<i_2<i_3<i_4\leq n}\frac{tr_{-}[i_1i_2i_3i_4]}{\langle 12\rangle\langle 23\rangle....\langle n1\rangle}.
\ee

\section{One loop same helicity four point amplitude} 
The same helicity Yang-Mills amplitudes vanish at tree-level, but become non-zero
at one-loop. The QCD one-loop amplitudes at four (and five) points were computed by
the field theory techniques in \cite{Bern:1991aq}, and via string-inspired technology in \cite{Bern:1993sx} (four-points)
and \cite{Bern:1993mq} (five-points). The result for same helicity five gluon amplitude was then used
to conjecture \cite{Bern:1993sx} an $n$-gluon formula. Supersymmetry implies that there is a relation
between same helicity one-loop amplitudes in theories with different spin particles (e.g.
spin 1 and spin 1/2) propagating in the loop, see [3]. This means that the same helicity
one-loop amplitudes in YM are related to those in massless QED. The later were computed
in \cite{Mahlon:1993fe} using recursive methods, thus proving the conjecture of [4]. This conjecture received
additional support from the consideration of the collinear limits in \cite{Bern:1993qk}. At four points,
which is the case of main interest for us in this paper, the same helicity amplitude takes
the following extremely simple form \cite{Bern:1991aq}
\be 
\mathcal{A}_{one-loop}(1^-,2^-,3^-,4^-)\approx\frac{[12][34]}{\langle 12\rangle\langle 34\rangle}.
\ee 
where the spinor helicity notations are used, see below, and the proportionality factor
contains a numerical coefficient as well as powers of the relevant coupling constant.
\\~\\ 
In \cite{Bardeen:1995gk} William A. Bardeen suggested that the integrability of the self-dual sector of YM
theory is behind the simplicity of the all-same helicity sector of the full YM. This paper
also conjectured that the non-vanishing of the same-helicity one-loop amplitudes should
be interpreted as the anomaly of the currents responsible for the integrability of the selfdual sector. The paper [8] explicitly confirmed that the four-point same helicity one-loop
amplitudes in self-dual YM are given by (1.1). This paper also discusses symmetries of the
self-dual YM.
More recently, the non-vanishing of the one-loop same helicity amplitude in YM and
gravity was shown to be linked to the UV divergence of the two-loop quantum gravity, as we explained earlier. This
is very clear from the calculations \cite{Bern:2015xsa, Bern:2017puu} that use unitarity methods and directly link the
two-loop divergence to the one-loop amplitude. It thus
becomes more pressing to revisit the possible anomaly interpretation of the one-loop same
helicity amplitude. Indeed, if this amplitude’s non-vanishing is the signal of an anomaly, it
may be made to vanish by appropriately canceling the anomaly. It is thus very important
to understand the anomaly interpretation, if any, of the same helicity one-loop amplitudes
of YM and gravity.\\~\\
The purpose of our computation is a modest step in this direction. The available calculations \cite{Bern:1993sx, Mahlon:1993fe} of the four-point one-loop same helicity amplitude are not transparent. The
first of these uses string theory inspired methods. The second calculates all $n$-point amplitudes (in massless QED) and uses usual Feynman diagrams but resorts to dimensional
regularisation to extract the final result. Given that the amplitude one calculates is non-divergent, this makes it hard to understand where the result is coming from.

\subsection{One loop four point amplitude from shifts}
The four point same helicity amplitude in Yang-Mills is captured by the simpler theory of self-dual Yang-Mills. However, because of the similarity between the Feynman rules of this theory and of massless QED, one can perform the exact same calcualtion in either of them. The only difference is an overall numerical factor which arises because the self-dual Yang-Mills Feynman rules contain some additional spinor metrics which get contracted in the one-loop diagram to produce an overall factor. We omit such numerical factors in the calculation. The first diagram we consider is the box. As we will show subsequently, the bubbles are zero to begin with as this can be attributed to a Lorentz invariance argument by noticing the form of the integrand. The triangles on the other hand are in general not vanishing. We thus compute them using the same technique as is described here. To get the colour ordered four point amplitude, we add the box diagram with four distinct triangles and this reproduces the correct result. 
\subsubsection{Box}
Let us start with the box. Using the Feynman rules of self-dual Yang-Mills, the box diagram is constructed as follows
\vspace{1em}
\vspace{1em}
\vspace{1em}
\vspace{1em}
\be
\begin{fmfgraph*}(110,80)
\fmfbottom{i1,i2}
\fmftop{o1,o2}
\fmf{photon,tension=3}{i1,v1}
\fmf{photon,tension=3}{i2,v2}
\fmf{photon,tension=3}{o1,v3}
\fmf{photon,tension=3}{o2,v4}
\fmf{photon}{v3,v3'}
\fmf{vanilla}{v3',v1}
\fmf{photon}{v1,v1'}
\fmf{vanilla}{v1',v2}
\fmf{photon}{v2,v2'}
\fmf{vanilla}{v2',v4}
\fmf{photon}{v4,v4'}
\fmf{vanilla}{v4',v3}
\fmflabel{$1^-$}{o1}
\fmflabel{$4^-$}{o2}
\fmflabel{$2^-$}{i1}
\fmflabel{$3^-$}{i2}
\fmflabel{$l+1+2$}{v1'}
\fmflabel{$l+1$}{v3'}
\fmflabel{$l-4$}{v2'}
\fmflabel{$l$}{v4'}
\end{fmfgraph*}\nonumber
\ee\\
In index free notations, the box diagram is given by
\be 
\label{amp!}
i\mathcal{A}=\int \frac{d^4l}{(2\pi)^4}\frac{\langle q|l|4]\langle q|l+1|1]\langle q|l+1+2|2]\langle q|l-4|3]}{l^2(l+1)^2(l+1+2)^2(l-4)^2\prod_{j=1}^4\langle qj\rangle}.
\ee
\\~\\
We start multiplying the numerator and denominator of the integral by $\langle 43\rangle$. With this, rearranging the numerator, we have
\be 
i\mathcal{A}=\int \frac{d^4l}{(2\pi)^4}\frac{\langle q|l\circ4\circ3\circ(l-4)|q\rangle\langle q|l+1|1]\langle q|l+1+2|2]}{l^2(l+1)^2(l+1+2)^2(l-4)^2\prod_{j=1}^4\langle qj\rangle\langle 43\rangle}.
\ee
We then replace $4=l-(l-4)$ and write the integral as
\be 
\label{amp12!}
i\mathcal{A}=\int \frac{d^4l}{(2\pi)^4}\frac{\langle q|l\circ (l-(l-4))\circ3\circ (l-4)|q\rangle\langle q|l+1|1]\langle q|l+1+2|2]}{l^2(l+1)^2(l+1+2)^2(l-4)^2\prod_{j=1}^4\langle qj\rangle\langle 43\rangle}
\\\nonumber=\frac{1}{2}\int \frac{d^4l}{(2\pi)^4}\frac{\langle q|3\circ (l-4)|q\rangle\langle q|l+1|1]\langle q|l+1+2|2]}{(l+1)^2(l+1+2)^2(l-4)^2\prod_{j=1}^4\langle qj\rangle\langle 43\rangle}\\\nonumber-\int \frac{d^4l}{(2\pi)^4}\frac{\langle q|l\circ(l-4)\circ3\circ(l-4)|q\rangle\langle q|l+1|1]\langle q|l+1+2|2]}{l^2(l+1)^2(l+1+2)^2(l-4)^2\prod_{j=1}^4\langle qj\rangle\langle 43\rangle}.
\ee
Next we use the spinor identity, 
\be
\label{id!}
A\circ B=-B\circ A+(A.B)\mathds{1}
\ee
for any arbitrary mixed spinors $A$ and $B$. We cancel denominators and write (\ref{amp12!}) as\\ 
\be 
\label{amp3}
2i\mathcal{A}=\int \frac{d^4l}{(2\pi)^4}\frac{\langle q|3\circ(l-4)|q\rangle\langle q|l+1|1]\langle q|l+1+2|2]}{(l+1)^2(l+1+2)^2(l-4)^2\prod_{j=1}^4\langle qj\rangle\langle 43\rangle}\\\nonumber+\int \frac{d^4l}{(2\pi)^4}\frac{\langle q|l\circ (l-4)|q\rangle\langle q|l+1|1]\langle q|l+1+2|2]}{l^2(l+1)^2(l-4)^2\prod_{j=1}^4\langle qj\rangle\langle 43\rangle}\\\nonumber+\int \frac{d^4l}{(2\pi)^4}\frac{\langle q|l\circ(l-3-4)|q\rangle\langle q|l+1|1]\langle q|l+1+2|2]}{l^2(l+1)^2(l+1+2)^2\prod_{j=1}^4\langle qj\rangle\langle 43\rangle}.
\ee
Eliminating one of the factors of $l$, we rewrite the three integrals 
\be
\begin{split}
\label{ammpli2}
2i\mathcal{A}_1&=\int \frac{d^4l}{(2\pi)^4}\frac{\langle q|3\circ(l-4)|q\rangle\langle q|l+1|1]\langle q|l+1+2|2]}{(l+1)^2(l+1+2)^2(l-4)^2\prod_{j=1}^4\langle qj\rangle\langle 43\rangle}\\\nonumber
2i\mathcal{A}_2&=\int \frac{d^4l}{(2\pi)^4}\frac{\langle q|l\circ (l-4)|q\rangle\langle q|l+1|1]\langle q|l+1+2|2]}{l^2(l+1)^2(l-4)^2\prod_{j=1}^4\langle qj\rangle\langle 43\rangle}\\\nonumber
2i\mathcal{A}_3&=\int \frac{d^4l}{(2\pi)^4}\frac{\langle q|l\circ(l-3-4)|q\rangle\langle q|l+1|1]\langle q|l+1+2|2]}{l^2(l+1)^2(l+1+2)^2\prod_{j=1}^4\langle qj\rangle\langle 43\rangle}
\end{split}
\ee
such that 
\be 
2i\mathcal{A}=2i\mathcal{A}_1+2i\mathcal{A}_2+2i\mathcal{A}
_3
\ee 
Next consider the first integral of (\ref{ammpli2}). We multiply the numerators and denominators by $\langle 23\rangle$. This yields
\be 
\label{amp13}
2i\mathcal{A}_{1}=\int \frac{d^4l}{(2\pi)^4}\frac{\langle q|l+1|1]\langle q|(l+1+2)\circ3\circ(l-4)|q\rangle}{(l+1)^2(l+1+2)^2(l-4)^2\langle 23\rangle\langle 43\rangle\langle q1\rangle\langle q2\rangle\langle q4\rangle}.
\ee
We use
\be 
3=-(l+1+2)+(l-4)\nonumber
\ee
and rewrite the integral as
\be 
\label{amplit1}
2i\mathcal{A}_{1}=\int \frac{d^4l}{(2\pi)^4}\frac{\langle q|l+1|1]\langle q|(l+1+2)\circ2\circ((l+1+2)-(l-4))\circ(l-4)|q\rangle}{(l+1)^2(l+1+2)^2(l-4)^2\langle 23\rangle\langle 43\rangle\langle q1\rangle\langle q2\rangle\langle q4\rangle}\\\nonumber=-\int \frac{d^4l}{(2\pi)^4}\frac{\langle q|l+1|1]\langle q|(l+1+2)\circ2\circ(l+1+2)\circ(l-4)|q\rangle}{(l+1)^2(l+1+2)^2(l-4)^2\langle 23\rangle\langle 43\rangle\langle q1\rangle\langle q2\rangle\langle q4\rangle}\\+\frac{1}{2}\nonumber\int \frac{d^4l}{(2\pi)^4}\frac{\langle q|l+1|1]\langle q|(l+1+2)\circ2|q\rangle}{(l+1)^2(l+1+2)^2\langle 23\rangle\langle 43\rangle\langle q1\rangle\langle q2\rangle\langle q4\rangle}.
\ee
Using (\ref{id}), we do a similar manipulation with the second integral of (\ref{amplit1}), giving
\be 
\label{amplit111}
4i\mathcal{A}_{1}=-\int \frac{d^4l}{(2\pi)^4}\frac{\langle q|l+|1]\langle q|(l+1+2)\circ(l-4)|q\rangle}{(l+1+2)^2(l-4)^2\langle 23\rangle\langle 43\rangle\langle q1\rangle\langle q2\rangle\langle q4\rangle}\\\nonumber+\int \frac{d^4l}{(2\pi)^4}\frac{\langle q|l+1|1]\langle q|(l+1)\circ(l-4)|q\rangle}{(l+1)^2(l-4)^2\langle 23\rangle\langle 43\rangle\langle q1\rangle\langle q2\rangle\langle q4\rangle}\\+\nonumber\int \frac{d^4l}{(2\pi)^4}\frac{\langle q|l+1|1]\langle q|(l+1+2)\circ2|q\rangle}{(l+1)^2(l+1+2)^2\langle 23\rangle\langle 43\rangle\langle q1\rangle\langle q2\rangle\langle q4\rangle}.
\ee
These integrals can be seen to vanish after we shift the loop momentum variable. Consider the first integral. We do a shift $l\rightarrow l-1$, which results into 
\be 
\int \frac{d^4l}{(2\pi)^4}\frac{\langle q|l|1]\langle q|l\circ (l-1-4)|q\rangle}{(l-1-4)^2l^2}=-\int \frac{d^4l}{(2\pi)^4}\frac{\langle q|l|1]\langle q|l\circ (1+4)|q\rangle}{(l-1-4)^2l^2};
\ee 
Clearly, the above integral can only depend on (1+4). This then vanishes due to the contraction of the auxiliary spinor variable. Let us then compute the shift. The linear part of the shift is given by 
\be 
-i\lim_{l\to\infty}\int \frac{d\Omega}{(2\pi)^4}1_{\mu}l^{\mu}\frac{\langle q|l+1|2]\langle q|(l+1)\circ(2+3)|3|q\rangle}{l^2}\Bigg(1+2\frac{(l.4)-(l.1)}{l^2}\Bigg).
\ee 
In the linear part of the shift which is quadratic in the loop momentum gives a vanishing result, while the quartic part gives $-(i/3.32\pi^2)\langle q|4|1]\langle q|4\circ 1|q\rangle$. Also, the quadratic part of the shift is zero. Then, the first integral gives a result for the shift 
\be 
-\frac{i}{3.32\pi^2}\frac{\langle q|4|1]\langle q|4\circ 1|q\rangle}{\langle 23\rangle\langle 43\rangle\langle q1\rangle\langle q2\rangle\langle q4\rangle}.
\ee 
Now consider the second integral of (\ref{amplit111}). If we shift the loop momentum $l\rightarrow l+4$, then we have the following loop integral 
\be 
\int \frac{d^4l}{(2\pi)^4}\frac{\langle q|l+4|1]\langle q|(l-3)\circ l|q\rangle}{(l-3)^2l^2}=-\int \frac{d^4l}{(2\pi)^4}\frac{\langle q|l+4|1]\langle q|3\circ l|q\rangle}{(l-3)^2l^2}.
\ee 
After the shift, the quadratically divergent part in the above integral vanishes due to spinor contraction $[33]=0$ and the linearly divergent part also vanishes similarly. Then, we compute the shift. We have the following integrand 
\be 
\frac{\langle q|l+1|1]\langle q|(l+1+2)\circ (l-4)|q\rangle}{(l+1+2)^2(l-4)^2}=-\frac{\langle q|l|1]\langle q|3\circ (l-4)|q\rangle}{(l+1+2)^2(l-4)^2}.
\ee 
We have for the linear part of the shift 
\be 
-i\lim_{l\to\infty}\int \frac{d\Omega}{(2\pi)^4}4_{\mu}l^{\mu}\frac{\langle q|l|1]\langle q|3\circ(l-4)|q\rangle}{l^2}\Bigg(1+2\frac{(l.4)+(l.3)}{l^2}\Bigg);
\ee 
The quadratic part of this shift gives $(i/32\pi^2)\langle q|4|1]\langle 3\circ 4|q\rangle$. The quartic part gives 
\be 
-\frac{i}{3.32\pi^2}(4\langle q|4|1]\langle q|3\circ 4|q\rangle + \langle q|3|1]\langle q|3\circ 4|q\rangle. 
\ee 
When we add the two contributions, we get 
\be 
-\frac{i}{3.32\pi^2}\langle q|4+3|1]\langle q|3\circ 4|q\rangle.
\ee 
The quadratic part of the shift is 
\be 
-\frac{i}{2}\lim_{l\to\infty}\int \frac{d\Omega}{(2\pi)^4}4_{\mu}4_{\nu}l^{\mu}l^2\frac{\partial}{\partial l_{\nu}}\frac{\langle q|l|1]\langle q|3\circ(l-4)|q\rangle}{(l+1+2)^2(l-4)^2}.
\ee 
When the derivative hits the denominator, we get 
\be 
2i\lim_{l\to\infty}\int \frac{d\Omega}{(2\pi)^4}4_{\mu}4_{\nu}l^{\mu}l^{\nu}\frac{\langle q|l|1]\langle q|3\circ l|q\rangle}{l^4}=\frac{2i}{3.32\pi^2}\langle q|4|1]\langle q|3\circ 4|q\rangle; 
\ee 
When the derivative hits the numerator, we get two equal contributions resulting in 
\be 
-\frac{i}{32\pi^2}\langle q|4|1]\langle q|3\circ 4|q\rangle.
\ee 
When we add the two contributions, we get for the quadratic part of the shift 
\be 
-\frac{i}{3.32\pi^2}\langle q|4|1]\langle 3\circ 4|q\rangle. 
\ee 
So overall we get for this shift 
\be 
-\frac{i}{3.32\pi^2}(2\langle q|4|1]+\langle q|3|1])\langle q|3\circ 4|q\rangle.
\ee 
We have checked this result by computing the $l\rightarrow l-1-2$ shift instead. This can be further simplified using momentum conservation. Writing $2k_4+k_3=k_4-(k_1+k_2)$ shows that we can rewrite this result as 
\be 
-\frac{i}{3.32\pi^2}\langle q|4-2|1]\langle 3\circ 4|q\rangle. 
\ee 
The integral is given by minus the shift. Thus we get the following result for the second term 
\be 
\frac{i}{3.32\pi^2}\frac{\langle q|4-2|1]\langle q|3\circ 4|q\rangle}{\langle 23\rangle\langle 43\rangle \langle q1\rangle\langle q2\rangle\langle q4\rangle}.
\ee 
We now consider the last term in (\ref{amplit111}). We do a shift $l\rightarrow l-1$. The relevant integrand is \be 
\frac{\langle q|l|1]\langle q|(l+1)\circ 2|q\rangle}{(l+1)^2(l+1+2)^2}. 
\ee 
The linear part of the shift is 
\be 
-i\lim_{l\to\infty}\int \frac{d\Omega}{(2\pi)^4}1_{\mu}l^{\mu}\frac{\langle q|l|1]\langle q|(l+1)\circ2|q\rangle}{l^2}\Bigg(1-2\frac{(l.1)+(l.2)}{l^2}\Bigg).
\ee 
The quadratic part gives no contribution, while the quartic part gives $(i/3.32\pi^2)\langle q|2|1]\langle q|1\circ 2|q\rangle$. There is no contribution from the quadratic part of the shift. The full answer for the last term in (\ref{amplit111}) is then 
\be 
-\frac{i}{3.32\pi^2}\frac{\langle q|2|1]\langle q|1\circ 2|q\rangle}{\langle 23\rangle\langle 43\rangle\langle q1\rangle\langle q2\rangle\langle q4\rangle}. 
\ee 
When we add all the three contributions, we get 
\be 
\mathcal{A}_1=-\frac{1}{12.32\pi^2}\frac{\langle q|4|1]\langle q|4\circ 1|q\rangle+\langle q|2-4|1]\langle q|3\circ 4|q\rangle+\langle q|2|1]\langle q|1\circ 2|q\rangle}{\langle 23\rangle\langle 43\rangle\langle q1\rangle\langle q2\rangle\langle q4\rangle}. 
\ee 
The result for $\mathcal{A}_1$ is quite long and it is desirable to simplify it. To this end, we multiply the numerator and denominator with $\langle 1q\rangle$, which converts the numerator into 
\be 
\langle q|4\circ 1|q\rangle\langle q|4\circ 1|q\rangle+\langle q|(2-4)\circ 1|q\rangle\langle q|3\circ 4|q\rangle + \langle q|2\circ 1|q\rangle\langle q|1\circ 2|q\rangle.
\ee 
We can simplify this using momentum conservation. We have relations of the type 
\be 
1\circ 2+1\circ3+1\circ4=-4\circ4=-\frac{1}{2}4^2=0.
\ee 
There are four such relations among six different momentum products. This makes it possible to choose any three of them as independent three of them as independent and write the other three in terms of the basis chosen. Let us choose as independent the products $1\circ 2, 1\circ 4, 2\circ 3$. The choice is motivated by the fact that $1\circ 2, 1\circ 4$ appears in $\mathcal{A}_1$. We then get 
\be 
3\circ 4=-1\circ 2-1\circ 4+2\circ 3. 
\ee 
Substituting this into the numerator of the amplitude we notice that there are multiple cancellations, and the result is 
\be 
\mathcal{A}_1=\frac{1}{12.32\pi^2}\frac{\langle q|1\circ (4-2)|q\rangle\langle 2\circ 3|q\rangle}{\langle 23\rangle\langle 43\rangle\langle q1\rangle^2\langle q2\rangle\langle q4\rangle}. 
\ee 

Next we consider $\mathcal{A}_2$. Multiplying numerator and denominator by $\langle 41\rangle$ yields
\be 
\label{amp234}
2i\mathcal{A}_{2}=-\int \frac{d^4l}{(2\pi)^4}\frac{\langle q|l+1+2|2]\langle q|l\circ4\circ1\circ(l+1)|q\rangle}{l^2(l+1)^2(l-4)^2\langle 41\rangle\langle 43\rangle\langle q1\rangle\langle q2\rangle\langle q3\rangle}.
\ee
Using $1=l+1-l$ we write the integral as
\be 
\label{amp2311}
2i\mathcal{A}_{2}=\int \frac{d^4l}{(2\pi)^4}\frac{\langle q|l+1+2|2]\langle q|l\circ4\circ(l-(l+1))\circ(l+1)|q\rangle}{l^2(l+1)^2(l-4)^2\langle 41\rangle\langle 43\rangle\langle q1\rangle\langle q2\rangle\langle q3\rangle}\\\nonumber=-\int \frac{d^4l}{(2\pi)^4}\frac{\langle q|l+1|2]\langle q|l\circ4\circ l\circ(l+1)|q\rangle}{l^2(l+1)^2(l-4)^2\langle 41\rangle\langle 43\rangle\langle q1\rangle\langle q2\rangle\langle q3\rangle}\\\nonumber+\frac{1}{2}\int \frac{d^4l}{(2\pi)^4}\frac{\langle q|l+1|2]\langle q|l\circ4|q\rangle}{l^2(l-4)^2\langle 41\rangle\langle 43\rangle\langle q1\rangle\langle q2\rangle\langle q3\rangle}.
\ee
Using (\ref{id!}), we do a similar manipulation with the second integral of (\ref{amp2311}), giving
\be 
\label{amp2322}
4i\mathcal{A}_{2}=\int \frac{d^4l}{(2\pi)^4}\frac{\langle q|l+1|2]\langle q|l\circ1|q\rangle}{l^2(l+1)^2\langle 41\rangle\langle 43\rangle\langle q1\rangle\langle q2\rangle\langle q3\rangle}\\\nonumber-\int \frac{d^4l}{(2\pi)^4}\frac{\langle q|l+1|2]\langle q|(l-4)\circ(l+1)|q\rangle}{(l+1)^2(l-4)^2\langle 41\rangle\langle 43\rangle\langle q1\rangle\langle q2\rangle\langle q3\rangle}\\\nonumber+\int \frac{d^4l}{(2\pi)^4}\frac{\langle q|l+1|2]\langle q|l\circ4|q\rangle}{l^2(l-4)^2\langle 41\rangle\langle 43\rangle\langle q1\rangle\langle q2\rangle\langle q3\rangle}.
\ee
The first and last integrals can only depend on the momenta $1$ and $4$ respectively and then vanish as causing a $\langle qq\rangle=0$ contraction.\\~\\
The second integral vanishes after a shift. First, using $l+1=l-4-2-3$ we rewrite it as 
\be 
\int \frac{d^4l}{(2\pi)^4}\frac{\langle q|l+1|2]\langle q|(l-4)\circ (2+3)|q\rangle}{(l+1)^2(l-4)^2\langle 41\rangle\langle 43\rangle\langle q1\rangle\langle q2\rangle\langle q3\rangle}, 
\ee 
to make it clear that it is at most quadratically divergent. A reasonably good shift is then $l\rightarrow l-1$ which renders the integral 
\be 
\int \frac{d^4l}{(2\pi)^4}\frac{\langle q|l|2]\langle q|l\circ (2+3)|q\rangle}{l^2(l+2+3)^2\langle 41\rangle\langle 43\rangle\langle q1\rangle\langle q2\rangle\langle q3\rangle}.
\ee 
This vanishes by the already familiar argument. 
\\
We thus need to compute the shift. The relevant integrand is 
\be 
\frac{\langle q|l+1|2]\langle q|(l-4)\circ(2+3)|q\rangle}{(l+1)^2(l-4)^2}.
\ee 
The linear part of the shift is 
\be 
-i\lim_{l\to\infty}\int \frac{d\Omega}{(2\pi)^4}1_{\mu}l^{\mu}\frac{\langle q|l+1|2]\langle q|(l-4)\circ(2+3)|q\rangle}{l^2}\Bigg(1-2\frac{(l.4)-(l.1)}{l^2}\Bigg).
\ee 
The quadratic part in $l$ gives the following two terms 
\be 
-\frac{i}{32\pi^2}(-\langle q|1|2]\langle q|4\circ(2+3)|q\rangle+\langle q|1|2]\langle q|1\circ (2+3)|q\rangle)\nonumber\\
=-\frac{i}{32\pi^2}\langle q|1|2]\langle q|(1-4)\circ (2+3)|q\rangle.
\ee 
We now use $(2+3)=-(1+4)$ to rewrite the above as 
\be 
\frac{2i}{32\pi^2}\langle q|1|2]\langle q|1\circ 4|q\rangle.
\ee 
Next, the quartic part in $l$ gives, after simplification 
\be 
-\frac{i}{3.32\pi^2}(3\langle q|1|2]-\langle q|4|2])\langle q|1\circ 4|q\rangle.
\ee 
Therefore the complete linear part of the shift is given by 
\be 
\frac{i}{32\pi^2}\langle q|1|2]\langle q|1\circ 4|q\rangle +\frac{i}{3.32\pi^2}\langle q|4|2]\langle q|1\circ 4|q\rangle.
\ee 
The quadratic part of the shift is 
\be 
\frac{i}{2}\lim_{l\to\infty}\int \frac{d\Omega}{(2\pi)^4}1_{\mu}1_{\nu}l^{\mu}l^2\frac{\partial}{\partial l_{\nu}}\frac{\langle q|l+1|2]\langle q|(l-4)\circ (2+3)|q\rangle}{(l+1)^2(l-4)^2};
\ee 
When the derivative hits the denominator we get 
\be 
-\frac{2i}{3.32\pi^2}\langle q|1|2]\langle q|1\circ (2+3)|q\rangle. 
\ee 
When the derivative hits the numerator we get two terms, which are however equal with the result 
\be 
\frac{i}{32\pi^2}\langle q|1|2]\langle q|1\circ (2+3)|q\rangle. 
\ee 
Thus we get for the quadratic part of the shift 
\be 
\frac{i}{3.32\pi^2}\langle q|1|2]\langle q|1\circ (2+3)|q\rangle=-\frac{i}{3.32\pi^2}\langle q|1|2]\langle q|1\circ 4|q\rangle. 
\ee 
We then combine the quadratic and linear parts of the shift, obtaining the full shift 
\be 
\frac{i}{3.32\pi^2}(2\langle q|1|2]+\langle q|4|2])\langle q|1\circ 4|q\rangle =\frac{i}{32\pi^2}\langle q|1-3|2]\langle q|1\circ 4|q\rangle, 
\ee 
where we wrote $1+4=-2-3$. The integral is minus the shift. This therefore gives us the result for $\mathcal{A}_2$ 
\be 
\mathcal{A}_2=\frac{1}{12.32\pi^2}\frac{\langle q|(3-1)\circ 2|q\rangle\langle q|1\circ 4|q\rangle}{\langle 14\rangle\langle 43\rangle\langle q1\rangle\langle q2\rangle^2\langle q3\rangle},
\ee 
where we wrote the result by multiplying the numerator and denominator by $\langle 2q\rangle$
On similar lines, we compute the third integral in (\ref{ammpli2}). Multiplying numerator and denominator by $\langle 12\rangle$ yields
\be 
\label{amp33!}
i\mathcal{A}_{3}=\int \frac{d^4l}{(2\pi)^4}\frac{\langle q|l+1|2]\langle q|l+1|2|1|l|q\rangle}{l^2(l+1)^2(l+1+2)^2\langle 12\rangle\langle 34\rangle\langle q1\rangle\langle q2\rangle\langle q3\rangle\langle q4\rangle}.
\ee
Using $1=(l-1)-l$ we rewrite the integral as
\be 
\label{amp331!}
i\mathcal{A}_{3}=\int \frac{d^4l}{(2\pi)^4}\frac{\langle q|l|3+4|q\rangle\langle q|l+1|2|l+1-l|l|q\rangle}{l^2(l+1)^2(l+1+2)^2\langle 12\rangle\langle 34\rangle\langle q1\rangle\langle q2\rangle\langle q3\rangle\langle q4\rangle}\\\nonumber=\int \frac{d^4l}{(2\pi)^4}\frac{\langle q|l|3+4|q\rangle\langle q|l+1|2|l+1|l|q\rangle}{l^2(l+1)^2(l+1+2)^2\langle 12\rangle\langle 34\rangle\langle q1\rangle\langle q2\rangle\langle q3\rangle\langle q4\rangle}\\-\nonumber\int \frac{d^4l}{(2\pi)^4}\frac{\langle q|l|3+4|q\rangle\langle q|l+1|2|q\rangle}{(l+1)^2(l+1+2)^2\langle 12\rangle\langle 34\rangle\langle q1\rangle\langle q2\rangle\langle q3\rangle\langle q4\rangle}.
\ee
Using (\ref{id!}), we do a similar manipulation with the second integral of (\ref{amp331!}), giving
\be 
\label{amp332!}
4i\mathcal{A}_{3}=\int \frac{d^4l}{(2\pi)^4}\frac{\langle q|l\circ(3+4)|q\rangle\langle q|(1+2)\circ l|q\rangle}{l^2(l+1+2)^2\langle 12\rangle\langle 43\rangle\langle q1\rangle\langle q2\rangle\langle q3\rangle\langle q4\rangle}\\\nonumber+\int \frac{d^4l}{(2\pi)^4}\frac{\langle q|l\circ(3+4)|q\rangle\langle q|(l+1)\circ 1|q\rangle}{l^2(l+1)^2\langle 12\rangle\langle 43\rangle\langle q1\rangle\langle q2\rangle\langle q3\rangle\langle q4\rangle}\\+\nonumber\int \frac{d^4l}{(2\pi)^4}\frac{\langle q|l\circ(3+4)|q\rangle\langle q|(l+1)\circ2|q\rangle}{(l+1)^2(l+1+2)^2\langle 12\rangle\langle 43\rangle\langle q1\rangle\langle q2\rangle\langle q3\rangle\langle q4\rangle}.
\ee
We now follow the similar set of arguments like the ones we used earlier to argue that some of the terms are zero. Indeed the first integral with $l_{\mu}l_{\nu}$ in the numerator can only give $\eta_{\mu\nu}$, which causes $\langle qq\rangle=0$ contraction or $1_{\mu}1_{\nu}$ which produces $1^2=0$. In the second term, similarly $l_{\mu}l_{\nu}$ in the numerator can only give $(1+2)_{\mu}(1+2)_{\nu}$. This gives a $\langle qq\rangle=0$ contraction. In the third term, however, we argue that the integral is zero only after a shift. The relevant shifts which does the job are $l\rightarrow l-1-2$ and $l\rightarrow l-1$. Consider the shift $l\rightarrow l-1-2$. This converts the integral to 
\be 
\int \frac{d^4l}{(2\pi)^4}\frac{\langle q|(l-1-2)\circ (3+4)|q\rangle\langle q|(l-2)\circ 2|q\rangle}{l^2(l-2)^2}.
\ee 
We now use $-1-2=3+4$ to see that the above integral equals 
\be 
\int \frac{d^4l}{(2\pi)^4}\frac{\langle q|l\circ (3+4)|q\rangle\langle q|l\circ 2|q\rangle}{l^2(l-2)^2}.
\ee 
Then the argument as before shows that this integral should be proportional to $2^2=0$. Therefore the part $\mathcal{A}_3$ reduces to a shift. Let us compute the shift. The linear part of the shift is given by 
\be 
-i\lim_{l\to\infty}\int \frac{d\Omega}{(2\pi)^4}(1+2)_{\mu}l^{\mu}\frac{\langle q|(l+1)\circ2|q\rangle\langle q|l\circ(3+4)|q\rangle}{l^2}\Bigg(1-2\frac{(l.1)+(l.2)}{l^2}\Bigg).
\ee 
We can see that there is no contribution coming from the quadratic in $l$ part of the numerator. This is because this part would give a factor $\langle q|(1+2)\circ (3+4)|q\rangle$ which would vanish owing to $\langle qq\rangle=0$ contraction. Then, from the quartic in $l$ part, we get the contribution 
\be 
\label{totalr}
-\frac{i}{3.32\pi^2}\langle q|1\circ2|q\rangle^2. 
\ee 
The quadratic part of the shift is 
\be 
\frac{i}{2}\lim_{l\to\infty}\int \frac{d\Omega}{(2\pi)^4}(1+2)_{\mu}l^{\mu}l^2(1+2)_{\nu}\frac{\partial}{\partial l_{\nu}}\frac{\langle q|l\circ 3+4|q\rangle\langle q|(l+1)\circ 2|q\rangle}{(l+1)^2(l+1+2)^2}.
\ee 
When the derivative hits the denominator we get, in the large $l$ limit 
\be 
-2i\lim_{l\to\infty}\int \frac{d\Omega}{(2\pi)^4}(1+2)_{\mu}l^{\mu}(1+2)_{\nu}l_{\nu}\frac{\langle q|l\circ 3+4|q\rangle\langle q|(l+1)\circ 2|q\rangle}{l^4}.
\ee 
which can be seen to be zero using $\langle q|(1+2)\circ(3+4)|q\rangle=0$. When the derivative hits the numerator it must hit the second term, but then the integration causes $(1+2)\circ(3+4)$ contraction again, rendering $\langle qq\rangle=0$. So we do not get any quadratic part of the shift and the total result is given by (\ref{totalr}).\\~\\
We can check this result by doing the second shift mentioned earlier, $l\rightarrow l-1$. This converts the integral to 
\be 
\int \frac{d^4l}{(2\pi)^4}\frac{\langle q|(l-1)\circ (3+4)|q\rangle\langle q|l\circ 2|q\rangle}{l^2(l+2)^2}.
\ee 
Once again, we use the same argument as before, which then shows that the integral is proportional to $2^2=0$. Thus the part $\mathcal{A}_3$ reduces to a shift. We now compute the shift. The linear part of the shift is given by 
\be 
-i\lim_{l\to\infty}\int \frac{d\Omega}{(2\pi)^4}1_{\mu}l^{\mu}\frac{\langle q|(l+1)\circ2|q\rangle\langle q|l\circ(3+4)|q\rangle}{l^2}\Bigg(1-2\frac{(l.1)+(l.2)}{l^2}\Bigg).
\ee 
We see that all the terms are zero for the linear part of the shift. Next, for the quadratic part of the shift, we have 
\be 
\frac{i}{2}\lim_{l\to\infty}\int \frac{d\Omega}{(2\pi)^4}1_{\mu}l^{\mu}l^21_{\nu}\frac{\partial}{\partial l_{\nu}}\frac{\langle q|l\circ (3+4)|q\rangle\langle q|(l+1)\circ 2|q\rangle}{(l+1)^2(l+1+2)^2}.
\ee 
When the derivative hits the denominator, we get in the large $l$ limit 
\be 
-2i\lim_{l\to\infty}\int \frac{d\Omega}{(2\pi)^4}1_{\mu}l^{\mu}1_{\nu}l_{\nu}\frac{\langle q|l\circ (3+4)|q\rangle\langle q|(l+1)\circ 2|q\rangle}{l^4}
\ee  
which gives $2i/3.32\pi^2\langle q|1\circ 2|q\rangle^2$. Adding everything up we get the same result for the shift like before. This establishes that the two shifts are consistent with each other. \\~\\ 
The result for $\mathcal{A}_3$ is given by minus the shift. Taking other factors into consideration, we get the result 
\be 
\mathcal{A}_3=\frac{1}{12.32\pi^2}\frac{\langle q1\rangle\langle q2\rangle}{\langle q3\rangle\langle q4\rangle}\frac{[12]^2}{\langle 12\rangle\langle 43\rangle}. 
\ee 
\subsubsection{Collecting the results} 
Let us collect the results for $\mathcal{A}_{1,2,3}$ for a single box diagram. We get 
\be 
\begin{split} 
12.32\pi^2\mathcal{A}&=-\frac{[12][23]}{\langle 23\rangle\langle 43\rangle}\frac{\langle q2\rangle\langle q3\rangle}{\langle q1\rangle\langle q4\rangle}+\frac{[14][23]}{\langle 23\rangle\langle 43\rangle}\frac{\langle q3\rangle}{\langle q1\rangle}
\\
&-\frac{[12][14]}{\langle 14\rangle\langle 43\rangle}\frac{\langle q1\rangle\langle q4\rangle}{\langle q2\rangle\langle q3\rangle}-\frac{[14][23]}{\langle 14\rangle\langle 43\rangle}\frac{\langle q4\rangle}{\langle q2\rangle}+\frac{[12]^2}{\langle 12\rangle\langle 43\rangle}\frac{\langle q1\rangle\langle q2\rangle}{\langle q3\rangle\langle q4\rangle}.
\end{split} 
\ee 
We note that the replacement $1\leftrightarrow2$ gives the same result as $3\leftrightarrow 4$. This is because the amplitude is symmetric under the simultaneous transformation $1\leftrightarrow 2$ and $3\leftrightarrow 4$. The amplitude above is not manifestly invariant under the simultaneous $2\leftrightarrow 4$ and $3\leftrightarrow 1$exchange. To rectify this, we first rewrite the two terms that have single powers of $q$in the numerator and denominator. We have 
\be 
\begin{split} 
&\frac{[14][23]}{\langle 23\rangle\langle 43\rangle}\frac{\langle q3\rangle}{\langle q1\rangle}-\frac{[14][23]}{\langle 14\rangle\langle 43\rangle}\frac{\langle q4\rangle}{\langle q2\rangle} \\
&=\frac{[14][23]}{\langle 14\rangle\langle 23\rangle\langle 43\rangle\langle q1\rangle\langle q2\rangle}(\langle 14\rangle\langle q2\rangle\langle q3\rangle-\langle 23\rangle\langle q1\rangle\langle q4\rangle).
\end{split} 
\ee 
Next we use the Schouten identities for the terms inside the bracket
\be 
\langle q4\rangle\langle 23\rangle=\langle q2\rangle\langle 43\rangle -\langle q3\rangle \langle 42\rangle, ~~~~~~\langle q2\rangle\langle 14\rangle=\langle q1\rangle\langle 24\rangle -\langle q4\rangle \langle 21\rangle.
\ee 
This gives 
\be 
\langle 14\rangle\langle q2\rangle\langle q3\rangle-\langle 23\rangle\langle q1\rangle\langle q4\rangle=\langle q1\rangle\langle q2\rangle\langle 34\rangle+ \langle q3\rangle\langle q4\rangle\langle 12\rangle. 
\ee 
This means that the amplitude can be written as 
\be 
\begin{split} 
12.32\pi^2\mathcal{A}&=\frac{[12][23]}{\langle 23\rangle\langle 43\rangle}\frac{\langle q2\rangle\langle q3\rangle}{\langle q1\rangle\langle q4\rangle}-\frac{[34]^2}{\langle 12\rangle\langle 34\rangle}\frac{\langle q3\rangle\langle q4\rangle}{\langle q1\rangle\langle q2\rangle}
\\
&\frac{[12][14]}{\langle 14\rangle\langle 43\rangle}\frac{\langle q1\rangle\langle q4\rangle}{\langle q2\rangle\langle q3\rangle}-\frac{[12]^2}{\langle 12\rangle\langle 43\rangle}\frac{\langle q1\rangle\langle q2\rangle}{\langle q3\rangle\langle q4\rangle}-\frac{[14][23]}{\langle 14\rangle\langle 23\rangle}.
\end{split} 
\ee 
where we also used $[23]=[34]\langle 14\rangle/\langle 12\rangle$ and $[14]-[34]\langle 23\rangle/\langle 12\rangle$ in the second term on the second line. Note that the last term is $q$-independent and thus gauge invariant. \\~\\
The invariances are now manifest. The gauge invariant term is manifestly invariant under $1\leftrightarrow 2$ and $3\leftrightarrow 4$ as well as $2\leftrightarrow 4$ and $3\leftrightarrow 1$ symmetry. The terms in the first line go into one another under both of these symmetries. So there are also invariant. The first two terms in the second line are invariant under the exchange $1\leftrightarrow 2$ and $3\leftrightarrow 4$  and go into each under the exchanges $2\leftrightarrow 4$ and $3\leftrightarrow 1$. 
\\~\\
Next there are ghost contribution to the amplitude. With the Feynman rules, we construct the ghost box diagram, with ghosts running in the loop and the external legs are projected to negative helicity gluons. The contribution is given by 
\vspace{1em}
\be
\begin{fmfgraph*}(140,90)
\fmftop{i1,i2}
\fmfbottom{o1,o2}
\fmf{photon}{i1,v1}
\fmf{photon}{i2,v2}
\fmf{dots,label=$l$}{v2,v1}
\fmf{photon}{o1,v3}
\fmf{photon}{o2,v4}
\fmf{dots,label=$l+1+2$}{v4,v3}
\fmf{dots,label=$l-4$}{v4,v2}
\fmf{dots,label=$l+1$}{v1,v3}
\fmflabel{$1^-$}{i1}
\fmflabel{$4^-$}{i2}
\fmflabel{$2^-$}{o1}
\fmflabel{$3^-$}{o2}
\end{fmfgraph*}\nonumber
\ee
Upto numerical and colour factors, this is 
\be 
\label{amp}
i\mathcal{A}_{ghost-box}=\int \frac{d^4l}{(2\pi)^4}\frac{\langle q|l|4]\langle q|l+1|1]\langle q|l+1+2|2]\langle q|l-4|3]}{l^2(l+1)^2(l+1+2)^2(l-4)^2\prod_{j=1}^4\langle qj\rangle}.
\ee
This integral is exactly the same as the original box diagram and therefore is a multiple of the previous  result.
\subsubsection{Triangles} 
Consider the triangle diagrams in SDYM. We evaluate one of them and then permute the legs to obtain others. We omit any numerical and colour factors arising since the colour decomposition allows us to just compute and add the colour-stripped part of the diagrams, which should be gauge invariant.
\be 
\begin{fmfgraph*}(140,90)
\fmfbottom{i1,i2}
\fmftop{o1,o2}
\fmf{photon,tension=3}{i1,v1,o1}
\fmf{photon,tension=3}{i2,v3}
\fmf{photon,tension=3}{o2,v4}
\fmf{vanilla}{v1,v1'}
\fmf{photon}{v1',v2}
\fmf{photon,label=$l$}{v2',v2}
\fmf{vanilla}{v2',v4}
\fmf{photon,label=$l-4$}{v4',v4}
\fmf{vanilla}{v4',v3}
\fmf{photon,label=$l+1+2$}{v3',v3}
\fmf{vanilla}{v3',v2}
\fmflabel{$1^-$}{o1}
\fmflabel{$4^-$}{o2}
\fmflabel{$2^-$}{i1}
\fmflabel{$3^-$}{i2}
\fmflabel{}{v1'}
\fmflabel{}{v2'}
\fmflabel{}{v3'}
\fmflabel{}{v4'}
\end{fmfgraph*}\nonumber 
\ee
Using the Feynman rules, this diagram takes the particular form \be 
i\mathcal{A}_{triangle}=\frac{[12]}{[12]\langle 12\rangle}\int \frac{d^4l}{(2\pi)^4}\frac{\langle q|l|4]\langle q|l-4|3]\langle q|(l+1+2)\circ(1+2)|q\rangle}{l^2(l-4)^2(l+1+2)^2\prod_{j=1}^4\langle qj\rangle}.
\ee 
Let us start by multiplying the numerator and denominator by $\langle 34\rangle$. This allows to write the integral as 
\be 
i\mathcal{A}_{triangle}=-\frac{[12]}{[12]\langle 12\rangle\langle 34\rangle}\int \frac{d^4l}{(2\pi)^4}\frac{\langle q|l|4\circ 3\circ(l-4)|q\rangle\langle q|l|1+2|q\rangle}{l^2(l-4)^2(l+1+2)^2\prod_{j=1}^4\langle qj\rangle},
\ee 
where we used $\langle q|(1+2)\circ(1+2)|q\rangle=0$ to get rid of $(1+2)$ from the last factor in the numerator. 
We now replace $4=l-(l-4)$ and use the identity $l\circ l=\frac{1}{2}l^2\mathds{1}$ to write the integral as
\be 
i\mathcal{A}_{triangle}=-\frac{[12]}{[12]\langle 12\rangle\langle 34\rangle}\Bigg[-\frac{1}{2}\int \frac{d^4l}{(2\pi)^4}\frac{\langle q|(l-4)\circ 3|q\rangle\langle q|l|1+2|q\rangle}{(l-4)^2(l+1+2)^2\prod_{j=1}^4\langle qj\rangle}\nonumber\\-\int \frac{d^4l}{(2\pi)^4}\frac{\langle q|l\circ(l-4)\circ 3\circ(l-4)|q\rangle\langle q|l|1+2|q\rangle}{l^2(l-4)^2(l+1+2)^2\prod_{j=1}^4\langle qj\rangle}\Bigg].
\ee 
Next we use the spinor identity 
\be 
A\circ B=-B\circ A+(A.B)\mathds{1},
\ee 
where $(A.B)$ is the metric pairing. This identity holds for any two arbitrary rank two mixed spinors $A$ and $B$. Using this, we have 
\be 
\label{id}
(l-4)\circ 3\circ(l-4)=-(l-4)\circ(l-4)\circ3+(l-4)((l-4).3)\nonumber\\=-\frac{1}{2}(l-4)^2 3+(l-4)((l-4).3).
\ee 
The second term in the last line of (\ref{id}) can be written as 
\be 
(l-4).3=\frac{1}{2}((l-4)^2-(l-4-3)^2)
\ee 
Therefore, this gives us 
\be 
(l-4)\circ 3\circ(l-4)=\frac{1}{2}(l-4)^2 (l-4-3)-\frac{1}{2}(l-4-3)^2(l-4).
\ee 
We now replace $l-3-4=l+1+2$ and cancel denominators to get 
\be 
i\mathcal{A}_{triangle}=\frac{[12]}{[12]\langle 12\rangle\langle 34\rangle}\Bigg[\frac{1}{2}\int \frac{d^4l}{(2\pi)^4}\frac{\langle q|(l-4)\circ 3|q\rangle\langle q|l|1+2|q\rangle}{(l-4)^2(l+1+2)^2\prod_{j=1}^4\langle qj\rangle}\nonumber\\+\frac{1}{2}\int \frac{d^4l}{(2\pi)^4}\frac{\langle q|l\circ(l+1+2)|q\rangle\langle q|l|1+2|q\rangle}{l^2(l+1+2)^2\prod_{j=1}^4\langle qj\rangle}\nonumber\\-\frac{1}{2}\int \frac{d^4l}{(2\pi)^4}\frac{\langle q|l\circ(l-4)|q\rangle\langle q|l|1+2|q\rangle}{l^2(l-4)^2\prod_{j=1}^4\langle qj\rangle}\Bigg].
\ee 
The first integral is quadratically divergent while the second and third are cubic divergent. However, one of the factors of $l$ in the second and third integrals can be immediately eliminated owing to the identity $\langle q|l\circ l|q\rangle=0$. Thus we have for this diagram 
\be 
\label{tri}
i\mathcal{A}_{triangle}=\frac{[12]}{[12]\langle 12\rangle\langle 34\rangle}\Bigg[\frac{1}{2}\int \frac{d^4l}{(2\pi)^4}\frac{\langle q|(l-4)\circ 3|q\rangle\langle q|l|1+2|q\rangle}{(l-4)^2(l+1+2)^2\prod_{j=1}^4\langle qj\rangle}\nonumber\\+\frac{1}{2}\int \frac{d^4l}{(2\pi)^4}\frac{\langle q|l\circ(1+2)|q\rangle\langle q|l\circ(1+2)|q\rangle}{l^2(l+1+2)^2\prod_{j=1}^4\langle qj\rangle}\nonumber\\+\frac{1}{2}\int \frac{d^4l}{(2\pi)^4}\frac{\langle q|l\circ4|q\rangle\langle q|l|1+2|q\rangle}{l^2(l-4)^2\prod_{j=1}^4\langle qj\rangle}\Bigg].
\ee 
Let us consider the second integral. In any Lorentz invariant regularization, the numerator $l_{\mu}l_{\nu}$ can be either proportional to $\eta_{\mu\nu}$ in which case, there will be $\langle qq\rangle=0$ contraction, making the integral vanish or it can be proportional to $(1+2)_{\mu}(1+2)_{\nu}$ which yields the numerator to be $(\langle q|(1+2)\circ(1+2)|q\rangle)^2$. This again vanishes due to $\langle qq\rangle$ contraction. Thus the integral does not contribute. A similar argument applies to the third integral of (\ref{tri}) rendering it to vanish due to $\langle qq\rangle=0$ contraction. We then consider the first integral and argue that it vanishes under a shift. The shift is given by $l\rightarrow l-1-2$. This converts the integral to   
\be 
\label{tri2}
\int \frac{d^4l}{(2\pi)^4}\frac{\langle q|l\circ 3|q\rangle\langle q|l|1+2|q\rangle}{l^2(l+3)^2\prod_{j=1}^4\langle qj\rangle}.
\ee 
Again, in any Lorentz invariant regularization, the $l_{\mu}l_{\nu}$ part of the numerator can either be proportional to $\eta_{\mu\nu}$, in which case the $q$s contract, making it vanish or it can be proportional to $3_{\mu}3_{\nu}$ rendering $[33]=0$. Therefore, this integral vanishes under this shift. We then compute the shift.\\~\\ 
The linear part of the shift is given by 
\be 
i\lim_{l\to\infty}\int\frac{d\Omega}{(2\pi)^4}(1+2)_{\mu}l^{\mu}\frac{\langle q|(l-4)\circ 3|q\rangle\langle q|l\circ(1+2)|q\rangle}{l^2}\nonumber\\\times\Bigg(1-\frac{2(l.4)-l.(1+2)}{l^2}\Bigg).
\ee
The non-zero contribution can only come from the quadratic and quartic in $l$ terms of this shift, after taking an average over all directions. The quadratic term is 
\be 
-(1+2)_{\mu}l^{\mu}\langle q|4\circ 3|q\rangle\langle q|l\circ(1+2)|q\rangle.
\ee 
This vanishes owing to $\langle q| (1+2)\circ(1+2)|q\rangle=0$. \\~\\There are two quartic terms. One of the terms is 
\be 
2i\lim_{l\to\infty}\int\frac{d\Omega}{(2\pi)^4}(1+2)_{\mu}l^{\mu}(1+2)_{\nu}l^{\nu}\frac{\langle q|l\circ 3|q\rangle\langle q|l\circ(1+2)|q\rangle}{l^4}.
\ee 
The integral is computed using 
\be
\label{quart}
\int\frac{d\Omega}{(2\pi)^4}\frac{l_{\mu}l_{\nu}l_{\rho}l_{\sigma}}{l^4}=\frac{1}{32.6\pi^2}(\eta_{\mu\nu}\eta_{\rho\sigma}+\eta_{\mu\rho}\eta_{\nu\sigma}+\eta_{\mu\sigma}\eta_{\rho\nu}).
\ee 
All possible contractions of $l$ vanish and hence this part does not contribute. The other quartic term is given by 
\be 
-2i\lim_{l\to\infty}\int\frac{d\Omega}{(2\pi)^4}(1+2)_{\mu}l^{\mu}4_{\nu}l^{\nu}\frac{\langle q|l\circ 3|q\rangle\langle q|l\circ(1+2)|q\rangle}{l^4}.
\ee 
Using (\ref{quart}), we find only one of the contractions is non-zero. Thus the linear part of the shift is given by
\be 
-\frac{i}{32.3\pi^2}(\langle q||3|4|q\rangle)^2.
\ee 
For the quadratic part of the shift, the integral is given by 
\be 
\label{quadratic}
\lim_{l\to\infty}\int\frac{d\Omega}{(2\pi)^4}(1+2)_{\mu}(1+2)_{\nu}l^{\mu}\frac{\partial}{\partial l_{\nu}}\frac{\langle q|(l-4)\circ 3|q\rangle\langle q|l\circ(1+2)|q\rangle}{(l-4)^2(l+1+2)^2}.
\ee 
When the derivative hits the denominator, it produces a factor proportional to 
\be 
\lim_{l\to\infty}\int\frac{d\Omega}{(2\pi)^4}(1+2)_{\mu}(1+2)_{\nu}l^{\mu}l^{\nu}\frac{\langle q|(l-4)\circ 3|q\rangle\langle q|l\circ(1+2)|q\rangle}{l^4}.
\ee 
Note that only the quartic in $l$ part of the numerator can contribute after averaging over all directions. Using (\ref{quart}) we find that all possible contractions lead to $\langle qq\rangle=0$. Hence this part vanishes. When the derivative hits the numerator of (\ref{quadratic}), it can hit either of the two factors. When it hits $\langle q|(l-4)\circ 3|q\rangle$, the other factor contracts with $l_\mu$ giving $\langle q|(1+2)\circ(1+2)|q\rangle$, which vanishes because of $\langle qq\rangle$ contraction. On the other hand, hitting $\langle q|l\circ(1+2)|q\rangle$ produces the same factor $\langle q|(1+2)\circ(1+2)|q\rangle$ rendering it to vanish again. So, the quadratic part of the shift is zero.
\\~\\ 
Overall, the triangle diagram gives the result 
\be
\label{result}
\mathcal{A}_{triangle}=\frac{1}{32.3\pi^2}\frac{[12](\langle q||3|4|q\rangle)^2}{[12]\langle 12\rangle\langle 34\rangle\prod_{j=1}^4\langle qj\rangle}=\frac{1}{32.3\pi^2}\frac{[12][34]^2}{[12]\langle 12\rangle\langle 34\rangle}\frac{\langle q3\rangle\langle q4\rangle}{\langle q1\rangle\langle q2\rangle}.
\ee 
Let us write it in two parts, one which is $q$-independent and another which is q-dependent. We omit numerical factors here
\be 
\mathcal{A}_{triangle}=\frac{[12][34]^2}{[12]\langle 12\rangle\langle 34\rangle}\frac{\langle q3\rangle\langle q4\rangle}{\langle q1\rangle\langle q2\rangle}=\frac{[12][34]\Big([32]\langle 2q\rangle+[31]\langle 1q\rangle\Big)\langle q3\rangle}{[12]\langle 12\rangle\langle 34\rangle\langle q1\rangle\langle q2\rangle}\nonumber\\=-\frac{[12][34][32]\langle q3\rangle}{[12]\langle 12\rangle[34]\langle q1\rangle}-\frac{[12][34][31]\langle q3\rangle}{[12]\langle 12\rangle[34]\langle q2\rangle}\nonumber\\=\frac{[12][34]}{\langle 12\rangle\langle 34\rangle}-\frac{[12][34][24]\langle q4\rangle}{[12]\langle 12\rangle\langle 34\rangle\langle q1\rangle}-\frac{[12][34][31]\langle q3\rangle}{[12]\langle 12\rangle\langle 34\rangle \langle q2\rangle}.
\ee
where we used momentum conservation of the form $4=-(1+2+3)$ in the first line and $3=-(1+2+4)$ in the second line. 
Let us now add to it, all the possible cyclic permutations of the external legs for this diagram. The distinct permutations are given by $(1\leftrightarrow 3,4\leftrightarrow 2)$, $(1\leftrightarrow 3)$, $(2\leftrightarrow 4)$. Adding these, we get for the triangle diagram, upto numerical factors 
\be
\label{full}
\mathcal{A}_{triangle}(1^-,2^-,3^-,4^-)=\frac{[12][34]}{\langle 12\rangle\langle 34\rangle}-\frac{[12][34][24]\langle q4\rangle}{[12]\langle 12\rangle\langle 34\rangle\langle q1\rangle}-\frac{[12][34][31]\langle q3\rangle}{[12]\langle 12\rangle\langle 34\rangle \langle q2\rangle}\nonumber\\+
\frac{[34][12]}{\langle 34\rangle\langle 12\rangle}-\frac{[34][12][42]\langle q2\rangle}{[34]\langle 34\rangle\langle 12\rangle\langle q3\rangle}-\frac{[34][12][13]\langle q1\rangle}{[34]\langle 34\rangle\langle 12\rangle \langle q4\rangle}\nonumber\\+
\frac{[14][32]}{\langle 14\rangle\langle 32\rangle}-\frac{[14][32][42]\langle q2\rangle}{[14]\langle 14\rangle\langle 32\rangle\langle q1\rangle}-\frac{[14][32][31]\langle q3\rangle}{[14]\langle 14\rangle\langle 32\rangle \langle q4\rangle}\nonumber\\+
\frac{[32][14]}{\langle 32\rangle\langle 14\rangle}-\frac{[32][14][24]\langle q4\rangle}{[32]\langle 32\rangle\langle 14\rangle\langle q3\rangle}-\frac{[32][14][13]\langle q1\rangle}{[32]\langle 32\rangle\langle 14\rangle \langle q2\rangle}.\\\nonumber
\ee
Next we have the triangle diagram with ghosts propagating in the loop
\vspace{1em}
\vspace{1em}
\be 
\begin{fmfgraph*}(140,90)
\fmfbottom{i1,i2}
\fmftop{o1,o2}
\fmf{photon,tension=3}{i1,v1,o1}
\fmf{photon,tension=3}{i2,v3}
\fmf{photon,tension=3}{o2,v4}
\fmf{vanilla}{v1,v1'}
\fmf{photon}{v1',v2}
\fmf{dots,label=$l$}{v2,v4}
\fmf{dots,label=$l-4$}{v4,v3}
\fmf{dots,label=$l+1+2$}{v3,v2}
\fmflabel{$1^-$}{o1}
\fmflabel{$4^-$}{o2}
\fmflabel{$2^-$}{i1}
\fmflabel{$3^-$}{i2}
\fmflabel{}{v2}
\fmflabel{}{v2}
\fmflabel{}{v3}
\fmflabel{}{v4}
\end{fmfgraph*}\nonumber 
\ee 
\be 
i\mathcal{A}_{ghost-triangle}=\frac{[12]}{[12]\langle 12\rangle}\int \frac{d^4l}{(2\pi)^4}\frac{\langle q|l|1+2|q\rangle\langle q|l+1+2|3]\langle q|l-4|4]}{l^2(l-4)^2(l+1+2)^2\prod_{j=1}^4\langle qj\rangle}.
\ee 
Using momentum conservation, this can be written as
\be 
i\mathcal{A}_{ghost-triangle}=\frac{[12]}{\langle 12\rangle}\int \frac{d^4l}{(2\pi)^4}\frac{\langle q|l+1+2|1+2|q\rangle\langle q|l-4|3]\langle q|l|4]}{l^2(l-4)^2(l+1+2)^2\prod_{j=1}^4\langle qj\rangle}.
\ee 
This integral is again similar to the one we have computed earlier. Thus, when evaluated and permuted accordingly, the ghost diagrams contribute an overall factor to the already evaluated ones. 
\\~\\
\newpage
\subsection{Result}
\textbf{Full amplitude}
\\~\\
We now obtain the full amplitude by adding the box and the triangles. It is then easy to check that the $q$-dependent terms from the box cancels the corresponding ones in the triangles after using momentum conservation identities. What is left is a sum of $q$-independent terms 
\be 
-12.32\pi^2\mathcal{A}_{one-loop}(1^-,2^-,3^-,4^-)=\frac{[14][32]}{\langle 14\rangle\langle 32\rangle}+\frac{[14][23]}{\langle 14\rangle\langle 23\rangle}+\frac{[24][13]}{\langle 24\rangle\langle 13\rangle}.
\ee 
All the three terms are equal to one another due to momentum conservation identities. We finally get 
\be 
\label{res}
\mathcal{A}_{one-loop}(1^-,2^-,3^-,4^-)=-\frac{1}{128\pi^2}\frac{[12][34]}{\langle 12\rangle\langle 34\rangle}.
\ee 

We have computed the same helicity one-loop 4-point amplitude in self-dual
Yang-Mills (SDYM) theory, relating it to a set of integrals that vanish after a shift of the
loop momentum, and thus reducing the computation to shifts. The same helicity one-loop
amplitudes in self-dual YM coincide with such amplitudes in full YM, as can be seen by
comparing the Feynman rules of the two theories. Thus, our computation actually gives
the full YM same helicity 4-point amplitude.
\\~\\
The difference between the present computation and those available in the literature \cite{Bern:1991aq, Mahlon:1993fe}, as well as \cite{Ossola:2008xq, Brandhuber:2005jw} is that the computation was carried out in four dimensions.
No dimensional regularisation, common to papers \cite{Mahlon:1993fe}, \cite{Brandhuber:2005jw, Binoth:2006hk, Ossola:2008xq} was used. Doing the computation in a way that avoids the use of dimensional regularisation was an important part
of our motivation. This is because the SDYM is a chiral theory, and there are well-known
difficulties reconciling chiral objects with the dimensional regularisation. One could even
suspect that the answer arises only by using a certain prescription for chiral objects
in $4-2\epsilon$ dimensions, and the result vanishes if a different prescription is used. Reproducing the answer by a computation that avoids introducing extra dimensions shows that the
result is independent of a way used to compute it, as it should be.
Our computation related the amplitude of interest to nominally quadratically divergent
integrals, which however vanish after a shift. So, in a sense, the finite result comes from a
divergent and at the same time vanishing expression. This is similar to the computations \cite{Mahlon:1993fe, Bern:1991aq, Brandhuber:2005jw} that use the dimensional regularisation and get the result as the coefficient of $\epsilon/\epsilon$.
So, it may be that the trick that relates the amplitude of interest to shifts can be used for
other amplitudes, to extract their rational parts. The technology of extracting the rational
parts of amplitudes by combining generalised unitarity with dimensional regularisation is
by now well-established \cite{Binoth:2006hk, Ossola:2008xq}. It is possible that a variant of this technology is possible
by relating amplitudes to shifts, but we will not attempt to demonstrate this here. \\~\\
The computation performed above also holds for massless QED. This is because in spinorial notations, the Feynman rules of massless QED is similar to that of SDYM. The massless QED Lagrangian is chiral, in the sense that it only incorporates one chiral half of the fermion. Let us write it here. 
\be 
\mathcal{L}_{\textrm{massless~QED}}=i(\psi^{\dagger})^{B'}(\partial^{~~B}_{B'}+iA_{B'}^{~~B})\psi_{B}.
\ee 
Here $\psi_B$ is the two-component unprimed spinor (Weyl spinor) and $(\psi^{\dagger})_{B'}$ is its Hermitian conjugate. The object $A^{~~B}_{B'}$ is the spinor version of the electromagnetic potential $A_{\mu}$. The Lagrangian is real modulo a surface term. The spinor propagator and the vertex which results from this Lagrangian reads 
\be 
\label{qed}
\langle \psi_B(-k)(\psi^{\dagger})^{B'}(k)\rangle=\frac{2}{ik^2}k^{~~B'}_B\nonumber\\
\langle (\psi^{\dagger})_{B'}\psi^{B}A^{~~C}_{C'}\rangle=i\epsilon_{B'C'}\epsilon^{BC}.
\ee 
Thus, the only difference in the propagator of SDYM from that in (\ref{qed})
is that there is an extra Kronecker delta for the colour, a phase factor and an extra factor of the spinor
metric for the additional unprimed index present in the SDYM as compared to the case of QED. The same thing is true for the vertex. Thus the only change one has to do in the computation for the QED box diagram is to mod out some of the unprimed spinor metrics and the colour factors. 
However, in the absence of any colour ordering in QED, the box itself gives the complete gauge invariant result, unlike in SDYM where in addition to the box, we needed triangle diagrams. Further, using supersymmetric Ward identities, it is easy to see that the result of scalar QED is a multiple of massless QED. Thus, overall the four point same helicity amplitude in all these theories is related to a shift computation. \\~\\
From the next section, we take a different approach to understand the same helicity amplitudes in SDYM. This can be coined as the interpretation of the amplitudes in terms of self-energy bubbles.
\newpage
 \section{Sum of integrands at one loop} 
We have till now focused on the box and triangle diagrams and computed the result of the amplitude entirely from those. The bubble graphs are zero to begin with, where we used Lorentz invariance arguments. There are no tadpole contributions to the four point amplitude. However, getting motivated from the earlier works in \cite{Chakrabarti:2005ny, Brandhuber:2007vm}, we want to inspect the sum of one loop integrand of the complete amplitude. In the previous works, the lightcone formalism is used to show that the complete integrand vanishes. In our case, we intend to perform the computation in the covariant formalism of SDYM. It is different from the other works in that there is no regularization used here and as we will see, the non-vanishing of the bubbles is not an artifact of a regulator, but instead of shifts of loop momentum. We want to sum all the integrands of the arising geometries, i.e box, triangles and bubbles, where only cyclic permutations of legs are allowed. Thus there will be just one box diagram since the box remains unchanged under cyclic permutations. There will be four distinct permutations of the triangle, two for the internal bubble and four for the external bubble. The idea is then to decompose the box and triangle integrals into quadratically divergent integrals, following what we have done earlier. Then there are bubble graphs which are by themselves quadratically divergent. We then add all the integrands of these quadratically divergent contributions. It is important to note that we use a particular choice of loop momentum variable for each graph. To do so, we employ region momenta, see the Review section for more details. There are four external lines for this particular amplitude. Thus there are four distinct external regions for any graph and so there are four different region momenta. There is one enclosed region for the loop and we associate a momenta for that region. We discuss about these subsequently.
\subsection{Box} 
Let us start with the box diagram.
\vspace{1em}
\vspace{1em}
\be
\begin{fmfgraph*}(110,80)
\fmfbottom{i1,i2}
\fmftop{o1,o2}
\fmf{photon,tension=3}{i1,v1}
\fmf{photon,tension=3}{i2,v2}
\fmf{photon,tension=3}{o1,v3}
\fmf{photon,tension=3}{o2,v4}
\fmf{photon}{v3,v3'}
\fmf{vanilla}{v3',v1}
\fmf{photon}{v1,v1'}
\fmf{vanilla}{v1',v2}
\fmf{photon}{v2,v2'}
\fmf{vanilla}{v2',v4}
\fmf{photon}{v4,v4'}
\fmf{vanilla}{v4',v3}
\fmflabel{$1^-$}{o1}
\fmflabel{$4^-$}{o2}
\fmflabel{$2^-$}{i1}
\fmflabel{$3^-$}{i2}
\fmflabel{$l+1+2$}{v1'}
\fmflabel{$l+1$}{v3'}
\fmflabel{$l-4$}{v2'}
\fmflabel{$l$}{v4'}
\end{fmfgraph*}\nonumber
\ee\\
We already described the box integral in the previous section. It is given by
\be 
\label{amp}
i\mathcal{M}=\int \frac{d^4l}{(2\pi)^4}\frac{\langle q|l|4]\langle q|l+1|1]\langle q|l+1+2|2]\langle q|l-4|3]}{l^2(l+1)^2(l+1+2)^2(l-4)^2\prod_{j=1}^4\langle qj\rangle}.
\ee
We now use the previous results obtained when the box integral was decomposed into quadratically divergent integrals. There are a total of nine such integrals, each of which are obtained from three linearly divergent integrals. We refer (\ref{amplit111}) to write the first set of quadratically divergent integrals
\be 
\label{amp111}
4i\mathcal{A}_{1}=-\int \frac{d^4l}{(2\pi)^4}\frac{\langle q|l+|1]\langle q|(l+1+2)\circ(l-4)|q\rangle}{(l+1+2)^2(l-4)^2\langle 23\rangle\langle 43\rangle\langle q1\rangle\langle q2\rangle\langle q4\rangle}\\\nonumber+\int \frac{d^4l}{(2\pi)^4}\frac{\langle q|l+1|1]\langle q|(l+1)\circ(l-4)|q\rangle}{(l+1)^2(l-4)^2\langle 23\rangle\langle 43\rangle\langle q1\rangle\langle q2\rangle\langle q4\rangle}\\+\nonumber\int \frac{d^4l}{(2\pi)^4}\frac{\langle q|l+1|1]\langle q|(l+1+2)\circ2|q\rangle}{(l+1)^2(l+1+2)^2\langle 23\rangle\langle 43\rangle\langle q1\rangle\langle q2\rangle\langle q4\rangle}.
\ee
We similarly consider the next two sets using (\ref{amp2322}) and (\ref{amp332!}). They are given by
\be 
\label{amp232}
4i\mathcal{A}_{2}=\int \frac{d^4l}{(2\pi)^4}\frac{\langle q|l+1|2]\langle q|l\circ1|q\rangle}{l^2(l+1)^2\langle 41\rangle\langle 43\rangle\langle q1\rangle\langle q2\rangle\langle q3\rangle}\\\nonumber-\int \frac{d^4l}{(2\pi)^4}\frac{\langle q|l+1|2]\langle q|(l-4)\circ(l+1)|q\rangle}{(l+1)^2(l-4)^2\langle 41\rangle\langle 43\rangle\langle q1\rangle\langle q2\rangle\langle q3\rangle}\\\nonumber+\int \frac{d^4l}{(2\pi)^4}\frac{\langle q|l+1|2]\langle q|l\circ4|q\rangle}{l^2(l-4)^2\langle 41\rangle\langle 43\rangle\langle q1\rangle\langle q2\rangle\langle q3\rangle}
\ee
and 
\be 
\label{amp332}
4i\mathcal{A}_{3}=\int \frac{d^4l}{(2\pi)^4}\frac{\langle q|l\circ(3+4)|q\rangle\langle q|(1+2)\circ l|q\rangle}{l^2(l+1+2)^2\langle 12\rangle\langle 43\rangle\langle q1\rangle\langle q2\rangle\langle q3\rangle\langle q4\rangle}\\\nonumber+\int \frac{d^4l}{(2\pi)^4}\frac{\langle q|l\circ(3+4)|q\rangle\langle q|l+1\circ 1|q\rangle}{l^2(l+1)^2\langle 12\rangle\langle 43\rangle\langle q1\rangle\langle q2\rangle\langle q3\rangle\langle q4\rangle}\\+\nonumber\int \frac{d^4l}{(2\pi)^4}\frac{\langle q|l\circ(3+4)|q\rangle\langle q|(l+1)\circ2|q\rangle}{(l+1)^2(l+1+2)^2\langle 12\rangle\langle 43\rangle\langle q1\rangle\langle q2\rangle\langle q3\rangle\langle q4\rangle}.
\ee
We now implement the use of region momenta, following the convention of the world-sheet formulation. We direct the reader to the Review section for further details. In a particular diagram, if we have four external lines of momenta $k_1,k_2,....k_4$, such that there are four distinct regions outside the diagram, with region momenta $p_1,p_2....p_{4}$, then we make the following convention of the assignment of region momenta.
\be 
\begin{split} 
k_1&=p_1-p_4,\\
k_2&=p_2-p_1,\\
k_3&=p_3-p_2,\\
k_4&=p_4-p_3.
\end{split}
\ee 
In terms of these region momenta, momentum conservation is automatically satisfied as can be checked from the assignment above. For the one-loop diagrams, we also assign the region momenta inside the loop and re-express the loop momentum in terms of region momenta. Since the loop momentum has to be integrated, there is an arbitrariness in its assignment. Indeed, we can add any momenta $q$ to the loop. We thus make specific choices of the loop momenta for the various diagrams in such a way that we can assemble terms with common denominators and no single term is left away. For the box, we make the choice $l=x+p_4$, where $x$ is the region momentum inside the loop and $p_4$ is one of the external region momenta. The momentum $x$ is now to be integrated over. Note that we only use the dual momentum variables for terms involving loop momenta, and for all other places, the usual momenta.  We then re-write the quadratically divergent integrals in terms of the dual momentum variables and usual momenta, given this particular choice of loop momentum. Implementing this in (\ref{amp332}), we get
\be 
4i\mathcal{A}_3=4i\mathcal{A}_{31}+4i\mathcal{A}_{32}+4i\mathcal{A}_{33}
\ee 
where
\be 
\begin{split}
\label{amp3321}
4i\mathcal{A}_{31}&=\int \frac{d^4x}{(2\pi)^4}\frac{\langle q|(x+p_4)\circ(3+4)|q\rangle\langle q|(1+2)\circ (x+p_2)|q\rangle}{(x+p_4)^2(x+p_2)^2\langle 12\rangle\langle 43\rangle\langle q1\rangle\langle q2\rangle\langle q3\rangle\langle q4\rangle},\\
4i\mathcal{A}_{32}&=\int \frac{d^4x}{(2\pi)^4}\frac{\langle q|(x+p_4)\circ(3+4)|q\rangle\langle q|(x+p_1)\circ 1|q\rangle}{(x+p_4)^2(x+p_1)^2\langle 12\rangle\langle 43\rangle\langle q1\rangle\langle q2\rangle\langle q3\rangle\langle q4\rangle},\\
4i\mathcal{A}_{33}&=-\int \frac{d^4x}{(2\pi)^4}\frac{\langle q|(x+p_2)\circ(1+2)|q\rangle\langle q|(x+p_1)\circ2|q\rangle}{(x+p_1)^2(x+p_2)^2\langle 12\rangle\langle 43\rangle\langle q1\rangle\langle q2\rangle\langle q3\rangle\langle q4\rangle}.
\end{split}
\ee
Next, we have
\be 
4i\mathcal{A}_2=4i\mathcal{A}_{21}+4i\mathcal{A}_{22}+4i\mathcal{A}_{23},
\ee 
where
\be 
\begin{split}
\label{amp2321}
4i\mathcal{A}_{21}&=\int \frac{d^4x}{(2\pi)^4}\frac{\langle q|x+p_1|2]\langle q|(x+p_4)\circ1|q\rangle}{(x+p_4)^2(x+p_1)^2\langle 41\rangle\langle 43\rangle\langle q1\rangle\langle q2\rangle\langle q3\rangle},\\\nonumber
4i\mathcal{A}_{22}&=\int \frac{d^4x}{(2\pi)^4}\frac{\langle q|x+p_1|2]\langle q|(x+p_3)\circ(3+2)|q\rangle}{(x+p_1)^2(x+p_3)^2\langle 41\rangle\langle 43\rangle\langle q1\rangle\langle q2\rangle\langle q3\rangle},\\\nonumber
4i\mathcal{A}_{23}&=\int \frac{d^4x}{(2\pi)^4}\frac{\langle q|x+p_1|2]\langle q|(x+p_4)\circ4|q\rangle}{(x+p_4)^2(x+p_3)^2\langle 41\rangle\langle 43\rangle\langle q1\rangle\langle q2\rangle\langle q3\rangle}.
\end{split}
\ee
and the remaining one
\be 
4i\mathcal{A}_1=4i\mathcal{A}_{11}+4i\mathcal{A}_{12}+4i\mathcal{A}_{13},
\ee 
where
\be
\begin{split}
\label{amp111}
4i\mathcal{A}_{11}&=\int \frac{d^4x}{(2\pi)^4}\frac{\langle q|x+p_1|1]\langle q|(x+p_2)\circ2|q\rangle}{(x+p_2)^2(x+p_1)^2\langle 23\rangle\langle 43\rangle\langle q1\rangle\langle q2\rangle\langle q4\rangle},\\\nonumber
4i\mathcal{A}_{12}&=\int \frac{d^4x}{(2\pi)^4}\frac{\langle q|x+p_1|1]\langle q|(x+p_3)\circ(1+4)|q\rangle}{(x+p_1)^2(x+p_3)^2\langle 23\rangle\langle 43\rangle\langle q1\rangle\langle q2\rangle\langle q4\rangle},\\\nonumber
4i\mathcal{A}_{13}&=\int \frac{d^4x}{(2\pi)^4}\frac{\langle q|x+p_1|1]\langle q|(x+p_2)\circ3|q\rangle}{(x+p_3)^2(x+p_2)^2\langle 23\rangle\langle 43\rangle\langle q1\rangle\langle q2\rangle\langle q4\rangle}.
\end{split}
\ee
\subsection{Triangles}
There are four in-equivalent triangle diagrams to consider. We start by considering the one where the legs $3$ and $4$ are attached to two of the vertices.
\vspace{1em}
\vspace{1em}

~~~~~~~~~~~~~~~~~~~~~~~~~~~~~~~~~~~~~~~~~\begin{fmfgraph*}(150,90)
\fmfbottom{i1,i2}
\fmftop{o1,o2}
\fmf{photon,tension=3}{i1,v1,o1}
\fmf{photon,tension=3}{i2,v3}
\fmf{photon,tension=3}{o2,v4}
\fmf{vanilla}{v1,v1'}
\fmf{photon}{v1',v2}
\fmf{photon,label=$l$}{v2',v2}
\fmf{vanilla}{v2',v4}
\fmf{photon,label=$l-4$}{v4',v4}
\fmf{vanilla}{v4',v3}
\fmf{photon,label=$l+1+2$}{v3',v3}
\fmf{vanilla}{v3',v2}
\fmflabel{$1^-$}{o1}
\fmflabel{$4^-$}{o2}
\fmflabel{$2^-$}{i1}
\fmflabel{$3^-$}{i2}
\fmflabel{}{v1'}
\fmflabel{}{v2'}
\fmflabel{}{v3'}
\fmflabel{}{v4'}
\end{fmfgraph*}\\~\\
\be 
2i\mathcal{A}_{t1}=\frac{1}{\langle 12\rangle}\int \frac{d^4l}{(2\pi)^4}\frac{\langle q|l|4]\langle q|l-4|3]\langle q|(l+1+2)\circ(1+2)|q\rangle}{l^2(l-4)^2(l+1+2)^2\prod_{j=1}^4\langle qj\rangle}.
\ee 
Using the same techniques, the reduction to quadratically divergent integrals is straightforward. We refer (\ref{tri}) and write these integrals. 
\be 
\label{tri2}
i\mathcal{A}_{t1}=\frac{1}{4\langle 12\rangle\langle 34\rangle}\Bigg[\int \frac{d^4l}{(2\pi)^4}\frac{\langle q|(l-4)\circ 3|q\rangle\langle q|l|1+2|q\rangle}{(l-4)^2(l+1+2)^2\prod_{j=1}^4\langle qj\rangle}\nonumber\\+\int \frac{d^4l}{(2\pi)^4}\frac{\langle q|l\circ(1+2)|q\rangle\langle q|(l+1+2)\circ(1+2)|q\rangle}{l^2(l+1+2)^2\prod_{j=1}^4\langle qj\rangle}\nonumber\\+\int \frac{d^4l}{(2\pi)^4}\frac{\langle q|l\circ4|q\rangle\langle q|(l+1+2)\circ(1+2)|q\rangle}{l^2(l-4)^2\prod_{j=1}^4\langle qj\rangle}\Bigg].
\ee 
Let us then rewrite this in the dual momentum variables. As before, we choose the loop momentum variable $l=p_4-x$. We have  
\be 
4i\mathcal{T}^{12}=4i\mathcal{T}^{12}_1+4i\mathcal{T}^{12}_2+4i\mathcal{T}^{12}_3.
\ee 
where 
\be 
\begin{split}
\label{tri2}
4i\mathcal{T}^{12}_1&=\int \frac{d^4l}{(2\pi)^4}\frac{\langle q|(p_3-x)\circ 3|q\rangle\langle q|(p_2-x)\circ(1+2)|q\rangle}{(p_3-x)^2(p_2-x)^2\langle 12\rangle\langle 34\rangle\prod_{j=1}^4\langle qj\rangle},\\
4i\mathcal{T}^{12}_2&=\int \frac{d^4l}{(2\pi)^4}\frac{\langle q|(p_4-x)\circ(1+2)|q\rangle\langle q|(p_2-x)\circ(1+2)|q\rangle}{(p_4-x)^2(p_2-x)^2\langle 12\rangle\langle 34\rangle\prod_{j=1}^4\langle qj\rangle},\\
4i\mathcal{T}^{12}_3&=\int \frac{d^4l}{(2\pi)^4}\frac{\langle q|(p_4-x)\circ4|q\rangle\langle q|(p_2-x)\circ(1+2)|q\rangle}{(p_4-x)^2(p_3-x)^2\langle 12\rangle\langle 34\rangle\prod_{j=1}^4\langle qj\rangle}.
\end{split}
\ee 

The diagram with legs 1 and 2 inserted to two of the vertices can be obtained from (\ref{tri2}) by substituting
\be
1\to 3, \quad 2\to 4,\quad 3\to 1,\quad 4\to 2.
\ee
We need to make a choice for the loop momentum parameter. There are three different choices as there are three possible external region momenta adjacent to the region momenta inside the loop. Motivated by the vanishing of the sum of integrands, we choose the loop momentum in terms of region momenta to be $l=p_2-x$. We have
\be 
4i\mathcal{T}^{34}=4i\mathcal{T}^{34}_1+4i\mathcal{T}^{34}_2+4i\mathcal{T}^{34}_3,
\ee 
where 
\be 
\begin{split}
\label{tri31}
4i\mathcal{T}^{34}_1&=\int \frac{d^4x}{(2\pi)^4}\frac{\langle q|(p_1-x)\circ 1|q\rangle\langle q|(p_4-x)\circ (3+4)|q\rangle}{(p_1-x)^2(p_4-x)^2\langle 12\rangle\langle 34\rangle\langle q1\rangle\langle q2\rangle\langle q3\rangle\langle q4\rangle},\\
4i\mathcal{T}^{34}_2&=\int \frac{d^4x}{(2\pi)^4}\frac{\langle q|(p_2-x)\circ(3+4)|q\rangle\langle q|(p_4-x)\circ (3+4)|q\rangle}{(p_2-x)^2(p_4-x)^2\langle 12\rangle\langle 34\rangle\langle q1\rangle\langle q2\rangle\langle q3\rangle\langle q4\rangle},\\
4i\mathcal{T}^{34}_3&=\int \frac{d^4x}{(2\pi)^4}\frac{\langle q|(p_2-x)\circ 2|q\rangle\langle q|(p_4-x)\circ(3+4)|q\rangle}{(p_2-x)^2(p_1-x)^2\langle 12\rangle\langle 34\rangle\langle q1\rangle\langle q2\rangle\langle q3\rangle\langle q4\rangle}.
\end{split}
\ee 
We note that the terms arising here cancel precisely the terms in ${\mathcal M}_3$. 

The diagram with legs 2 and 3 inserted to two of the vertices can be obtained from (\ref{tri2})  by substitutions
\be
1\to 4, \quad 4\to 3,\quad 3\to 2,\quad 2\to 1.
\ee
Let us choose the loop momentum in terms of region momenta to be $l=p_3-x$. We get
\be 
4i\mathcal{T}^{41}=4i\mathcal{T}^{41}_1+4i\mathcal{T}^{41}_2+4i\mathcal{T}^{41}_3,
\ee 
where 
\be 
\begin{split}
\label{tri3}
4i\mathcal{T}^{41}_1&=\int \frac{d^4x}{(2\pi)^4}\frac{\langle q|(p_2-x)\circ 2|q\rangle\langle q|(p_1-x)\circ(1+4)|q\rangle}{(p_2-x)^2(p_1-x)^2\langle 41\rangle\langle 23\rangle\prod_{j=1}^4\langle qj\rangle}\\
4i\mathcal{T}^{41}_2&=\int \frac{d^4x}{(2\pi)^4}\frac{\langle q|(p_3-x)\circ(1+4)|q\rangle\langle q|(p_1-x)\circ(1+4)|q\rangle}{(p_3-x)^2(p_1-x)^2\langle 41\rangle\langle 23\rangle\prod_{j=1}^4\langle qj\rangle}\\
4i\mathcal{T}^{41}_3&=\int \frac{d^4x}{(2\pi)^4}\frac{\langle q|(p_3-x)\circ3|q\rangle\langle q|(p_1-x)\circ(1+4)|q\rangle}{(p_2-x)^2(p_3-x)^2\langle 41\rangle\langle 23\rangle\prod_{j=1}^4\langle qj\rangle}
\end{split}
\ee

The diagram with legs 1 and 4 inserted to two of the vertices can be obtained from (\ref{tri2}) by substitutions
\be
1\to 2, \quad 2\to 3,\quad 3\to 4,\quad 4\to 1.
\ee
The loop momentum in terms of region momenta is $l=p_1-x$. This gives
\be 
4i\mathcal{T}^{23}=4i\mathcal{T}^{23}_1+4i\mathcal{T}^{23}_2+4i\mathcal{T}^{23}_3,
\ee 
where 
\be 
\begin{split}
\label{tri31}
4i\mathcal{T}^{23}_1&=\int \frac{d^4x}{(2\pi)^4}\frac{\langle q|(p_4-x)\circ 4|q\rangle\langle q|(p_3-x)\circ(2+3)|q\rangle}{(p_4-x)^2(p_3-x)^2\langle 23\rangle\langle 41\rangle\prod_{j=1}^4\langle qj\rangle}\\
4i\mathcal{T}^{23}_2&=\int \frac{d^4x}{(2\pi)^4}\frac{\langle q|(p_1-x)\circ (2+3)|q\rangle\langle q|(p_3-x)\circ(2+3)|q\rangle}{(p_1-x)^2(p_3-x)^2\langle 23\rangle\langle 41\rangle\prod_{j=1}^4\langle qj\rangle}\\
4i\mathcal{T}^{23}_3&=\int \frac{d^4x}{(2\pi)^4}\frac{\langle q|(p_1-x)\circ 1|q\rangle\langle q|(p_3-x)\circ(2+3)|q\rangle}{(p_1-x)^2(p_4-x)^2\langle 23\rangle\langle 41\rangle\prod_{j=1}^4\langle qj\rangle}.
\end{split}
\ee 
\subsection{Bubble Diagrams}
\subsubsection{Internal Bubbles}
There are two in-equivalent permutations of the bubbles on internal lines. In one diagram, the legs 1 and 2 join on one side of the internal line while 3 and 4 on the other side. This is given by  \\~\\

~~~~~~~~~~~~~~~~~~~~~~~~~~~~~~~~~\begin{fmfgraph*}(140,50)
     \fmftop{i1,i3,i2}
     \fmfbottom{o1,o3,o2}
     \fmf{photon}{i1,v2,o1}
     \fmf{vanilla}{v2,v1'}
     \fmf{photon}{v1',v1}
     \fmf{photon}{v1,i3}
     \fmf{vanilla}{i3,v4}
     \fmf{photon}{v4,o3}
     \fmf{vanilla}{o3,v1}
     \fmf{photon}{v4,v3}
     \fmf{vanilla}{v3,v3'}
     \fmf{photon}{i2,v3',o2}
     \fmflabel{l}{i3}
     \fmflabel{l+1+2}{o3}
     \fmflabel{$1^-$}{i1}
     \fmflabel{$2^-$}{o1}
     \fmflabel{$3^-$}{o2}
     \fmflabel{$4^-$}{i2}
\end{fmfgraph*}
\\~\\

 \be 
4i\mathcal{B}_{11}=\frac{1}{\langle 12\rangle\langle 34\rangle}\int \frac{d^4l}{(2\pi)^4}\frac{\langle q|l\circ (1+2)|q\rangle\langle q|(l+1+2)\circ(3+4)|q\rangle}{l^2(l+1+2)^2\prod_{j=1}^4\langle qj\rangle}
\ee 
We use the choice $l=x+p_4$ for the loop momentum and write the above in dual momentum variables
 \be 
 \label{bubb1}
4i\mathcal{\tilde{B}}_{11}=\frac{1}{\langle 12\rangle\langle 34\rangle}\int \frac{d^4x}{(2\pi)^4}\frac{\langle q|(x+p_4)\circ (1+2)|q\rangle\langle q|(x+p_2)\circ(3+4)|q\rangle}{(x+p_4)^2(x+p_2)^2\prod_{j=1}^4\langle qj\rangle}
\ee 
The other internal bubble diagram is
\\~\\

~~~~~~~~~~~~~~~~~~~~~~~~~~~~~~~~~\begin{fmfgraph*}(140,50)
     \fmftop{i1,i3,i2}
     \fmfbottom{o1,o3,o2}
     \fmf{photon}{i1,v2,o1}
     \fmf{vanilla}{v2,v1'}
     \fmf{photon}{v1',v1}
     \fmf{photon}{v1,i3}
     \fmf{vanilla}{i3,v4}
     \fmf{photon}{v4,o3}
     \fmf{vanilla}{o3,v1}
     \fmf{photon}{v4,v3}
     \fmf{vanilla}{v3,v3'}
     \fmf{photon}{i2,v3',o2}
     \fmflabel{l}{i3}
     \fmflabel{l+1+4}{o3}
     \fmflabel{$4^-$}{i1}
     \fmflabel{$1^-$}{o1}
     \fmflabel{$2^-$}{o2}
     \fmflabel{$3^-$}{i2}
\end{fmfgraph*}
\\~\\
\be 
4i\mathcal{B}_{12}=\frac{1}{\langle 41\rangle\langle 23\rangle}\int \frac{d^4l}{(2\pi)^4}\frac{\langle q|l\circ (4+1)|q\rangle\langle q|(l+1+4)\circ(3+2)|q\rangle}{l^2(l+1+4)^2\prod_{j=1}^4\langle qj\rangle}
\ee 
We use the choice $l=x+k_3$ for the loop momentum, re-writing the above 
\be 
\label{bubb2}
4i\mathcal{\tilde{B}}_{12}=\frac{1}{\langle 41\rangle\langle 23\rangle}\int \frac{d^4x}{(2\pi)^4}\frac{\langle q|(x+k_3)\circ (1+4)|q\rangle\langle q|(x+k_1)\circ(3+2)|q\rangle}{(x+k_3)^2(x+k_1)^2\prod_{j=1}^4\langle qj\rangle}
\ee 

\subsubsection{Bubbles on External Lines}

There are four distinct diagrams here, corresponding to the different ways of insertion of bubbles to four external lines. One can write each contribution as an insertion of a current to one of the legs of the bubble. For example, we have for the insertion of the bubble between the particle 1 and the $J(2,3,4)$ current
\\~\\ 
\be
\begin{gathered}
~~~\begin{fmfgraph*}(130,30)
     \fmfleft{i}
     \fmfright{o}
     \fmfbottom{v}
     \fmftop{u}
     \fmf{photon}{i,v1}
     \fmf{photon}{v1,u}
     \fmf{vanilla}{u,v2}
     \fmf{photon}{v2,v}
     \fmf{vanilla}{v,v1}
     \fmf{photon}{v2,v3}
     \fmf{vanilla}{v3,o}
     \fmfblob{.15w}{o}
     \fmflabel{l}{u}
     \fmflabel{l+1}{v}
     \fmflabel{$1^-$}{i}
\end{fmfgraph*}
\end{gathered}\nonumber 
\ee 
\\
\be 
4i\mathcal{B}^1=\frac{\langle q3\rangle}{\langle 23\rangle\langle 34\rangle}\int \frac{d^4l}{(2\pi)^4}\frac{\langle q|(l+1)\circ 1|q\rangle\langle q|l|1]}{l^2(l+1)^2\prod_{j=1}^4\langle qj\rangle}
\ee 
Rewriting this in terms of the dual momentum variables with $l=p_4-x$ we get
\be 
\label{b1}
4i\mathcal{B}^1=\frac{\langle q3\rangle}{\langle 23\rangle\langle 34\rangle}\int \frac{d^4x}{(2\pi)^4}\frac{\langle q|(p_1-x)\circ 1|q\rangle\langle q|p_4-x|1]}{(p_4-x)^2(p_1-x)^2\prod_{j=1}^4\langle qj\rangle}.
\ee 

The other similar contributions are obtained by cyclic permutations. We have
\be 
\label{b2}
4i\mathcal{B}^2=\frac{\langle q4\rangle}{\langle 34\rangle\langle 41\rangle}\int \frac{d^4x}{(2\pi)^4}\frac{\langle q|(p_2-x)\circ 2|q\rangle\langle q|p_1-x|2]}{(p_2-x)^2(p_1-x)^2\prod_{j=1}^4\langle qj\rangle},
\ee 
\be 
\label{b3}
4i\mathcal{B}^3=\frac{\langle q1\rangle}{\langle 41\rangle\langle 12\rangle}\int \frac{d^4x}{(2\pi)^4}\frac{\langle q|(p_3-x)\circ 3|q\rangle\langle q|p_2-x|3]}{(p_2-x)^2(p_3-x)^2\prod_{j=1}^4\langle qj\rangle},
\ee 
\be 
\label{b4}
4i\mathcal{B}^4=\frac{\langle q2\rangle}{\langle 12\rangle\langle 23\rangle}\int \frac{d^4x}{(2\pi)^4}\frac{\langle q|(p_4-x)\circ 4|q\rangle\langle q|p_3-x|4]}{(p_4-x)^2(p_3-x)^2\prod_{j=1}^4\langle qj\rangle}
\ee

\subsection{Sum of Integrands}

As we previously noticed, the terms in ${\mathcal A}_3$ get cancelled by one of the triangle diagrams ${\mathcal T}^{34}$. Our aim is now to show that all other terms get cancelled agains each other as well. To this end, we will group the terms according to their denominators of the form 
$(p_i-x)^2(p_j-x)^2$ for some $i,j$. 
\\~\\
We start with the denominator factor $(p_1-x)^2(p_2-x)^2$. The contributions to this come from ${\mathcal A}_{11}, {\mathcal T}^{41}_1$ as well as one of the bubbles ${\mathcal B}^2$. 
The sum of the corresponding numerators is given by 
\be 
\label{n1}
\begin{split} 
&\frac{\langle q|(p_2-x)\circ 2|q\rangle}{\langle 23\rangle\langle 43\rangle \langle 14\rangle}\Big(\langle q|(p_1-x)\circ (1+4)|4\rangle\langle q3\rangle
-\langle q|(p_1-x)\circ (1+4)|q\rangle\langle 43\rangle
\\&~~~~~~~~~~~~~~~~~~~~~~~~~~
+\langle q|(p_1-x)\circ 2|3\rangle\langle q4\rangle\Big),
\end{split}
\ee 
where we rewrote the first term in a suggestive way.
We now use the momentum conservation $2=-(1+3+4)$ in the last term, and then Schouten identity 
$|4\rangle \langle q3\rangle = |q\rangle \langle 43\rangle + |3\rangle \langle q4\rangle$ to see that the sum in brackets is zero. 
\\~\\
Next consider the denominator factor $(p_1-x)^2(p_3-x)^2$. The contributions come from $\mathcal{A}_{12},\mathcal{A}_{22}, {\mathcal T}^{41}_2, {\mathcal T}^{23}_2$ and the bubble 
$\mathcal{B}^{23}$. The triangle contributions double each other, and the sum of these numerators is 
\be 
\label{n2}
\begin{split} 
&\frac{\langle q|(p_3-x)\circ(1+4)|q\rangle}{\langle 23\rangle\langle 43\rangle \langle 14\rangle}\Big(\langle q|(p_1-x)\circ (1+4) |4\rangle\langle q3\rangle+\langle q|(p_1-x)\circ (2+3)| 3\rangle\langle q4\rangle\\&
-2\langle q|(p_1-x)\circ (1+4)|q\rangle\langle 43\rangle
+\langle q|(p_1-x)\circ (1+4)|q\rangle\langle 43\rangle\Big),
\end{split}
\ee 
where again we rewrote the first terms in a suggestive way. Relacing $2+3=-(1+4)$ in the second term and using the same Schouten identity as above we see the cancellation.

Next consider the denominator factor $(p_1-x)^2(p_4-x)^2$. The contributions come from $\mathcal{A}_{21},\mathcal{T}^{23}_3$ and ${\mathcal B}^1$. The sum of these numerators is given by 
\be 
\label{n3}
\begin{split} 
&\frac{\langle q|(p_4-x)\circ1|q\rangle}{\langle 23\rangle\langle 43\rangle \langle 41\rangle}\Big(
\langle q|(p_1-x)\circ 2|3\rangle\langle q4\rangle
+\langle q|(p_3-1)\circ (2+3)|q\rangle\langle 43\rangle
\\&~~~~~~~~~~~~~~~~~~~~~~~~~~
+\langle q|(p_4-x)\circ 1|4\rangle\langle q3\rangle\Big).
\end{split}
\ee 
We now use the fact that $p_3=p_1+2+3$, and so $p_3$ can be replaced by $p_1$ in the second term. Similarly, $p_4=p_1-1$, and so we can replace $p_4$ with $p_1$ in the last term. We then similarly replace $2$ by $(2+3)$ in the first term, and $1$ by $(1+4)$ in the last. Then again the same Schouten identity implies the cancellation. 

Let us now consider the denominator factor $(p_2-x)^2(p_3-x)^2$. The situation is somewhat more interesting here. There are 4 integrands that contribute, namely ${\mathcal A}_{13}, {\mathcal T}^{12}_1, {\mathcal T}^{41}_3$ and the bubble ${\mathcal B}^3$. Let us start by considering the sum of ${\mathcal T}^{41}_3$ and ${\mathcal B}^3$. This can be written as
\be
\frac{\langle q| (p_3-x)\circ 3|q\rangle}{ \langle 41\rangle \langle 12\rangle \langle 23\rangle}
\left( \langle q|(p_1-x)\circ (1+4)|q\rangle \langle 12\rangle - \langle q|(p_2-x)\circ 3|2\rangle\langle q1\rangle \right).
\ee
Using $p_3-p_1=3+2$ in the first term, as well as $1+4=-(2+3)$, we can replace $p_1$ there by $p_3$. Similarly, in the second term we can use $p_3-p_2=3$ to replace $p_2$ with $p_3$. The expression in brackets is then
\be
-\langle q|(p_3-x)\circ (2+3)|q\rangle \langle 12\rangle - \langle q|(p_3-x)\circ (2+3)|2\rangle\langle q1\rangle\nonumber\\= -\langle q|(p_3-x)\circ (2+3)|1\rangle \langle q2\rangle 
\ee
by Schouten identity. This can be written as
\be\label{app-1}
-\langle q|(p_3-x)\circ (2+3)|1\rangle \langle q2\rangle = -\langle q|(p_3-x)\circ (1+2+3)|1\rangle \langle q2\rangle\nonumber\\ = \langle q|(p_3-x)|4] \langle 41\rangle \langle q2\rangle.
\ee

On the other hand, the sum of ${\mathcal A}_{13}$ and ${\mathcal T}^{12}_1$ is given by
\be
\frac{\langle q| (p_3-x)\circ 3|q\rangle}{ \langle 43\rangle \langle 12\rangle \langle 23\rangle}
\left( \langle q|(p_1-x)\circ 1|2\rangle \langle q3\rangle - \langle q|(p_2-x)\circ (1+2)|q\rangle\langle 23\rangle \right).
\ee
We can replace $p_1$ by $p_4$ and $1$ by $(1+2)$ in the first term, and $p_2$ by $p_4$ in the second term. This gives for the expression in the brackets
\be
\langle q|(p_4-x)\circ (1+2)|2\rangle \langle q3\rangle - \langle q|(p_4-x)\circ (1+2)|q\rangle\langle 23\rangle\nonumber\\ = \langle q|(p_4-x)\circ (1+2)|3\rangle \langle q2\rangle.
\ee
This can be written as
\be\label{app-2}
\langle q|(p_4-x)\circ (1+2)|3\rangle \langle q2\rangle= \langle q|(p_4-x)\circ (1+2+3)|3\rangle \langle q2\rangle
\nonumber\\= -\langle q|(p_4-x)|4] \langle 43\rangle \langle q2\rangle.
\ee
Given that we can replace here $p_4$ by $p_3$, it is clear that the terms ${\mathcal A}_{13}, {\mathcal T}^{12}_1, {\mathcal T}^{41}_3, {\mathcal B}^3$ cancel each other.

For the denominator $(p_2-x)^2(p_4-x)^2$ there are only two contributing terms ${\mathcal T}^{12}_2$ and ${\mathcal B}^{12}$, which directly cancel each other. For the denominator $(p_3-x)^2(p_4-x)^2$ we have the terms ${\mathcal A}_{23}, {\mathcal T}^{12}_3, {\mathcal T}^{23}_1, {\mathcal B}^4$ contributing, and the cancelation here is similar to the one encountered in the case $(p_2-x)^2(p_3-x)^2$.

Therefore the total integrand, as a sum of box, four triangles, two internal bubbles and eight external bubbles vanishes. Pictorially, this can be represented as 
\be
\begin{gathered}
\begin{fmfgraph*}(60,45)
\fmfbottom{i1,i2}
\fmftop{o1,o2}
\fmf{photon,tension=3}{i1,v1}
\fmf{photon,tension=3}{i2,v2}
\fmf{photon,tension=3}{o1,v3}
\fmf{photon,tension=3}{o2,v4}
\fmf{photon}{v3,v3'}
\fmf{vanilla}{v3',v1}
\fmf{photon}{v1,v1'}
\fmf{vanilla}{v1',v2}
\fmf{photon}{v2,v2'}
\fmf{vanilla}{v2',v4}
\fmf{photon}{v4,v4'}
\fmf{vanilla}{v4',v3}
\fmflabel{}{o1}
\fmflabel{}{o2}
\fmflabel{}{i1}
\fmflabel{}{i2}
\fmflabel{}{v1'}
\fmflabel{}{v3'}
\fmflabel{}{v2'}
\fmflabel{}{v4'}
\end{fmfgraph*}\nonumber
\end{gathered}~~+~~
4\times\begin{gathered}
\begin{fmfgraph*}(70,40)
\fmfbottom{i1,i2}
\fmftop{o1,o2}
\fmf{photon,tension=3}{i1,v1,o1}
\fmf{photon,tension=3}{i2,v3}
\fmf{photon,tension=3}{o2,v4}
\fmf{vanilla}{v1,v1'}
\fmf{photon}{v1',v2}
\fmf{photon}{v2,v2'}
\fmf{vanilla}{v2',v4}
\fmf{photon}{v4',v4}
\fmf{vanilla}{v4',v3}
\fmf{photon}{v3',v3}
\fmf{vanilla}{v3',v2}
\fmflabel{}{o1}
\fmflabel{}{o2}
\fmflabel{}{i1}
\fmflabel{}{i2}
\fmflabel{}{v1'}
\fmflabel{}{v2'}
\fmflabel{}{v3'}
\fmflabel{}{v4'}
\end{fmfgraph*}
\end{gathered}~~+~~
2\times \begin{gathered}
\begin{fmfgraph*}(70,30)
     \fmftop{i1,i3,i2}
     \fmfbottom{o1,o3,o2}
     \fmf{photon}{i1,v2,o1}
     \fmf{vanilla}{v2,v1'}
     \fmf{photon}{v1',v1}
     \fmf{photon}{v1,i3}
     \fmf{vanilla}{i3,v4}
     \fmf{photon}{v4,o3}
     \fmf{vanilla}{o3,v1}
     \fmf{photon}{v4,v3}
     \fmf{vanilla}{v3,v3'}
     \fmf{photon}{i2,v3',o2}
\end{fmfgraph*}
\end{gathered}~~+~~
4\times \begin{gathered} 
\begin{fmfgraph*}(70,30)
     \fmfleft{i}
     \fmfright{o}
     \fmfbottom{v}
     \fmftop{u}
     \fmf{photon}{i,v1}
     \fmf{photon}{v1,u}
     \fmf{vanilla}{u,v2}
     \fmf{photon}{v2,v}
     \fmf{vanilla}{v,v1}
     \fmf{photon}{v2,v3}
     \fmf{vanilla}{v3,o}
     \fmfblob{.15w}{o}
\end{fmfgraph*}
\end{gathered}~~=0
\ee

\newpage
\section{Shift dependence of self energy bubble}
Consider the two point one loop self energy graph

\vspace{1em} 
\vspace{1em} 
\be 
\begin{fmfgraph*}(150,60)
     \fmfleft{i}
     \fmfright{o}
     \fmfbottom{v}
     \fmftop{u}
     \fmf{photon,tension=3,label=-k}{i,v1}
     \fmf{photon}{v1,u}
     \fmf{vanilla}{u,v2}
     \fmf{photon}{v2,v}
     \fmf{vanilla}{v,v1}
     \fmf{photon,tension=3,label=k}{v2,o}
     \fmflabel{l}{u}
     \fmflabel{l+k}{v}
\end{fmfgraph*}\nonumber 
\ee 
\\~\\
In the diagram as pictured above, the external lines are projected to two negative helicity states and the convention being all external momenta incoming. Using the Feynman rules reviewed in the previous section, the amplitude can be written as 
\begin{equation}
\label{se}
i\Pi^{--}=\int \frac{d^4l}{(2\pi)^4}\frac{\langle q|l|k]\langle q|l+k|k]}{l^2(l+k)^2\langle qk\rangle^2}.
\end{equation}
In the form it is written this integral can be argued to vanish. Indeed, 
there is no linear in $l$ part of the integrand as it is proportional to $\langle q|k|k]=\langle qk\rangle[k k]=0$. The only non-vanishing contribution thus comes from 
\be 
\int \frac{d^4l}{(2\pi)^4}\frac{l_{\mu}l_{\nu}}{l^2(l+k)^2}.
\ee
Any Lorentz invariant regularisation of this will yield $x_{\mu}x_{\nu}$ to be proportional to $k^2 \eta_{\mu\nu}$ which is zero because $k$ is null or to $k_{\mu}k_{\nu}$, which gives the numerator factor $\langle q|k|k]=0$ by using $[kk]=0$. Thus, the self-energy diagram (\ref{se}) can be argued to be zero. 

However, (\ref{se}) is a quadratically divergent integral, and so one must be careful in reaching the conclusion that this object is zero. Let us consider shifting the loop momentum as in $l= x+ \tilde{s}$, where $\tilde{s}$ is some momentum and $x$ is the new integration variable. The argument above depends on the specific form of the integrand and is no longer applicable to the shifted integrand. In fact, below we shall compute the effect of the shift by $\tilde{s}$ and see that the shift is non-vanishing. What this means is that the self-energy diagram projected onto two negative helicity states cannot in general be assumed to vanish. Instead, it is given by a finite quantity, depending on the shift parameter. 

Let us compute the shift dependence of the self-energy diagram. We will use the region momenta so that $k=s-\tilde{s}$, and 
\be 
\begin{split} 
\label{rel}
l&=x+\tilde{s}\\
l+k&=x + s.
\end{split}
\ee 
We then have
\be 
\label{dse}
i\Pi^{--}=\int \frac{d^4x}{(2\pi)^4}\frac{\langle q|x+\tilde{s}|k]\langle q|x+s|k]}{(x+\tilde{s})^2(x+s)^2\langle qk\rangle^2}.
\ee
We have already seen that this integral vanishes after the shift $x\rightarrow x-\tilde{s}$. We then compute the result of the shift. This is done using the standard techniques, which are reviewed in the Appendix of \cite{Chattopadhyay:2020oxe}. The linear part of the shift is given by 
\be 
-i\lim_{x\to\infty}\int\frac{d\Omega}{(2\pi)^4}\tilde{s}_{\mu}x^{\mu}\frac{\langle q|x+\tilde{s}|k]\langle q|x+s|k]}{x^2}\Bigg(1-\frac{2x.(\tilde{s}+s)}{x^2}\Bigg).
\ee 
The non-zero contribution can only come from the quadratic and quartic in $x$ terms. The quadratic term is 
\be 
\tilde{s}_{\mu}x^{\mu}\Big(\langle q|x|k]\langle q|s|k]+\langle q|x|k]\langle q|\tilde{s}|k]\Big).
\ee 
Integrating over the directions of $x^\mu$ produces
\be 
-\frac{i}{32\pi^2}\Big(\langle q|\tilde{s}|k]\langle q|s|k]+\langle q|\tilde{s}|k]\langle q|\tilde{s}|k]\Big).
\ee 
The quartic in $x$ part is given by 
\be 
2i\lim_{x\to\infty}\int\frac{d\Omega}{(2\pi)^4}\tilde{s}_{\mu}x^{\mu}(\tilde{s}+s)_{\nu}x^{\nu}\frac{\langle q|x|k]\langle q|x|k]}{x^4}.
\ee 
The integral is computed using 
\be
\label{quart-YM}
\int\frac{d\Omega}{(2\pi)^4}\frac{x_{\mu}x_{\nu}x_{\rho}x_{\sigma}}{x^4}=\frac{1}{32.6\pi^2}(\eta_{\mu\nu}\eta_{\rho\sigma}+\eta_{\mu\rho}\eta_{\nu\sigma}+\eta_{\mu\sigma}\eta_{\rho\nu}).
\ee 
This results in the following two contributions
\be
\begin{split} 
\frac{i}{32.3\pi^2}\Big(\langle q|\tilde{s}|k]\langle q|\tilde{s}+s|k]+\langle q|\tilde{s}+s|k]\langle q|\tilde{s}|k]\Big)
\\=\frac{i}{16.3\pi^2}\Big(\langle q|\tilde{s}|k]\langle q|\tilde{s}|k]+\langle q|\tilde{s}|k]\langle q|s|k]\Big).
\end{split}
\ee 
For the quadratic part of the shift, the integral is given by 
\be 
\label{quadratic}
\frac{i}{2}\lim_{x\to\infty}\int\frac{d\Omega}{(2\pi)^4}\tilde{s}_{\mu}\tilde{s}_{\nu}x^{\mu}x^2\frac{\partial}{\partial x_{\nu}}\frac{\langle q|x+\tilde{s}|k]\langle q|x+s|k]}{(x+\tilde{s})^2(x+s)^2};
\ee 
When the derivative hits the denominator, it produces a factor proportional to 
\be 
\frac{i}{2}(-4)\lim_{x\to\infty}\int\frac{d\Omega}{(2\pi)^4}\tilde{s}_{\mu}\tilde{s}_{\nu}x^{\mu}x^{\nu}\frac{\langle q|x+\tilde{s}|k]\langle q|x+s|k]}{x^4};
\ee 
The quartic in $x$ part of the numerator is the only which contributes. Using (\ref{quart-YM}) we find one of the contractions vanish and the other two contractions are equal, giving
\be 
-\frac{2i}{32.3\pi^2}\langle q|\tilde{s}|k]\langle q|\tilde{s}|k].
\ee 
When the derivative hits the numerator, in the large $x$ limit, we get 
\be 
\frac{i}{2}\lim_{x\to\infty}\int\frac{d\Omega}{(2\pi)^4}\tilde{s}_{\mu}x^{\mu}\frac{\langle q|x+\tilde{s}|k]\langle q|\tilde{s}|k]+\langle q|x+s|k]\langle q|\tilde{s}|k]}{x^2};
\ee 
Using the relevant contraction, this gives 
\be 
\frac{i}{32\pi^2}\langle q|\tilde{s}|k]^2.
\ee
Adding all the contributions, we have for this amplitude 
\be 
\label{shiftr-ym}
\Pi^{--}= -\frac{i}{32.3\pi^2}\frac{\langle q|\tilde{s}|k]\langle q|s|k]}{\langle qk\rangle^2}.
\ee
Using the fact that $k=s-\tilde{s}$ we could write this result in terms of only $s$ or $\tilde{s}$, but the form we chose will be most convenient below.
We see that the two point one loop self energy bubble projected to same helicity states is in general region momenta dependent. It is not possible to get rid of this dependence unless we set the shift parameter to zero. However, as we will show, this shift dependence helps us to compute the four point one loop amplitude by inserting different combination of states to the two point bubble diagram. The sum of all bubble diagrams lead to independence of region momenta. This is an interesting fact to note because the two point bubble is by itself not independent of region momenta and hence unphysical. While if we sum all the bubbles appropriately, we achieve region momenta independence.  
\\~\\
\newpage
\section{One loop four point amplitude from self energy} 
To compute the one loop four point amplitude, we need to sum over distinct bubble diagrams. There can be two categories of bubbles. One in which we insert the bubble on internal line and for the other we insert it on external lines. Recall, amplitudes in Yang Mills theory and in particular the self-dual sector admit colour decomposition. Thus any such amplitude can be expressed as a sum over colour ordered amplitudes where in each of these sums, a particular cyclic ordering is manifest. Thus we consider a cyclic permutation of external legs for each category of bubble diagrams and this will give us a gauge independent answer. It turns out that for the internal bubbles, there can only be two such distinct cyclic permutations while for the external bubbles, we need four different permutations. We start with the internal bubbles.   
\subsection{Shifts for internal bubbles}
There are two in-equivalent permutations of the bubbles on internal lines. In one diagram, the legs 1 and 2 join on one side of the internal line while 3 and 4 on the other side. Using (\ref{bubb1}), the result for this diagram is 
\be 
\mathcal{B}_{11}= \frac{1}{32.12\pi^2}\frac{\langle q|p_4\circ(1+2)|q\rangle\langle q|p_2\circ(1+2)| q\rangle}{\langle 12\rangle\langle 34\rangle\prod_{j=1}^4\langle qj\rangle}.
\ee
The result for the other internal bubble diagram, after using (\ref{bubb2}) is 
\be 
\mathcal{B}_{12}= \frac{1}{32.12\pi^2}\frac{\langle q|p_3\circ(1+4)|q\rangle\langle q|p_1\circ(1+4)| q\rangle}{\langle 41\rangle\langle 23\rangle\prod_{j=1}^4\langle qj\rangle}.
\ee
\subsection{Shifts for external bubbles}
There are four distinct diagrams here, corresponding to the different ways of insertion of bubbles to four external lines. It is feasible to write each contribution as an insertion of a current to one of the legs of the bubble. The current is the color ordered sum of all tree level diagrams with all but one leg on-shell.
Next we insert each of these currents into the bubbles such that for a particular graph, the internal line adjacent to a bubble is the off-shell leg of the current. Note that the pole resulting from the denominator in the propagator of the internal line,~i.e $1/k^2$, where $k=1,2,..,4$ gets cancelled by the corresponding $k^2$ factor in the numerator of the current, yielding a finite result in the on-shell limit. We now give the shift computed results for these diagrams after using (\ref{b1}) to (\ref{b4}). We have
\be 
\mathcal{B}_{21}= -\frac{1}{32.12\pi^2}\frac{\langle q3\rangle\langle q|p_4|1]\langle q|p_1|1| q\rangle}{\langle 23\rangle\langle 34\rangle\prod_{j=1}^4\langle qj\rangle},
\ee
\be 
\mathcal{B}_{22}= -\frac{1}{32.12\pi^2}\frac{\langle q4\rangle\langle q|p_1|2]\langle q|p_2|1| q\rangle}{\langle 14\rangle\langle 23\rangle\prod_{j=1}^4\langle qj\rangle},
\ee
\be 
\mathcal{B}_{23}= -\frac{1}{32.12\pi^2}\frac{\langle q1\rangle\langle q|p_3|3]\langle q|p_2|3| q\rangle}{\langle 41\rangle\langle 12\rangle},
\ee
\be 
\mathcal{B}_{24}= -\frac{1}{32.12\pi^2}\frac{\langle q2\rangle\langle q|p_3|4]\langle q|p_4|4| q\rangle}{\langle 32\rangle\langle 21\rangle\prod_{j=1}^4\langle qj\rangle}.
\ee
\subsection{Summing the results} 
Let us now collect and add all the results for the bubble diagrams. We need to get rid of the dual momentum variables since the physical amplitude should be independent of them. Thus, under any arbitrary shift of these variables, say $p_i\rightarrow p_i+b$, for any $b$, the amplitude should be invariant, although each individual bubbles are not. We use this important fact and first solve for each dual momenta $p_i,i=1,2,3$ in terms of $p_4$ and some combination of usual line momenta. We have 
\be
\label{dualm}
\begin{split} 
p_1&=p_4+1,\\
p_2&=p_4+1+2,\\
p_3&=p_4+1+2+3.
\end{split} 
\ee 
Then each bubble contribution is a function of $p_4$ and line momenta. Let us first write the sum total of all the contributions
\be 
\label{amplitude}
\begin{split} 
\mathcal{A}_4&=\frac{1}{32.12\pi^2\prod_{j=1}^4\langle qj\rangle}\Bigg[\frac{\langle q|p_4\circ(1+2)|q\rangle\langle q|p_4\circ(1+2)| q\rangle}{\langle 12\rangle\langle 34\rangle}\\&+\frac{\langle q|(p_4+1)\circ(2+3)|q\rangle\langle q|(p_4+1)\circ(2+3)| q\rangle}{\langle 41\rangle\langle 23\rangle}\\
&-\frac{\langle q3\rangle\langle q|p_4|1]\langle q|p_4|1| q\rangle}{\langle 23\rangle\langle 34\rangle}-\frac{\langle q4\rangle\langle q|p_4+1|2]\langle q|p_4+2|1| q\rangle}{\langle 14\rangle\langle 23\rangle}\\
&-\frac{\langle q1\rangle\langle q|p_4-4|3]\langle q|p_4-4|3| q\rangle}{\langle 41\rangle\langle 12\rangle}-\frac{\langle q2\rangle\langle q|p_4|4]\langle q|p_4|4| q\rangle}{\langle 32\rangle\langle 21\rangle}\Bigg].
\end{split}
\ee 
Now let us consider the shift $p_i\rightarrow p_i-p_4$. This means that terms with a factor of $p_4$ vanish identically, while the others contribute. We then have for the amplitude in (\ref{amplitude}) 
\be 
\label{amplitude2}
\begin{split} 
\mathcal{A}_4=\frac{1}{32.12\pi^2\prod_{j=1}^4\langle qj\rangle}\Bigg[\frac{\langle q|1\circ4|q\rangle\langle q|1\circ4| q\rangle}{\langle 41\rangle\langle 23\rangle}
-\frac{\langle q4\rangle\langle q|1|2]\langle q|2|1| q\rangle}{\langle 14\rangle\langle 23\rangle}
\\-\frac{\langle q1\rangle\langle q|4|3]\langle q|4|3| q\rangle}{\langle 41\rangle\langle 12\rangle}\Bigg]\\
=\frac{1}{32.12\pi^2}\Bigg[\frac{[14]^2\langle q1\rangle\langle q4\rangle}{\langle 41\rangle\langle 23\rangle\langle q2\rangle\langle q3\rangle}+\frac{[12]^2\langle q1\rangle}{\langle 14\rangle\langle 43\rangle\langle q3\rangle}+\frac{[43]^2\langle q4\rangle}{\langle 41\rangle\langle 12\rangle\langle q2\rangle}\Bigg].
\end{split}
\ee 
Let us simplify the second line in (\ref{amplitude2}). We first add the last two terms, giving
\be 
\label{simplify}
\frac{[12]^2\langle q1\rangle}{\langle 14\rangle\langle 43\rangle\langle q3\rangle}+\frac{[43]^2\langle q4\rangle}{\langle 41\rangle\langle 12\rangle\langle q2\rangle}=\frac{[12][34]\Big(\langle 34\rangle\langle q2\rangle\langle q1\rangle+\langle 12\rangle\langle q4\rangle\langle q3\rangle\Big)}{\langle 14\rangle\langle 12\rangle\langle 43\rangle\langle q2\rangle\langle q3\rangle}.
\ee 
Next we use the Schouten identity for the terms in brackets on the right hand side of (\ref{simplify})
\be 
\begin{split} 
\langle q1\rangle\langle 34\rangle=\langle q3\rangle\langle 14\rangle-\langle q4\rangle\langle 13\rangle, \\
\langle q3\rangle\langle 12\rangle=\langle q1\rangle\langle 32\rangle-\langle q2\rangle\langle 31\rangle .
\end{split} 
\ee 
Plugging this back into (\ref{simplify}), we get 
\be 
\frac{[12]^2\langle q1\rangle}{\langle 14\rangle\langle 43\rangle\langle q3\rangle}+\frac{[43]^2\langle q4\rangle}{\langle 41\rangle\langle 12\rangle\langle q2\rangle}=-\frac{[12][34]}{\langle 12\rangle\langle 34\rangle}+\frac{[12][34]\langle 32\rangle}{\langle 14\rangle\langle 12\rangle\langle 43\rangle}\frac{\langle q1\rangle\langle q4\rangle}{\langle q2\rangle\langle q3\rangle}.
\ee 
Then we have for the amplitude in (\ref{amplitude2}),
\be 
\mathcal{A}_4=\frac{1}{32.12\pi^2}\Bigg[-\frac{[12][34]}{\langle 12\rangle\langle 34\rangle}+\Bigg(\frac{[12][34]\langle 32\rangle}{\langle 14\rangle\langle 12\rangle\langle 43\rangle}+\frac{[14]^2}{\langle 41\rangle\langle 23\rangle}\Bigg)\frac{\langle q1\rangle\langle q4\rangle}{\langle q2\rangle\langle q3\rangle}\Bigg].
\ee 
It is now easy to see that the $q$-dependent term vanish, owing to momentum conservation, giving the result
\be 
\label{result}
\mathcal{A}_4=-\frac{1}{32.12\pi^2}\frac{[12][34]}{\langle 12\rangle\langle 34\rangle}.
\ee 
Thus we reproduce the four point amplitude from a bubble computation. 
\newpage 
\section{General $n$-point amplitude}
The analysis of the four point one-loop diagrams tells us that the box amplitude can be expressed as a sum over diagrams with bubble insertions. The bubble diagrams are simpler to compute and thus the complicated box integral can be avoided in order to extract the result. However, unlike the box, the bubble diagrams are quadratically divergent to start with. Thus, shifting the loop momentum variable in any one bubble diagram changes the result of the computation. This however is not in contradiction with the amplitude result, because one needs to sum over different bubble diagrams to get the final amplitude and the final sum is independent of any shift of the loop variable. This is an interesting fact to note because any single bubble diagram is divergent and sensitive to shifts. However, when we take an appropriate sum of bubbles, the shift dependence disappears and the obtained result is unambiguous. \\~\\
The technique presented here in the covariant formalism was used by several authors earlier in the light-cone formalism \cite{Chakrabarti:2005ny, Brandhuber:2007vm}. This has been carried out first in QCD in the light cone gauge and subsequently in Yang Mills and chiral high spin gravity \cite{Skvortsov:2020gpn}. In QCD and Yang Mills, the analysis is mainly done for the particular case of four points. In their case, the proof for the sum of integrands being zero is carried out by using standard results in complex analysis. Whereas in our case, we use two specific identities, in particular momentum conservation and the Schouten identity to show that the sum vanishes. Also, they use dimensional regularization to compute the self-energy bubble whereas in our case, it is the method of shifts that gives us the result for the self-energy. We know that there are obvious difficulties to use dimensional regularization when we are dealing with two component spinors and this is why it is reasonable to avoid it. Another important point to emphasize is they interpret the shift dependence of the self-energy bubble as an artifact of a very specific regulator. However, in our case, in the absence of any such regulators, it is evident that this is not an artifact of some regularization scheme but it is true for a divergent graph solely because of the ambiguity of assigning a loop momentum variable to its internal lines, independent of whatever dimensions one is computing in. The analysis in \cite{Chakrabarti:2005ny} is limited to four point amplitude. In \cite{Brandhuber:2007vm}, the MHV rules were used to represent the four point amplitude as a sum over two point insertions. They also argued about generalising it to $n$-point one loop amplitudes. However, their analysis is once carried out in the light cone gauge and thus does not take into account the role played by currents. Thus, their formula for the one-loop amplitude is different from what we are going to present.
\\~\\
It is not unreasonable to extend this technique for amplitudes with higher number of external legs. Indeed, to obtain the four point amplitude, we simply inserted pairs of all possible combinations of lower order Berends-Giele currents to the bubble and summed them up. Likewise, we can generalise this for any higher leg amplitude. In particular, for the $n$-point one-loop amplitude, we first form pairs of currents of orders $a<n$ and $b<n$, such that $a+b=n$ and compute the sum of all such possible combinations. Our diagrams will be expressed in terms of region momenta. However, the amplitude must be translation invariant in the region momentum space. Thus any shift of all the region momenta by the same amount should not change the result of the amplitude. In such a case, we need to ensure region momentum independence for the sum over bubbles. After we show that such a sum is independent of region momenta, we can start computing the sum and extract the $q$-independent result. Let us then explain our new formula for the series of same helicity one-loop amplitudes and subsequently give explicit computations for the three, four and five point cases. To do so, we start with the self-energy bubble and interpret it as an effective propagator.

\subsection{Bubble as an effective propagator} 
\be 
\begin{fmfgraph*}(200,130)
     \fmfleft{i}
     \fmfright{o}
     \fmf{vanilla}{i,v1}
     \fmf{vanilla,left,tension=0.6,tag=1,label=$l+k$}{v2,v1}
     \fmf{vanilla,left,tension=0.6,tag=1,label=$l$}{v1,v2}
     \fmf{vanilla}{o,v2}
     \fmflabel{-k}{i}
     \fmflabel{k}{o}
\end{fmfgraph*}\nonumber
\ee 
\\~\\
Consider the bubble diagram above, where the external lines are projected to two positive helicity states and the convention being all external momenta incoming.
A direct computation of shifts gives the result for the amplitude, upto overall numerical factors,
\be 
\label{shift}
\Pi^{--}_{SDYM}\approx\frac{\langle q|s|k]\langle q|\tilde{s}|k]}{\langle qk\rangle^2}.
\ee
At this stage, we keep the parameter $s$ to be arbitrary. The above diagram can be a part of a sub-diagram in a 2 to 2 scattering process and we will later choose the shift parameter accordingly. Alternatively, we can write the one loop correction to the propagator in the form of an operator 
\be
\label{ope}
\begin{gathered}
\begin{fmfgraph*}(140,70)
     \fmfleft{i}
     \fmfright{o}
     \fmf{vanilla}{i,v1}
     \fmfdot{v1}
     \fmf{vanilla}{o,v1}
     \fmflabel{$MM'$}{i}
     \fmflabel{$NN'$}{o}
     \end{fmfgraph*}
     \end{gathered}~~~~~~~~~~:=~~~~~s^{~M}_{N'} \tilde{s}_{N}^{~M'}
\ee 
where the dot signify that it is a one-loop corrected graph.
We can then start gluing currents to each of the legs of this graph. This implies contracting the indices of the operator with that of the current(s). For instance, when polarization states of the form $\epsilon_{MM'}=\frac{q_Mk_{M'}}{\langle qk\rangle}$ are glued to both the legs, it results in (\ref{shift}).
\subsection{Gluing BG currents to bubble}
The effective one-loop propagator (\ref{ope}) gives a non-vanishing expectation value of the product of two
connection states. This means that it can be used to glue together the off-shell legs of the BG currents.
Our main claim is that this gives the correct one-loop same helicity amplitude of SDYM theory.
To see how this arises, we consider a colour ordered diagram with n external lines. Because the
diagram is colour ordered we can adopt the use of region momenta $p_i$ so that the null momenta on the
external legs are given by the difference of the region momenta
\be 
k_i=p_i-p_{i-1},~~p_n\equiv p_0.
\ee 
This ensures momentum conservation, but introduces indeterminacy in that all region momenta can
be shifted by an arbitrary amount without changing the external momenta.
The amplitude can be written as a sum over partitions of the set of external states into two groups.
Each group of states then produces a current, and two such currents are glued by the effective propagator (\ref{ope}). The amplitude is colour ordered, and so the states in one of the groups can be numbered
as those starting with the external particle $i$ and ending with particle $j$. This gives the following
amplitude
\\~\\
\be 
\label{YM-general1}
\mathcal{A}=\sum_{part} \begin{gathered}
~~~\begin{fmfgraph*}(120,70)
     \fmfleft{i,i1,i2,i3,i4}
     \fmfright{o1,o4,o5,o6,o7}
     \fmf{vanilla}{i,v}
     \fmf{vanilla}{i4,v}
     \fmfblob{.15w}{v}
     \fmf{vanilla,label=$p_j$,label.distance=1cm,label.side=left}{v,v1}
     \fmfdot{v1}
     \fmf{vanilla,label=$p_{i-1}$,label.distance=1cm,label.side=right}{v1,o}
     \fmfblob{.15w}{o}
     \fmf{vanilla}{o,o1}
     \fmf{vanilla}{o,o7}
     \fmflabel{$i$}{i}
      \fmflabel{$.$}{i1}
     \fmflabel{$.$}{i2}
      \fmflabel{$.$}{i3}
      \fmflabel{$j$}{i4}
     \fmflabel{$i-1$}{o1}
      \fmflabel{$.$}{o4}
      \fmflabel{$.$}{o5}
      \fmflabel{$.$}{o6}
     \fmflabel{$j+1$}{o7}    
     \end{fmfgraph*}
     \end{gathered}
     \quad\quad
\ee 
\be 
=\sum_{i=1}^{n/2}J(i,..j)J(j+1,..i-1)\langle q|p_j\circ (k_i+..+k_j)|q\rangle\langle q|(k_{i+1}+...+k_{i-1})\circ p_{i-1}|q\rangle.\nonumber 
\ee 
Using 
\be 
k_i+...+k_j=p_j-p_{i-1}=-(k_{j+1}+...+k_{i-1})
\ee 
we can write this compactly as 
\be 
\label{BG}
\mathcal{A}=\sum_{part}J(i,..,j)J(j+1,..,i-1)\langle q|p_j\circ p_{i-1}|q\rangle^2.
\ee 
\subsection{Collinear limit}
For a digression, let us consider the MHV tree amplitude in YM. This is the $n$-point tree amplitude where $(n-2)$ gluons have the same helicity and the remaining two gluons have the opposite helicity. The representation for this amplitude is given by the famous Park-Taylor formula 
\be 
\label{Park}
A(1^-,2^-,....,j^+,l^+,...,n^-)=\frac{\langle jl\rangle^4}{\langle 12\rangle\langle 23\rangle\langle 34\rangle....\langle n1\rangle}. 
\ee 
Next, consider the limit where any two adjacent gluons becomes collinear or parallel to each other. Without loss of generality, we take the gluons $k_2$ and $k_3$ to be collinear. This limit is singular because the intermediate momentum $P=k_2+k_3$ becomes null in the collinear limit 
\be 
P^2=(k_2+k_3)^2=2k_2.k_3\rightarrow 0.
\ee 
Let us then specify the fraction of the total momentum carried by the individual momenta $k_2$ and $k_3$. Where confusions do not arise, we write $k_i\equiv i$ for each of the momenta. 
\be 
2\approx zP, ~~~~ 3\approx (1-z)P
\ee 
where $z\in [0,1]$. We can then construct the corresponding relations between the spinors. In particular, the above relations imply 
\be 
\label{rel} 
\lambda_2\approx \sqrt{z}\lambda_P, ~~~~\lambda_3\approx \sqrt{1-z}\lambda_P\nonumber,\\
\lambda'_2\approx \sqrt{z}\lambda'_P, ~~~~\lambda'_3\approx \sqrt{1-z}\lambda'_P,
\ee 
where $\lambda_i$ is the unprimed spinor corresponding to the null momentum $i$ and $\lambda'_i$ is the primed spinor.
Using (\ref{rel}), we can write (\ref{Park}) in the limit when $3\parallel 4$ as  
\be 
\label{new park} 
A(1^-,2^-,....,j^+,l^+,...,n^-)\xrightarrow{3\parallel 4}\Bigg(\frac{1}{\sqrt{z(1-z)}\langle 23\rangle}\Bigg)\frac{\langle jl\rangle^4}{\langle 1P\rangle\langle P4\rangle....\langle n1\rangle}\nonumber \\
=\Bigg(\frac{1}{\sqrt{z(1-z)}\langle 23\rangle}\Bigg)A(1^-,P^-,....,j^+,l^+,...,n^-).
\ee

At one-loop, we have the following collinear limit of colour ordered amplitudes in YM
\be 
\label{coloop}
A^{\textrm{loop}}_{n;1}\xrightarrow{a\parallel b}\sum_{\lambda=\pm}\textrm{Split}^{\textrm{tree}}_{-\lambda}(a^{\lambda_a},b^{\lambda_b}) A^{\textrm{loop}}_{n-1;1}(....(a+b)^{\lambda},...)\nonumber\\+\textrm{Split}^{\textrm{loop}}_{-\lambda}(a^-,b^-)A^{\textrm{tree}}_{n-1}(....,(a+b)^{
\lambda},...)
\ee 
in the limit where two of the momenta, say $a$ and $b$ become collinear, such that $a\rightarrow zP$ and $b\rightarrow (1-z)P$. In particular, in the same helicity sector, the second term in (\ref{coloop}) drops out because the split factor in this case multiples a tree amplitude with all same helicity, which vanishes. Then, for the case of the all same helicity one loop amplitudes, the collinear limit takes the form 
\be 
\label{coloop2}
A^{\textrm{one-loop}}_n(1^-,2^-,...,n^-)\xrightarrow{a\parallel b}\textrm{Split}^{\textrm{tree}}_-(a^-,b^-) A^{\textrm{loop}}_{n-1}(....(a+b)^-,...),
\ee 
where the splitting amplitude in this case is given by
\be 
\textrm{Split}_-^{\textrm{tree}}(a^-,b^-)=\frac{1}{\sqrt{z(1-z)}\langle ab\rangle}. 
\ee 
Note, it resembles the collinear factorization of the Park-Taylor formula. Let us now write the current representation of the one-loop all same helicity amplitude. 
\\~\\
\be 
\mathcal{A}_n(1^-,2^-,...,n^-)
=\sum_{part}J(i,..,j)J(j+1,..,i-1)\langle q|p_j\circ p_{i-1}|q\rangle^2.
\ee 
where $p_{j}$ is the region momentum bounded by the lines $j$ and $j-1$.
In the sum over permutations of external legs, there are two currents which are glued to the bubble in each term. Let us now choose a pair of adjacent momenta $(a,b)$ which become collinear to each other. Then there will be two cases depending upon which of the two currents do the momenta belong.
\subsubsection{Case I}
In the first case, both the momenta can belong to the same current. The current however has a similar denominator structure like the Park-Taylor formula and therefore admits a similar splitting behaviour. In particular, if we consider the pair of momenta $(a,b)$ belongs to the current $J(1,...i)$, then in the limit of these two momenta becoming collinear, we have 
\be 
J(1,...a,b,..i)\xrightarrow{a\parallel b}\Bigg(\frac{1}{\sqrt{z(1-z)}\langle ab\rangle}\Bigg)\frac{1}{\langle q1\rangle\langle 12\rangle...\langle jP\rangle\langle P(j+3)\rangle..\langle (i-1)i\rangle}\nonumber\\
=\Bigg(\frac{1}{\sqrt{z(1-z)}\langle ab\rangle}\Bigg)J(1,...P,...i).
\ee 
If we relabel the arguments of the current on the right hand side of the above equation, we recover that 
\be 
J(1,...a,b,..i)\xrightarrow{a\parallel b}\Bigg(\frac{1}{\sqrt{z(1-z)}\langle ab\rangle}\Bigg)J(1,...(i-1)).
\ee 
Thus the $i^{th}$ current goes to the $(i-1)^{th}$ current times the usual split factor. A similar thing happens for the other current, i.e in some of the terms of the amplitude, when the pair $(a,b)$ belongs to the current $J(i,...n)$, then in the limit when $a\parallel b$, the current goes like $J(i,..a,b,..n)\rightarrow J(i,..P,..(n-1))$. 
\subsubsection{Case II}
The next case is when the two momenta belong to two different currents. Thus, we can have momentum $'a'$ belonging to the current $J(1,...,i)$ while momentum $'b'$ belonging to the current $J(i,...,n)$ or vice versa. Let us denote the region momentum bounded by the two line momenta $a,b$ be $p_{ab}$. Since the momenta $a,b$ are adjacent, the term in the amplitude in such a case should have the following structure 
\be 
\label{term}
J(i,..,a)J(b,..,i-1)\langle q|p_{ab}\circ p_{i-1}|q\rangle^2.
\ee 
However, we can use region momentum independence and thus can choose the region momentum bounded by the lines $a,b$ to be zero. In particular, we have
\be 
p_{ab}\xrightarrow{a\parallel b}0.
\ee 
It is then easy to see that all such terms analogous to the one in (\ref{term}) vanish in the collinear limit. The remaining non-zero terms are entirely the ones which fall in the first case and can be written in the following way, after some relabelling of the momentum indices
\be
\label{eqn1}
\mathcal{A}_n\xrightarrow{a\parallel b}\textrm{Split}^{\textrm{tree}}_-(a^-,b^-)\sum_{part}J(i-1,..,j-1)J(j,..,i)\langle q|p_{j-1}\circ p_{i}|q\rangle^2.
\ee 
However, the right hand side of (\ref{eqn1}) is the relevant expression for the $(n-1)$ amplitude $\mathcal{A}_{n-1}(1^-,...,(n-1)^-)$. Thus, we showed that the current representation of the general $n$-point amplitude follows the correct collinear behaviour like the one in (\ref{coloop2}) as two of the momenta become collinear.

\subsection{Alternative ways of writing the formula}

Using the fact that 
\be
p_j = p_{i-1} + \sum_{l=i}^j k_l,
\ee
we can rewrite the formula (\ref{YM-general1}) lowering the power of the region momenta in it
\be
{\cal A}= \sum_{part} J(i,\ldots, j) J(j+1,\ldots, i-1) \langle q| p_j \circ (\sum_{l=i}^j k_l) |q\rangle^2.
\ee

We can also rewrite the sum over partitions as a sum over cyclic permutations of the set $1,\ldots, n$. Indeed, it is easy to check that
\be\label{formula-cyclic}
2 {\cal A} = \sum_{i=1}^{n-1} J(1,\ldots, i) J(i+1,\ldots,n) \langle q| p_i \circ ( k_{i+1} + \ldots + k_n)| q\rangle^2 + {\rm cyclic},
\ee
where we need to add all cyclic permutations of the set $(1,\ldots, n)$. This form of the formula is particularly convenient for establishing the region momentum independence. Written in this way the amplitude formula is very similar to the one that appears in \cite{Skvortsov:2020gpn} in the light-cone gauge

\subsection{Region momentum independence}

The purpose of this subsection is to argue that the amplitude is invariant under shifts of all region momenta by the same amount. Combined with our collinear limit argument, this gives a proof of the formula (\ref{YM-general1}).
.

\subsubsection{Quadratic part of the dependence}

When we shift all region momenta by some value $x$ there are both quadratic and linear in $x$ terms that appear. Using (\ref{formula-cyclic}), the part quadratic in the shift  can be written as 
\be
{\cal A}^{x^2}= \sum_{i=1}^{n-1} J(1,\ldots, i) J(i+1,\ldots,n) \langle q| x \circ ( k_{i+1} + \ldots + k_n)| q\rangle^2 + {\rm cyclic}.
\ee
Because $x$ here is an arbitrary vector, so is the primed spinor $\langle q|x|:=\mu$. This means that we must consider
\be
\sum_{i=1}^{n-1} J(1,\ldots, i) J(i+1,\ldots,n) [\mu |  ( k_1 + \ldots + k_i)| q\rangle  [\mu | k_{i+1} + \ldots + k_{n}|q\rangle \nonumber\\ + {\rm cyclic},
\ee
where we wrote the expression more symmetrically. This can be computed using the identity
\be
\sum_{i=1}^{n-1} J(1,\ldots, i) J(i+1,\ldots, n) [ \mu | k_1+ \ldots + k_i | q\rangle [\mu | k_{i+1} + \ldots + k_{n}|q\rangle \nonumber\\= 
\frac{ [\mu | \sum_{i<j} i\circ j| \mu]}{\langle q1\rangle \langle 12\rangle \ldots \langle (n-1) n\rangle \langle nq\rangle},
\ee
which holds for arbitrary momenta $1,\ldots n$. The momenta in this formula are not assumed to add up to zero. This formula is proven analogously to how the recursive formula for the Berends-Giele currents is established. 

We will also need the identity
\be\label{identity}
\sum_{i=1}^{n-1} \frac{\langle i(i+1)\rangle}{\langle iq\rangle \langle (i+1)q\rangle} =\frac{\langle 1n\rangle}{\langle 1q\rangle \langle nq\rangle},
\ee
which is a simple consequence of Schouten identity. It can also be written as
\be
\sum_{i=1}^{n} \frac{\langle i(i+1)\rangle}{\langle iq\rangle \langle (i+1)q\rangle} =0,
\ee
with the convention that $n+1=1$. Using this identity we have
\be
\begin{split} 
&\sum_{i=1}^{n-1} J(1,\ldots, i) J(i+1,\ldots,n) [\mu |  ( k_1 + \ldots + k_i)| q\rangle  [\mu | k_{i+1} + \ldots + k_{n}|q\rangle  + {\rm cyclic} \\&=
\frac{2}{ \langle 12\rangle \ldots \langle (n-1)n\rangle \langle n1\rangle}
\left( \sum_{i<j} \frac{ [\mu| i\circ j|\mu] (ij)}{\langle iq\rangle \langle jq\rangle}\right).
\end{split} 
\ee
No momentum conservation has yet been used. It is then easily checked that when the momentum conservation e.g. in the form $-n=1+\ldots + (n-1)$ is used, the coefficients in front of independent $[\mu|i\circ j|\mu]$ factors with $i,j = 1,\ldots, n-1$ vanish. Thus, the quadratic in $x$ part of dependence of the amplitude on the region momentum vanishes.

\subsubsection{Linear part of the dependence}

Taking the first variation of the amplitude as all region momenta vary, and denoting $\langle q| x|=[\mu|$ as before, we get a multiple of
\be
\begin{split} 
&\sum_{i=1}^{n-1} J(1,\ldots, i) J(i+1,\ldots,n) [\mu| ( k_{i+1} + \ldots + k_n)| q\rangle 
\langle q| p_i \circ ( k_{i+1} + \ldots + k_n)| q\rangle
 \\&+{\rm cyclic}.
\end{split} 
\ee
The idea is again to compute the sum here explicitly, similar to what one does in the check of the Berends-Giele formula for the all same helicity currents. This is an exercise in applying Schouten identity. The result is
\be
\begin{split}
\label{proof-1}
&\sum_{i=1}^{n-1} J(1,\ldots, i) J(i+1,\ldots,n) [\mu| ( k_{i+1} + \ldots + k_n)| q\rangle 
\langle q| p_i \circ ( k_{i+1} + \ldots + k_n)| q\rangle\\&=
\frac{1}{ \langle 1q\rangle \langle 12\rangle \ldots \langle (n-1) n\rangle \langle nq\rangle} 
\left( \sum_{i<j} [\mu| i\circ j\circ p_i|q\rangle - \sum_{i=1}^{n-2} [\mu | i| q\rangle\sum_{j>i}^n\sum_{l>j}^n s_{jl}  \right),
\end{split}
\ee
where the last terms contain Mandelstam variables $s_{ij}:= \langle ij\rangle [ij]$ and arise from relating the region momenta to each other via relations of the type $p_{i+1}=p_i + k_{i+1}$. 

It remains to add the cyclic permutations, and then apply the momentum conservation. Taking the cyclic permutation of the first set of terms in brackets in (\ref{proof-1}), and using (\ref{identity}) gives
\be
\begin{split}
&-\frac{1}{  \langle 12\rangle \ldots \langle (n-1) n\rangle \langle n1\rangle  } 
\left( \sum_{i=1}^{n-1}\sum_{j>i} \frac{\langle ij\rangle}{\langle iq\rangle \langle jq\rangle} ( [\mu|i\circ j\circ p_i|q\rangle - [\mu|j\circ i\circ p_j|q\rangle)\right).
\end{split} 
\ee
We now use the momentum conservation to express the last momentum $k_n$ in terms of all the rest. After this, we collect the terms in front of similar $[\mu| i\circ j\circ p|q\rangle$ expressions. Using (\ref{identity}) one more time we get for these terms
\be
\frac{1}{  \langle 12\rangle \ldots \langle (n-1) n\rangle \langle n1\rangle  }  \sum_{i=1}^{n-1} \sum_{j=1}^{n-1} \frac{\langle in\rangle}{\langle iq\rangle \langle nq\rangle} [\mu| j\circ i\circ (p_n-p_j)|q\rangle. 
\ee
This is now written in terms of differences of region momenta, and so depends just on the external momenta $p_n-p_j =  -(k_1+ \ldots + k_j)$. Using identities of the form $i\circ l=-l\circ i-s_{il}\mathds{1}$ for some arbitrary momentum null momentum $l$, we can write the above in terms of Mandelstam variables. The compact formula is given by 
\be 
\label{lin}
-\frac{1}{  \langle 12\rangle \ldots \langle (n-1) n\rangle \langle n1\rangle  }  \sum_{i=1}^{n-1}[\mu|i|q\rangle \sum_{j=1}^{n-1} \frac{\langle jn\rangle}{\langle jq\rangle \langle nq\rangle} \sum_{l=1}^is_{jl} 
\ee
Let us add the cyclic permutations of the second group of terms. Using (\ref{identity}) repeatedly, we extract the coefficients of $[\mu|i|q\rangle s_{jl}$ factors. Then we use momentum conservation and write the momentum $k_n$ in terms of the others. We get the exact similar terms as in (\ref{lin}), but with opposite signs. Thus all these terms cancel each other and the linear part vanishes. 
 \subsection{3-point amplitude} 
Using (\ref{BG}) we have for the 3-point amplitude
\be 
\mathcal{A}_3 = J(1, 2)J(3)\langle q|p_2 \circ p_3|q\rangle^2 + J(3, 1)J(2)\langle q|p_1\circ p_2|q\rangle^2\nonumber\\ + J(2, 3)J(1)\langle q|p_3\circ p_1|q\rangle^2.
\ee 
We would like to understand the dependence of this on the region momenta. To this end, it is helpful
to reduce the power of the region momenta in the expression. Thus, we write the amplitude in the
form (28)
\be 
\mathcal{A} = J(1, 2)J(3)\langle q|p_2\circ k_3|q\rangle^2 + J(3, 1)J(2)\langle q|p_1\circ k_2|q\rangle^2 \nonumber\\+ J(2, 3)J(1)\langle q|p_3\circ k_1|q\rangle^2
\ee 
We now substitute the explicit form of the currents, and simplify the arising expression to get
\be 
\mathcal{A} =\frac{\langle q|p_2|3]^2}{\langle 1q\rangle\langle 12\rangle\langle 2q\rangle}+\frac{\langle q|p_1|2]^2}{\langle 3q\rangle\langle 31\rangle\langle 1q\rangle}+\frac{\langle q|p_3|1]^2}{\langle 2q\rangle\langle 23\rangle\langle 3q\rangle}.
\ee
If we now parametrise all region momenta in terms of one of them and the external momenta
\be 
p_1=x,~ p_2=x+k_2,~p_3=x+k_2+k_3=x-k_1
\ee 
we get
\be 
\mathcal{A} =\frac{\langle q|x+2|3]^2}{\langle 1q\rangle\langle 12\rangle\langle 2q\rangle}+\frac{\langle q|x|2]^2}{\langle 3q\rangle\langle 31\rangle\langle 1q\rangle}+\frac{\langle q|x|1]^2}{\langle 2q\rangle\langle 23\rangle\langle 3q\rangle}.
\ee 
Bringing the last two terms to the common denominator we have for them
\be 
\label{form}
\frac{1}{\langle 1q\rangle\langle 2q\rangle\langle 3q\rangle\langle 31\rangle\langle 23\rangle}(\langle 23\rangle\langle 2q\rangle\langle q|x|2]^2+\langle 31\rangle\langle 1q\rangle\langle q|x|1]^2).
\ee 
We now have the momentum conservation that can be written in the form
\be 
\label{mom}
\langle q|x|1]1 + \langle q|x|2]2 + \langle q|x|3]3 = 0.
\ee 
Projecting onto $|3\rangle$ we get
\be 
\langle q|x|1]\langle 13\rangle + \langle q|x|2]\langle 23\rangle = 0
\ee 
which means we can write (\ref{form}) as
\be 
\frac{1}{\langle 1q\rangle\langle 2q\rangle\langle 3q\rangle\langle 31\rangle\langle 23\rangle}(\langle 23\rangle\langle 2q\rangle\langle q|x|2]^2+\langle 31\rangle\langle 1q\rangle\langle q|x|1]^2)
=\frac{\langle q|x|3]^2}{\langle 1q\rangle\langle 2q\rangle\langle 21\rangle},
\ee 
where we used (\ref{mom}) another time (projected onto $(|q\rangle)$ to get the first equality, and projected onto
$|1\rangle$ to obtain the last expression. This means that
\be 
\mathcal{A} = \frac{1}{\langle 1q\rangle\langle 2q\rangle\langle 12\rangle}(\langle q|x+2|3]^2-\langle q|x|3]^2).
\ee 
The quadratic in $x$ part cancels in the above expression. The linear part in $x$ can be written in terms of Mandelstam variables $s_{ij}$. However, all such Mandelstam variables vanish at three points. Thus the amplitude vanishes.  
\subsection{4-point amplitude} 
We substitute the expressions for the currents in this case and we get the four point amplitude 
\be 
\begin{split} 
\label{4pt}
\mathcal{A}_4&=\frac{1}{\langle q1\rangle\langle 1q\rangle}\frac{1}{\langle q2\rangle\langle 23\rangle\langle 34\rangle\langle 4q\rangle}\langle q|p_1\circ p_4|q\rangle^2+ 
\frac{1}{\langle q2\rangle\langle 2q\rangle}\frac{1}{\langle q3\rangle\langle 34\rangle\langle 41\rangle\langle 1q\rangle}\langle q|p_2\circ p_1|q\rangle^2\\&+
\frac{1}{\langle q3\rangle\langle 3q\rangle}\frac{1}{\langle q4\rangle\langle 41\rangle\langle 12\rangle\langle 2q\rangle}\langle q|p_3\circ p_2|q\rangle^2+
\frac{1}{\langle q4\rangle\langle 4q\rangle}\frac{1}{\langle q1\rangle\langle 12\rangle\langle 23\rangle\langle 3q\rangle}\langle q|p_4\circ p_3|q\rangle^2\\&+
\frac{1}{\langle q1\rangle\langle 12\rangle\langle 2q\rangle}\frac{1}{\langle q3\rangle\langle 34\rangle\langle 4q\rangle}\langle q|p_4\circ p_2|q\rangle^2\\&+
\frac{1}{\langle q2\rangle\langle 23\rangle\langle 3q\rangle}\frac{1}{\langle q4\rangle\langle 41\rangle\langle 1q\rangle}\langle q|p_3\circ p_1|q\rangle^2.
\end{split} 
\ee 
The amplitude is written with explicit region momentum dependence. But it can be shown to be region
momentum independent, i.e. invariant under the shift of all region momenta. Assuming that this is
the case it is easy to extract the answer for this amplitude. We parametrise all region momenta in
terms of one of them, e.g. $p_1=x$, and the external momenta. We have 
\be 
p_1=x,~p_2=2+x,~p_3=3+2+x,~p_4=x-1.
\ee 
We can then drop all terms containing $x$ because these terms must vanish to render the result region momentum independent. This collapses the result to 
\be 
\begin{split} 
\mathcal{A}_4&=\frac{1}{\langle q3\rangle\langle 3q\rangle}\frac{1}{\langle q4\rangle\langle 41\rangle\langle 12\rangle\langle 2q\rangle}\langle q|3\circ 2|q\rangle^2+\frac{1}{\langle q4\rangle\langle 4q\rangle}\frac{1}{\langle q1\rangle\langle 12\rangle\langle 23\rangle\langle 3q\rangle}\langle q|1\circ 4|q\rangle^2\\&+\frac{1}{\langle q1\rangle\langle 12\rangle\langle 2q\rangle}\frac{1}{\langle q3\rangle\langle 34\rangle\langle 4q\rangle}\langle q|1\circ 2|q\rangle^2\\&=\frac{\langle 2q\rangle[32]^2}{\langle 4q\rangle\langle 12\rangle\langle 41\rangle}+\frac{\langle 1q\rangle[14]^2}{\langle 3q\rangle\langle 12\rangle\langle 23\rangle}+\frac{\langle 1q\rangle\langle 2q\rangle[12]^2}{\langle 3q\rangle\langle 12\rangle\langle 34\rangle\langle
4q\rangle}.
\end{split}
\ee 
We now eliminate $q$ dependence using the momentum conservation 
\be 
\frac{[32]}{\langle 41\rangle}=\frac{[12]}{\langle 34\rangle},~~\frac{[14]}{\langle 23\rangle}=\frac{[12]}{\langle 34\rangle} 
\ee 
This gives 
\be 
\begin{split} 
\mathcal{A}_4&=\frac{[12]}{\langle 12\rangle\langle 34\rangle\langle 3q\rangle\langle 4q\rangle}(\langle 3q\rangle\langle 2q\rangle[32]+\langle 1q\rangle\langle 4q\rangle[14]+\langle 1q\rangle\langle 2q\rangle[12])\\
&=-\frac{[12][34]}{\langle 12\rangle\langle 34\rangle}.
\end{split} 
\ee 
This is the expected result (modulo sign and numerical factors).

\subsection{5-point amplitude}

The expression for the amplitude in terms of currents is
\be
{\mathcal A}_5=J(1) J(2,3,4,5) \langle q| p_1\circ p_5 |q\rangle^2 + J(2) J(3,4,5,1) \langle q| p_2\circ p_1 |q\rangle^2 \nonumber \\  + J(3) J(4,5,1,2) \langle q| p_3\circ p_2 |q\rangle^2 
+J(4) J(5,1,2,3) \langle q| p_4\circ p_3 |q\rangle^2 \nonumber \\ + J(5) J(1,2,3,4) \langle q| p_5\circ p_4 |q\rangle^2 
+J(1,2) J(3,4,5)  \langle q| p_2\circ p_5 |q\rangle^2 \nonumber \\+ J(2,3) J(4,5,1)  \langle q| p_3\circ p_1 |q\rangle^2
+J(3,4) J(5,1,2)  \langle q| p_4\circ p_2 |q\rangle^2 \\ \nonumber
+J(4,5) J(1,2,3)  \langle q| p_5\circ p_3 |q\rangle^2+J(5,1) J(2,3,4)  \langle q| p_1\circ p_4 |q\rangle^2.
\ee
Again, we know that it must reproduce the known answer, but would like to see explicitly how this happens. This requires much more work as compared to the 4-point case.

\subsubsection{Extracting the region momentum independent result}

We again parametrise the region momenta in terms of one of them, and the external momenta
\be
p_1=x, \quad p_2=2+x, \quad p_3=3+2 +x, \quad p_4=4+3+2+x, \quad p_5=x-1.
\ee
All terms containing $x$ must drop out by region momentum independence. This gives the following expression
\be
 {\mathcal A}_5=J(3) J(4,5,1,2) \langle q| 3\circ 2 |q\rangle^2 + 
J(4) J(5,1,2,3) \langle q| 4\circ (3+2) |q\rangle^2 \nonumber\\ + J(5) J(1,2,3,4) \langle q| 1\circ 5 |q\rangle^2 
+J(1,2) J(3,4,5)  \langle q| 2\circ 1 |q\rangle^2 
\nonumber\\+J(3,4) J(5,1,2)  \langle q| (4+3)\circ 2 |q\rangle^2+J(4,5) J(1,2,3)  \langle q| (4+5)\circ 1 |q\rangle^2.
\ee
Substituting the expressions for the currents we get
\be
 {\mathcal A}_5=\frac{[23]^2\langle 2q\rangle}{ \langle 4q\rangle \langle 45\rangle \langle 51\rangle \langle 12\rangle} + \frac{([41]\langle 1q\rangle + [45]\langle 5q\rangle)^2}{ \langle 5q\rangle \langle 51\rangle \langle 12\rangle \langle 23\rangle \langle 3q\rangle}
 \nonumber\\+\frac{[15]^2\langle 1q\rangle}{ \langle 12\rangle \langle 23\rangle \langle 34\rangle \langle 4q\rangle}
 + \frac{[12]^2\langle 1q\rangle\langle 2q\rangle}{ \langle 3q\rangle  \langle 12\rangle \langle 34\rangle \langle 45\rangle \langle 5q\rangle}\nonumber\\
 + \frac{([21]\langle 1q\rangle + [25]\langle 5q\rangle)^2 \langle 2q\rangle}{ \langle 3q\rangle \langle 4q\rangle \langle 5q\rangle\langle 34\rangle \langle 51\rangle \langle 12\rangle}
 + \frac{([14]\langle 4q\rangle + [15]\langle 5q\rangle)^2 \langle 1q\rangle}{\langle 3q\rangle  \langle 4q\rangle \langle 5q\rangle \langle 45\rangle \langle 12\rangle \langle 23\rangle}.
 \ee
 Let us start by bringing it all to the common denominator
 \be
 \begin{split} 
  {\mathcal A}_5&= \frac{1}{\langle 3q\rangle  \langle 4q\rangle \langle 5q\rangle \langle 12\rangle  \langle 23\rangle  \langle 34\rangle \langle 45\rangle  \langle 51\rangle}\times\\ \nonumber
  &\Big(
  [23]^2\langle 2q\rangle\langle 3q\rangle\langle 5q\rangle\langle 23\rangle  \langle 34\rangle
  + [15]^2\langle 1q\rangle \langle 3q\rangle\langle 5q\rangle  \langle 45\rangle  \langle 51\rangle
  + [12]^2\langle 1q\rangle\langle 2q\rangle\langle 4q\rangle \langle 23\rangle\langle 51\rangle
  \\ 
&+([41]\langle 1q\rangle + [45]\langle 5q\rangle)^2 \langle 4q\rangle \langle 34\rangle \langle 45\rangle
   + ([23]\langle 3q\rangle + [24]\langle 4q\rangle)^2 \langle 2q\rangle \langle 23\rangle \langle 45\rangle
  \\&+([14]\langle 4q\rangle + [15]\langle 5q\rangle)^2 \langle 1q\rangle \langle 34\rangle \langle 51\rangle
  \Big).
  \end{split} 
  \ee
 We then expand the squares and collect the terms next to common square bracket factors. One then notices that such terms can be rewritten  more compactly using Schouten identity
 \be
 \begin{split} 
 &[15]^2\langle 1q\rangle \langle 3q\rangle\langle 5q\rangle  \langle 45\rangle  \langle 51\rangle 
 +  [15]^2 \langle 5q\rangle^2 \langle 1q\rangle \langle 34\rangle \langle 51\rangle
 \\&= [15]^2 \langle 1q\rangle \langle 4q\rangle \langle 5q\rangle \langle 51\rangle \langle 35\rangle
 \\
 &[23]^2\langle 2q\rangle\langle 3q\rangle\langle 5q\rangle\langle 23\rangle  \langle 34\rangle
 + [23]^2\langle 3q\rangle^2  \langle 2q\rangle \langle 23\rangle \langle 45\rangle\\&=
 [23]^2\langle 2q\rangle  \langle 3q\rangle  \langle 4q\rangle \langle 23\rangle \langle 35\rangle
 \\
 &[14]^2\langle 1q\rangle^2 \langle 4q\rangle \langle 34\rangle \langle 45\rangle+
 [14]^2 \langle 4q\rangle^2 \langle 1q\rangle \langle 34\rangle \langle 51\rangle
 =[14]^2\langle 1q\rangle \langle 4q\rangle \langle 5q\rangle\langle 34\rangle \langle 41\rangle .
 \end{split} 
 \ee
 This gives for the amplitude
 \be\label{A5-result-q}
 \begin{split} 
   {\mathcal A}_5&= \frac{1}{\langle 3q\rangle  \langle 4q\rangle \langle 5q\rangle \langle 12\rangle  \langle 23\rangle  \langle 34\rangle \langle 45\rangle  \langle 51\rangle}\times\\
  &\Big(
 s_{23} [23]\langle 2q\rangle  \langle 3q\rangle  \langle 4q\rangle  \langle 35\rangle
  + s_{15} [15] \langle 1q\rangle \langle 4q\rangle \langle 5q\rangle  \langle 53\rangle
  + [12]^2\langle 1q\rangle\langle 2q\rangle\langle 4q\rangle \langle 23\rangle\langle 51\rangle
  \\ 
&+ s_{45} [45]\langle 5q\rangle^2 \langle 4q\rangle \langle 34\rangle 
   +  [24]^2 \langle 4q\rangle^2 \langle 2q\rangle \langle 23\rangle \langle 45\rangle
  +s_{14} [14] \langle 1q\rangle \langle 4q\rangle \langle 5q\rangle\langle 43\rangle 
    \\ 
    &+2 s_{45} [41]\langle 1q\rangle  \langle 5q\rangle \langle 4q\rangle \langle 34\rangle 
   + 2 s_{23} \langle 3q\rangle  [24]\langle 4q\rangle \langle 2q\rangle  \langle 45\rangle
  +2 s_{15} [14]\langle 4q\rangle \langle 5q\rangle \langle 1q\rangle \langle 43\rangle 
  \Big),
  \end{split} 
  \ee
  where we used the notation $s_{ij} := \langle ij\rangle [ij]$.
  
  To understand the steps that follow we start by writing the amplitude that we want to reproduce. Our starting point is the form (\ref{A5-result-q}) of the amplitude. It is clear that in this expression there are terms that can be written in terms of Mandelstam variables, but there is always a remainder that cannot be written in this way. In the formula (\ref{A5-result-q}) this is the last term. This term, however, can be written in many different ways. Let us first massage it into the form that will be useful later.

We use the momentum conservation in the form $-\langle 23\rangle [34] = \langle 21\rangle [14] +\langle 25\rangle [54]$ to rewrite
\be
\langle 12\rangle  \langle 23\rangle  \langle 34\rangle \langle 45\rangle  \langle 51\rangle  {\mathcal A}_5 =  s_{12} s_{23}  +s_{45} s_{51} + s_{25}s_{45}+ \langle 21\rangle [14] \langle 45\rangle [52].
\ee
Finally, we use Schouten identity in the last term to rewrite the amplitude as
 \be\label{A5-full}
\langle 12\rangle  \langle 23\rangle  \langle 34\rangle \langle 45\rangle  \langle 51\rangle  {\mathcal A}_5 =  s_{12} s_{23}  +s_{45} s_{51} + s_{25}s_{45}+ s_{25} s_{14} \nonumber\\+\langle 24\rangle [14] \langle 15\rangle [52].
\ee

The idea now is to see which of the terms in the amplitude (\ref{A5-result-q}) can reproduce the last term in (\ref{A5-full}). Most of the terms in (\ref{A5-result-q})  already contain factors of Mandelstam variables, and so cannot be responsible for this term. The only terms that can be responsible are the ones containing $[12]^2$ and $[24]^2$. To massage these terms into the desired form we use 
\be
- [21] \langle 1q\rangle  = [23] \langle 3q\rangle  + [24] \langle 4q\rangle + [25] \langle 5q\rangle.
\ee
This gives, using the momentum conservation in terms proportional to $[24]$
\be
[12]^2\langle 1q\rangle\langle 2q\rangle\langle 4q\rangle \langle 23\rangle\langle 51\rangle
+ [24]^2 \langle 4q\rangle^2 \langle 2q\rangle \langle 23\rangle \langle 45\rangle \nonumber\\
=s_{23} [21] \langle 2q\rangle \langle 3q\rangle \langle 4q\rangle    \langle 15\rangle
+[21] [25] \langle 2q\rangle  \langle 4q\rangle \langle 5q\rangle \langle 23\rangle  \langle 15\rangle
\nonumber\\- s_{23} [24] \langle 4q\rangle^2 \langle 2q\rangle \langle 35\rangle .
\ee
We then use for the middle term
\be\nonumber
[21]\langle 23\rangle = [14] \langle 43\rangle + [15] \langle 53\rangle, 
\ee 
to get
\be
\begin{split} 
&[12]^2\langle 1q\rangle\langle 2q\rangle\langle 4q\rangle \langle 23\rangle\langle 51\rangle
+ [24]^2 \langle 4q\rangle^2 \langle 2q\rangle \langle 23\rangle \langle 45\rangle \\= 
&s_{23} [21] \langle 2q\rangle \langle 3q\rangle \langle 4q\rangle    \langle 15\rangle
- s_{23} [24] \langle 4q\rangle^2 \langle 2q\rangle \langle 35\rangle
\\&+ [25]  [14] \langle 43\rangle  \langle 2q\rangle  \langle 4q\rangle \langle 5q\rangle \langle 15\rangle+ s_{15} [25]  \langle 53\rangle \langle 2q\rangle  \langle 4q\rangle \langle 5q\rangle .
\end{split} 
\ee
As the last step, we extract the $q$-independent part of the third term using Schouten identity. 
\be\label{A5-interm-1}
[12]^2\langle 1q\rangle\langle 2q\rangle\langle 4q\rangle \langle 23\rangle\langle 51\rangle
+ [24]^2 \langle 4q\rangle^2 \langle 2q\rangle \langle 23\rangle \langle 45\rangle \nonumber\\= 
[52]  [14] \langle 24\rangle \langle 15\rangle \langle 3q\rangle  \langle 4q\rangle \langle 5q\rangle 
+s_{23} [21] \langle 2q\rangle \langle 3q\rangle \langle 4q\rangle    \langle 15\rangle
\nonumber\\- s_{23} [24] \langle 4q\rangle^2 \langle 2q\rangle \langle 35\rangle
+ s_{15} [25]  \langle 53\rangle \langle 2q\rangle  \langle 4q\rangle \langle 5q\rangle 
 \nonumber\\+ [25]  [14] \langle 23\rangle    \langle 4q\rangle^2 \langle 5q\rangle \langle 15\rangle.
\ee
The term on the right-hand side of the first line (after dividing by the $q$-dependent terms in the denominator) is precisely the last term in (\ref{A5-full}) that can not be written in terms of Mandelstam variables. The term on the last line can also be written in terms of Mandelstam variables. Indeed, we first use Schouten identity $ \langle 15\rangle \langle 4q\rangle =\langle 14\rangle \langle 5q\rangle-\langle 1q\rangle \langle 54\rangle$ to write
\be
[25]  [14] \langle 23\rangle    \langle 4q\rangle^2 \langle 5q\rangle \langle 15\rangle=
s_{14} [25]  \langle 23\rangle    \langle 4q\rangle \langle 5q\rangle^2 \nonumber\\- [25]  [14] \langle 23\rangle    \langle 1q\rangle \langle 4q\rangle \langle 5q\rangle \langle 54\rangle.
\ee
We then use $- [25]\langle 54\rangle = [21]\langle 14\rangle+[23]\langle 34\rangle$ to finally get
\be
[25]  [14] \langle 23\rangle    \langle 4q\rangle^2 \langle 5q\rangle \langle 15\rangle=
s_{14} [25]  \langle 23\rangle    \langle 4q\rangle \langle 5q\rangle^2 
+ s_{14} [21] \langle 23\rangle \langle 1q\rangle \langle 4q\rangle \langle 5q\rangle 
\nonumber\\+ s_{23} [14] \langle 34\rangle \langle 1q\rangle \langle 4q\rangle \langle 5q\rangle.
\ee

It thus remains to reproduce the other $s$-containing terms in the formula (\ref{A5-full}) for the amplitude. We substitute the terms in the second and third line of (\ref{A5-interm-1}) into (\ref{A5-result-q}) instead of the $[12]^2, [24]^2$ terms. This gives a part of the amplitude that is supposed to contain all terms with Mandelstam variables
\be
\begin{split} 
 \langle 3q\rangle  \langle 4q\rangle \langle 5q\rangle \langle 12\rangle  \langle 23\rangle  \langle 34\rangle \langle 45\rangle  \langle 51\rangle   {\mathcal A}'_5 = s_{23} [23] \langle 35\rangle \langle 2q\rangle  \langle 3q\rangle  \langle 4q\rangle  
 \\ \nonumber+ s_{15} [15] \langle 53\rangle \langle 1q\rangle \langle 4q\rangle \langle 5q\rangle 
+ s_{45} [45]\langle 34\rangle   \langle 4q\rangle \langle 5q\rangle^2
  +s_{14} [14] \langle 43\rangle \langle 1q\rangle \langle 4q\rangle \langle 5q\rangle
    \\  \nonumber+   2 s_{45} [14] \langle 43\rangle \langle 1q\rangle  \langle 4q\rangle \langle 5q\rangle 
   + 2 s_{23}   [24] \langle 45\rangle \langle 2q\rangle \langle 3q\rangle \langle 4q\rangle 
  +2 s_{15} [14] \langle 43\rangle\langle 1q\rangle \langle 4q\rangle \langle 5q\rangle  
 \\ \nonumber+s_{23} [21] \langle 15\rangle\langle 2q\rangle \langle 3q\rangle \langle 4q\rangle    
- s_{23} [24] \langle 35\rangle \langle 2q\rangle\langle 4q\rangle^2  
 + s_{15} [25]  \langle 53\rangle \langle 2q\rangle  \langle 4q\rangle \langle 5q\rangle 
 \\ \nonumber+ s_{14} [25]  \langle 23\rangle    \langle 4q\rangle \langle 5q\rangle^2 
+ s_{14} [21] \langle 23\rangle \langle 1q\rangle \langle 4q\rangle \langle 5q\rangle 
- s_{23} [14] \langle 43\rangle \langle 1q\rangle \langle 4q\rangle \langle 5q\rangle.
\end{split} 
\ee
There are some immediate simplifications. The terms containing $s_{23}  \langle 2q\rangle  \langle 3q\rangle  \langle 4q\rangle $ simplify using $[23]\langle 35\rangle + [21]\langle 15\rangle=- [24]\langle 45\rangle$. The terms containing $[14] \langle 43\rangle \langle 1q\rangle \langle 4q\rangle \langle 5q\rangle$ simplify using $s_{14}+2s_{45} + 2s_{15} -s_{23}=s_{23}-s_{14}$, and so
\be
\begin{split} 
 \langle 3q\rangle  \langle 4q\rangle \langle 5q\rangle \langle 12\rangle  \langle 23\rangle  \langle 34\rangle \langle 45\rangle  \langle 51\rangle   {\mathcal A}'_5 &= 
    s_{23}   [24] \langle 45\rangle \langle 2q\rangle \langle 3q\rangle \langle 4q\rangle    
   \\ \nonumber
  &+ s_{15} [15] \langle 53\rangle \langle 1q\rangle \langle 4q\rangle \langle 5q\rangle 
+ s_{45} [45]\langle 34\rangle   \langle 4q\rangle \langle 5q\rangle^2  
 \\ \nonumber
&- s_{23} [24] \langle 35\rangle \langle 2q\rangle\langle 4q\rangle^2  
 + s_{15} [25]  \langle 53\rangle \langle 2q\rangle  \langle 4q\rangle \langle 5q\rangle 
 \\ \nonumber
 &+ s_{14} [25]  \langle 23\rangle    \langle 4q\rangle \langle 5q\rangle^2 
+ s_{14} [21] \langle 23\rangle \langle 1q\rangle \langle 4q\rangle \langle 5q\rangle \\ \nonumber
&+ (s_{23} -s_{14})[14] \langle 43\rangle \langle 1q\rangle \langle 4q\rangle \langle 5q\rangle.
\end{split} 
\ee
The two terms in the last line containing $s_{14} \langle 1q\rangle \langle 4q\rangle \langle 5q\rangle$ simplify using $-[12]\langle 23\rangle - [14]\langle 43\rangle = [15]\langle 53\rangle$, and so
\be
 \langle 3q\rangle  \langle 4q\rangle \langle 5q\rangle \langle 12\rangle  \langle 23\rangle  \langle 34\rangle \langle 45\rangle  \langle 51\rangle   {\mathcal A}'_5 = 
    s_{23}   [24] \langle 45\rangle \langle 2q\rangle \langle 3q\rangle \langle 4q\rangle    
   \\ \nonumber
  + (s_{15}+s_{14})  [15] \langle 53\rangle \langle 1q\rangle \langle 4q\rangle \langle 5q\rangle 
+ s_{45} [45]\langle 34\rangle   \langle 4q\rangle \langle 5q\rangle^2  
 \\ \nonumber
- s_{23} [24] \langle 35\rangle \langle 2q\rangle\langle 4q\rangle^2  
 + s_{15} [25]  \langle 53\rangle \langle 2q\rangle  \langle 4q\rangle \langle 5q\rangle 
 \\ \nonumber
 + s_{14} [25]  \langle 23\rangle    \langle 4q\rangle \langle 5q\rangle^2 
+ s_{23} [14] \langle 43\rangle \langle 1q\rangle \langle 4q\rangle \langle 5q\rangle.
\ee
We then again use the same momentum conservation formula on the very last term to get
\be
 \langle 3q\rangle  \langle 4q\rangle \langle 5q\rangle \langle 12\rangle  \langle 23\rangle  \langle 34\rangle \langle 45\rangle  \langle 51\rangle   {\mathcal A}'_5 = 
    s_{23}   [24] \langle 45\rangle \langle 2q\rangle \langle 3q\rangle \langle 4q\rangle    
   \\ \nonumber
  -s_{45}  [15] \langle 53\rangle \langle 1q\rangle \langle 4q\rangle \langle 5q\rangle 
+ s_{45} [45]\langle 34\rangle   \langle 4q\rangle \langle 5q\rangle^2  
 \\ \nonumber
- s_{23} [24] \langle 35\rangle \langle 2q\rangle\langle 4q\rangle^2  
 + s_{15} [25]  \langle 53\rangle \langle 2q\rangle  \langle 4q\rangle \langle 5q\rangle 
 \\ \nonumber
 + s_{14} [25]  \langle 23\rangle    \langle 4q\rangle \langle 5q\rangle^2 
- s_{23} [12] \langle 23\rangle \langle 1q\rangle \langle 4q\rangle \langle 5q\rangle.
\ee
We can now use Schouten identity to extract the $q$-invariant pieces, and match these to those in (\ref{A5-full}). We have
\be
- s_{23} [12] \langle 23\rangle \langle 1q\rangle \langle 4q\rangle \langle 5q\rangle=
s_{23} s_{12} \langle 3q\rangle \langle 4q\rangle \langle 5q\rangle +
s_{23} [12] \langle 31\rangle \langle 2q\rangle \langle 4q\rangle \langle 5q\rangle, 
\\ \nonumber
-s_{45}  [15] \langle 53\rangle \langle 1q\rangle \langle 4q\rangle \langle 5q\rangle =
s_{45} s_{51} \langle 3q\rangle \langle 4q\rangle \langle 5q\rangle +
s_{45} [15] \langle 31\rangle  \langle 4q\rangle \langle 5q\rangle^2, 
\\ \nonumber
 s_{14} [25]  \langle 23\rangle    \langle 4q\rangle \langle 5q\rangle^2 =
 s_{14} s_{25} \langle 3q\rangle \langle 4q\rangle \langle 5q\rangle +
s_{14} [25] \langle 53\rangle \langle 2q\rangle \langle 4q\rangle \langle 5q\rangle.
\ee
These gives three of the four Mandelstam variable containing terms in (\ref{A5-full}). The remainder, which is supposed to give the last $s_{25} s_{45}$ term is 
\be
  s_{23}   [24] \langle 45\rangle \langle 2q\rangle \langle 3q\rangle \langle 4q\rangle 
  +s_{45} [15] \langle 31\rangle  \langle 4q\rangle \langle 5q\rangle^2
  + s_{45} [45]\langle 34\rangle   \langle 4q\rangle \langle 5q\rangle^2  
 \\ \nonumber
- s_{23} [24] \langle 35\rangle \langle 2q\rangle\langle 4q\rangle^2  
 + s_{15} [25]  \langle 53\rangle \langle 2q\rangle  \langle 4q\rangle \langle 5q\rangle 
 \\ \nonumber
 +s_{14} [25] \langle 53\rangle \langle 2q\rangle \langle 4q\rangle \langle 5q\rangle
 +s_{23} [12] \langle 31\rangle \langle 2q\rangle \langle 4q\rangle \langle 5q\rangle.
 \ee
The second and third terms here, using the momentum conservation, give
\be
s_{45} [25] \langle 23\rangle  \langle 4q\rangle \langle 5q\rangle^2 =
s_{45} s_{25} \langle 3q\rangle \langle 4q\rangle \langle 5q\rangle +
s_{45} [25] \langle 53\rangle \langle 2q\rangle \langle 4q\rangle \langle 5q\rangle.
\ee
The first term gives the last term in (\ref{A5-full}). Thus, we have the remainder which is
\be
  s_{23}   [24] \langle 45\rangle \langle 2q\rangle \langle 3q\rangle \langle 4q\rangle 
  +s_{45} [25] \langle 53\rangle \langle 2q\rangle \langle 4q\rangle \langle 5q\rangle
 \\ \nonumber
- s_{23} [24] \langle 35\rangle \langle 2q\rangle\langle 4q\rangle^2  
 + s_{15} [25]  \langle 53\rangle \langle 2q\rangle  \langle 4q\rangle \langle 5q\rangle 
 \\ \nonumber
 +s_{14} [25] \langle 53\rangle \langle 2q\rangle \langle 4q\rangle \langle 5q\rangle
 +s_{23} [12] \langle 31\rangle \langle 2q\rangle \langle 4q\rangle \langle 5q\rangle.
 \ee
 Applying Schouten identity once more on the first term in the first and second lines
 \be
s_{23}   [24] \langle 45\rangle \langle 2q\rangle \langle 3q\rangle \langle 4q\rangle
-s_{23} [24] \langle 35\rangle \langle 2q\rangle\langle 4q\rangle^2
=s_{23}   [24] \langle 43\rangle \langle 2q\rangle \langle 4q\rangle \langle 5q\rangle,
\ee
we get a set of terms all proportional to $\langle 2q\rangle \langle 4q\rangle \langle 5q\rangle$
\be\nonumber
s_{23}   [24] \langle 43\rangle+s_{23} [21] \langle 13\rangle + s_{45} [25] \langle 53\rangle+s_{15} [25] \langle 53\rangle +s_{14} [25] \langle 53\rangle \\
=[25] \langle 53\rangle ( s_{45} + s_{15} + s_{14} - s_{23})=0,
\ee
where we applied momentum conservation to the first two terms. Thus, the correct expression (\ref{A5-full}) for the 5-point amplitude is reproduced. 
\\~\\
The main ingredient of our formula is that the self-energy bubble as an operator connects the lower order Berends-Giele currents in all possible ways. Thus it is a new variant of the Berends-Giele recursion relation which generates the series of one-loop amplitudes in the theory of SDYM. In the tree level case, the currents connect using the cubic vertex of the SDYM and thus lower order pieces combine to build higher order ones. The recursion is valid off-shell and thus one needs to amputate the final leg propagator to construct the on-shell tree amplitudes which are then trivially zero. In this case however, the off-shell legs of the currents connect via the effective propagator and all the external legs are thus on-shell, resulting in the construction of the amplitude rather directly. As we have seen, such a construction is motivated from the fact that the sum of all possible geometries vanish for the one-loop amplitude. We have explicitly demonstrated it for the four point case and expect it to be valid for the case of arbitrary numbers of external gluons of the same helicity. The simple formula for the series of same helicity amplitudes raises the question of why the classically conserved currents play such an important in an otherwise quantum computation. Bardeen conjectured that the amplitude at four points might be related to an anomaly of the currents responsible for integrability of SDYM. In our case we see that this notion is more general and may apply to amplitudes at all points. The shift behaviour of the self-energy bubble coupled with the fact that the currents get inserted to it can be possibly interpreted as the quantum expectation value of the divergence of these currents, which if non-zero would result in an anomaly. It is however not clear how to achieve this result and it might be that there are some missing gaps left in the analysis. We expect to understand this further in a future work. 

\part{Aspects of self-dual gravity}
\newpage
\newpage 
\chapter{Introduction}
The metric formulation of Einstein gravity is the most studied one and in fact the best understood. However, the perturbative treatment of this formulation becomes complicated as we go to high loop orders or increase the number of external legs in Feynman diagrams. To give an example, when the action is expanded around flat space, the quartic order term in the Lagrangian occupies half a page. Thus, computing any physical quantity of interest becomes a daunting task in this prescription. One of the major motivations of alternative formulations of General Relativity is to simplify computations at the perturbative level. In almost all such formulations, the role played by the metric becomes secondary while the connection becomes the main object of interest. Another motivation to consider alternative formulations is about coupling matter to gravity in a consistent way. In the usual metric formulation, it is not possible to couple fermions because the group of diffeomorphisms does not admit spinor representations. One thus constructs orthonormal frames at each point in spacetime and the metric can then be expressed in terms of the basis vectors of these frames, which are called tetrads. The orthonormal frames over the spacetime manifold constitute what is called the frame bundle. In the frame bundle, one associates a connection such that covariant derivatives can be defined. In the tetradic Palatini action, the tetrad and spin connection thus become independent variables and the metric is understood as a second order construct. The interesting aspect of the Palatini formulation is that the action becomes first order in the independent variables, in contrast to the Einstein-Hilbert case where the action is second order in the metric. Also, the action becomes polynomial in the fields in the case of zero cosmological constant. However, in the case of non-zero cosmological constant, the action no longer stays polynomial. Thus, a better first order formulation is required which keeps the action polynomial even in the case when the cosmological constant is non-zero. The Einstein-Cartan first order formulation is the one which precisely does the same. The action of this formulation reads 
\be 
\label{non-chiral}
S_{EC}(e,\omega)=\frac{1}{32\pi G}\int \epsilon_{ijkl}e^ie^j\Big(F^{kl}(\omega)-\frac{\Lambda}{6}e^ke^l\Big)
\ee 
The action is a function of the tetrad (e)
and the spin connection ($\omega$). If one varies this action with respect to the connection, one obtains the zero-torsion condition, $d^\omega e^i=0$. Substituting the solution of this equation into (\ref{non-chiral}) gives us the Einstein-Hilbert action, spelled in terms of the tetrad variables. Even though the Einstein-Cartan action gives a better formulation for perturbative computations, it is not very economical. This is because in addition to tetrad components, the Lagrangian depends on 24 connection components per spacetime points. The gauge fixing of the theory becomes complicated and in addition to the propagator of the tetrad with itself, there exists other propagators involving the connection, which makes computations difficult. The resolution to all this is provided by the chiral formulation. The basic reason behind the chiral formulations of four dimensional GR is that in this many dimensions, there happens to be some 'accidental' isomorphisms in the Lie algebra of four dimensional Lorentz groups. Let us write two such isomorphisms, first in the Euclidean signature and then in the Lorentzian case 

\be 
so(4)=su(2)\oplus su(2),\nonumber\\
so(1,3)=sl(2,\mathbb{C})\oplus \overline{sl(2,\mathbb{C})}.
\ee 
The underlying reason for these isomorphisms to exist lie in the fact that the Hodge dual operator in four dimensions maps any 2-form to another 2-form and thus decomposes the space of 2-forms into self-dual (SD) and anti-self-dual (ASD) parts. 
\be 
\star: \Lambda^2\rightarrow \Lambda^2.
\ee 
The eigenvalues of the Hodge dual operator ($\star$) are $\pm 1$ in the case of Euclidean or split signatures, whereas $\pm i$ in the case of Lorentzian signature. The space of such 2-forms thus gets split into the eigenspaces of this operator, which is just the SD and ASD decomposition. 
\be 
\Lambda^2= \Lambda^+\oplus \Lambda^-.
\ee 
This leads to an elegant decomposition of the Riemann curvature, which we now describe. The Riemann curvature is a symmetric $\Lambda^2\otimes \Lambda^2$-valued matrix. We can decompose it into the SD and ASD parts and get the following block 
\be 
R=\begin{pmatrix}
X & Y \\
Y^T & Z 
\end{pmatrix}.
\ee 
where $X$ is the self-dual self-dual component, $Z$ is the anti-self-dual anti-self dual component and $Y$ is the self-dual anti-self-dual component. These components can also be obtained by applying appropriate SD/ASD projectors $P_{\pm}$ to the Riemann curvature. The main point of the chiral formulations is that in view of the above decomposition, the Einstein condition $R_{\mu\nu}=\Lambda g_{\mu\nu}$ is equivalent to  
\be 
Y=0,~~~~ \textrm{Tr}(X)=\Lambda.
\ee 
which implies that for a metric to be Einstein, it is enough to have access to just half of the Riemann curvature. Indeed, in the above perspective we find that the ASD-ASD part $Z$ is not constrained by the Einstein equations. Thus, in four spacetime dimensions, the full dynamical theory of gravity can be analyzed by just accessing the half (either SD or ASD) of the curvature. Let us then apply this to the Einstein-Cartan formulation. The Riemann curvature as we said, is encoded in the curvature of the spin connection. It is then possible to impose SD or ASD projectors to this curvature and build an action which just contains the SD part of the full curvature. This is what is done in the chiral Einstein-Cartan action as we will describe in details in the next chapter.
\\~\\
Much of what has been described here is analogous to the YM story. In that case, one uses a self-dual auxiliary field as a projector on the full curvature and this makes the action to be chiral. A beautiful gauge fixing procedure then gives rise to very simple Feynman rules in spinor notations. Also, it becomes convenient to pass to SDYM by a truncation of the full YM. As we will see, a analogous thing is going to happen in the case of gravity. First, there exists a nice gauge fixing procedure which is detailed in \cite{Krasnov:2020bqr} for the case of a flat background. In this case, one of the propagators, namely the connection with itself vanishes. Also, the spinor structure of the Feynman rules becomes very simple. We will employ a similar variant of such a gauge fixing in the case of an Einstein background and develop the ghost Lagrangian using the BRST formalism. It is then straightforward to pass to the flat background case by replacing covariant derivatives with partial derivatives everywhere. This will give us the ghost Feynman rules in addition to the existing rules for the tetrad and the connection. Also, it is convenient to pass from full gravity to self-dual gravity (SDGR) in flat space using the chiral Einstein-Cartan action as has been described in \cite{Krasnov:2020bqr}. 
\\~\\
The same helicity amplitudes in gravity are correctly captured by the simpler SDGR Feynman rules. Thus, the flat space covariant formulation of SDGR is most relevant for amplitude computations. The same helicity tree amplitudes in SDGR vanish, analogous to SDYM. The vanishing is related to the fact that to get such amplitudes, one has to remove the final leg propagator of the currents. Thus one needs to multiply the current by a factor of $k^2$ and take the on-shell limit $k^2\rightarrow 0$. However, there is no such pole to cancel this propagator and hence the amplitudes vanish. The structure of the BG currents is however more complicated in this case. In particular, to construct the current one has to take all possible permutations of the insertion of legs to the cubic vertex, instead of just cyclic permutations. However, it is possible to write down a general form of the current using the Berends-Giele recursion \cite{Krasnov:2016emc}.
\\~\\
Self-dual gravity is finite, despite possessing a negative dimension coupling constant. The reason being that all possible one-loop divergences are proportional to topological invariants of the underlying manifold and thus does not contribute to the S-matrix. Higher loop diagrams do not exist as will be evident from the action and therefore the theory is one-loop finite. The one-loop amplitudes are however non-trivial. They are interesting on their own right because of the simplicity in the structure of them. In particular, all such one-loop amplitudes are rational functions of the momenta involved and are cut-free. The only singularities are those of 2-particle poles. The structure of these amplitudes is very similar to the YM case and a possible anomaly interpretation is conjectured in \cite{Krasnov:2016emc}. It remains to be understood what kind of anomaly may give rise to the non-vanishing of such amplitudes. Further, these amplitudes have recently been linked to the 2-loop divergence in quantum gravity \cite{Bern:2017puu}. It thus becomes much more interesting to investigate the reason behind the finiteness of such amplitudes.
A general ansatz of these amplitudes was given by Bern and collaborators \cite{Bern:1998sv} using the soft and collinear limit arguments. It is possible to compute them using supersymmetry, by replacing a graviton propagating in the loop by a scalar. However, there are no direct computations of these amplitudes from self-dual gravity Feynman rules till date. In this part of the thesis, we will attempt to compute this amplitude at four points and show that they behave in an analogous way to their YM cousin.

\newpage
\chapter{Chiral Einstein-Cartan Gravity}
In the first order Einstein-Cartan formulation of General Relativity, the action becomes polynomial in the fields but at the expense of introducing an auxiliary connection (spin-connection) variable. The Riemann curvature is then encoded into the curvature of this spin connection. However, even being polynomial, the perturbative treatment of this formulation becomes complicated because in addition to the metric/tetrad propagator, there exists two other propagators, namely the metric-connection and the connection-connection. This gives rise to too many Feynman diagrams and thus the algebraic complexity in any amplitude computation increases. The chiral first order formulation resolves this in an interesting way. In the chiral Einstein-Cartan, one considers self-dual projection of the curvature of the spin-connection. One can then construct an action with just the self-dual projection instead of the full curvature. The difference between the two is a total derivative term (Holst term), which does not change the dynamics of the theory. Thus this action is equivalent to the non-chiral first order Einstein-Cartan action. Further, one can rewrite the self-dual part of the curvature of the spin connection as the curvature of the self-dual part of the spin connection. This then gives rise to a better perturbation theory where one of the propagators, namely the propagator of the connection with itself vanishes. Thus the algebraic complexity gets reduced significantly in Feynman diagram computations. 
\\~\\
It is to be noted that this chiral action is quite analogous to its YM counterpart. In YM, as we described in the previous part, one uses self-dual projections on the curvature 2-form and constructs a first order chiral formulation where the propagator of the auxiliary field with itself vanishes. This not only gives rise to a very nice perturbation theory but makes YM an extension of SDYM. As we will see, the same holds true in gravity. In particular, it is straightforward to construct the covariant action of self-dual gravity (SDGR) from the chiral Einstein-Cartan just by removing the $\omega \omega$ term from the curvature. 
\section{Chiral Einstein Cartan action}
We start with the chiral Einstein Cartan action for gravity, with zero cosmological constant. In units $32\pi G=1$ we have
\be
\label{EC}
\mathcal{S}_{chiral}[h,A]=\frac{4}{i}\int\Sigma^iF^i,
\ee
where
\be 
F^i=d\omega^i+\epsilon^{ijk}\omega^j\wedge \omega^k
\ee 
is the field strength of the corresponding chiral part of the spin connection, i.e $\omega^j$ and the constraint on self dual two forms is imposed as \be 
\Sigma^i\Sigma^j=\delta^{ij}.
\ee 
This constraint guarantees that the self dual two forms are constructed from the metric, denoted as $h$ which is one of the arguments in the action. Therefore, $\Sigma^i$ is not an independent variable and only variations of $\Sigma^i$ upto second order is non zero. The SO(3) indices are lowered and raised with the Kronecker delta metric and following convention, we keep them in the upper position. The relation between the metric and the tetrad is given by 
\be 
g_{\mu\nu}=e^{AA'}_{\mu}e_{\nu AA'}.
\ee 
The indices $A,A'$ are a pair of spinor indices for the Lorentz degree of freedom and are raised or lowered by the epsilon spinors $\epsilon^{AB}, \epsilon_{A'B'}$. The indices $\mu,\nu$ are curvy indices and are raised or lowered by the metric $h_{\mu\nu}$. When all the indices are converted into spinorial ones, this reads 
\be 
g_{MM'NN'}=e^{AA'}_{~~MM'}e_{NN' AA'}.
\ee 
\subsection{Action in spinor notation}
The chiral Einstein-Cartan action requires chiral projections, which are easiest to describe in spinor notations. The action in (\ref{EC}) then reads
\be 
\label{ECs}
S[\theta, \omega]=2i\int \Sigma^{AB}\wedge F_{AB}
\ee 
where $A,B=1,2$ are unprimed 2-component spinor indices and the self-dual 2-forms are 
\be 
\Sigma^{AB}=\frac{1}{2}e^A_{C'}\wedge e^{BC'},
\ee 
where $e^{AA'}$ is the soldering form and the curvature 2-form $F^{AB}$is given by 
\be 
F^{AB}=d\omega^{AB}+\omega^{AC}\wedge \omega^{~B}_C.
\ee 
The object $\omega^{AB}$ is the self-dual part of the spin connection. Locally, it takes values in the spin bundle of symmetric second rank unprimed spinors. The action in (\ref{ECs}) is obtained by applying the chiral self-dual projection to the first order Einstein-Cartan action in terms of the tetrad $e^{AA'}$ and the full spin connection.
 \subsection{Spin connection and covariant derivative}
The internal space is equipped with the spin connection given by $\omega^{~i}_{\mu~j}$, which acts on the internal indices of a given tensor. The covariant derivative acting on a Lorentz vector with respect to this connection is then given by 
\be 
D_{\mu}Y^{i}=\partial_{\mu}Y^i+\omega^{~i}_{\mu~k}Y^k.
\ee 
Similarly, when we have a Lorentz vector with lower index, the covariant derivative acts as 
\be 
D_{\mu}Y_{j}=\partial_{\mu}Y_j-\omega^{~i}_{\mu~j}Y_i.
\ee 
With the tetrad field, we can establish a relation between the spin connection variables in the internal bundle and the Christoffel connections which act on spacetime. The Christoffel connection on spacetime is given as follows 
\be 
\label{relation}
\Gamma^{\alpha}_{\mu\nu}=e^{\alpha}_{~j}\partial_{\mu}e^j_{~\nu}+\omega^{~j}_{\mu~i}e^i_{~\nu}e^{\alpha}_{~j}.
\ee 
Let us now multiply by $e^{~k}_{\alpha}$ on both sides, which gives 
\be 
\begin{split}
\Gamma^{\alpha}_{\mu\nu}e^{~k}_{\alpha}&=e^{\alpha}_{~j}\partial_{\mu}e^j_{~\nu}e^{~k}_{\alpha}+\omega^{~j}_{\mu~i}e^i_{~\nu}e^{\alpha}_{~j}e^{~k}_{\alpha}\\
&=\partial_{\mu}e^k_{~\nu}+\omega^{~k}_{\mu~j}e^j_{~\nu}.
\end{split}
\ee 
We thus arrive at the tetrad postulate by some rearrangement of terms in the above, which implies that the total covariant derivative of the tetrad vanishes
\be 
\nabla^T_{\mu}e^{~k}_{\nu}=\partial_{\mu}e^k_{~\nu}-\Gamma^{\alpha}_{\mu\nu}e^{~k}_{\alpha}+\omega^{~k}_{\mu~j}e^j_{~\nu}=0.
\ee 
This is reminiscent of the fact that the metric is covariantly conserved. Indeed, using the relation in (\ref{relation}), and the antisymmetry of the Lorentz indices of the spin connection, it is not hard to check that $\nabla_{\gamma}g_{\mu\nu}=0$.
The total covariant derivative can now act on both the internal bundle indices and on spacetime indices. Let us clarify this by some examples. Consider first a pair of vector fields possessing an upper and a lower spacetime index respectively. The space-time covariant derivative then acts as 
\be 
\begin{split}
\nabla_{\mu}Y^{\nu}&=\partial_{\mu}Y^{\nu}+\Gamma^{\nu}_{\mu\lambda}Y^{\lambda},\\
\nabla_{\mu}X_{\nu}&=\partial_{\mu}X_{\nu}-\Gamma^{\lambda}_{\mu\nu}X_{\lambda}.
\end{split}
\ee 
Next, we take a pair of Lorentz vectors which have either an upper or a lower Lorentz index. The covariant derivative with respect to the connection in this case acts as follows 
\be 
\begin{split}
D_{\mu}A^{i}&=\partial_{\mu}A^{i}+\omega^{~i}_{\mu~k}A^{k},\\
D_{\mu}B_{j}&=\partial_{\mu}B_{j}-\omega^{~k}_{\mu~j}B_{k}.
\end{split}
\ee 
Finally, when we have an object with both spacetime and internal indices, the total covariant derivative is given as 
\be 
\label{cov}
\nabla^T_{\mu}Z^{i}_{~\nu}&=\partial_{\mu}Z^{i}_{~\nu}-\Gamma^{\lambda}_{\mu\nu}Z^i_{\lambda}+\omega^{~i}_{\mu~k}Z^{k}_{~\nu}. 
\ee 
It is instructive to rewrite all this by converting to spinor indices. The covariant derivative $\nabla^T_{\mu}$ in spinor indices becomes an object $\nabla^T_{MM'}$, while the internal and spacetime indices of any general tensor field are all converted to spinor indices henceforth. We will use only the self dual part of the spin connection from now as this is what is relevant for the chiral action in (\ref{EC}). Then the $SL(2,C)$ indices of the self-dual spin connection becomes a pair of symmetrised unprimed indices $(AB)$, while the spacetime index $\mu$ becomes a pair $MM'$. So in spinor notations, the spin connection becomes an object $\omega^{AB}_{~~~MM'}$. A general object which has Lorentz index and another spacetime index in spinorial notations become 
\be 
Z^{i}_{~\nu}\rightarrow Z^{BB'}_{~~~NN'}.
\ee 
The spinor form of the covariant derivative in (\ref{cov}) is
\be 
\label{cov2}
\nabla^T_{MM'}Z^{AA'}_{~~~NN'}&=\partial_{MM'}Z^{AA'}_{~~~NN'}-\Gamma^{LL'}_{~~~MM'NN'}Z^{AA'}_{~~~LL'}+\omega^{~A}_{~~BMM'}Z^{BA'}_{~~~NN'}.
\ee 
where the Christoffel connection acts on the spinor version of the spacetime indices and the self dual spin connection acts on the unprimed Lorentz index of the general tensor.
\subsection{Symmetries}
The action in (\ref{EC}) is invariant under two classes of transformations. One is the diffeomorphisms which is similar to usual Einstein gravity. Another is the local $SL(2,C)$ gauge transformations. When the background is not flat, both these transformations act on the independent fields i.e, the tetrad and the spin connection. As we will see, it is possible to correct the diffeomorphism by a gauge transformation and this results in a simpler set of rules for the transformation of the fields.

\subsubsection{Diffeomorphisms}
Under the infinitesimal coordinate transformation 
\be 
x'^{\mu}=x^{\mu}+\epsilon\xi^{\mu}
\ee 
the tetrad field transforms (at order $\epsilon$) as follows
\be 
\begin{split} 
\delta_{\xi}e^{i}_{\nu}&=\xi^{\mu}\partial_{\mu}e^{i}_{\nu}+e^{i}_{\mu}\partial_{\nu}\xi^{\mu}\\
&=\xi^{\mu}\partial_{\mu}e^{i}_{\nu}-\xi^{\mu}\Gamma^{\lambda}_{\mu\nu}e^i_{\lambda}+e^{i}_{\mu}\partial_{\nu}\xi^{\mu}+e^i_{\mu}\Gamma^{\mu}_{\nu\lambda}\xi^{\lambda}\\
&=\xi^{\mu}\nabla_{\mu}e^{i}_{\nu}+e^{i}_{\mu}\nabla_{\nu}\xi^{\mu}.
\end{split}
\ee 
where in the second line of the above equation, we added and subtracted a term with the Levi-Civita connection, which allows us to write the variation of the tetrad in terms of the spacetime covariant derivative. We can also add and subtract a term with the spin connection. It is then possible to arrange some of the terms in such a way that by the tetrad postulate they vanish. The remaining terms are the ones in which the total covariant derivative operator acts on the vector field and this is corrected by a gauge transformation. We thus have 
\be 
\begin{split} 
\delta_{\xi}e^{i}_{\nu}&= e^{i}_{\mu}\nabla^T_{\nu}\xi^{\mu}-\tilde{\phi}^i_{~j}e^j_{~\nu}.
\end{split}
\ee 
Thus, we arrive at a simple transformation rule for the tetrad under diffeomorphisms. The first term is the covariant derivative of the parameter $\xi^{\mu}$ and this is corrected by a gauge transformation. As we will see, this lets us to get simpler rules for the total transformation when we add the local Lorentz to it. Let us now write the transformation rule for the connection under diffeomorphisms. The connection is a one-form like the tetrad and thus it will have a similar set of transformations. In particular,
\be 
\begin{split} 
\delta_{\xi}\omega^{~i}_{\mu~j}&=\xi^{\nu}\partial_{\nu}\omega^{~i}_{\mu~j}+\omega^{i}_{\nu~j}\partial_{\mu}\xi^{\nu}.\\
\end{split}
\ee 
We can also write the diffeomorphism transform of the connection in such a way so that it does not contain explicit derivatives of the vector field which generates it. In an analogous way like the tetrad case, we add and subtract pair of terms with a spin connection. We then arrange terms in a way that the vector field is inserted into the curvature of the connection. The remaining term is a total covariant derivative. However, the total derivative term can be matched with the corresponding local Lorentz transform of the connection and this results again in a simpler transformation rule, as we will see. The diffeomorphism of the connection thus reads
\be 
\label{diffconn}
\begin{split} 
\delta_{\xi}\omega^{~i}_{\mu~j}&=\xi^{\nu}F^{i}_{~\nu\mu j}+\nabla^T_{\mu}(\xi^{\nu}\omega^{i}_{\nu~j})
\end{split}.
\ee 
where the last term in (\ref{diffconn}) is of the form of a gauge transformation, which we can add to the local Lorentz part of the transformation of the connection. It is computationally simple to use this version of the transformation. Next, we write the local Lorentz transformations of both the tetrad and the connection fields. We consider one chiral half of the Lorentz to act on the fields, the other half being set to zero from the beginning.
\subsubsection{Local Lorentz transformations}
$SL(2,C)$ transformations act on both the tetrad and the connections fields. They are given by 
\be 
\begin{split}
\delta_{\phi}e^{i}_{~\mu}&=\phi^{i}_{~j}e^{j}_{~\mu},\\
\delta_{\phi}\omega^{i}_{\mu~j}&=D_{\mu}\phi^{i}_{~j}\\
&=\nabla^T_{\mu}\phi^{i}_{~j}.
\end{split}
\ee 
Thus the action of the $SL(2,C)$ transformation on the tetrad amounts to a Lorentz rotation given by the intfinitesimal parameter $\phi^i_{~j}$. The action on the connection is given by the covariant derivative on the parameter $\phi^i_{~j}$. 
\subsubsection{Total transformation}
Note that the diffeomorphism of the tetrad is corrected by a gauge transformation, with some parameter $\tilde{\phi}^{ij}$. It is then possible to absorb this into the local Lorentz transformation parameter $\phi^{ij}$ and just set the total gauge transformation parameter to be $\phi^{ij}$. Then the full transformation of the tetrad reads
\be 
\delta e^{i}_{~\mu}=e^{i}_{~\mu}\nabla^T_{\nu}\xi^{\mu}-\phi^i_{~j}e^j_{~\nu}.
\ee 
Let us also write the full transformation for the spin connection. We add the diffeomorphism and local Lorentz to get 
\be 
\delta_{\phi}\omega^{i}_{\mu~j}=\nabla^T_{\mu}\phi^{i}_{~j}+\xi^{\nu}F^i_{\nu\mu j}.
\ee

\newpage 
\section{Quantization and BRST formalism}
The next step is to quantize the Einstein-Cartan action on a general Einstein background. The system has two types of gauge symmetries, diffeomorphism and local Lorentz. Thus the system falls under the realm of constrained dynamics. It is essential to fix the gauges in order to get the physical states of the theory. We will follow the well developed BRST formalism to gauge fix the theory. The BRST formalism was developed many years ago in order to systematically renormalize non-abelian gauge theories. One invokes anti-commuting (fermionic) Grassmann valued fields and enlarges the Hilbert space of the theory. Then the notion of gauge transformations is generalised to the more powerful notion of BRST transformation which mixes bosonic and fermionic fields. The BRST transformation $'s'$ which acts on the fields is nilpotent, i.e, $s^2=0$. The usual gauge transformations are then written in an analogous way but with bosonic parameters replaced by the anti-commuting ones. These anti-commuting parameters are interpreted as fictitious fields in the theory. The important point is that the Lagrangian which is originally gauge invariant is now BRST invariant by definition because the gauge parameters (c-numbers) are simply replaced by anti-commuting variables. It is this BRST symmetry that remains after we gauge fix the theory.
\\~\\
In order to gauge fix, one invokes a suitable gauge fixing fermion such that the action of the BRST transformation returns a gauge fixing term along with the so called ghost term. The ghost term in general has ghost fields coupled to the usual fields of the theory, in addition to kinetic terms. Thus the ghosts participate in scattering processes in a non trivial way. However, in any particular process, the ghost fields are not present as asymptotic states but only propagate off-shell and thus they are not physical. The physical states of the theory lie in the cohomology class of the BRST operator.\\~\\
In the present case, we have a system which has two kinds of gauge symmetries, local $SL(2,C)$ and diffeomorphisms. Thus, we need to introduce two kinds of ghost fields which enter the BRST transformations. As we will see, the transformation for the local Lorentz ($SL(2,C)$) ghost will have a term which contains the diffeomorphism ghosts and thus the two ghosts are coupled to generate the BRST complex. This is a  non-trivial feature of our formalism. We will gauge fix our theory on a general Einstein background and this makes the BRST closure property more non-trivial to verify. Once we have gauge fixed the theory, we can get the one-loop effective action using the heat kernel methods. Let us then begin to describe all this.
\\~\\
\subsection{BRST complex}
Consider two pairs of anti-commuting fields $c^{\mu},\bar{c}^{\mu}$ for diffeomorphisms and $b^{ij},\bar{b}^{ij}$ for $SL(2,C)$ transformations. The total BRST transformation operator is  
\be 
s=s_D +s_L,
\ee 
where $s_D$ is the operator corresponding to diffeomorphisms and $s_L$ is for local $SL(2,C)$ transformations. The transformation we use however is most conveniently written in the condensed form using the total transformation operator. Thus the splitting we described above is just to illustrate that it is the total transformation parameter that is nilpotent, i.e $s^2=0$. The individual transformation parameters are not nilpotent and this is the non triviality of the complex in the present case. 
Let us now define the BRST transformations
\be
\label{brstn}
\begin{split}
se^{i}_{~\mu}&=e^{i}_{~\nu}\nabla^T_{\mu}c^{\nu'}-b^{i}_{~j}e^{j}_{~\mu},\\ 
sc^{\mu}&=c^{\nu}\nabla^T_{\nu}c^{\mu},\\
s\bar{c}^{\mu}&=\lambda^{\mu},\\ 
s\lambda^{\mu}&=0,\\
s\omega^{ij}_{~~\mu}&=\nabla^T_{\mu}b^{ij}+c^{\nu}F^{ij}_{~~\nu\mu},\\
sb^{ij}&=-\frac{1}{2}[b,b]^{ij}+\frac{1}{2}c^{\mu}c^{\nu}F^{ij}_{~~\mu\nu},\\
s\bar{b}^{ij}&=\beta^{ij},\\
s\beta^{ij}&=0.
\end{split}
\ee 
As we can see, the transformation for the tetrad has two terms on it, one for diffeomorphism ghost and another for the Lorentz ghost. The relative sign is minus which is purely a convention. In a similar way, the spin connection transformation admits two terms, one of which follows from its Lorentz transformation and another gets added owing to diffeomorphisms. We now verify that the BRST transformation $s$ is nilpotent, i.e $s^2=0$. This is required to generate the BRST complex, such that physical states of the theory lie in the cohomology class of the BRST operator. For the third, fourth, seventh, eighth transformations in (\ref{brstn}), this condition trivially follows because $s^2\lambda^{\mu}=s^2\beta^{ij}=0$ and $s^2\bar{c}^{\mu}=s\lambda^{\mu}=0$, $s^2\bar{b}^{ij}=s\beta^{ij}=0$. For the rest of the transformations, we detail out in Appendix (3). Let us now write the BRST transformation in spinor notations.
In these notations, the total covariant derivative $\nabla^T_{\mu}$ becomes an object $\nabla^T_{MM'}$ which acts on the spinor version of both Lorentz and spacetime indices an $\omega^{~i}_{\mu~j}$ becomes and object $\omega^{AB}_{~~MM'}$ where $(AB)$ is a pair of symmetrised Lorentz indices and $MM'$ corresponds to the spacetime index $\mu$. We then write the following BRST transformations
\be
\label{brstnew}
\begin{split}
se^{AA'}_{~~MM'}&=\nabla^T_{MM'}c^{AA'}-b^{A}_{~~B}e^{BA'}_{~~MM'}\\
sc^{MM'}&=c^{LL'}\nabla^T_{LL'}c^{MM'},\\
s\bar{c}^{MM'}&=\lambda^{MM'},\\ 
s\lambda^{LL'}&=0,\\
s\omega^{AB}_{~~~CC'}&=c^{MM'}F^{AB}_{~~~MM'CC'}+\nabla^T_{CC'}b^{AB},\\
sb^{AB}&=-\frac{1}{2}[b,b]^{AB}+\frac{1}{2}c^{MM'}c^{NN'}F^{AB}_{~~~MM'NN'}\\
s\bar{b}^{AB}&=\beta^{AB},\\
s\beta^{A'B'}&=0.
\end{split}
\ee 
\subsection{Linearised action}
The main objective of the development of BRST complex is to deduce the ghost Feynman rules and also to understand the ghost contribution to the one-loop effective action in our chiral formalism. We will first analyze the later and subsequently describe the former. The one loop computation is efficiently done using the heat kernel methods. To employ it one needs to expand the action around an arbitrary background and then compute the regularised determinant of the differential operator which arises. However, in an arbitrary background, the computation becomes very complicated to suitably apply heat kernel techniques. Thus, let us try to expand the action in (\ref{EC}) around an Einstein background. We decompose the tetrad and the spin connection into a fixed background and consider small fluctuations around the background 
\be
\label{dec}
e^{AA'}_{~~~MM'}=\tilde{e}^{AA'}_{~~~MM'}+h^{AA'}_{~~~MM'},\ee 
where $\tilde{e}^{AA'}_{~~~MM'}$ is the background tetrad corresponding to an arbitrary Einstein background, so that the background metric is 
\be 
\tilde{g}_{\mu\nu}=\tilde{e}^{AA'}_{\mu}\tilde{e}_{\nu AA'}.
\ee 
The background metric satisfies Einstein equations. The fluctuations around this metric are then considered and the full metric is decomposed as
\be 
g_{\mu\nu}=\tilde{g}_{\mu\nu}+h_{\mu\nu}. 
\ee 
Similarly, the connection can be split into 
\be 
\label{split}
\omega^{AB}_{T~~CC'}= \omega^{AB}_{0~CC'}+w^{AB}_{~~~CC'}.
\ee 
where $\omega^{AB}_{0~CC'}$ is the background connection.
As a result, the decomposition of the curvature is given by 
\be
\begin{split} 
F_{T}&=F_{0}+\delta F,\\
F^{AB}_{0}&=d\omega^{AB}_0 +\omega_0^{AC}\wedge\omega_{0C}^B,\\
\delta F&=D_{\omega_0}\delta\omega=D_{\omega_0}w,
\end{split}
\ee \\
where in the above, $F_0$ is the background curvature constructed from the background connection, $\delta F$ is the linear order fluctuations around this baackground. The fluctuation can be expressed as a covariant derivative of the connection fluctuation with respect to the background connection. Let us now write the self-dual two form in terms of the splitting form of the tetrad. This will then be used to decompose the action into kinetic and interaction terms in a convenient way.
The self-dual two form can be written as 
\be 
\label{sd}
\begin{split} 
\Sigma^{AB}&=\frac{1}{2}\Big(\tilde{e}^{A}_{~A'}+h^{A}_{~A'}\Big)\wedge\Big(\tilde{e}^{BA'}+h^{BA'}\Big)\\
&=\frac{1}{2}\Big(\tilde{e}^A_{~A'}\wedge \tilde{e}^{BA'}+\tilde{e}^{A}_{~A'}\wedge h^{BA'}+h^{A}_{~A'}\wedge \tilde{e}^{BA'}+h^{A}_{~A'}\wedge h^{BA'}\Big).
\end{split}
\ee
Linearisation of the action in (\ref{EC}) through the decomposition in (\ref{dec}) is given by 
\be 
\begin{split} 
\label{linear}
S&=i\int \Big(\tilde{e}^A_{~A'}\wedge \tilde{e}^{BA'}+\tilde{e}^{A}_{~A'}\wedge h^{BA'}+h^{A}_{~A'}\wedge \tilde{e}^{BA'}+h^{A}_{~A'}\wedge h^{BA'}\Big)\\&\wedge\Big(d\omega_{TAB}+\omega_{TAC}\wedge \omega^{C}_{T~B}\Big)\\
&=i\int \Big(\tilde{e}^A_{~A'}\wedge \tilde{e}^{BA'}+\tilde{e}^{A}_{~A'}\wedge h^{BA'}+h^{A}_{~A'}\wedge \tilde{e}^{BA'}+h^{A}_{~A'}\wedge h^{BA'}\Big)\\&\wedge\Big(d\omega_{0AB}+\omega_{0AC}\wedge \omega^{C}_{0~B}\Big)\\
&+i\int \Big(\tilde{e}^A_{~A'}\wedge \tilde{e}^{BA'}+\tilde{e}^{A}_{~A'}\wedge h^{BA'}+h^{A}_{~A'}\wedge \tilde{e}^{BA'}+h^{A}_{~A'}\wedge h^{BA'}\Big)\wedge\Big(dw_{AB}+\omega_{0AC}\wedge w^{C}_{~B}\\&+w_{AC}\wedge w^{C}_{~B}\Big).
\end{split}
\ee 
where in the above steps we have decomposed the tetrad and the connection fields and written the action in terms of this decomposition. We now expand the wedge product between the tetrad field and the connection and collect all individual terms which can arise. The terms which comprise of just the background fields do not participate in the dynamical equations because they are fixed. Thus we ignore such terms. The rest of the terms have both the background and fluctuations in them. Let us first write the free part of the action. It is given by three terms respectively. 
\be
\begin{split} 
\label{action}
S_{free}&=i\int \tilde{e}^A_{~A'}\wedge \tilde{e}^{BA'}\wedge \Big(w_{AC}\wedge w^{C}_{~B}\Big)+i\int h^{A}_{~A'}\wedge h^{BA'}\wedge\Big(d\omega_{0AB}+\omega_{0AC}\wedge \omega^{C}_{0~B}\Big)\\
&+i\int \Big(\tilde{e}^{A}_{~A'}\wedge h^{BA'}+h^{A}_{~A'}\wedge \tilde{e}^{BA'}\Big)\wedge dw_{AB}.
\end{split} 
\ee 
The first term is quadratic in the connection fluctuations. The appearance of this term makes the tetrad propagator non-zero. The second term is quadratic in the tetrad fluctuations, while the third term is a cross term where metric/tetrad fluctuations is wedged with the connection fluctuation. Let us now write the interaction part of the action.
\be 
\begin{split} 
S_{interaction}&=2i\int\tilde{e}^{A}_{~A'}\wedge h^{BA'}\wedge w_{AC}\wedge w^{C}_{~B}\\&+i\int h^{A}_{~A'}\wedge h^{BA'}\wedge\Big(d\omega_{AB}+\omega_{0AC}\wedge \omega^{C}_{~B}+\omega_{AC}\wedge \omega^{C}_{~B}\Big).
\end{split}
\ee 
\\
We next expand the tetrad perturbation and the connection in terms of the background tetrad. This is done to parameterise our tetrad and connection fields in terms of the background and it appears convenient to write the action in terms of the coefficients of the expansion. Since the background is fixed once and for all, we can use it as a basis to expand the fluctuating fields. This is quite similar in spirit to that of the background field method for perturbation theory. The fluctuations of the tetrad and connection then reads
\be 
\begin{split}
\label{expand}
h^{AA'}&=h^{AA'}_{~~~MM'}\tilde{e}^{MM'},\\
\omega^{AB}&=\omega^{AB}_{~~~MM'}\tilde{e}^{MM'},\\
D\omega^{AB}&=\tilde{\nabla}_{MM'}\omega^{AB}_{~~~NN'}\tilde{e}^{MM'}\wedge \tilde{e}^{NN'}.
\end{split}
\ee 
where $\tilde{\nabla}_{MM'}$ is the background covariant derivative. Using the above expansion, we write the kinetic part of the free action as follows 
\be 
\begin{split}
\label{kfree}
S_{free}&=i\int\Big(\tilde{e}^{A}_{~A'}\wedge h^{BA'}+h^{A}_{~A'}\wedge \tilde{e}^{BA'}\Big)\wedge d\omega_{AB}\\
&=i\int(\tilde{e}^{A}_{A'}\wedge \tilde{e}^{MM'}\wedge \tilde{e}^{KK'}\wedge \tilde{e}^{JJ'})h^{BA'}_{~~~MM'}\tilde{\nabla}_{KK'}\omega_{ABJJ'}
\\
&+i\int(\tilde{e}^{MM'}\wedge \tilde{e}^{BA'}\wedge \tilde{e}^{KK'}\wedge \tilde{e}^{JJ'})h^A_{MM'A'}\tilde{\nabla}_{KK'}\omega_{ABJJ'}.
\end{split}
\ee 
In the above action, we have replaced the derivative on the spin connection by the appropriate background covariant derivative, which acts on the coefficient of the expansion of the connection on the basis of the background tetrads. This then couples with the coefficient of the fluctuation of the tetrad field, thus giving the resultant kinetic terms of the form $hd\omega$ with the background covariant derivatives everywhere in place of the ordinary derivatives.
Next we identify the wedge product of four copies of tetrad as the oriented volume form
\be 
\label{wedge}
\tilde{e}^{AA'}\wedge \tilde{e}^{BB'}\wedge \tilde{e}^{CC'}\wedge \tilde{e}^{DD'}=\epsilon^{AA'BB'CC'DD'}v.
\ee 
and use the spinor representation of the totally anti-symmetric tensor. 
\be 
\label{anti}
\epsilon^{AA'BB'CC'DD'}=i\Big(\epsilon^{AD}\epsilon^{BC}\epsilon^{A'C'}\epsilon^{B'D'}-\epsilon^{AC}\epsilon^{BD}\epsilon^{A'D'}\epsilon^{B'C'}\Big).
\ee 
The validity of the spinor representation can be easily checked. For instance, if we swap any pair of indices, say $AA'\leftrightarrow BB'$, the right hand side picks up an overall minus sign, because the first term of the right side goes to the second under the swap. In a similar way, for all such pairs, an overall sign factor comes up. This shows the totally anti-symmetric nature of the tensor $\epsilon^{AA'BB'CC'DD'}$. Let us then use it and write the kinetic part in (\ref{kfree}) as
\be 
\begin{split} 
\label{kinetic1} 
S_{free}&=\int d^4x\Big[ h^{BJ'JK'}\tilde{\nabla}_{KK'}\omega^K_{~BJJ'}- h^{BK'KJ'}\tilde{\nabla}_{KK'}\omega^J_{~BJJ'}\Big]\\
&+\int d^4x\Big[h^{AKJ'K'}\tilde{\nabla}_{KK'}\omega_{A~JJ'}^{~J}-h^{AJK'J'}\tilde{\nabla}_{KK'}\omega^{~K}_{~AJJ'}\Big]\\
&=-2\int d^4x \Big[ h^{BJ'JK'}\Big(\tilde{\nabla}_{JJ'}\omega^{K}_{~BKK'}-\tilde{\nabla}_{KK'}\omega^K_{~BJJ'}\Big)\Big].
\end{split} 
\ee 
We now decompose the tetrad perturbation into its irreducible components. This is reminiscent of the decomposition of a tensorial quantity, for instance the Riemann tensor, in which we treat the tetrad field as an object with four spinor indices (analogous to a rank four tensor). The decomposition reads
\be 
\label{tetrad dec}
h^{AA'MM'}=h^{(AM)(A'M')}+h^{(AM)}\epsilon^{A'M'}+h^{(A'M')}\epsilon^{AM}+\epsilon^{AM}\epsilon^{A'M'}h.
\ee 
The first object on the right hand side above is a tetrad field with its primed and unprimed indices symmetrised. The object with two spinor indices in the second and third terms are the part of the tetrad perturbation is a symmetric field which only propagate off-shell.
It is then convenient to combine the second and fourth terms and define a new field $h^{AM}=h^{(AM)}+h\epsilon^{AM}$. This field is thus no more symmetric in its pair of unprimed indices. This results in 
\be 
h^{AA'MM'}=h^{(AM)(A'M')}+h^{AM}\epsilon^{A'M'}+h^{(A'M')}\epsilon^{AM}.
\ee 
We get rid of the $h^{A'M'}$ part of the perturbation by setting one chiral half of the Lorentz gauge to zero. This can be done because this part of the perturbation does not appear in the free Lagrangian. Thus it simplifies the computation to some extent. We put the decomposition 
\be 
h^{AA'MM'}=h^{(AM)(A'M')}+h^{AM}\epsilon^{A'M'}
\ee 
in (\ref{kinetic1}) to get for the kinetic part of the Lagrangian 
\be 
\begin{split}
\label{Lk}
\mathcal{L}_{kinetic}&=2h^{JKJ'K'}\Big(\tilde{\nabla}_{KJ'}\omega^M_{~JMK'}-\tilde{\nabla}_{MK'}\omega_{J~KJ'}^{~M}\Big)\\&+2h^{JK}\Big(\tilde{\nabla}_{KJ'}\omega^{M~~J'}_{~JM}-\tilde{\nabla}_{MK'}\omega_{J~K}^{~M~K'}\Big).
\end{split}
\ee 
The kinetic term above can be written in a compact way. It is possible to combine the parts of the tetrad perturbation into a single object and the covariant derivative can then act on an appropriate combination of connection perturbations. This is helpful because it gives rise to a familiar kinetic term of the form $h\nabla\omega$ in a general background which is the one suitable for a chiral first order formulation. Thus, re-written in this way, we get  
\be 
\label{compactke}
\begin{split}
\mathcal{L}_{kinetic}&=-2\Big(h^{JKJ'K'}-h^{(JK)}\epsilon^{J'K'}\Big)\tilde{\nabla}_{MJ'}\Big(\omega_{J~KK'}^{~M}+\epsilon^{M}_{~~K}\omega^N_{~JNK'}\Big)\\&+4h^{JK}\tilde{\nabla}_{KK'}\omega^{M~~K'}_{~JM}.
\end{split}
\ee 
We now redefine some fields as follows 
\be 
\begin{split} 
\omega^{JJ'}&=\omega^{MJ~~J'}_{~~~M},\\
\Omega^{JKMJ'}&=\omega^{JMKJ'}+\epsilon^{M}_{~~K}\omega^{JJ'}.
\end{split}
\ee 
Thus the Lagrangian is written in terms of the new variables 
\be 
\label{compactke}
\begin{split}
\mathcal{L}_{kinetic}&=-2\Big[h^{JKJ'K'}-h^{(JK)}\epsilon^{J'K'}\Big]\tilde{\nabla}_{MJ'}\Omega^{~M}_{J~KK'}+4h^{JK}\tilde{\nabla}_{KJ'}\omega^{~J'}_{J}.
\end{split}
\ee 
As we see, there are two terms in the kinetic part above. However, the kinetic term is degenerate. This is because the connection field $\Omega^{JKMJ'}$ has twelve independent components, while the other field $\omega^{JJ'}$ has four components. The tetrad perturbation $h^{JKJ'K'}$ has nine components, which is conjugate to $\Omega^{JKMJ'}$ and therefore there is a mismatch in the number of components, which makes it degenerate. To remove the degeneracy we need to completely fix the gauge. We want to add a specific gauge fixing and ghost term. This will be done by introducing a gauge fixing fermion in the BRST formalism. Thus we now introduce the linearised BRST which we subsequently use to fix our gauges. 
\subsection{Linearised BRST}
The main purpose of this section is to disentangle the background from the perturbations while defining the BRST transformations. This will allow us to systematically use the transformations for the perturbations themselves. Let us note how the perturbations of the metric and the connection transform under diffeomorphisms and local $SL(2,C)$ transformations. We begin with the transformation for the general tetrad. It admits the following decomposing into background and quantum fluctuations
\be 
e^{AA'}_{~~~BB'}=\tilde{e}^{AA'}_{~~~BB'}+h^{AA'}_{~~~BB'}.
\ee 
where the perturbation $h^{AA'}_{~~~BB'}$is of order $\epsilon$ and is very small compared to the background. 
Then, we get for the total transformation of the tetrad perturbation
\be 
\begin{split}
\label{trans}
\delta_{\xi}h^{AA'}_{~~~BB'}&=\tilde{e}^{AA'}_{~~~NN'}\tilde{\nabla}_{BB'}\xi^{NN'}-\phi^A_{~C}\tilde{e}^{CA'}_{~~BB'}.
\end{split}
\ee
where $\tilde{\nabla}_{BB'}$ is the background total covariant derivative.
For the connection field, we similarly decompose it into the background and perturbations
\be 
\label{pert}
\omega^{AB}_{T~~CC'}= \omega^{AB}_{0~CC'}+w^{AB}_{~~~CC'}.
\ee 
where $\omega^{AB}_{0~CC'}$ is the background and $w^{AB}_{~~~CC'}$ is the perturbation of order $\epsilon$.
We get for the transformation of the perturbation, under diffeomorphisms,
\be 
\label{omegavar}
\delta_{\xi}w^{AB}_{~~CC'}&=\xi^{NN'}F^{AB}_{0~~NN'CC'}+\tilde{\nabla}_{CC'}\phi^{AB}.
\ee
where $F^{AB}_{0~~NN'CC'}$ is the background curvature.
With these linearised versions of the diffeomorphisms and gauge transformations, we are now ready to write the BRST transformations of these fields. As usual, we have two pairs of ghost fields, each for diffeomorphisms and local Lorentz transformations. Our linearised BRST is then given by
\be
\label{brstlin}
\begin{split}
sh^{AA'}_{~~MM'}&=\tilde{\nabla}_{MM'}c^{AA'}-b^{A}_{~~B}\tilde{e}^{BA'}_{~~MM'},\\
sc^{MM'}&=c^{LL'}\nabla_{LL'}c^{MM'},\\
s\bar{c}^{MM'}&=\lambda^{MM'},\\ 
s\lambda^{LL'}&=0,\\
sw^{AB}_{~~~CC'}&=c^{MM'}F^{AB}_{0~~MM'CC'}+\tilde{\nabla}_{CC'}b^{AB},\\
sb^{AB}&=-\frac{1}{2}[b,b]^{AB}+\frac{1}{2}c^{MM'}c^{NN'}F^{AB}_{0~~MM'NN'},\\
s\bar{b}^{AB}&=\beta^{AB},\\
s\beta^{A'B'}&=0.
\end{split}
\ee 
\subsection{Gauge fixing fermion and ghosts}
We follow the gauge fixing procedure which is already described in \cite{Krasnov:2020bqr}. The main difference is that we are dealing with a general Einstein background as opposed to the flat background in \cite{Krasnov:2020bqr}. Thus, instead of linear gauges, there will be non-linearities present in our formalism. We also adopt the BRST gauge fixing procedure, in which we write the gauge fixing fermion and take BRST variations to produce the gauge fixing and ghost terms. This will help us to generate the ghost Lagrangian in a general Einstein background. Later, in the context of amplitudes, we can take its flat space limit and get the corresponding Feynman rukes. Let us then begin to fix the gauges. One chiral half of the Lorentz was fixed by making the $h^{A'B'}$ part of the perturbation vanish. The other chiral half of the Lorentz can be fixed by imposing the non-linear version of the Lorentz gauge fixing condition. Consider then the following gauge fixing fermion 
\be 
\label{Lgf}
\begin{split}
\psi_{Lorentz}&=\bar{b}^{JK}\tilde{\nabla}_{MM'}\Omega^{~~~MM'}_{JK}.
\end{split}
\ee 
Using (\ref{brstlin}), the BRST variation of the gauge fixing fermion gives
\be
\begin{split}
s\psi_{Lorentz}&=s\bar{b}^{JK}\tilde{\nabla}_{MM'}\Omega^{~~~MM'}_{JK}-\bar{b}^{JK}\tilde{\nabla}_{MM'}s\Omega^{~~~MM'}_{JK}\\
&=\beta^{JK}\tilde{\nabla}_{MM'}\Omega^{~~~MM'}_{JK}-\bar{b}^{JK}\tilde{\nabla}_{MM'}\tilde{\nabla}^{MM'}b_{JK}\\&-\bar{b}^{JK}\tilde{\nabla}_{MM'}\Big(c^{CC'}F^{MM'}_{0~~JKCC'}\Big).
\end{split}
\ee 
The variation of the Lorentz gauge fixing fermion results in a Lorentz gauge fixing term of the form
\be 
\mathcal{L}_{Lorentz~g.f}=-2\beta^{JK}\tilde{\nabla}_{MM'}\Omega_{JK}^{~~~MM'}.
\ee 
The above term can be written as $-2\beta^{JK}\epsilon^{J'K'}\nabla_{MJ'}\Omega_{J~KK'}^{~M}$, which when added to the Lagrangian in (\ref{compactke}) along with the ghost terms results in
\be 
\label{lorentzgf}
\begin{split}
   \mathcal{L}&=-2\Big[h^{JKJ'K'}+(\beta^{JK}-h^{(JK)})\epsilon^{J'K'}\Big]\tilde{\nabla}_{MJ'}\Omega^{~M}_{J~KK'}+4h^{JK}\tilde{\nabla}_{KJ'}\omega^{~J'}_{J} \\&-\bar{b}^{JK}\tilde{\nabla}_{MM'}\tilde{\nabla}^{MM'}b_{JK}-\bar{b}^{JK}\tilde{\nabla}_{MM'}\Big(c^{CC'}F^{MM'}_{0~~JKCC'}\Big).
\end{split}
\ee
The remaining diffeomorphism gauge can be fixed by implementing a variant of the de-Donder gauge fixing condition as follows. First we introduce a new name for the combination of the terms in the bracketed expression of the first line in (\ref{lorentzgf})
\be 
\label{redf}
H^{JKJ'K'}:=h^{JKJ'K'}+(\beta^{JK}-h^{(JK)})\epsilon^{J'K'}.
\ee 
Note that since the field $\beta^{JK}$ is independent, this results in the two fields $H^{JKJ'K'}$ and $h^{JK}$ being independent of each other. This then leads to decoupling of the kinetic part of the Lagrangian into two sectors, $(H,\Omega)$ and $(h,\omega)$. We now construct the diffeomorphism gauge-fixing fermion to be 
\be 
\label{gff}
\psi_{diffeo}=2\bar{c}^{JJ'}\Big[\tilde{\nabla}_{KK'}\Big(h^{~K'K}_{J~~~J'}+h^{K~~~K'}_{~J'J}\Big)-2\tilde{\nabla}_{KJ'}\beta^{~K}_J-2\beta_{JJ'}\Big].
\ee 
After using the redefinition in (\ref{redf}) we take the BRST variation of it. We have 
\be 
\begin{split} 
\label{brstdiffeo}
s\psi_{diffeo}&=4\beta^{JJ'}\Big[\tilde{\nabla}^{KK'}H_{JKJ'K'}-\tilde{\nabla}^K_{~J'}H^{~~M'}_{JK~M'}+\tilde{\nabla}^K_{~J'}h_{JK}-\beta_{JJ'}\Big]\\
&-4\bar{c}^{JJ'}\Big[\tilde{\nabla}^{KK'}\Big(\tilde{\nabla}_{KK'}c_{JJ'}-b_{JK}\epsilon_{J'K'}\Big)\Big].
\end{split}
\ee 
The variation produces a diffeomorphism gauge fixing term and a ghost term. The gauge fixing term is an analogue of the variant of de-Donder gauge in curved spacetime. The ghost term consists of a kinetic part for the diffeomorphism ghost and a mixed part where the diffeomorphism and Lorentz ghosts couple.
Then, the total gauge fixing term becomes 
\be
\begin{split} 
\label{totalgf}
\mathcal{L}_{g.f}&=4\beta^{JJ'}\Big[\tilde{\nabla}^{KK'}H_{JKJ'K'}-\tilde{\nabla}^{K}_{~J'}H^{~~M'}_{JK~M'}+\tilde{\nabla}^{K}_{~J'}h_{JK}-\beta_{JJ'}\Big]\\&-2\beta^{JK}\tilde{\nabla}_{MM'}\Omega_{JK}^{~~~MM'}.
\end{split} 
\ee 
and the ghost Lagrangian is given by 
\be 
\label{ghost}
\begin{split} 
\mathcal{L}_{ghost}&=-4\bar{c}^{JJ'}\tilde{\nabla}^{KK'}\tilde{\nabla}_{KK'}c_{JJ'}+4\bar{c}^{JK'}\tilde{\nabla}_{KK'}b^{K}_{~J}-\bar{b}^{JK}\tilde{\nabla}_{MM'}\tilde{\nabla}^{MM'}b_{JK}\\&-\bar{b}^{JK}\tilde{\nabla}_{MM'}\Big(c^{CC'}F^{MM'}_{0~~JKCC'}\Big).
\end{split}
\ee 
The Lagrangian above has terms with Lorentz and diffeomorphism ghosts mixed. To compute the ghost contribution to the one-loop effective action, we need to diagonalize and rid of the mixed terms. First, we re-scale the fields 
\be 
\begin{split} 
\label{rescale} 
c^{AA'}&\rightarrow 2c^{AA'},~~\bar{c}^{AA'}\rightarrow 2\bar{c}^{AA'}.
\end{split} 
\ee 
and rewrite the Lagrangian 
\be 
\label{ghost2}
\begin{split} 
\mathcal{L}_{ghost}&=-\bar{c}^{JJ'}\tilde{\nabla}^{KK'}\tilde{\nabla}_{KK'}c_{JJ'}+2\bar{c}^{JK'}\tilde{\nabla}_{KK'}b^{K}_{~J}-\bar{b}^{JK}\tilde{\nabla}_{MM'}\tilde{\nabla}^{MM'}b_{JK}\\&-\frac{1}{2}\bar{b}^{JK}\tilde{\nabla}_{MM'}\Big(c^{CC'}F^{MM'}_{0~~JKCC'}\Big).
\end{split}
\ee 
We now use the background field equation and write 
\be
\begin{split}
\label{psi}
F_{0~~MM'NN'}^{AB}&= \psi^{AB}_{~~CD}\Sigma^{CD}_{0~~MM'NN'}
=\psi^{AB}_{~~MN}\epsilon_{M'N'}.
\end{split}
\ee 
where $\psi^{ABCD}$ is the self-dual part of the Weyl tensor and the self dual two-form $\Sigma_0$ has the particular representation in terms of the epsilons, given in the second line of the above equation. When we contract all the indices of $F_{0}^{AB}$, we end up with contracting all the indices of $\psi^{ABCD}$, which implies vanishing of the self dual part of the Weyl tensor. 
\be 
\begin{split} 
F_0&=F_{0~~MM'NN'}^{AB}\epsilon^{N}_{~A}\epsilon^{M}_{B}\epsilon^{M'N'}\\
&=\psi^{AB}_{~~MN}\epsilon_{M'N'}\epsilon^{N}_{~A}\epsilon^{M}_{~B}\epsilon^{M'N'}=0.
\end{split}
\ee
where we used the fact that the self-dual part of the Weyl tensor vanishes upon index contraction.
The Lagrangian in (\ref{ghost2}) can now be written by replacing $F_0$ with the self-dual part of the Weyl tensor. We have  
\be 
\label{ghost3!}
\begin{split} 
\mathcal{L}_{ghost}&=-\bar{c}^{JJ'}\tilde{\nabla}^{KK'}\tilde{\nabla}_{KK'}c_{JJ'}+2\bar{c}^{JK'}\tilde{\nabla}_{KK'}b^{K}_{~J}-\bar{b}^{JK}\tilde{\nabla}_{MM'}\tilde{\nabla}^{MM'}b_{JK}\\&-\frac{1}{2}\psi^{M}_{~~CJK}\bar{b}^{JK}\tilde{\nabla}_{MM'}c^{CM'}.
\end{split}
\ee 
\subsubsection{Setup for heat kernel computation}
The Lagrangian in (\ref{ghost3!}) is not diagonal in the  fields. It has mixed terms which arise from the coupling of diffeomorphism and Lorentz ghosts. To do a heat kernel computation, it is necessary to obtain a diagonalised Lagrangian such that the differential operator which arises is of Laplace type. Thus, we now proceed to diagonalize it. Let us first give a outline of the diagonalization procedure and then we will implement it in our problem. Consider a differential operator of the form 
\be 
\partial^\mu \partial_\mu \mathds{1} + \partial_\mu A^\mu + B,
\ee 
where $A^{\mu}$ and $B$ are matrices in some matrix space. This operator can in turn be written as a matrix which then acts on the column vector comprising of fields $b,c$. The conjugate fields $\bar{b},\bar{c}$ are contracted from the left and this produces the Lagrangian. It is then possible to absorb the linear parts of the differential operator by a suitable redefinition such that it results in an operator of the form 
\be 
\mathcal{D}^{\mu} \mathcal{D}_{\mu} + E,
\ee 
where $E$ is some endomorphism on the field space and $\mathcal{D}_{\mu}$ is an appropriate covariant derivative operator. It is then customary to first write the Lagrangian in (\ref{ghost3!}) in a matrix form, so that one can diagonalize the matrix by suitable field re-definitions. We have
\be 
\label{ghost3}
\mathcal{L}_{ghost}&=-\begin{pmatrix}
    \bar{b}^{JK} & \bar{c}^{BB'}  
  \end{pmatrix}
  \begin{pmatrix}
   X &  Y\\
   Z & X
  \end{pmatrix}
  \begin{pmatrix}
    b_{JK} \\
    c_{BB'}\\
   \end{pmatrix},
\ee 
where 
\be 
\begin{split} 
X&=\tilde{\nabla}_{MM'}\tilde{\nabla}^{MM'},\\
Y&=\frac{1}{2}\psi^{MB}_{~~~JK}\tilde{\nabla}^{~B'}_{M},\\
Z&=-\tilde{\nabla}^{K}_{~B'}\epsilon^{J}_{~B}.\\
\end{split} 
\ee 
where in writing $Y$, we used the Bianchi identity $\nabla_{MB'}\psi^{M}_{~~BJK}=0$ and got rid of the other term.
The matrix in (\ref{ghost3}) can be split in the following way \be
\begin{split}
\label{matrix}
 \begin{pmatrix}
   X &  Y\\
   Z & X
  \end{pmatrix}&=
\tilde{\nabla}_{MM'}\tilde{\nabla}^{MM'}\begin{pmatrix}
   1 &  0\\
   0 & 1
  \end{pmatrix}+
 \begin{pmatrix}
   0 & \frac{1}{2}\epsilon^{M'B'} \psi^{MB}_{~~~JK}\\
  -\epsilon^{J}_{~B}\epsilon^{KM}\epsilon^{M'}_{~B'} & 0
  \end{pmatrix}\tilde{\nabla}_{MM'} \\~\\&+
  \tilde{\nabla}^{MM'}\begin{pmatrix}
    0 & \frac{1}{2}\epsilon_{M'B'} \psi^{B}_{~~MJK} \\
  -\epsilon^{J}_{~B}\epsilon^{K}_{~M}\epsilon_{~M'B'} & 0
  \end{pmatrix}.
  \end{split}
  \ee 
This can now be re-written by absorbing the first order derivative parts and expressing it in the form of a Laplace type operator 
\be 
\begin{pmatrix}
   X &  Y\\
   Z & X
  \end{pmatrix}&=\mathcal{D}^{MM'}\mathcal{D}_{MM'}+E,
\ee 
where we define the new connection
\be 
\label{connnew}
\mathcal{D}^{MM'}=\begin{pmatrix}
   \tilde{\nabla}^{MM'} &  \frac{1}{2}\epsilon^{M'B'} \psi^{MB}_{~~~JK} \\~\\
  \epsilon^{J}_{~B}\epsilon^{KM}\epsilon^{M'}_{~B'}  & \tilde{\nabla}^{MM'}
  \end{pmatrix}
 \ee 
and 
\be
E=-\begin{pmatrix}
   \psi^{KJ}_{~~~JK} &  0 \\~\\
  0 & \psi^{KJ}_{~~~JK} 
  \end{pmatrix}.
 \ee 
Upon index contractions, the self dual part of the Weyl tensor
vanishes and therefore we can set $E=0$ henceforth. 
We thus have the relevant Laplace type operator, which in this case with $E=0$ is 
\be 
\Delta :=\mathcal{D}_{MM'}\mathcal{D}^{MM'}.
\ee
with the new connection defined in (\ref{connnew}).
\section{Heat Kernel computation}
Let us consider a Laplace type operator of the form $\Delta=D^{\mu}D_{\mu}+E$. The operator acts on some vector bundle, $E$ is an endomorphism on the fibre and $D_{\mu}$ is an appropriate covariant derivative. The interesting object to study is the determinant $\textrm{det}(\Delta)$. We use the identity log det($\Delta$)=Tr log($\Delta$) and then rewrite the logarithm of the determinant in terms of an integral  
\be 
\label{logdet}
\textrm{log det}(\Delta)=-\int^{\infty}_0\frac{dt}{t}Tr(e^{-t\Delta}).
\ee 
Then we have the well known expansion of the trace under the integral, in powers of the auxiliary variable $t$
\be 
\label{trace}
\textrm{Tr}(e^{-t\Delta})=\int d^4x \sqrt{g}\frac{1}{(4\pi t)^2}\sum_{n=0}^{\infty} t^n a^R_n(E).
\ee 
Our task is to compute the heat kernel coefficients $a^R_n(E)$. In the chiral Einstein-Cartan theory around some Einstein background, we have in consideration a manifold $M$ without a boundary, over which we have a vector bundle $V$. The Laplace operator $\mathcal{D}_{\mu}\mathcal{D}^{\mu}$ acts on $V$. Then there exists an expansion of the heat kernel coefficients, such that all the odd indexed coefficients vanish and the even indexed coefficients are given by geometric invariants. Particularly, the UV divergent behaviour is controlled by the coefficient $a^R_2(E)$, which is given by 
\be
\label{coeff}
\begin{split} 
a^\mathcal{R}_2(E)&=Tr_{\mathcal{R}}\Big[\frac{1}{6}\mathcal{D}^2E+\frac{1}{2}E^2+\frac{1}{6}RE+\frac{1}{12}\Omega_{\mu\nu}\Omega^{\mu\nu}\\
&+\frac{1}{30}\mathcal{D}^2R+\frac{1}{72}R^2-\frac{1}{180}R_{\mu\nu}R^{\mu\nu}+\frac{1}{180}R_{\mu\nu\rho\sigma}R^{\mu\nu\rho\sigma}\Big].
\end{split}
\ee 
where $\Omega_{\mu\nu}$ is the curvature of the resulting new covariant derivative operator, defined as 
\be 
\begin{split}
\Omega_{\mu\nu}&=[\mathcal{D}_{\mu},\mathcal{D}_{\nu}]\\
&=\begin{pmatrix}
   \tilde{\nabla}_{\mu} &   Y_{1\mu}\\
   Y_{2\mu} & \tilde{\nabla}_{\mu}
  \end{pmatrix}
  \begin{pmatrix}
   \tilde{\nabla}_{\nu} &   Y_{1\nu}\\
   Y_{2\nu} & \tilde{\nabla}_{\nu}
  \end{pmatrix}
  -\begin{pmatrix}
   \tilde{\nabla}_{\nu} &   Y_{1\nu}\\
   Y_{2\nu} & \tilde{\nabla}_{\nu}
  \end{pmatrix}
  \begin{pmatrix}
   \tilde{\nabla}_{\mu} &   Y_{1\mu}\\
   Y_{2\mu} & \tilde{\nabla}_{\mu}
  \end{pmatrix}\\
&=\begin{pmatrix}
   \tilde{\nabla}_{[\mu}\tilde{\nabla}_{\nu]}+Y_{1[\mu}Y_{2\nu]} &  \tilde{\nabla}_{[\mu}Y_{1\nu]} +Y_{1[\mu}\tilde{\nabla}_{\nu]}\\~\\
   Y_{2[\mu}\tilde{\nabla}_{\nu]}+\tilde{\nabla}_{[\mu}Y_{2\nu]} & Y_{2[\mu}Y_{1\nu]}+\tilde{\nabla}_{[\mu}\tilde{\nabla}_{\nu]}
  \end{pmatrix}.
 \end{split}
\ee
Let us compute the matrix elements of $\Omega_{\mu\nu}$ using spinor notations, which will yield simpler expressions. The diagonal elements are equal to the commutator of the total covariant derivative. This commutator can act on the Lorentz ghost which has just the internal indices or the diffeomorphism ghost which has only the spacetime indices. Whatever the ghost field, the action of the commutator can only be proportional to the curvature of the spin connection. This is because the curvature of the spin connection is related to the curvature of the Levi-civita connection via the tetrad one-forms and thus they are equivalent descriptions of the same entity. Further, the background field equation expresses the curvature in terms of the Weyl spinor. We already arrived at a simple form of the curvature in spinor notations when expressed in terms of the Weyl spinor. Thus, overall the diagonal part of $\Omega_{\mu\nu}$ in spinor notations contributes  
\be 
\begin{split} 
\label{matrixelements}
\tilde{\nabla}_{[MM'}\tilde{\nabla}_{NN']}+Y_{1[MM'}Y_{2NN']}
&=\epsilon_{M'N'}\psi^{AB}_{~~MN},
\end{split} 
\ee 
where we used the fact that when two of the indices of the Weyl spinor are contracted, it vanishes, i.e $\psi^{J}_{~MJK}=0$. Thus the contribution from the second term of the left hand side above is zero. Let us now compute the off-diagonal parts. The lower off-diagonal element when evaluated in spinor notations give 
\be
\begin{split} 
\tilde{\nabla}_{[\mu}Y_{2\nu]} +Y_{2[\mu}\tilde{\nabla}_{\nu]}&=\tilde{\nabla}_{MM'}\epsilon^{J}_{~B}\epsilon^{K}_{~N}\epsilon_{N'B'}-\tilde{\nabla}_{NN'}\epsilon^{J}_{~B}\epsilon^{K}_{~M}\epsilon_{M'B'}\\&+\epsilon^{J}_{~B}\epsilon^{K}_{~M}\epsilon_{M'B'}\tilde{\nabla}_{NN'}-\epsilon^{J}_{~B}\epsilon^{K}_{~N}\epsilon_{N'B'}\tilde{\nabla}_{MM'}\\
&=0.
\end{split} 
\ee 
Let us also compute the upper right off-diagonal part. This is given by 
\be
\begin{split} 
\tilde{\nabla}_{[\mu}Y_{1\nu]} +Y_{1[\mu}\tilde{\nabla}_{\nu]}&=\epsilon_{N'C'}\nabla_{MM'}\psi^{JK}_{~~NC}-\epsilon_{M'C'}\nabla_{NN'}\psi^{JK}_{~~MC}\\&+\psi^{JK}_{~~MC}\epsilon_{M'C'}\tilde{\nabla}_{NN'}-\psi^{JK}_{~~NC}\epsilon_{N'C'}\tilde{\nabla}_{MM'}.
\end{split} 
\ee 
We now need to compute $\Omega^2$. Let us first see the structure of the off-diagonal elements which arise in it. Indeed, it is easy to check that the lower left off-diagonal element must vanish and the upper left will be given by
\be 
\begin{split} 
&\psi^{MN}_{~~~AB}\epsilon^{M'N'}\Big(\epsilon_{N'C'}\nabla_{MM'}\psi^{JK}_{~~NC}-\epsilon_{M'C'}\nabla_{NN'}\psi^{JK}_{~~MC}\\&+\psi^{JK}_{~~MC}\epsilon_{M'C'}\tilde{\nabla}_{NN'}-\psi^{JK}_{~~NC}\epsilon_{N'C'}\tilde{\nabla}_{MM'}\Big).
\end{split}
\ee 
Due to index contractions of the Weyl spinor, the above term vanishes. Then, the square of the curvature is given entirely by diagonal terms, which are however equal and is given by 
\be 
\begin{split} 
\Omega_{MM'NN'}\Omega^{MM'NN'}&=\begin{pmatrix}
  2\psi_{ABMN}\psi^{ABMN}  &  0 \\~\\
 0 & 2\psi_{ABMN}\psi^{ABMN}
  \end{pmatrix}.
\end{split}
\ee
We thus have the relevant heat kernel coefficient as 
\be
\label{coeff}
\begin{split} 
a^\mathcal{R}_2&=Tr_{\mathcal{R}}\Bigg[\frac{19}{90}\psi_{MNPS}\psi^{MNPS}\Bigg].
\end{split}
\ee 
We thus see that the heat kernel coefficient depends on just the self-dual part of the Weyl tensor. The one-loop ghost contribution to the effective action thus also depends entirely on it. We do not compute the complete one-loop effective action here. This is because the contribution from the bosonic part of the action is left to be worked out and we do not compute this in our thesis. We postpone this remaining work and plan to include it in a future publication. For completeness and brevity, we outline the basic technology behind extracting the effective action and the relavant $\beta$-function from the heat kernel computation. 
\section{One-loop effective action} 
In the background field method, the effective action is defined by considering the quantum fluctuations around a fixed background field. The generating functional is given by 
\be 
\label{pathint}
e^{iW(\xi_0, J)}=\int \mathcal{D}\xi e^{i(S(\xi+\xi_0)+J\xi)}.
\ee 
So, the background field effective action is 
\be 
\Gamma[\xi_0, \xi]=W(\xi_0, J)-J\tilde{\xi}, 
\ee 
where 
\be 
\tilde{\xi}=\frac{\delta W(\xi_0,J)}{\delta J}.
\ee 
Under the Wick rotation, we go to the Euclidean signature and then define the Euclidean path integral analogous to (\ref{pathint}). In the absence of sources, we put $J=0$ and obtain 
\be
\begin{split} 
e^{-\Gamma(\xi_0)}&=\int \mathcal{D}\xi e^{-S(\xi+\xi_0)}\\
&=e^{-S(\xi_0)}\int\mathcal{D}\xi e^{-\int\xi\Delta\xi}.
\end{split} 
\ee 
Using the Gaussian integration formula 
\be 
\int\Pi_j dx_je^{-\frac{1}{2}x_a M_{ab}x_b}=[\textrm{det}(M)]^{-1/2}
\ee 
we find 
\be
\begin{split} 
e^{-\Gamma(\xi_0)}&=e^{-S(\xi_0)}[\textrm{det}(\Delta)]^{-1/2}.
\end{split} 
\ee 
where $\Delta$ is generalized Laplace operator obtained after linearizing the action around the background field. We then have the one-loop effective action 
\be 
\Gamma_{1-loop}=\frac{1}{2}\textrm{log det}(\Delta).
\ee 
The relevant expression for the one-loop effective action is given using (\ref{logdet}) and (\ref{trace})
\be 
\Gamma_{1-loop}=-\frac{1}{2(4\pi)^2}\int_0^{\infty}dt~ t^{n-3}\sum_{n=0}^{\infty}\int d^4x \sqrt{g}~a^R_n(E).
\ee 
In a renormalization scale, which is free of any mass scale, the running of the coupling constants is obtained from the logarithmically divergent part. This part is extracted from the third term in the above expansion, particularly 
\be 
\gamma=\frac{1}{(4\pi)^2}\int d^4x \sqrt{g}~a_2.
\ee 
which allows to write 
\be 
\Gamma^{\textrm{log}}_{1-loop}=-\frac{1}{2}\gamma\int_0^{\infty}\frac{dt}{t}=\gamma~\textrm{log}\frac{\delta}{\delta_0}.
\ee 
where we have regularized the integral of $t$ by introducing some cut-offs $t_{max}\approx 1/\delta^2$, $t_{min}\approx 1/\delta_0^2$. Therefore, we can write 
\be 
\frac{\partial \Gamma_{1-loop}^{\textrm{log}}}{\partial log \delta}=\gamma.
\ee 
\newpage 
\section{Chiral Einstein-Cartan in flat space} 
In the previous section, we developed the ghost Lagrangian for the chiral action in (\ref{EC}) in a general Einstein background. Using a non-linear gauge fixing fermion, we arrived at this Lagrangian and computed the ghost contribution to the one-loop effective action. In this section, we sketch the perturbation theory of the same action on Minkowski background. Thus, we expand the chiral action around the Minkowski space configuration, with zero background connection. We will not detail the gauge fixing procedure here. The interested reader may refer \cite{Krasnov:2020bqr}, where a detailed gauge fixing procedure for the flat background case is outlined. The gauge fixing fermion in this case can be recovered in the flat space limit where we replace all covariant derivatives in (\ref{totalgf}) by partial derivatives.
The perturbation theory in flat space is best suited for amplitude computations. The chiral Einstein-Cartan action after gauge fixing leads to very simple Feynman rules. In particular, amongst the three propagators, namely tetrad-tetrad, tetrad-connection and connection-connection, the propagator of the connection to itself vanishes. These aspects are detailed in \cite{Krasnov:2020bqr} and we will only quote the main results here. Our new result here is the ghost Lagrangian and the corresponding ghost Feynman rules at the linearised level, which we explain in some details. The interaction part of the linearised gauge fixed Lagrangian contains three kinds of terms. But the only relevant one for amplitude computations in our case is $hh\partial \omega$. We will explain some simplifications which occur for this interaction in light of the gauge fixing procedure.

\subsection{Linearised action}
The action in (\ref{EC}) can be expanded around the Minkowski background. Thus, we start with the zero connection configuration and denote the perturbation of the connection as $\omega^{CD}$. The Minkowski background tetrad is denoted by $e^{BB'}$ and its corresponding perturbation is $h^{BB'}$. Where confusion do not arise, we use the same symbols for the perturbations in the flat background case, analogous to the general Einstein background. The free part of the linearised action reads
\be
\begin{split} 
\label{action1}
S_{free}&=i\int e^A_{~A'}\wedge e^{BA'}\wedge \Big(w_{AC}\wedge w^{C}_{~B}\Big)+i\int \Big(e^{A}_{~A'}\wedge h^{BA'}+h^{A}_{~A'}\wedge e^{BA'}\Big)\\&\wedge dw_{AB}.
\end{split} 
\ee 
The first term is quadratic in the connection fluctuations. The appearance of this term makes the tetrad propagator non-zero. The second term is of the form $h d\omega$. Let us now write the interaction part of the action.
\be 
\begin{split} 
\label{action2}
S_{interaction}&=2i\int e^{A}_{~A'}\wedge h^{BA'}\wedge w_{AC}\wedge w^{C}_{~B}\\&+i\int h^{A}_{~A'}\wedge h^{BA'}\wedge\Big(d\omega_{AB}+\omega_{AC}\wedge \omega^{C}_{~B}\Big).
\end{split}
\ee 
The interaction Lagrangian in this case does not have the $hh$ term, which was present in the case of a general Einstein background. This leads to simpler Feynman rules. Let us also quickly summarise the symmetries at the linearised level. 
\subsection{Symmetries} 
The linearised action above can be seen to be invariant under both the diffeomorphisms and the local $SL(2,C)$ transformations. The diffeomorphisms are generated by the vector field $\xi^{\mu}=\xi^{BB'}e^{\mu}_{BB'}$. It acts on the tetrad perturbations but does not act on the connection perturbation for the flat background. The local $SL(2,C)$ gauge transformations act on both the fields. It is generated by the infinitesimal parameters $\phi^{CD}, \bar{\phi}^{C'D'}$ which belong to the two chiral halves of the Lorentz group. Let us explicitly write these transformations.
\be
\begin{split} 
\delta_{\xi}\omega^{CD}_{~~~BB'}&=0,\\ \delta_{\xi}h^{CC'}_{~~~DD'}&=\partial_{DD'}\xi^{CC'},\\
\delta_{\phi}\omega^{CD}_{~~~BB'}&=\partial_{BB'}\phi^{CD},\\ \delta_{\phi}h^{CC'}_{~~~DD'}&=-\phi^C_{~D}\epsilon^{C'}_{~D'}-\phi^{C'}_{~D'}\epsilon^{C}_{~D}.
\end{split} 
\ee 
\subsection{Kinetic and potential terms}
The linearised action in (\ref{action1}) can be expanded by using the following decomposition of all the forms in terms of the background 1-forms $e^{BB'}$. Let us write the decomposition first. We have 
\be 
\begin{split} 
h^{BB'}&=h^{BB'}_{~~~NN'}e^{NN'},\\
\omega^{CD}&=\omega^{CD}_{~~~NN'}e^{NN'},\\
d\omega^{CD}&=\partial_{NN'}\omega^{CD}_{~~~MM'}e^{NN'}\wedge e^{MM'}.
\end{split} 
\ee 
One can then use the relation for the wedge product of the forms, which we outlined in (\ref{wedge}) and then the decomposition for the tetrad perturbation into its irreducible components as in (\ref{tetrad}) to expand the free part. The final gauge fixed kinetic part of the Lagrangian is 
\be 
\label{kinetic}
L_{kinetic}=-H^{(MN)A'B'}\partial_{AA'}\Omega_{MN~~~B'}^{~~~~~A}+ 2h^{AC}\partial_{CD'}\omega_A^{~~D'},
\ee 
where $\Omega^{ABCD'}$ and $\omega^{AA'}$ are the redefined components of the connection field. The new components are related to the self-dual part of the spin connection as
\be 
\begin{split} 
\omega^{MM'}&=\omega^{JM~M'}_{~~~J},\\
\Omega^{MNJM'}&=\omega^{MJNM'}+\epsilon^{JN}\omega^{MM'}=\omega^{MNJM'}-\epsilon^{NJ}\omega^{P~~~MM'}_{~~P}
\end{split}.
\ee
The spinorial object in the second equation is symmetric in its first two indices. Thus, it has twelve independent components. The spinor in the first equation has a total of four components. Thus, together they have sixteen components, appropriate for the spin connection. In terms of the new variables, the potential term can be written in a diagonalized form. The diagonalization is possible after we have split the wedge product form of the potential term in the Wald form, see (3.3) in \cite{Krasnov:2020bqr}. Without further ado, let us write down the potential in terms of the new variables.
\be 
\label{pot}
L_{pot}= -\frac{1}{2}(\omega^{ACBA'})^2+\omega^{AC~~A'}_{~~~A}\omega_{C~~~BA'}^{~~B}\nonumber\\
=-\frac{1}{2}(\Omega^{MNJM'})^2+\omega^{MM'}\omega_{MM'}.
\ee 
As we can see, the potential decouples into two sectors. One in which the field $\Omega$ is quadratically coupled to itself and another in which the other component $\omega$ is coupled again to itself. There are no mixed terms. Thus, it makes the deduction of propagators trivial.
We now use the generating functional method to derive the propagators. Since the kinetic part is decoupled into two sectors, we demonstrate this method for one sector and the other will be similar.

\subsection{Propagators} 
To derive the propagators, we use the machinery of coupling the fields to currents and using the field equations to eliminate the fields from the Lagrangian. When the generating Lagrangian is written down completely in terms of the currents, we read off the propagators from it directly. To this end, let us write the kinetic and potential terms of the Lagrangian with currents coupled to fields
\be 
\begin{split} 
\label{l1}
\mathcal{L}&=-\Omega^{MNJM'}\Omega_{MNJM'}-2\Omega_{MNJM'}\partial^J_{~N'}H^{MNN'M'}+2\omega^{MM'}\omega_{MM'}+4\omega_{MM'} \partial_N^{~M'}h^{MN}\\
&+J_{MNJM'}\Omega^{MNJM'}+J_{MM'}\omega^{MM'}+J_{MNM'N'}H^{MNM'N'}+J_{MN}h^{MN}.
\end{split} 
\ee 
We now obtain the field equations from the Lagrangian in (\ref{l1}). The field equation for the redefined tetrad variable $H$ is 
\be 
\label{H}
H:~~~~\partial^J_{~M'}\Omega_{MNJN'}+\frac{1}{2}J_{MNM'N'}=0.
\ee 
This is obtained by applying the partial derivative $\frac{\partial}{\partial H}$ to the Lagrangian. Clearly, all other terms are killed except two of them. In one of the terms, we use integration by parts to shift the derivative on the left hand side of $\Omega$, using the fact that the total derivative term is irrelevant in the action and thus can be taken to vanish. Let us also write down the other field equations. The field equation for the other metric component is given by 
\be 
\label{h}
h:~~~~\partial_{NM'}\omega^{~M'}_{M}+\frac{1}{4}J_{MN}=0.
\ee 
While for the two redefined connection fields, the equations of motion are as follows 
\be 
\label{omega}
\Omega:~~~~\Omega^{MNJM'}=\frac{1}{2}J^{MNJM'}-\partial^J_{~N'}H^{MNN'M'}.
\ee 
\be 
\omega:~~~~\omega^{MM'}=\partial^{NM'}h^M_{~N}-\frac{1}{4}J^{MM'}.
\ee 
Let us now substitute the solution of the connection fields in (\ref{h}) and (\ref{H}). We get 
\be 
\label{eq1}
\partial^J_{~M'}J_{MNJN'}+\square H_{MNM'N'}+J_{MNM'N'}=0,
\ee 
\be 
\label{eq2}
2\square h_{MN}+\partial_{NM'}J^{~M'}_M-J_{MN}=0.
\ee 
The laplacian operator can be written in the spinor notations as $\square=\partial^{M}_{~M'}\partial^{~~M'}_M$. Now we can solve for all the fields completely in terms of the currents. Each field equation is a laplacian on the field which gets equated to some function of the currents and derivatives. Let us write down each of these equations 
\be 
\label{eq3}
\square H_{MNM'N'}=-\partial^J_{~M'}J_{MNJN'}-J_{MNM'N'},
\ee 
\be 
\label{eq4}
\square h_{MN}=\frac{1}{2}\Big(J_{MN}-\partial_{NM'}J^{~M'}_M\Big),
\ee 
\be 
\label{eq5}
\square \Omega_{MNJM'}=\partial_{JJ'}J^{~~~J'}_{MN~M'},
\ee 
\be 
\label{eq6}
\square \omega^{MM'}=\frac{1}{2}\partial^{NM'}J^M_{~N}.
\ee 
Using the solutions for the fields in terms of currents, we can now directly substitute them in the part of the Lagrangian in (\ref{l1}), where the currents are coupled to fields. Upon doing this, we get the generating Lagrnagian written completely in terms of the currents. We have 
\be 
\label{gen-l}
\mathcal{L}_W=J_{MNJM'}\square^{-1}\partial_{JJ'}J^{~~~J'}_{MN~M'}+J_{MM'}\square^{-1}\frac{1}{2}\partial^{NM'}J^M_{~N}\nonumber\\-J_{MNM'N'}\square^{-1}\Big(\partial^J_{~M'}J_{MNJN'}-J_{MNM'N'}\Big)\nonumber\\+\frac{1}{2}J_{MN}\square^{-1}\Big(J_{MN}-\partial_{NM'}J^{~M'}_M\Big).
\ee 
Let us rearrange these terms and write it in a convenient way. 
\be 
\label{gen-l1}
\mathcal{L}_W=-\frac{1}{2}J_{MNM'N'}\square^{-1}J^{MNM'N'}+\frac{1}{4}J_{MN}\square^{-1}J^{MN}\nonumber\\
+J^{MNJM'}\square^{-1}\partial_{JN'}J^{~~~N'}_{MN~M'}-
\frac{1}{2}J^{MM'}\square^{-1}\partial_{NM'}J^{~N}_M.
\ee 
This is the final form of the generating Lagrangian. The propagators can be obtained by hitting it with appropriate derivatives. For instance, to get the $\langle HH\rangle$ propagator, one needs to apply partial derivatives w.r.t the field $J^{PQP'Q'}$ twice on the generating function above. However, the form of the generating Lagrangian is quite simple such that the propagators can be directly read off from here. We have for the $HH$ propagator
\be 
\begin{gathered} 
\begin{fmfgraph*}(120,0)
\fmfleft{i1}
\fmfright{o1}
\fmf{dbl_wiggly}{i1,v1,o1}
\fmflabel{}{i1}
\fmflabel{}{o1}
\end{fmfgraph*}
\end{gathered}\nonumber\\~\nonumber\\
\langle H^{PQP'Q'}(k)H_{STS'T'}(-k)\rangle=\frac{1}{ik^2}\epsilon^{(P}_{~S}\epsilon^{Q)}_{~T}\epsilon^{P'}_{~S'}\epsilon^{Q'}_{~T'}.
\ee 
\\
where we use the spinor translation for the Minkowski metric as a product of two spinor metrics. There is no copy of momentum in the numerator whereas there is a factor of $k^2$ in the denominator and this makes the tetrad propagator trivial. The symmetrization of the unprimed indices of the tetrad perturbation is captured by the symmetric product of two $\epsilon$ on the right hand side. Let us now write the other propagator in this theory, which connects the tetrad to the redefined connection field. 

\be 
\label{Homega}
\begin{gathered} 
\begin{fmfgraph*}(140,0)
\fmfleft{i1,i2}
\fmfright{o1,o2}
\fmf{dbl_wiggly}{i1,v1}
\fmf{dbl_dashes}{v1,o1}
\fmflabel{}{i1}
\fmflabel{}{o1}
\end{fmfgraph*}
\end{gathered}\nonumber\\~\nonumber\\
\langle H^{MN~Q'}_{~~~Z'}(k)~\Omega_{JLZG'}(-k)\rangle= \frac{1}{k^2}\epsilon^M_{~~J}\epsilon^{N}_{~~L}\epsilon^{Q'}_{~~G'}~k_{ZZ'}.
\ee 
In the $\langle H\Omega\rangle$ propagator, we find that there is both a factor of the spinor metric and a single copy of momentum sitting in the numerator. The spinor metric connects two unprimed indices, which do not have any symmetry between them while the momentum connects the third primed index of the tetrad field to the third unprimed index of the connection.  The $k^2$ factor appears as usual in the denominator. This is quite analogous to the $\langle ba\rangle$ propagator in the chiral Yang Mills. It is interesting to observe the fact that the $\langle H\Omega\rangle$ propagator can be directly related to the derivative of the $\langle HH\rangle$ propagator. This can be understood from (\ref{omega}), where if we omit the current, we have 
\be 
\label{omega2}
\Omega^{MNJM'}=-\partial^J_{~N'}H^{MNN'M'},
\ee 
which then leads to the relation 
\be 
\langle \Omega_{PQCC'} (k) H^{ABA'B'}(-k)\rangle= -ik_{CD'}\langle H^{~~~D'}_{PQ~~C'}(k)H^{ABA'B'}(-k)\rangle.
\ee 
\vspace{1em}
\be 
\begin{gathered} 
\begin{fmfgraph*}(140,0)
\fmfleft{i1}
\fmfright{o1}
\fmf{zigzag}{i1,v1}
\fmf{dashes}{v1,o1}
\end{fmfgraph*}
\end{gathered}\nonumber\\~\nonumber\\
\langle h^{B}_{~D}(k)~\omega_{AD'}(-k)\rangle= \frac{1}{k^2}\epsilon^B_{~~A}~k_{DD'}.
\ee 
The $\langle h\omega\rangle$ propagator is quite similar to the $\langle H\Omega\rangle$ except that in the later case there are few more unprimed and primed index contractions. The propagator is non-trivial in one of the primed and the other unprimed indices because the momentum connects these two. In the rest of the work, as we will demonstrate, this propagator does not contribute to the amplitudes. 
\\
\vspace{1em}
\be 
\begin{gathered} 
\begin{fmfgraph*}(140,0)
\fmfleft{i1}
\fmfright{o1}
\fmf{zigzag}{i1,o1}
\end{fmfgraph*}
\end{gathered}\nonumber\\~\nonumber\\
\langle h^{AB}(k)~h_{MN}(-k)\rangle= \frac{1}{k^2}\epsilon^{A}_{~M}\epsilon^{B}_{~N}.
\ee 
Similar to that of $\langle HH\rangle$ propagator, this propagator is trivial. There are no momentum factors in the numerator and only the unprimed spinors are contracted. This too will not contribute in the computation of amplitudes as we will see in the subsequent chapters. 
\subsubsection{Ghost propagators} 
The BRST quantization which we described in the previous sections leads to a gauge fixed Lagrangian with ghost terms. There are two kinds of ghosts: diffeomorphisms and local Lorentz. The Lagrangian consists of mixed terms between these ghosts besides the usual kinetic terms. Thus there are a total of three such terms to which we can add the relevant currents coupled with the ghosts and deduce the generating Lagrangian as before. The free ghost Lagrangian with currents coupled is given by 
\be 
\label{ghost4}
\begin{split} 
\mathcal{L}_{ghost}&=-\bar{c}^{JJ'}\partial^{KK'}\partial_{KK'}c_{JJ'}+2\bar{c}^{JK'}\partial_{KK'}b^{K}_{~J}-\bar{b}^{JK}\partial_{MM'}\partial^{MM'}b_{JK}\\&+J_{1MM'}c^{MM'}+\bar{J}_{1NN'}\bar{c}^{NN'}+J_{2}^{PQ}b_{PQ}+\bar{J}_2^{MN}\bar{b}_{MN}.
\end{split}
\ee 
We follow the method outlined for the tetrad and connection propagators. The propagators can be obtained by hitting the generating Lagrangian with appropriate derivatives. However, the form of the generating Lagrangian is quite simple such that the propagators can be directly read off from there. We do not repeat the calculations here. Let us first write down the propagator for the diffeomorphism ghost field. We have
\be 
\begin{gathered} 
\begin{fmfgraph*}(120,0)
\fmfleft{i1}
\fmfright{o1}
\fmf{ghost}{i1,o1}
\fmflabel{}{i1}
\fmflabel{}{o1}
\end{fmfgraph*}
\end{gathered}\nonumber\\~\nonumber\\
\langle \bar{c}^{MM'}(k)c_{NN'}(-k)\rangle=\frac{1}{ik^2}\epsilon^{M}_{~N}\epsilon^{M'}_{~N'}.
\ee 
\\
There is no copy of momentum in the numerator as should be the case, whereas there is a factor of $k^2$ in the denominator. Further, there is no symmetrization of any indices. This makes the diffeormphism ghost field propagator trivial. This is quite analogous to that of the ghost field propagator in chiral YM theory. Next, we have the propagator for the Lorentz ghost field
\be 
\begin{gathered} 
\begin{fmfgraph*}(140,0)
\fmfleft{i1,i2}
\fmfright{o1,o2}
\fmf{dbl_dots_arrow}{i1,o1}
\fmflabel{}{i1}
\fmflabel{}{o1}
\end{fmfgraph*}
\end{gathered}\nonumber\\~\nonumber\\
\langle b^{MN}(k)~b_{PQ}(-k)\rangle= \frac{1}{ik^2}\epsilon^{M}_{~P}\epsilon^{N}_{~Q}.
\ee 
This propagator connects the Lorentz ghost with its anti-ghost. There is no symmetrization in the indices. The $k^2$ factor appears as usual in the denominator. There is no factor of momentum in the numerator and this makes the propagator trivial as before. The remaining propagator is the one which connects the Lorentz ghost with the diffeomorphism anti-ghost. It is given by \\
\vspace{1em}
\be 
\begin{gathered} 
\begin{fmfgraph*}(140,0)
\fmfleft{i1}
\fmfright{o1}
\fmf{dots}{i1,v1}
\fmf{dbl_dots}{v1,o1}
\end{fmfgraph*}
\end{gathered}\nonumber\\~\nonumber\\
\langle \bar{c}^{A}_{~B'}(k)~b_{PQ}(-k)\rangle= \frac{1}{k^2}\epsilon^A_{~Q} k_{PB'}.
\ee 
The $\langle \bar{c}b\rangle$ propagator is quite similar to the $\langle h\omega\rangle$ one except that in the latter case there are few more unprimed and primed index contractions.

\subsection{Interaction}
There are three kinds of interaction terms. One is the single derivative cubic term, which will play the main role in amplitudes. The other two are the non-derivative cubic interaction and the non-derivative quartic interaction. In the single derivative cubic interaction, there are again three different kinds of terms. One is of the form $HH\partial\Omega$ and the other two are $hh\partial\Omega$ and $hH\partial\Omega$ respectively.
It is the $HH\partial\Omega$ interaction which is relevant in the same helicity amplitude computations. Despite being a single derivative interaction, as we will show, it is possible to write the connection variable in terms of the tetrad perturbation, resulting in the two derivative interaction vertex, as is relevant for gravity. Let us then first write the complete set of interactions. We write this in two parts. The single derivative cubic interaction reads 
\be 
\label{int21}
L_{hh\partial\omega}=-\frac{1}{2}\Big(H^{AR}_{~~M'N'}H^{BSM'N'}\partial_{RR'}\Omega_{ABS}^{~~~~R'}-H^{A~~~~R'}_{~MM'}H^{BMM'S'}\partial_{RR'}\Omega_{AB~S'}^{~~~R}\Big)\nonumber\\-h^{AR}h^{BS}\partial_{RR'}\Omega_{ABS}^{R'}-h^{AN}H^{B~~R'S'}_{~~N}\partial_{RR'}\Omega_{AB~~S'}^{~~~R}.
\ee 
Note that the above interaction only contains one of the components of the connection field. The other component, namely $\omega^{BB'}$ does not enter the interaction. We can write this interaction in a better way. In particular, it is possible to combine the first two terms into a single effective term and rewrite the complete interaction. This then gives 
\be 
\label{int22}
L_{hh\partial\omega}=-2H^{AR}_{~~(R'N')}H^{BS(M'N')}\partial_{RM'}\Omega_{ABS}^{~~~~R'}-2h^{AR}h^{BS}\partial_{RR'}\Omega_{ABS}^{R'}\nonumber\\+2h^{AN}H^{B~~(R'S')}_{~~N}\partial_{RR'}\Omega_{AB~~S'}^{~~~R}.
\ee 
The only change in the tetrad components is that the primed spinors are now symmetrised. Let us also write the other components of the interaction. The $h\omega\omega$ and $hh\omega\omega$ parts of the interaction read 
\be 
L_{h\omega\omega}=2h^{CC'DD'}(\omega^A_{~CDC}\omega^{B}_{~ABD'}-\omega^A_{~CBD'}\omega^{B}_{~ADC'}),
\ee 
\be 
L_{hh\omega\omega}=2h^{CC'DD'}h^{A~~BA'}_{~C'}(\omega^{~E}_{C~BD'}\omega_{EADA'}-\omega^{~E}_{C~DA'}\omega_{EABD'}).
\ee 
\subsection{Vertices} 
There are both cubic and quartic interaction terms in the linearised Lagrangian. However, as we mentioned earlier, the only relevant interaction for our purposes is the $HHd\Omega$. It is possible to use the relation of the connection variable in terms of the tetrad, as is shown in (\ref{omega2}) and write this interaction term with just the tetrad variable. This leads to an effective Feynman rule, where one of the legs of the tetrad is marked special. Two such special legs can never contract because the $\langle \Omega \Omega\rangle$ propagator vanishes in our theory. Let us first write the vertex factor in the form which is there in the interaction Lagrangian. 
\be 
\begin{gathered} 
\begin{fmfgraph*}(140,100)
\fmftop{i1,i2}
\fmfbottom{o1}
\fmf{dbl_wiggly}{i1,v1,i2}
\fmf{dbl_dashes}{v1,o1}
\fmflabel{1}{i1}
\fmflabel{3}{o1}
\fmflabel{2}{i2}
\end{fmfgraph*}
\end{gathered}\nonumber\\~\nonumber\\~\nonumber\\
\langle H^{PQ~~S'}_{~~~N'} H^{BSM'Q'} \partial_{SS'} \Omega_{JRDD'} \rangle=-2i\epsilon^P_{~J}\epsilon^{Q}_{~D}\epsilon^{Q'}_{~D'}\epsilon^{M'}_{~N'}\epsilon^B_{~R}k^S_{~S'}.
\ee 
Now we replace the connection variable and put a copy of the tetrad with a derivative in front of it. This then becomes an effective $\langle HHH\rangle$ vertex, but with the rule that the particular leg which is replaced is marked special. For convenience, we put a cilia in this leg and correspondingly in the expression of the vertex. When this leg is external, we project it to the positive helicity state of the connection. We have for the effective $\langle HHH\rangle$ vertex
\be 
\begin{gathered} 
\begin{fmfgraph*}(140,100)
\fmftop{i1,i2}
\fmfbottom{o1}
\fmf{dbl_wiggly}{i1,v1,i2}
\fmf{dbl_wiggly}{v1,v2}
\fmf{dbl_wiggly}{v2,o1}
\fmfdot{v2}
\fmflabel{1}{i1}
\fmflabel{3}{o1}
\fmflabel{2}{i2}
\end{fmfgraph*}
\end{gathered}\nonumber\\~\nonumber\\~\nonumber\\
\langle H^{PQ~~S'}_{~~~N'} H^{BSM'Q'} \partial_{SS'}\partial_{DE'} H^{\bullet~~E'}_{JR~~D'} \rangle=2i\epsilon^P_{~J}\epsilon^{Q}_{~D}\epsilon^{Q'}_{~D'}\epsilon^{M'}_{~N'}\epsilon^B_{~R}k_{SS'}k_{DE'}.
\ee 
\subsection{Helicity states}  
Let us introduce the helicity spinors which we will use subsequently in the amplitude computation. The fields $H,\Omega$ can be on-shell. The fields $h,\omega$ are not physical and thus they vanish off-shell. The metric perturbation $H$ is supported by both the positive and negative helicities, given by the usual states
\be
\label{helicity}
\epsilon^{HH'MM'}_-(k)=\frac{q^Hk^{H'}q^Mk^{M'}}{\langle qk\rangle^2},\nonumber\\
\epsilon^{HH'MM'}_+(k)=\frac{k^Hq^{H'}k^Mq^{M'}}{ [qk]^2}.
\ee 
where $q^A$ and $q^{A'}$ are auxiliary 2-spinors. \\~\\
For the field $\Omega$, we solve the gauge fixed Lagrangian from (\ref{kinetic}) and (\ref{pot}) on-shell. This gives 
\be 
\Omega_{PQRS'}=\partial_{RJ'}H_{PQ~~S'}^{~~~J'}.
\ee 
The corresponding helicity state for the field $\Omega$ is  
\be 
\label{connps}
\epsilon^{PQRS'}_+(k)=i\frac{k^Pk^{Q}k^Rq^{S'}}{[qk]}.
\ee 
which is obtained by using the helicity states of the tetrad perturbation and going to the momentum space with the substitution $\partial_{MM'}=-ik_{M}k_{M'}$. The spinor $k_{M'}$ contracts with its copy in the negative helicity state of $H$ and thus it makes it vanish. The only non-vanishing helicity supported by the connection field is the positive one in (\ref{connps}).

\newpage 

\chapter{Self-dual gravity}
The purpose of this chapter is to describe a recently developed covariant formulation of self-dual gravity (SDGR) in flat space \cite{Krasnov:2021cva}, relevant for computing scattering amplitudes. SDGR is a theory in four dimensions, whose solutions are Einstein metrics with either the self-dual or the anti self-dual part of the Weyl curvature vanishing. Such solutions of the field equations of gravity are known as gravitational instantons. When appropriately linearised around an instanton background, the theory describes two propagating degrees of the graviton. However, the two polarization states are described on a different footing as we shall see. The interesting feature of the theory is the fact that despite having a negative dimension coupling constant, it is not just renormalizable but ultraviolet finite, unlike pure gravity in four dimensions. SDGR can only describe metrics of Euclidean or split signatures because in Lorentzian signature, the vanishing of either the self-dual or the anti self-dual part of the Weyl curvature means the vanishing of all of it. There are strong similarities between SDYM and SDGR. In particular, SDGR is one loop exact and there are no divergences as can be seen by computing the one-loop effective action. This is described in \cite{Krasnov:2016emc}. The tree amplitudes in the theory vanish in an analogous way like SDYM, whereas the one-loop amplitudes are rational functions of the momenta involved. Let us then start describing the flat space version of SDGR, which we will use to compute amplitudes in the subsequent chapter.  

\section{SDGR in flat space}
There are many non covariant formulations for SDGR which exist in the literature. For a generic reference, see Plebanski's description \cite{Plebanski:1975wn}. The well known covariant formulation is proposed in \cite{Siegel:1992wd} for zero scalar curvature case and then in \cite{Krasnov:2016emc} for non-zero scalar curvature. We are going to describe two such actions relevant for the zero scalar curvature. One of them can be obtained from the chiral Einstein-Cartan in (\ref{EC}) by removing the part of the curvature which is quadratic in the spin connection. This action is also contained in the bosonic sector of the corresponding supergravity action in \cite{Siegel:1992wd}. The resulting covariant action in flat space background is very similar to SDYM. In particular, it contains just a single propagator and a cubic vertex. The quartic vertex which is present in the full gravity action in (\ref{EC}) gets eliminated as a result of removing the $\omega\omega$ term from the curvature. However, the gauge fixing for this action is rather non-trivial and thus it complicates perturbative computations. The other action has a simple gauge fixing procedure and the Feynman rules are analogous to its YM cousin, in that there is just one propagator and a cubic vertex. One can also compute the BG currents in this theory and then it is easy to infer that all tree amplitudes vanish, in analogy with SDYM. The interesting part is the one-loop same helicity amplitude sector which is correctly captured by SDGR and we will expand on it in the next chapter. 
\subsection{Action}
To write the action in self-dual gravity, let us first consider the chiral Einstein-Cartan action once again in spinor notations with coupling factors included. It reads  \be
\label{EC2}
\mathcal{S}[e,\omega]=i\int e^{A}_{~C'}\wedge e^{BC'}\wedge (d\omega_{AB}+\kappa\omega_{AC}\wedge\omega^C_{~B}),
\ee 
where $e^{A}_{~C'}\wedge e^{BC'}$ is the self-dual 2-form constructed from the frame field and $\kappa=8\pi G$. The action in (\ref{EC2}) is equivalent to the action in (\ref{EC}) in that the connection field is just re-scaled by a factor of $\kappa$. We now construct the SDGR action by simply taking the $\kappa\rightarrow 0$ limit. This removes the $\omega\omega$ part of the curvature. The resulting field which appears after this is to be distinguished from the connection. Thus we call it $\xi_{AB}$. The SDGR action reads
\be 
\mathcal{S}_{SDGR}[e,\xi]=i\int e^{A}_{~C'}\wedge e^{BC'}\wedge d\xi_{AB}.
\ee
The action comprises of an exact 2-form $d\xi$ on which the self-dual 2-form acts and projects out its self dual component. This is because any arbitrary 2-form can be expanded into its self-dual and anti-self-dual parts. Then one can use the fact that the wedge product between a self-dual 2-form and an anti-self 2-form vanishes. Thus it projects out the self-dual part of $d\xi$. One can vary this action with respect to both the tetrad/frame field and the 1-form. If we vary with respect to the 1-form $\xi_{AB}$, we get an equation which says that the self-dual part of the spin connection vanishes. Thus, the curvature constructed from the self-dual part of the spin connection also vanish and this then implies that the metric obtained from the tetrad field is the one with only non-zero anti-self-dual part of the Riemann curvature. Also, varying with respect to the tetrad gives an equation which is equivalent to Einstein equation. This is explained in details in \cite{Abou-Zeid:2005zfo}. Overall, this gives the correct description of SDGR where the metric is Einstein and the self-dual part of the Riemann curvature vanishes. However, like the chiral Einstein-Cartan, the gauge fixing of this action is non-trivial and thus it is not so useful to do perturbative calculations with it.
\\~\\
We next explain another covariant formalism of SDGR which is recently proposed in \cite{Krasnov:2021cva}. It is relevant for metrics with zero scalar curvature and is thus suitable for perturbative computations. The action proposed can be motivated from another closely related formalism for SDGR \cite{Krasnov:2016emc} in the case of non-zero scalar curvature. In particular, the new action can be obtained by removing the quadratic in connection term from the latter. It is also illuminating because it is much closer to the SDYM action (\ref{asdym}) as we described in the previous part, in that there is only one propagator and just a cubic vertex, when expanded around an appropriate background. Another nice feature of the new formalism is that it has a simpler gauge fixing procedure and this is why it is most useful for amplitude computations. The action reads 
\be 
\label{SDGR2}
\mathcal{S}_{SDGR}[\psi, a]=\frac{1}{2}\int\psi^{mn}da^{m}\wedge da^{n},
\ee 
where $\psi^{mn}$ is a tracefree and symmetric field. Thus $\psi^{mn}\delta_{mn}=0$. The action consists of another field, namely the 1-form $a^m$ and the way this field enters in the Lagrangian is via an exact 2-form $da^m$. The action above can be understood to arise from another covariant action for SDGR, which is described in \cite{Krasnov:2016emc} by a process where the connection-connection term in the curvature 2-form is set to zero. Then the curvature 2-form becomes an exact 2-form, which is the same as $da^m$ and thus results in the action (\ref{SDGR2}). Next, we describe the perturbative expansion of this action around a flat background and construct the Feynman rules.

\subsection{Linearised action}
The starting point for doing perturbative calculations is to expand the above action around a flat background. The relevant background which describes flat space is 
\be 
da^m=M^2\Sigma^m,
\ee 
where $M^2$ is introduced for dimensional purposes and $\Sigma^m$ are the already described self-dual 2-forms. When we expand around this background, the action reads 
\be 
\label{lS}
\mathcal{S}_{1}=\int \psi_{mn}(M^2\Sigma^m\wedge da^n+\frac{1}{2}da^m\wedge da^n).
\ee 
The first term is the kinetic term for gravitons and the second term is a 2-derivative cubic interaction. It is interesting to see that the structure of the linearised Lagrangian is analogous to that in SDYM. Indeed, as we described in SDYM, the linearised Lagrangian contains a kinetic piece and a cubic term. The only difference is that in the case of SDYM, the interaction is a non-derivative one, while in the case of SDGR, the interaction contains two derivatives. To do perturbative computations, it is most useful to express the above action in spinor notations. Thus let us start describing everything in terms of spinors. The field $a^m_{\mu}$ is a 1-form and translating the index $m$ into spinor indices, it becomes $a^{MN}_{\mu}$. The unprimed indices $M,N,...$ and the primed ones $M',N',...$ are the $sl(2,\mathbb{C})$ spinor indices. We also translate the spacetime index $\mu$ and then we get the object $a^{MN}_{~~~~PP'}$. The field $\psi^{mn}$ gets translated to the spinor $\psi^{MNPQ}$, which is totally symmetric. The spinor translation of the derivatives are already described in the previous part. For completeness, let us state that the spinorial translation of the partial derivative is $\partial_{M'N'}$. The exterior derivative then becomes $\partial_{PP'}a^{MN}_{~~~~QQ'}$. Next, when the self dual 2-forms are wedged with the exterior derivative it results in the self-dual projection of the latter, which in spinor notation reads $\partial_{PP'}a^{MN~~P'}_{~~~~Q}$. Overall, the linearised kinetic term reads 
\be 
\label{lk}
\mathcal{L}_2= \psi_{MNPQ}\partial^{M}_{~M'}a^{NPQM'}.
\ee 
The kinetic term above is normalised, because we have absorbed the dimensionful parameter $M^2$ into it. Thus, the mass dimension of $\psi$ is two whereas that of $a$ is one. It is convenient to use a notation where all the symmetrised indices of a particular field is denoted by the same letter. Thus, we denote the completely symmetrised field $\psi^{MNPQ}$ as $\psi^{MMMM}$. The equations of motion are now immediate to write down. Variation with respect to the field $a$ gives 
\be 
\label{e1}
\partial_{N}^{~~M'}\psi^{MMMN}=0.
\ee 
The field $\psi$ thus describes the $+2$ helicity state of the graviton. Its equation can be understood as the Bianchi identity for the Weyl tensor. The variation of the kinetic part with respect to $\psi$ gives 
\be 
\label{e2}
\partial^{M}_{~~N'}a^{MMMN'}=0.
\ee 
This equation describes the negative helicity state of the graviton. Let us next write the interaction. There is just a single cubic vertex. The 2-form $\partial_{PP'}a^{MN}_{~~~QQ'}$ can be split into a self-dual part and an anti-self-dual part. The self dual part results in a spinor of the form which is described earlier. The anti-self-dual part in spinor notation reads $\partial_{SP'}a^{MN~~S}_{~~~~Q'}\epsilon_{PQ}$. Then, the two copies of $da$ are wedged. In this process, there is a contraction of the primed indices of two copies of the anti-self-dual parts and the unprimed indices of the corresponding self-dual parts. The wedge product of a self-dual and anti-self-dual part vanishes. Overall, the cubic interaction can be written as 
\be 
\label{vert222}
\mathcal{L}_3=\psi_{MMMM}\Big(\partial_{PS'}a^{MM~~S'}_{~~~~Q}\partial^{P}_{~G'}a^{MMQG'}+\partial_{SP'}a^{MN~~S}_{~~~~Q'}\partial^{~P'}_{G}a^{MMQ'G}\Big).
\ee 
\\~\\
\subsection{Gauge fixing and Feynman rules}
Let us now describe the gauge fixing of our kinetic term. The linearized action around the flat background in (\ref{lS}) is invariant under both diffeomorphisms and shifts of the 1-forms $a^m$ by exact 1-forms. Thus, the relevant transformations for the fields are 
\be 
\begin{split} 
\delta \psi^{mn}&=i_{\eta}d\psi^{mn},\nonumber\\
\delta a^m&=i_{\eta}da^m+i_{\eta}\Sigma^m+d\chi^{m}.
\end{split}
\ee 
where the Lie derivative of the 1-form $a^m$ along some vector field $\eta$ is given by\\ $\mathcal{L}_{\eta}a^m=d(i_{\eta}a^m)+i_{\eta}da^m$ and we have absorbed the $d(i_{\eta}a^m)$ part of it in the definition of $\chi^m$. It is convenient for our purposes to write the above in spinor notations. The spinor translation of the vector field results in a bi-spinor $\eta^{MM'}$ where the conversion is done via the background vierbien 1-form $e^{MM'}_{\mu}$. Then the symmetries take the following form 
\be 
\begin{split}
\label{symg}
\delta \psi^{MMMM}&=0,\\
\delta a^{MN}&=\eta^{M}_{~N'}e^{NN'}+d\chi^{MN}.
\end{split}
\ee 
The field $a^{MN}$, where we now suppress the world index, admits a decomposition into its irreducible components. The decomposition reads 
\be 
a^{MN}=e_{PP'}\Phi^{MNP,P'}+e^{M}_{~N'}\Phi^{NN'}.
\ee 
Now the second component can be gauged away using the diffeomorphism symmetry. This is because it can be reabsorbed into the definition of $\eta^{M}_{~N'}$ in (\ref{symg}) and thus gets eliminated. The kinetic term in (\ref{lk}) is independent of this component and hence the diffeomorphism gauge is fixed. The resulting fields in the theory admit the following linearised transformations
\be 
\delta\psi^{MMMM}=0, ~~~\delta\Phi^{MMM,M'}=\partial^{MM'}\xi^{MM}.
\ee 
It is clear that the only symmetry remaining to be fixed is the $\delta\Phi^{MMM,M'}=\partial^{MM'}\xi^{MM}$. It is now fixed by imposing a variant of the Lorentz gauge. To do so, one considers a gauge fixing fermion of the form 
\be 
\Psi=\bar{c}_{l(MN)}\partial^{C}_{~C'}\Phi^{MNCC'}\equiv\bar{c}_{l(MN}\epsilon_{P)Q}\partial^{Q}_{~C'}\Phi^{MNPC'} .
\ee 
where $\bar{c_l}$ is the Lorentz ghost. BRST variation of this gives the bosonic contribution $\psi_{(MN}\epsilon_{P)Q}\partial^{Q}_{~C'}\Phi^{MNPC'}$ where $\psi_{(MN)}$ is the auxiliary field. The fermionic contribution is just the kinetic term for the ghosts, which is $\bar{c}_{lMN}\partial^C_{~C'}\partial^{C'}_{~C}c_l^{MN}$. One can then notice the structure of the bosonic contribution and combine the $\psi_{(MN}\epsilon_{P)Q}$ part of it to the field $\psi_{MNPQ}$ to obtain a new field, which is symmetric in its first three indices. 
\be 
\tilde{\psi}^{MNPQ}:=\psi^{MNPQ}+\psi^{(MN}\epsilon^{P)Q}.
\ee 
The new field $\tilde{\psi}$ is symmetric in its first three indices. The final gauge fixed action then depends on two fields, namely $\tilde{\psi}^{MMMN}$ and $\Phi^{MMM,N}$ and each of them contains the same number of components. It reads
\be 
\int \tilde{\psi}^{MMMN}\partial^{~~N'}_{N}\Phi_{MMM,N'}.
\ee 
The operator that maps one field to the other in the above term can now be inverted and the propagator is immediate. Let us write it. \\
\subsubsection{Propagator} 
\be 
\label{pp}
\begin{gathered} 
\begin{fmfgraph*}(140,0)
\fmfleft{i1,i2}
\fmfright{o1,o2}
\fmf{double}{i1,v1}
\fmf{dbl_wiggly}{v1,o1}
\fmflabel{}{i1}
\fmflabel{}{o1}
\end{fmfgraph*}
\end{gathered}\nonumber\\~\nonumber\\
\langle \tilde{\psi}^{MNPQ}(k)~\Phi_{ABC,Q'}(-k)\rangle= \frac{1}{k^2}\epsilon^{(M}_{~~A}\epsilon^{N}_{~~B}\epsilon^{P)}_{~~C}~k^{Q}_{~~Q'}.
\ee 
where the indices $M,N,P$ belong to a single symmetrized group and the symmetrization of these indices is denoted by a round bracket in the right hand side of (\ref{pp}).
\subsection{Helicity states}
As we described, the physical fields which enter the gauge fixed Lagrangian are $\Phi^{MMM,N}$ and $\tilde{\psi}^{MMMN}$. These constitute the two helicities of the graviton. The convention we take is that the negative helicity is described by $\Phi$ and the positive helicity by $\tilde{\psi}$. We can then solve the momentum space version of the linearised equations in (\ref{e1}) and (\ref{e2}) and get the corresponding polarisation tensors. They are 
\be 
\label{hel}
\epsilon^{-}_{MMM,M'}(k)=M\frac{q_Mq_Mq_Mk_{M'}}{\langle qk\rangle^3},~~~\epsilon^{+}_{MMMM}(k)=M^{-1}k_Mk_Mk_Mk_M.
\ee 
where the momentum is null, i.e $k_{MM'}=k_Mk_{M'}$ and $q^M$ is the auxiliary spinor as usual. The negative polarisation tensor is dimensionless and this is why we introduced the dimensionful parameter $M$ in its definition. The positive polarisation tensor is dimensionful, with its mass dimension being one. As we can see, the two polarisation states of the graviton are treated on a different footing. The negative state consists of both primed and unprimed spinors while the positive state just has only unprimed ones. Let us then apply this to compute the BG current in this theory.
\subsection{Relation to metric helicity state}
Let us see how the usual negative metric helicity state arise from the first expression in (\ref{hel}). This has been described in details in \cite{Delfino:2012aj}. The metric perturbation is obtained by the operation of taking the anti-self-dual part of the two form $d\Phi^i$. In the spinorial notation, this then can be described by the following relation, upto numerical factors
\be 
\label{metconn}
h_{MNM'N'}\sim \partial^{B}_{~M'}\Phi_{BMN,N'}.
\ee 
Here $\partial_{MM'}=-e^{\mu}_{MM'}\partial_{\mu}$ is the Dirac operator. We can apply this to the negative helicity spinor (\ref{hel}). Modulo numerical factors, it is easy to see that the usual negative helicity is reproduced 
\be 
h^{-}_{MM'NN'}(k)\sim M\frac{q_Mk_{M'}q_Nk_{N'}}{\langle qk\rangle^2}.
\ee 
Let us explain how the negative helicity state arises. In the passage to momentum space, the partial derivative in (\ref{metconn}) becomes the momentum vector $k$. This then splits into a primed and an unprimed spinor due to on-shell condition. Contracting one of the $q$ spinor by the unprimed momentum spinor $k$ cancels one of the $\langle qk\rangle$ factor from the denominator. This results in two copies of primed momentum spinors in the numerator and two copies of the $q$ spinor, which is the usual expression for the negative helicity state of the metric. 

\subsection{Berends-Giele current}
The first computation of the currents in gravity was done by Berends and Giele using the recursion relations \cite{Berends:1987me}. However, unlike in YM, the recursion in this case is quite complicated to solve and thus the MHV amplitude for gravity was not obtained using this procedure. We will give a brief outline of the computation of the 2-current from the SDGR Feynman rules which we just described and suggest the generalization to $n$-current. Like usual, we define a current as the sum of all tree level Feynman graphs with all but one leg on-shell. Thus one can have different states inserted into the on-shell legs and the off-shell leg is taken with the propagator on that leg. However, the most interesting current arises if we take all the on-shell legs to be of same helicity. By convention, we take the negative helicity state inserted into these legs. Let us write the form the general $n$-current. 
\be 
\label{bg2}
\mathbb{J}_{MNPQ'}(1,2,...,n)=M^{2-n}q_Mq_Nq_{P}q^{Q}(1+2+...+n)_{QQ'}\mathbb{J}(1,2,...,n).
\ee 
The one-current is the negative helicity polarization state itself. The scalar part of the current can be written with its index structure stripped off  
\be 
\mathbb{J}(1)=\frac{1}{\langle q1\rangle^4}.
\ee 
The two-current is obtained by taking the $\tilde{\psi}\partial \Phi\partial \Phi$ vertex in (\ref{vert222}), projecting the two $d\Phi$ legs into negative helicity states and then applying the final leg propagator. Upon inserting the negative helicity state, one of the terms in the vertex vanishes. We then read off the result for the scalar part of this current
\be 
\label{gr-2-cur}
\mathbb{J}(1,2)=-\frac{[12]}{\langle q1\rangle^2\langle q2\rangle^2\langle 12\rangle},
\ee 
The three current can be obtained by taking the two current and insert the third polarisation state to it, with a sum over permutations. Three such diagrams arise, one is the attachment of the third gluon to the current $J(1,2)$, the other being the attachment of the second gluon to $J(1,3)$ and the third one is the attachment of the first gluon to the current $J(2,3)$. The combinatorics is already complicated to solve. The details of this calculation is given in \cite{Krasnov:2016emc}. We outline the final result for the scalar part of the 3-current. 
\be 
\mathbb{J}(1,2,3)= \frac{1}{\langle q1\rangle^2\langle q3\rangle^2}\frac{[12][23]}{\langle 12\rangle\langle 23\rangle}+\frac{1}{\langle q1\rangle^2\langle q2\rangle^2}\frac{[23][31]}{\langle 23\rangle\langle 31\rangle}+\frac{1}{\langle q2\rangle^2\langle q3\rangle^2}\frac{[31][12]}{\langle 31\rangle\langle 12\rangle}.
\ee 
The pattern is becoming clear. The scalar part of the general $n$th-order current can be written as 
\be 
\label{sc}
\mathbb{J}(1,2,...,n):=J(\mathcal{M})=\prod_{j\in\mathcal{M}}\frac{1}{\langle qj\rangle^4}\sum_{T\in\mathcal{T}(\mathcal{M})}\prod_{\langle im\rangle\in H(T)}\frac{\langle im\rangle}{[im]}\langle qi\rangle^2\langle qm\rangle^2.
\ee 
This is a compact way of writing the general formula for the current. Let us explain the notation. The first product stands for all elements of $\mathcal{M}$. The notation $\mathcal{T}(\mathcal{M})$ denotes all the tree graphs with elements of $\mathcal{M}$ as vertices. Thus the sum after the product is over all such tree graphs from the set $\mathcal{T}(\mathcal{M})$. Next, the notation $H(T)$ denotes the set of edges for a tree graph $T$. The product is then over all such spinor brackets $\langle im\rangle$ which are in $H(T)$. The $q$ in $\langle qi\rangle^2$ stands for the usual auxiliary momentum spinor. Although the formula is complicated, it can be verified for simple cases, like $n=1,2$, etc. For the case when $n=2$, we have the set $\mathcal{M}=\{1,2\}$. In this case, there is just one tree graph and thus the sum reduces to just one term, which is the result in (\ref{gr-2-cur}). For the case when n=3, we have $\mathcal{M}=\{1,2,3\}$. In this case, there are three terms contributing to the current as can be seen from the formula. This is true because there can be three different permutations for the three legs of the current and then all of this is summed over. The formula can also be proved in general, see \cite{Krasnov:2013wsa}. It therefore is the closed form expression for our current.
\newpage
\section{Gravitational Instantons}
In analogy with the Yang Mills case, one can also  construct Euclidean solutions for the Einstein field equations which minimize the action functional. Such solutions are Einstein metrics with half of the Weyl curvature vanishing. These are the so called gravitational instantons. Thus, we have the decomposition of the Riemann curvature (for zero cosmological constant)
\be 
R=\begin{pmatrix}
W^+ & \textrm{Ricci}|_{\textrm{tracefree}} \\
\textrm{Ricci}|_{\textrm{tracefree}}& W^- 
\end{pmatrix},
\ee 
where $W^+$ and $W^-$ are the self dual and anti-self-dual parts of the Weyl curvature. The instanton condition implies that the tracefree part of Ricci and half of the Weyl curvature vanish. Such solutions then automatically correspond to self-dual or anti-self dual gravity depending on which half of the Weyl curvature is non-zero. Thus, for a gravitational instanton which is described by the anti-self-dual part of the Weyl curvature, we have the following condition
\be 
\label{condi}
\textrm{Ricci}|_{\textrm{tracefree}}=0,\nonumber\\
W^+=0.
\ee 
The description of instantons with non-zero cosmological constant is given in \cite{Krasnov:2016emc}. In this reference, a connection description of instantons is explained. Here we briefly review the case for zero cosmological constant, which has already been described in details in \cite{Krasnov:2021cva}. Consider the flat SDGR Lagrangian once again 
\be 
\mathcal{S}_{SDGR}[\psi,a]=\frac{1}{2}\int \psi^{mn}da^m\wedge da^n,
\ee 
where we have a triple of two-forms $\xi^m=da^m$ on some manifold X. The two forms are assumed to be closed and the variation of the action with respect to the field $\psi^{mn}$ leads to the Euler Lagrange equation 
\be 
\label{gi}
\xi^m\wedge\xi^n=2\delta^{mn}\mu.
\ee 
Any triple of two forms which are closed and satisfy the above equation describes a gravitational instanton. Let us explain this. First, any closed two-form satisfying (\ref{gi}) define a Riemannian signature metric $g_{\xi}$, see \cite{Krasnov:2021cva}. Further, when the two-forms are closed, it implies that the self-dual part of the spin connection vanishes. This also means that the curvature composed of just the self-dual part of the spin connection too vanish. However, we know that the curvature of the self-dual part of the spin connection can be decomposed into its self-dual and anti-self-dual parts. The self-dual part contains one half of the Weyl curvature and the scalar curvature, whereas the anti-self-dual part contains the tracefree part of the Ricci tensor. Vanishing of the curvature thus implies vanishing of both these parts. This then corresponds to the condition (\ref{condi}) and therefore describes gravitational instantons. 

\newpage 
\chapter{Amplitudes in self-dual gravity}
We now arrive at the final chapter of the thesis. In this chapter, we describe our results on the construction and computation of same helicity one-loop gravity amplitudes, particularly the one at four points using the covariant formulation of flat space SDGR. Before expanding on the loop amplitudes, let us briefly comment on the the same helicity tree amplitudes. These amplitudes with $n>3$ points vanish, analogous to the ones in SDYM. This is because in order to construct such amplitudes, one has to remove the final leg propagator in the BG current and project it on-shell. This is done by multiplying the current with a factor of $k^2$ where $k$ is the momentum in the final leg. However, on-shell, the momentum is null and there is no pole to cancel. Thus it renders the amplitude to vanish. The only non-vanishing tree amplitude is at 3-points. We quote the result from \cite{Krasnov:2016emc}
\be 
M^{++-}\sim \frac{1}{M}\frac{[12]^6}{[13]^2[23]^2}.
\ee 
The same helicity one-loop amplitudes in full GR is correctly captured by SDGR Feynman rules. These amplitudes are exactly analogous to their YM cousin, in the sense that they are cut-free and are rational functions of the momenta involved. Thus, inspite of being loop amplitudes, their behaviour is very similar to tree amplitudes. These amplitudes have not been computed using SDGR Feynman rules before. We provide a construction of such an amplitude at four points using the SDGR Feynman rules and show that they are finite. We then explain that it is also possible to construct it using the chiral Einstein-Cartan Feynman rules and both of these give exactly the same set of Feynman diagrams. We then provide a partial attempt to compute it using the one-loop bubbles but our computation stops because unlike in YM, it is not clear how to interpret the shift parameters in this case. Let us begin to describe all these. First we review the literature expression for these amplitudes
\section{Literature expression for one-loop amplitudes} 
The same helicity one-loop amplitudes were conjectured on the basis of soft and collinear limit arguments in \cite{Bern:1998xc}. In the same paper, it was also explained that they can be computed explicitly by using supersymmetry where a graviton in the loop is replaced by a massless scalar. The general expression of these amplitudes are written in terms of soft functions. Let us introduce and explain these functions. 
\subsection{Soft functions} 
We define a set of functions which have some simple behaviour in the soft limits. Let us write the first three of a series of such functions and subsequently write their soft limits. 
\be 
\begin{split} 
h(x,\{1\},y)&=\frac{1}{\langle x1\rangle^2\langle 1y\rangle^2},\nonumber\\
h(x,\{1,2\},y)&=\frac{[12]}{\langle 12\rangle\langle x1\rangle\langle1y\rangle\langle x2\rangle\langle 2y\rangle},\nonumber\\
h(x,\{1,2,3\},y)&=\frac{[12][23]}{\langle 12\rangle\langle 23\rangle\langle x1\rangle\langle1y\rangle\langle x3\rangle\langle 3y\rangle}+\frac{[23][31]}{\langle 23\rangle\langle 31\rangle\langle x2\rangle\langle2y\rangle\langle x1\rangle\langle 1y\rangle}\\&+\frac{[31][12]}{\langle 31\rangle\langle 12\rangle\langle x3\rangle\langle3y\rangle\langle x2\rangle\langle 2y\rangle}.
\end{split}
\ee 
where $x,y$ are arbitrary momenta and $\langle xi\rangle, \langle yj\rangle$ for $i,j =1,2,3...$ are the usual unprimed spinor contractions. These functions satisfy the soft limits  
\be 
h(x,M,y)\xrightarrow[]{m\rightarrow 0} -\mathcal{S}_m(x,M,y)\times h(x,M-m,y), ~~~~~\text{for $m\in M$},
\ee 
where 
\be 
\mathcal{S}_m(x,M,y)\equiv \frac{-1}{\langle xm\rangle\langle my\rangle}\sum_{j\in M}\langle xj\rangle\langle jy\rangle
\frac{[jm]}{\langle jm\rangle}
\ee 
and $M$ is a set containing the remaining legs other than $x,y$. 
\subsubsection{One loop amplitudes}
Using these soft functions, we now explicitly write the literature expressions for the one loop amplitudes at 4, 5 and 6 points. 
\be 
\begin{split} 
\mathcal{M}_4(1^-,2^-,3^-,4^-)&=h(1,\{2\},3)h(3,\{4\},1)tr^3[1234]\\&+h(1,\{2\},4)h(4,\{3\},1)tr^3[1243]
\\&+h(1,\{3\},2)h(2,\{4\},1)tr^3[1324],
\end{split} 
\ee 
\be 
\mathcal{M}_5(1^-,2^-,3^-,4^-,5^-)=h(1,\{2\},3)h(3,\{4,5\},1)tr^3[123(4+5)] + \text{perms},
\ee 
\be 
\mathcal{M}_6(1^-,2^-,3^-,4^-,5^-,6^-)=h(1,\{2,3\},4)h(4,\{5,6\},1)tr^3[1(2+3)4(5+6)]\nonumber\\ +h(1,\{2\},3)h(3,\{4,5,6\},1)tr^3[123(4+5+6)]+ \text{perms}.
\ee 
where 
\be 
tr[1234]=\langle 12\rangle[23]\langle34\rangle[41].
\ee 
The general $n$-point gravity amplitude can be written as \be 
\mathcal{M}_n(1^-,2^-,...,n^-)=\sum_{1\leq x<y\leq n~M,N}h(x,M,y)h(y,N,x)tr^{3}[xMyN],
\ee 
where $M$ and $N$ are two sets forming a distinct partition of the remaining $(n-2)$ legs. $M$ and $N$ are both considered to be non-empty and $(M,N)=(N,M)$.
Interestingly, it can be seen that these soft functions coincide with the scalar part of the BG currents in (\ref{bg2}) when the two arguments $x,y$ are taken to be equal to the auxiliary spinor $q$. In this case,
\be 
\begin{split} 
h(q,\{1\},q)&=\mathbb{J}(1),\\ 
h(q,\{1,2\},q)&=\mathbb{J}(1,2),\\
h(q,\{1,2,3\},q)&=\mathbb{J}(1,2,3),\\
&.\\
&.\\
&.\\
h(q,\{1,2,3...n\},q)&=\mathbb{J}(1,2....n).
\end{split} 
\ee 
There is yet another representation of these one-loop amplitudes in terms of the double off-shell scalar currents. This is described in \cite{Bern:1998sv} and we quote the results here.
\subsubsection{Double off-shell scalar current} 
The double off-shell scalar current $S_q(l;C)$ contains $n$ on-shell positive helicity gravitons for $C={1,2...n}$ and a massless scalar line which have outgoing momenta $l$ and $-l-K_C$ at the two ends. One can then sew the off-shell ends of this scalar line and thereby reconstruct the same helicity amplitudes in gravity. The recurrence relation which is obeyed by the double off-shell currents is given by 
\be
\begin{split} 
S_q(l;C)&=\frac{-1}{(l+K_C)^2}\sum_{A\subset C~B\equiv C-A}S_q(l;A)(l+K_A)^{\mu}(l+K_A)^{\alpha}J^{\mu\alpha}_{+q}(B)\nonumber\\
&=\frac{-1}{(l+K_C)^2}\sum_{A\subset C~B\equiv C-A}S_q(l;A)h(q,B,q)\langle q^-|(\cancel{l}+\cancel{K}_A)\cancel{K}_B|q^+\rangle^2,
\end{split} 
\ee 
where we have used the relation between the half-soft functions $h(q,C,q)$ when both the arguments in $h(a,C,b)$ are taken to be equal to the auxiliary spinor $q$ and the current $\mathbb{J}^{\mu\alpha}_{+q}(C)$
\be 
\mathbb{J}^{\mu\alpha}_{+q}(C)=\langle q^-|\gamma^{\mu}\cancel{K}_C|q^+\rangle\langle q^-|\gamma^{\alpha}\cancel{K}_C|q^+\rangle\times h(q,C,q).
\ee 
In the following representation, we equate the momenta of the two off-shell end of the scalar line for the current $S_q$ and integrate over $L$, to obtain the scalar loop contribution to $M_n(+,+,...,+)$. One can then use the supersymmetric Ward identities to see that the scalar contribution is the same as that of a graviton in the loop, up to overall numerical factors. 
\\~\\
One also needs to insert the current $\mathbb{J}$ into the sewing. One focuses on a particular leg, say $n$, and defines a tree by which it is attached to the loop to be the current $\mathbb{J}^{\mu\alpha}(B)$ for some subset $B$. One then obtains for the same helicity $n$-point amplitude in gravity, the following representation 
\be 
M_n(1^-,2^-,...,n^-)=\int \frac{d^DL}{(2\pi)^D}\sum_{B\subset C;~n\in B}\frac{S_q(L;A)}{L^2}\langle q^-|\cancel{l}\cancel{K_B}|q^+\rangle^2h(q,B,q).
\ee 

\section{One-loop same helicity four point graviton amplitude}
\subsection{Setup}
Let us explain the construction of the amplitude. The amplitude constitutes of a sum of three box diagrams, six triangle diagrams and fourteen bubbles, but for computation purposes, one diagram for each of these topologies is sufficient since we later permute the external legs in the result to obtain the other ones. We first construct the diagrams from the SDGR Feynman rules and then outline a construction from the chiral Einstein-Cartan action. Both results in the same loop integrals. The relevant diagrams to consider are the box, triangle and bubble. In SDGR at one-loop, one cannot have $\tilde{\psi}$ on external lines. This is because the propagator takes $\tilde{\psi}$ to $\Phi$ and the vertex is linear in $\tilde{\psi}$. Thus, it is not possible to construct one loop diagrams with external $\tilde{\psi}$ legs. Then we insert $d\Phi$ on external legs. In our convention, this is the all negative helicity graviton amplitude. With this, we begin with the box. 
\subsubsection{Box}
\vspace{1em}
\vspace{1em}

~~~~~~~~~~~~~~~~~~~~~~~~~~~~~~~~~~~~~~~~~\begin{fmfgraph*}(130,100)
\fmfbottom{i1,i2}
\fmftop{o1,o2}
\fmf{dbl_wiggly,tension=3}{i1,v1}
\fmf{dbl_wiggly,tension=3}{i2,v2}
\fmf{dbl_wiggly,tension=3}{o1,v3}
\fmf{dbl_wiggly,tension=3}{o2,v4}
\fmf{dbl_wiggly}{v3,v3'}
\fmf{double}{v3',v1}
\fmf{dbl_wiggly}{v1,v1'}
\fmf{double}{v1',v2}
\fmf{dbl_wiggly}{v2,v2'}
\fmf{double}{v2',v4}
\fmf{dbl_wiggly}{v4,v4'}
\fmf{double}{v4',v3}
\fmflabel{$1^-$}{o1}
\fmflabel{$4^-$}{o2}
\fmflabel{$2^-$}{i1}
\fmflabel{$3^-$}{i2}
\fmflabel{$l+1+2$}{v1'}
\fmflabel{$l+1$}{v3'}
\fmflabel{$l-4$}{v2'}
\fmflabel{$l$}{v4'}
\end{fmfgraph*}\\~\\
In the above diagram, the negative helicity states from (\ref{hel}) are inserted on the external legs. Consider any single vertex in the above diagram. From the interaction in (\ref{vert222}) it is clear that only the $\Phi^{MN}$ part of the $a^{MN}$ contributes. We now want to insert negative helicity states (\ref{hel}) to the external legs. Upon insertion, the second term vanishes due to $\langle qq\rangle$ contraction. The non-zero contribution thus comes from the first term, which is the anti-self dual part of the wedge product of two copies of $d\Phi$. Another way to see this is that on negative helicity states, the self-dual part of $d\Phi$ vanishes because of its linearised field equation. The ASD part is non-vanishing and results in the polarisation spinor $Mq_{M}q_Nk_{M'}k_{N'}/\langle qk\rangle^2$. This is the usual negative helicity state of a graviton. The resulting loop integral, upto coupling factors is 
\be 
\label{box1}
i\mathcal{M}_{box}=\int \frac{d^4l}{(2\pi)^4}\frac{\Big(\langle q|l|4]\langle q|l+1|1]\langle q|l+1+2|2]\langle q|l-4|3]\Big)^2}{l^2(l+1)^2(l+1+2)^2(l-4)^2\Big(\prod_{j=1}^4\langle qj\rangle\Big)^2}.
\ee
It is worth mentioning the structure of the loop integrand which arises from our construction. Comparing with the YM box diagram in (\ref{amp!}), we can see that the numerator of the gravity box integrand is the square of the numerator of the corresponding YM one. This is the double copy property between gravity and gauge theory amplitudes and we see that it arises directly from the Feynman rules of SDYM and SDGR. The expression of the above integrand is remarkably compact. In particular, in the absence of quartic vertices, there is just one box diagram to consider and it is in parallel with the construction in SDYM. This is the main outcome of the new covariant formalism for flat space SDGR. 
\\~\\
Let us also discuss the setup of the box diagram from the Feynman rules in chiral Einstein-Cartan theory. In this theory, the on-shell fields are $H$ and $\Omega$. We want to construct the all negative helicity diagram and thus insert it on the external $H$ leg. The field $\omega^{AA'}$ does not enter the interaction. This eliminates the possibility of $\langle h\omega\rangle$ type propagators. Therefore, neither the $h$ field nor the $\omega$ field contributes to the amplitude. This fixes the structure of the diagram, composed of $\langle H\Omega\rangle$ internal lines and $\langle HH\partial\Omega\rangle$ vertex. The vertex factor consists of two interaction terms. Consider first the second interaction term in the bracketed piece in (\ref{int21}). Let us apply it to the vertex where leg $1$ is inserted. From the $\langle H\Omega\rangle$ propagator in (\ref{Homega}), it is clear that the loop momentum $l$ will carry the third unprimed index of the connection and the third primed index of the metric. The derivative in the second interaction term is on the connection line and consists of the third unprimed index of the connection. This then hits the derivative in the propagator. Each polarization consists of two auxiliary unprimed spinors as is clear from (\ref{helicity}), which we take identical. Thus there are a total of 8 auxiliary unprimed spinors. These can either get contracted to the unprimed indices of the loop momenta or to themselves. However, in the above case where two loop momenta have the same unprimed index and gets contracted according to $l^{AA'}l_{AB'}=\frac{1}{2}\epsilon^{A'}_{~B'}l^2$, there are fewer loop momenta left to make them contract with all the  8 $qs$. Therefore, at least two auxiliary spinors in this case will contract amongst themselves, thus giving a vanishing contribution $\langle qq\rangle=0$. \\~\\Therefore, the second interaction term in (\ref{int21}) cannot contribute, because in all such terms one or many derivatives in the interaction hit the derivatives in the propagator, leading to $\langle qq\rangle=0$ contraction. The diagram thus consists of entirely the first interaction term. With this, it takes the particular form (\ref{box1})
\subsubsection{Triangle}
There is another non-zero contribution at one-loop, given by the triangle diagram
\vspace{1em}
\vspace{1em}

~~~~~~~~~~~~~~~~~~~~~~~~~~~~~~~~~~~~~~~~~\begin{fmfgraph*}(150,100)
\fmfbottom{i1,i2}
\fmftop{o1,o2}
\fmf{dbl_wiggly,tension=3}{i1,v1,o1}
\fmf{dbl_wiggly,tension=3}{i2,v3}
\fmf{dbl_wiggly,tension=3}{o2,v4}
\fmf{double}{v1,v1'}
\fmf{dbl_wiggly}{v1',v2}
\fmf{dbl_wiggly}{v2,v2'}
\fmf{double}{v2',v4}
\fmf{dbl_wiggly}{v4,v4'}
\fmf{double}{v4',v3}
\fmf{dbl_wiggly}{v3,v3'}
\fmf{double}{v3',v2}
\fmflabel{$1^-$}{o1}
\fmflabel{$4^-$}{o2}
\fmflabel{$2^-$}{i1}
\fmflabel{$3^-$}{i2}
\fmflabel{}{v1'}
\fmflabel{}{v2'}
\fmflabel{}{v3'}
\fmflabel{}{v4'}
\end{fmfgraph*}\\~\\
Using the SDGR Feynman rules, it is constructed in a similar way. The external legs are projected to negative helicity states and it is the ASD part of one of the $d\Phi$ which gives rise to the usual graviton polarization spinor. Thus we get four such spinors and these are then contracted with the other factors in the loop diagram. The result is 
\be 
\label{trigr}
\mathcal{M}_{triangle}=\frac{[12]}{\langle 12\rangle}\int \frac{d^4l}{(2\pi)^4}\frac{\Big(\langle q|l|3]\langle q|l-3|4]\langle q|l+1+2|1+2|q\rangle\Big)^2}{l^2(l-3)^2(l+1+2)^2\Big(\prod_{j=1}^4\langle qj\rangle\Big)^2}.
\ee 
For completeness, let us also mention that we can equivalently use the full GR Feynman rules, and this results in the same integral as in (\ref{trigr}). The process of construction is exactly similar to the box. In particular, the second term in the interaction Lagrangian in (\ref{int21}) does not contribute and thus the complete one loop integral is constructed using the first term.
\subsubsection{Bubble}
Finally we consider the bubble contribution to this amplitude. There are two types, bubbles on internal lines and bubbles on external legs. Let us describe the internal bubble diagram.
\be 
\begin{fmfgraph*}(180,50)
     \fmftop{i1,i3,i2}
     \fmfbottom{o1,o3,o2}
     \fmf{dbl_wiggly}{i1,v2,o1}
     \fmf{double}{v2,v1'}
     \fmf{dbl_wiggly}{v1',v1}
     \fmf{dbl_wiggly}{v1,i3}
     \fmf{double}{i3,v4}
     \fmf{dbl_wiggly}{v4,o3}
     \fmf{double}{o3,v1}
     \fmf{dbl_wiggly}{v4,v3}
     \fmf{double}{v3,v3'}
     \fmf{dbl_wiggly}{i2,v3',o2}
     \fmflabel{l}{i3}
     \fmflabel{l+1+2}{o3}
     \fmflabel{$1^-$}{i1}
     \fmflabel{$2^-$}{o1}
     \fmflabel{$3^-$}{o2}
     \fmflabel{$4^-$}{i2}
\end{fmfgraph*}
\ee 
\\~\\
The loop integral for the diagram is
\be 
\label{bubbgr}
i\mathcal{M}_{bubble}=\frac{[12][34]}{\langle 12\rangle\langle 34\rangle}\int \frac{d^4l}{(2\pi)^4}\frac{\Big(\langle q|l\circ (1+2)|q\rangle\langle q|(l+1+2)\circ(3+4)|q\rangle\Big)^2}{l^2(l+1+2)^2\prod_{j=1}^4\langle qj\rangle^2}.
\ee 
This is a nominally divergent integral. The denominator of the integrand is exactly like the one in YM, while the numerator is the square of the corresponding one in YM. Thus, in the form it is written above, the same arguments using Lorentz invariant regularization can be applied here. Indeed, the non-vanishing contribution in the numerator of the integrand comes from the quadratic, cubic and quartic in $l$ parts. However, using momentum conservation, the factor $\langle q|(1+2)\circ(3+4)|q\rangle$ vanishes by $\langle qq\rangle$ contraction and thus the quadratic and cubic parts do not contribute. The only non-vanishing contribution comes from the quartic part of the integral. This is proportional to 
\be 
\int \frac{d^4l}{(2\pi)^4}\frac{l_{\mu}l_{\nu}l_{\alpha}l_{\beta}}{l^2(l+1+2)^2}.
\ee
With any Lorentz invariant regularization, this can only be proportional to $(1+2)_{\mu}(1+2)_{\nu}\eta_{\alpha\beta}$+perm or to $(1+2)^4(\eta_{\mu\nu}\eta_{\alpha\beta}+\textrm{perm})$. Whatever the combination, the presence of the metric causes causes $\langle qq\rangle$ contraction and the diagram therefore vanishes. A similar thing holds for the bubbles on external lines and they too vanish. As we can see, this kind of argument is dependent on the form of the integrand, in conjunction with the SDYM case.
However, the integral in (\ref{bubbgr}) is quartic divergent and thus one again needs to be careful before concluding it to be vanishing. Indeed, like the SDYM bubble, one can try to shift the loop momenta by some arbitrary momenta and then check if the argument applies. We will come to this in a subsequent section, where we discuss the self energy diagram in gravity.
\subsection{One-loop finiteness} 
The box and triangle diagrams in (\ref{box1}) and (\ref{trigr}) are nominally divergent to start with. However, we now show that they are actually convergent. Let us consider the box integral in (\ref{box1}) and the same argument will hold for the triangle. We use the Feynman parametrization technique to rewrite the integrand. Thus, we assemble the quadratic factors of the denominator and complete the square by shifting the loop momentum parameter $l$ by an appropriate amount. We then rewrite the integral in terms of the shifted momentum and Feynman parameters
\be 
\label{amp'}
i\mathcal{M}_{box}=\int dX \int\frac{d^4l}{(2\pi)^4}\frac{\Big(\langle q|l-k|4]\langle q|l-k|1]\langle q|l+1-k|2]\langle q|l-4-k|3]\Big)^2}{(l^2+\Tilde{D})^4\Big(\prod_{j=1}^4\langle qj\rangle\Big)^2},
\ee
where 
\be 
\begin{split}
\int dX & = \int_{0}^{1}dx_1\int_{0}^{1-x_1}dx_2\int_0^{1-x_1-x_2}dx_3,\nonumber\\ 
\Tilde{D} & = -(x_31-x_14+x_2(1+2))^2+
x_2(1.2),\nonumber\\
k & = x_31-x_14+x_2(1+2).
\end{split}
\ee 
Let us consider the numerator of (\ref{amp'}). Note that the terms which have an odd number of loop momenta vanish when integrated over all momenta, since the denominator is symmetric under the exchange $l\rightarrow -l$, while the numerator is anti-symmetric. Next we collect the terms with an even number of loop momenta. In such terms, the Lorentz structure is carried only by the loop momenta and therefore can only be proportional to the metric tensor or sum of products of metric tensors. Let us consider the simplest cases of two and four loop momenta in the numerator. Such terms, by virtue of Lorentz invariance, can be written as 
\be 
\begin{split}
\int \frac{d^4l}{(2\pi)^4}\frac{l_{\mu_1}l_{\mu_2}}{(l^2+\Tilde{D})^4}&=\eta_{\mu_1\mu_2}\int \frac{d^4l}{(2\pi)^4}\frac{l^2}{(l^2+\Tilde{D})^4},\nonumber\\
\int \frac{d^4l}{(2\pi)^4}\frac{l_{\mu_1}...l_{\mu_4}}{(l^2+\Tilde{D})^4}&=(\eta_{\mu_1\mu_2}\eta_{\mu_3\mu_4}+\eta_{\mu_1\mu_3}\eta_{\mu_2\mu_4}+\eta_{\mu_1\mu_4}\eta_{\mu_2\mu_3})\int \frac{d^4l}{(2\pi)^4}\frac{l^4}{(l^2+\Tilde{D})^4}.
\end{split}
\ee 
The generalization to higher powers of loop momenta is now obvious. 
In 4-dimensions, the metric tensor can be written as a product of spinor metrics, $\eta_{\mu\nu}=\epsilon_{MN}\epsilon_{M'N'}$. Therefore, the combination of metric tensors produce combinations of a pair of spinor metrics, primed and unprimed. The unprimed spinor metrics contract the $q$s coming from the polarizations leading to $\langle qq\rangle=0$. Thus, we conclude that the terms with an even number of loop momenta also vanish when integrated over all momentum space. The remaining terms in the numerator do not contain any loop momentum factor and are therefore power counting finite. We have thus established that the integral in (\ref{box1}) is convergent. A similar argument renders the triangle integral to be power counting finite. 
\\~\\ 
Having established that the diagrams we consider are finite, it is reasonable to proceed towards computing these and extract the result for the amplitude. However, we encounter a problem at this point. The box and triangle integrals as in (\ref{box1}) and (\ref{trigr}) are quartic divergent to start with. The propagator factors in the denominator of these integrals are exactly similar to the ones in SDYM. Thus, it may seem reasonable to cancel propagators and decompose these into bubble like integrals with two propagator factors in the denominator. However, in this process, the degree of divergence of the individual integrals will increase as a result of the decrement of loop momentum factors in the denominator. In particular, the degree of divergence of the resulting bubble-like integrals will be more than or equal to five. Moreover, it is not clear if these bubble like integrals will vanish after shifts, since the numerator in this case will be a complicated function of the loop momentum and external momenta and it is not guaranteed that one can use Lorentz invariant arguments in this case. Even if the integrals vanish after shifting the loop momentum parameters, the computation of shifts will be a daunting task. It is then, if not impossible, almost impenetrable to do a shift computation and extract the result of the amplitude in this route.\\~\\ 
We thus follow a different route, namely a possible generalization of the bubble computation in the case of gravity. The motivation behind this follows from the close relations between SDYM and SDGR. As we have seen, the structure of one-loop diagrams in both these theories are on similar footing. The numerator of the integrand of the self-energy in this case is again the square of the one in SDYM, as we shall see below. Also, analogous to SDYM, the notion of self-dual BG currents is present in SDGR. Recall, it is the gluing of these currents to the self energy which leads to the construction of the amplitudes in the former case. We thus expect a similar thing will follow in the latter. Indeed, as we shall see below, it is possible to perform a shift computation of the self energy, except that one needs to calculate shifts upto quartic order in this case. However, it is possible to extract the result by arguing on general grounds. Once the self-energy is computed, one can glue BG currents to it and recover all possible bubble diagrams which can contribute to the amplitude. Let us describe all this  
\section{Self energy bubble}
Let us now consider the self-energy diagram in the theory of SDGR. As we shall see, the property of this diagram is in parallel to the one in SDYM. In particular, the diagram is sensitive to shifts and is in general non-vanishing. 
\\~\\
\be 
\begin{fmfgraph*}(150,80)
     \fmfleft{i}
     \fmfright{o}
     \fmfbottom{v}
     \fmftop{u}
     \fmf{dbl_wiggly,tension=3,label=-p}{i,v1}
     \fmf{dbl_wiggly}{v1,u}
     \fmf{double}{u,v2}
     \fmf{dbl_wiggly}{v2,v}
     \fmf{double}{v,v1}
     \fmf{dbl_wiggly,tension=3,label=p}{v2,o}
     \fmflabel{l}{u}
     \fmflabel{l+p}{v}
\end{fmfgraph*}\nonumber 
\ee 
\\~\\
In the diagram above, the external lines are projected to two negative helicity states and the convention being all external momenta incoming. The amplitude can be written as 
\begin{equation}
\label{segr}
i\Pi^{--}_{SDGR}=\int \frac{d^4l}{(2\pi)^4}\frac{(\langle q|l|p]\langle q|l+p|p])^2}{l^2(l+p)^2\langle qp\rangle^4}.
\end{equation}
In this form, the amplitude can be argued to vanish. Indeed, this amplitude is similar to the bubble diagrams, in that the form of the integrand as it is written is the square of the corresponding SDYM self-energy. Also, the diagram is quartic divergent. The only difference in this case is, the propagator which originates from the two vertices of the bubble is truncated and the external lines are projected to helicity states. Thus, the momenta $p$ flowing inside the loop is null. A similar argument like in the case of the bubble diagrams still applies and it is easy to see that with any Lorentz invariant regularisation, the diagram vanishes. However, as we explained earlier, the diagram is quartic divergent and thus one should be careful in analyzing it. Let us shift the loop momentum $l= x+ s'$, where $s'$ is some momentum and $x$ is the new integration variable. The shifted integral is 
\be 
\label{shifted}
i\Pi^{--}_{SDGR-shifted}=\int \frac{d^4x}{(2\pi)^4}\frac{(\langle q|x+\tilde{s}|p]\langle q|x+p+\tilde{s}|p])^2}{(x+\tilde{s})^2(x+\tilde{s}+p)^2\langle qp\rangle^4}.
\ee
The argument above depends on the specific form of the integrand in (\ref{segr}) and is no longer applicable to the shifted integrand. Let us then compute the difference between the shifted and the unshifted integral, analogous to the SDYM case. The difference with SDYM however is that, the integral in SDGR is quartic divergent and thus we have to compute shifts upto quartic order. The computation is complicated and we refer the reader to Appendix E for details. Here, we instead give an argument and give the general form of the result. 
\subsection{Result}
Let us understand the form of the result which can be obtained by a shift computation. The
denominator in the result can only have a factor of $\langle qp\rangle^4$. After the shift $x\rightarrow x-\tilde{s}$, the numerator can be a possible combination of two factors $\langle q|\tilde{s}|p]^m\langle q|p+\tilde{s}|p]^n$ with $m+n=4$. Let us give a name to the combination $p+\tilde{s}=s$. Then, it is easy to see that the combination of factors $\langle q|\tilde{s}|p]^m\langle q|p+\tilde{s}|p]^n$ can always be written in the form $(\langle q|s|p]\langle q|\tilde{s}|p])^2$, owing to $[pp]=0$ contraction. Overall, the amplitude takes the form
\be 
\label{shiftgr}
\Pi^{--}_{SDGR}\sim\Bigg(\frac{\langle q|s|p]\langle q|\tilde{s}|p]}{\langle qp\rangle^2}\Bigg)^2.
\ee
Thus, the self energy bubble projected to negative helicity states in SDGR is the square of the one in SDYM. This is again reminiscent of the double copy relation in the self-dual sector of these theories. Here we have analysed the bubble using general arguments and obtained the particular form (\ref{shiftgr}). Alternatively, the same result can be obtained from a direct shift calculation (\ref{shiftgr2}). The shift in the gravity case is calculated upto quartic order, in contrast with the YM case which was upto second order. One can now interpret the result in (\ref{shiftgr}) as an insertion of negative helicity states to an effective propagator.
\subsection{Effective gravitational propagator}
The effective propagator in SDGR is the one-loop correction to the $\langle\tilde{\psi}\Phi\rangle$ propagator. Thus, it is the bubble graph with two external $\Phi$ legs which are truncated. The 2-form $d\Phi$ on the external leg can be split into a self-dual part and the anti-self-dual part. We would like to insert negative helicity states to these legs. Such an insertion removes the self-dual part of $d\Phi$ in each of these legs. The anti-self-dual part which is non-vanishing reproduces the usual negative metric helicity. One can then insert negative helicity states to these legs and this results in (\ref{shiftgr}). In terms of the shifts, the effective propagator reads
\be
\label{grprop}
\begin{gathered}
\begin{fmfgraph*}(140,70)
     \fmfleft{i}
     \fmfright{o}
     \fmf{double}{i,v1}
     \fmfdot{v1}
     \fmf{double}{o,v1}
     \fmflabel{$MM'NN'$}{i}
     \fmflabel{$PP'QQ'$}{o}
     \end{fmfgraph*}
     \end{gathered}~~~~~~~~~~~~~~~~~~:=~~~~~s^{~M}_{P'} \tilde{s}_{P}^{~M'}s^{~N}_{Q'}\tilde{s}^{~Q}_{N'}.
\ee 
Comparing with the SDYM case, the effective propagator in SDGR can be seen to be the double copy of the former. We can now insert two copies of the negative helicity polarization spinor of the graviton (\ref{grprop}) and this results in (\ref{shiftgr}). Overall, the shift in gravitational effective propagator is the most natural generalisation of the effective propagator in YM from spin 1 to spin 2 case.
\\~\\
\subsection{BG current insertion into effective propagator} 
Similar to YM, we can construct bubble diagrams in gravity as a result of insertion of BG currents to the effective propagator. Let us explain this. 
\subsubsection{Internal Bubbles}
There are two in-equivalent permutations of the bubbles on internal lines. In one diagram, the legs 1 and 2 join on one side of the internal line while 3 and 4 on the other side. In the other diagram, the legs $1, 4$ join on one side while the legs $2,3$ join on the other end. When written in terms of insertion of 2-currents to the effective propagator in (\ref{grprop}), the result of these diagrams read 
\be 
\mathcal{M}_{11}=\mathbb{J}(1,2)\mathbb{J}(3,4)(\langle q|s_1\circ(1+2)|q\rangle\langle q|\tilde{s}_1\circ(1+2)| q\rangle)^2,
\ee
\be 
\mathcal{M}_{12}= \mathbb{J}(1,4)\mathbb{J}(2,3)(\langle q|s_{2}\circ(1+4)|q\rangle\langle q|\tilde{s}_2\circ(1+4)| q\rangle)^2,
\ee
where $s_i,\tilde{s}_i$ for $i=1, 2$ are arbitrary shifts of the loop momentum variable in the bubble integrals. Let us also write down the results for the shift in external bubbles.
\subsubsection{Bubbles on External Lines}
There are four distinct contributions here, each of which can be written as an insertion of the 3-current and 1-current to the legs of the bubble. Let us first denote a generic 3-current by a diagram
\vspace{1em}
\vspace{1em}
\be
\mathbb{J}_{MNPQ'}(2,3,4)~~~~=~~~~~\begin{gathered}
\begin{fmfgraph*}(60,50)
\fmfleft{i1}
\fmfright{o1,o2,o3}
\fmf{double}{i1,o}
\fmfblob{.15w}{o}
\fmf{dbl_wiggly}{o,o1}
     \fmf{dbl_wiggly}{o,o2}
     \fmf{dbl_wiggly}{o,o3}
\fmflabel{1}{i1}
\fmflabel{$2^-$}{o1}
\fmflabel{$3^-$}{o2}
\fmflabel{$4^-$}{o3}
\end{fmfgraph*}
\end{gathered}\nonumber
\ee 
\\~\\
On the left hand side of the diagram, the dashed line indicates the field $\tilde{\psi}$ with the convention that the propagator is included as a part of the definition of the current and the blob indicates the sum of three tree diagrams. 
Next we insert each of these currents into the bubbles such that for a particular graph, the internal line adjacent to a bubble is the off-shell leg of the current. The pole resulting from the denominator in the propagator of the internal line,~i.e $1/p^2$, where $p=1,2,..4$ gets cancelled yielding a finite result in the on-shell limit. The external bubble diagrams can then be expressed in terms of the gluing of BG currents to the effective propagator as\\~\\
\be
\begin{gathered}
~~~\begin{fmfgraph*}(80,60)
     \fmfleft{i}
     \fmfright{o1}
     \fmftop{o2}
     \fmfbottom{o3}
     \fmf{dbl_wiggly}{i,v1}
     \fmf{vanilla,left,tension=0.6,tag=1}{v1,v2}
     \fmf{vanilla,left,tension=0.6,tag=1}{v2,v1}
     \fmf{dbl_wiggly}{v2,o4}
     \fmf{double}{o4,o}
     \fmfblob{.15w}{o}
     \fmf{dbl_wiggly}{o,o1}
     \fmf{dbl_wiggly}{o,o2}
     \fmf{dbl_wiggly}{o,o3}
     \fmflabel{$1^-$}{i}
     \fmflabel{$2^-$}{o1}
     \fmflabel{$3^-$}{o2}
     \fmflabel{$4^-$}{o3}
\end{fmfgraph*}
\end{gathered}\nonumber 
\ee 
\be 
\mathcal{M}_{21}=\mathbb{J}(1)\mathbb{J}(2,3,4)(\langle q|s_3|1]\langle q|\tilde{s}_3|1| q\rangle)^2,
\ee
\be 
\mathcal{M}_{22}= \mathbb{J}(2)\mathbb{J}(1,3,4)(\langle q|s_4|2]\langle q|\tilde{s}_4|1| q\rangle)^2,
\ee
\be 
\mathcal{M}_{23}=\mathbb{J}(3)\mathbb{J}(1,2,4) (\langle q|s_5|3]\langle q|\tilde{s}_5|3| q\rangle)^2,
\ee
\be 
\mathcal{M}_{24}=\mathbb{J}(4)\mathbb{J}(1,2,3) (\langle q|s_6|4]\langle q|\tilde{s}_6|4| q\rangle)^2.
\ee
Here, we summarised the list of all possible bubble graphs for the one loop same helicity four point amplitude. The structure of the bubbles mimicks the YM ones, in that they arise by gluing BG currents to the effective propagator. However, in this case, one considers all possible permutations of the external legs in contrast to YM, where only cyclic permutations are allowed. The number of in-equivalent bubble diagrams are thus higher in the case of gravity. This is correctly captured by the structure of the currents in SDGR. In particular, the 3-currents in SDGR contains a sum of three distinct terms, unlike the ones in SDYM where there is only a single term. Thus, it is this higher number of terms in the gravity currents which help to capture all possible permutations of the bubbles and assemble them in a condensed form which we wrote above. We want to compute them explicitly and extract the result for the amplitude. However, here we face a problem. It is not clear how to interpret the shift parameters in the gravity case. In the case of YM we could associate these with the region momenta. There is no cyclic ordering in the case of gravity, and so the use of region momenta does not seem justified any longer. Thus, while there is strong evidence that there must exist a gravitational analog of the bubble interpretation of the four point amplitude and its generalization to arbitrary multiplicity as in the formula (\ref{YM-general}), it is not clear how to give meaning to the momentum variables appearing in the gravitational effective propagator, and so it is not clear what form the GR analog of (\ref{YM-general}) should take.

\conclusion{Conclusion and Outlook} 
In this thesis, we have studied the series of all minus amplitudes in massless QED, self-dual YM and self-dual GR. These amplitudes are rational functions of the momenta involved and are devoid of any branch cut singularities. Firstly, we have analyzed such an amplitude at four points in SDYM and massless QED, establishing that the result comes out from a computation of shifts. The important feature of such a computation is that it is done in 4 dimensions and avoids any regularization scheme. Indeed, in the absence of any singularities, except for two particle poles, the result should not depend upon any regularization and this is what we were able to demonstrate. An important insight gained from this computation is that the technique of reducing one-loop amplitudes to shifts can be an alternative variant to extract rational parts of a wider class of amplitudes. 
\\~\\
However, the most important result of this thesis is the understanding that the all minus amplitudes in Yang-Mills can be described by a very simple formula consisting of the BG currents and an effective propagator in SDYM.
\\~\\
\be 
\label{YM-general}
\mathcal{A}=\sum_{part} \begin{gathered}
~~~\begin{fmfgraph*}(120,70)
     \fmfleft{i,i1,i2,i3,i4}
     \fmfright{o1,o4,o5,o6,o7}
     \fmf{vanilla}{i,v}
     \fmf{vanilla}{i4,v}
     \fmfblob{.15w}{v}
     \fmf{vanilla,label=$p_j$,label.distance=1cm,label.side=left}{v,v1}
     \fmfdot{v1}
     \fmf{vanilla,label=$p_{i-1}$,label.distance=1cm,label.side=right}{v1,o}
     \fmfblob{.15w}{o}
     \fmf{vanilla}{o,o1}
     \fmf{vanilla}{o,o7}
     \fmflabel{$i$}{i}
      \fmflabel{$.$}{i1}
     \fmflabel{$.$}{i2}
      \fmflabel{$.$}{i3}
      \fmflabel{$j$}{i4}
     \fmflabel{$i-1$}{o1}
      \fmflabel{$.$}{o4}
      \fmflabel{$.$}{o5}
      \fmflabel{$.$}{o6}
     \fmflabel{$j+1$}{o7}    
     \end{fmfgraph*}
     \end{gathered}
     \quad\quad
\ee  
\be 
\label{SDYM-form}
\label{BG}
=\sum_{part}J(i,..,j)J(j+1,..,i-1)\langle q|p_j\circ p_{i-1}|q\rangle^2.
\ee
We have also managed to give an explicit proof of our formula by establishing that it is independent of the region momentum variables that are needed to make sense of the effective propagator. To establish our formula, we used the technology of shifts to show that the self-energy bubble in SDYM is ambiguous and shift dependent. Thus, it is rather a special choice of shift variables, namely the region momentum variables which produces a finite result for the bubble. The amplitudes are then built by gluing BG currents to this bubble and adding all possible cyclic permutations. The non-vanishing of such amplitudes reduces to the non-vanishing of the bubble. The formula we proposed is important for several reasons. Let us explain these. 
\\~\\ 
It is well known that the CSW \cite{Cachazo:2004kj} tree-level formalism gives a prescription for how to compute tree-level YM
amplitudes with effective Feynman rules that use MHV amplitudes as vertices. An attempt to extend
this to loop amplitudes was made in \cite{Cachazo:2004zb}, where it was in particular suggested that one-loop same
helicity amplitudes that are the subject of this thesis can be added as new interaction vertices, in
addition to the tree level MHV vertices. The interpretation of these one-loop amplitudes that emerges
from our work is that they are built from more elementary tree-level blocks. It thus seems unnatural
to try to build more involved amplitudes from the amplitudes that are already composed. Our results
show that these amplitudes are made from simpler tree-level amplitudes connected by an effective
propagator. Thus, it appears that it is the region momenta-dependent effective propagator that should
be added into the set of CSW rules if one is to reproduce loop amplitudes.\\
Our new interpretation is also relevant for the problem of UV divergence of quantum gravity. As we have already discussed, this divergence was linked to the non-vanishing of the same helicity one-loop amplitude
in GR. Given that the story with GR is likely mirroring that in YM, the
interpretation of this work shows that the non-vanishing of the same helicity one-loop amplitudes can
in turn be linked to the non-vanishing of the one-loop bubble. It would thus be very interesting to
better understand the significance and interpretation of the non-vanishing of the bubbles.
This leads to the question of interpretation of such a simple construction as our formula. The fact that
region momentum variables play such a prominent role points to both the worldsheet interpretation as
in \cite{Chakrabarti:2005ny}, as well as the (momentum) twistor space interpretation as in \cite{Mason:2009qx}. In any case, it would be very
interesting to understand a deeper origin of our proposed formula.
\\~\\
In this work we have analysed only the same helicity amplitudes. These amplitudes vanish at tree level, and are non-zero at one loop. There is, however, another
amplitude at one-loop with exactly the same status, this is the all but one same helicity
amplitude. It vanishes at tree level, but is given by a purely rational expression at one
loop, see e.g. the formula (3.10) in \cite{Binoth:2006hk}. This amplitude, however, cannot be computed in
SDYM theory. This is because in SDYM only diagrams with external gauge field lines can
be constructed at one loop. In the case of the same helicity one loop amplitudes this is sufficient. Indeed, in the conventions used in SDYM, auxiliary field can only describe the
positive helicity. When one computes the all minus amplitude, the Feynman diagrams with
the gauge field on the external lines are the only ones to consider, and these are precisely
the diagrams that get generated in SDYM. However, in the case of a $(---+)$ amplitude,
in addition to diagrams with the gauge field on external lines, there is a new diagram where
the minus helicity state gets inserted into an auxiliary field external line. This diagram
is not present in SDYM, and one needs the full YM Feynman rules to construct it. It is
possible that the methods outlined in this thesis can still be applied to this situation and the sum
of all arising diagrams gets reduced to shifts of the effective propagator, but this is far from certain and we will leave
this computation to future research.
\\~\\
Another important aspect of this thesis is to give the first steps of the understanding of all minus amplitudes in gravity. The main issue in gravity is that the existing Feynman rules in the usual Einstein-Hilbert formulation is a mess to deal with. Thus, we considered a recently developed chiral formulation for GR \cite{Krasnov:2020bqr} which has very simple Feynman rules. The only missing parts in this formalism was the ghost Feynman rules. We thus developed them in this work by using the BRST formalism. Before discussing the all plus amplitudes in gravity, let us describe as an aside, our study of the one-loop effective action around a general Einstein background in the new formulation. We devised a generalization of the flat space gauge fixing Lagrangian which is described in \cite{Krasnov:2020bqr}, to a general Einstein background. We found that this introduces a novel $hh$ term in the kinetic part of the action. The novelty in the gauge fixing is that the kinetic part of the Lagrangian gets decoupled into two distinct sectors in an analogous fashion as in flat space. Another important aspect in the computation is that the BRST transform of the Lorentz and diffeomorphism ghosts are coupled and this results in a ghost Lagrangian where there are coupling terms between the two ghosts. However, using field re-definitions, we could get rid of the mixed terms and found the relevant Laplace type operator for the kinetic part. We then studied the ghost contribution to the effective action using the heat kernel methods.  
\\

It is possible to construct the all plus (or minus) amplitudes in gravity both from the full GR Feynman rules or the SDGR Feynman rules. The SDGR Feynman rules are obtained from the recently proposed action for flat SDGR in \cite{Krasnov:2021cva}, where a detailed gauge fixing is outlined. The diagrams contributing to the four point all minus amplitude is constructed using these Feynman rules. The diagrams are all quartic divergent to start with. However, using Lorentz invariance arguments, we showed that the divergent pieces vanish and the diagrams are actually power counting finite. This is reminiscent of the one-loop finiteness of pure gravity. The problem arises when we attempt to compute these diagrams. The shift technology if at all applicable, is not feasible to extract the answer. This is because it is not clear if the process to cancel denominators and reducing the resulting integrand to shifts will hold in this case. Thus, we take the alternative route to compute the amplitude using a bubble computation, in analogy with SDYM. Indeed, we computed the self-energy bubble, with the result being the square of the corresponding YM one. We then glue BG currents to this bubble and the sum over all possible permutations is expected to give the amplitude. The only issue here is that is that the interpretation of the shift parameters is not clear, unlike in YM where the shifts were interpreted as region momenta. The main difference in gravity is that, unlike in YM (or SDYM), everything is permutation symmetric and thus the whole group of permutation symmetry is at play, rather than just the smaller subset of cyclic symmetry. Thus, the arguments in SDYM which depend on cyclic ordering do not apply in gravity. Thus, while there is strong evidence of an analogous formula as in (\ref{SDYM-form}) for gravity, it is not clear what form the formula will take. We leave this for future research. Let us now summarise the main points and results of the different chapters in this thesis. \\~\\
\textbullet~ Chapter 4: The $\beta$-function in the chiral formulation of Yang-Mills is worked out. The Lagrangian and Feynman rules for SDYM has been sketched and a description of Yang-Mills instantons has been outlined. There are no new results in this chapter but some warm-up calculations to fix notations.\\~\\
\textbullet~ Chapter 5: The results in this chapter appeared in \cite{Chattopadhyay:2020oxe} and \cite{Chattopadhyay:2021udc}. The one-loop four point amplitude is reduced to a shift computation in SDYM and massless QED (with scalar QED). The sum of integrands for this amplitude is computed in the covariant formalism. It is observed that the sum vanishes with a particular assignment of region momentum variables in the graphs. The self-energy bubble is observed to be shift dependent and the shift is computed using standard techniques. The result of the shift is interpreted as an effective propagator in SDYM. The four point amplitude is then obtained by gluing BG currents to this effective propagator and summing over cyclic permutations. A new formula for all YM amplitudes (same helicity) at one-loop is proposed and proved. The collinear and soft limits are analyzed. As an illustration of our formula, the explicit computation of the amplitude at 3, 4 and 5 points are outlined. 
\\~\\
\textbullet~ Chapter 7: Some of the results of this chapter are new. The quantization of the chiral Einstein-Cartan gravity is studied in a general Einstein background. We use the BRST formalism to develop the linearised gauge fixed Lagrangian. The non-triviality in the BRST transformations is that the transformation for the Lorentz and diffeomorphism ghosts are mixed. This results in a ghost Lagrangian where there is a coupling term between the two fields. Nevertheless, we diagonalised the arising operator in the Lagrangian and then used the heat kernel to compute the ghost contribution to the one loop effective action. This part has not appeared elsewhere.\\
We also deduced the ghost Feynman rules in flat space. For the sake of completeness, we mentioned the remaining Feynman rules for chiral Einstein-Cartan and this part is mostly taken from \cite{Krasnov:2020bqr}. 
\\~\\
\textbullet~ Chapter 8: This chapter is mostly based on \cite{Krasnov:2021cva} and \cite{Krasnov:2016emc}. The action for flat space self-dual gravity is described. There are two relevant actions and we described one of them in brief and the other in details. The Feynman rules and BG currents are outlined for the SDGR theory. A short description of gravitational instantons is also sketched.
\\~\\
\textbullet~ Chapter 9: In this chapter, we give partial results for the one-loop same helicity amplitudes in self-dual gravity. This part is new and has not appeared elsewhere. The construction of the one-loop same helicity amplitude at four point is sketched.  The setup of the four point amplitude is outlined in terms of the box, triangle and bubble diagrams. An explanation for the finiteness at one-loop is given. The result for the self-energy bubble in SDGR is obtained, both by direct computation and general arguments. The truncated bubble is then interpreted as an effective gravitational propagator in the theory. BG currents are glued to this propagator in all possible ways and it is expected that the sum of all these bubble diagrams should result in the four point amplitude. However, this last statement remains to be seen explicitly. The only missing element is the interpretation of the shift parameters for the bubbles, which is not clear in the case of gravity. Thus, the bubble interpretation of the gravity amplitudes is incomplete.
\\~\\
A possible direction to approach the computation in the gravity case is as follows. Using the analogy of SDYM, we can first try to compute the sum of integrands for the one loop amplitude. This would require contribution of the box, triangles and bubbles. The technique of decomposing the box and triangles to "bubble"-like integrals is the necessary next step. This part is non-trivial because unlike the case of SDYM, in gravity we deal with quartically divergent integrals from the beginning. Once this reduction is achieved, it may be possible to consider some good choice of loop momentum variables such that the integrand vanishes. It remains to be seen what kind of variables these are. The formula for the amplitude at all multiplicity in gravity can then possibly be sketched. We however leave this for future research.
\begin{appendices} 
\appendixpage
\noappendicestocpagenum
\addappheadtotoc
\chapter{Feynman parameter integrals}
In this part of the Appendix, we give a list of some of the results for the Feynman parameter integrals which we use in Chapter 4.
\\~\\
\begin{equation}
\label{1}
    \int \frac{d^{4-\epsilon}q}{(2\pi)^{4}}\frac{1}{(q^2+D)^2}=\frac{i}{16\pi^2}\frac{\Gamma(\epsilon/2)}{\Gamma(2)}\frac{1}{(-D)^{\frac{\epsilon}{2}}},
\end{equation}
\begin{equation}
\label{2}
    \int \frac{d^{4-\epsilon}q}{(2\pi)^4}\frac{q^2}{(q^2+D)^2}=\frac{-2i}{16\pi^2}\frac{\Gamma(\epsilon/2-1)}{\Gamma(2)}\frac{1}{(-D)^{\frac{\epsilon}{2}-1}},
\end{equation}

\begin{equation}
\label{3}
    \int \frac{d^{4-\epsilon}q}{(2\pi)^{4}}\frac{q^2}{(q^2+D)^3}=\frac{2i}{16\pi^2}\frac{\Gamma(\epsilon/2)}{\Gamma(3)}\frac{1}{(-D)^{\frac{\epsilon}{2}}},
\end{equation}
\begin{equation}
\label{4}
    \int \frac{d^{4-\epsilon}q}{(2\pi)^4}\frac{1}{(q^2+D)^3}=\frac{2i}{16\pi^2}\frac{\Gamma(\epsilon/2)}{\Gamma(3)}\frac{1}{(-D)^{\frac{\epsilon}{2}}},
\end{equation}
\begin{equation}
    \frac{1}{(-D)^{\epsilon/2}}=1-(\epsilon/2)log(-D)+.....
\end{equation}
The Gamma function can be expanded near its poles as follows:
\be
   \Gamma(x)=\frac{1}{x}-\gamma+O(x).
\ee
\chapter{Extracting the shift}
In this part, we extract the shift using standard technology, but in spinor notations. We use this result in Chapter 5.  
Let us consider the difference of two one dimensional quadratically divergent integrals of the form 
\be 
J(b)-J(0)\equiv\int_{-\infty}^{+\infty}dy [f(y+b)-f(y)],
\ee 
where $f(\pm\infty)=d_{\pm}$ are finite constants. If the integral were convergent, it is easy to see that $J(b)-J(0)$ would vanish by a quick change of integration variables $y+b\rightarrow y'$. In the present case, however, let us Taylor expand $f(y+b)$ in powers of $b$. Note that second and higher derivatives of $f(y)$ vanishes at $y=\pm\infty$.
\be 
J(b)=\int_{-\infty}^{+\infty}dy[f(y)+bf'(y)+\frac{1}{2}a^2f"(y)+...]\nonumber
\ee 
\be 
=J(0)+b(f(+\infty)-f(-\infty))+\frac{1}{2}b^2(f'(+\infty)-f'(-\infty)).
\ee 
We see that the difference gives a finite value in this case.
\be 
J(b)-J(0)=b(f(+\infty)-f(-\infty))+\frac{1}{2}b^2(f'(+\infty)-f'(-\infty)).
\ee 
\chapter{Check of region momenta independence in SDYM} 
In this part of the Appendix, we explicitly show that the four and five point amplitudes are region momenta independent. Thus, we use the shift symmetry of region momenta and write all region momenta in terms of one of them, say $x$. We then show that the terms containing $x$ vanish identically. We first do it for the four point case and subsequently for the five point case.
\section{Four point case} 
We demonstrate that the 4-point amplitude is region momentum independent. Let us start by analysing the terms linear in $x$. For instance, there is such term coming from $\langle q|p_3\circ p_2|q\rangle^2$. Indeed, we have 
\be 
\langle q|p_3\circ p_2|q\rangle^2= \langle q|(2 + x)\circ 3|q\rangle^2
= \langle q|2\circ 3|q\rangle^2 + 2\langle q|2\circ 3|q\rangle \langle q|x\circ 3|q\rangle\nonumber\\ + \langle q|x\circ 3|q\rangle^2.
\ee 
The second term on the right-hand side is linear in $x$. Keeping only such terms we get the linear in $x$
part of the amplitude
\be 
\begin{split} 
\mathcal{A}^x_4&=\frac{2[23]\langle q|x|3]}{\langle q4\rangle\langle 41\rangle\langle 12\rangle}-\frac{2[14]\langle q|x|4]}{\langle q3\rangle\langle 23\rangle\langle 12\rangle}+\frac{2[12](\langle q|x|3]\langle 3q\rangle+\langle q|x|4]\langle 4q\rangle)}{\langle q3\rangle\langle 34\rangle\langle 12\rangle\langle 4q\rangle}\\
&=\frac{2\langle q|x|3]}{\langle q4\rangle\langle 12\rangle}\Bigg(\frac{[23]}{\langle 41\rangle}+\frac{[12]}{\langle 34\rangle}\Bigg)-
\frac{2\langle q|x|4]}{\langle q3\rangle\langle 12\rangle}\Bigg(\frac{[14]}{\langle 23\rangle}-\frac{[12]}{\langle 34\rangle}\Bigg)\\
&=0.
\end{split}
\ee 
The last equality follows by momentum conservation as the terms in brackets are zero.
We now analyse the terms quadratic in $x$. There are such terms coming from all the terms in (\ref{4pt}).
We get
\be 
\begin{split} 
\mathcal{A}_4^{x^2}&=\frac{\langle q|x|1]^2}{\langle q2\rangle\langle 23\rangle34\rangle\langle q4\rangle}+
\frac{\langle q|x|2]^2}{\langle q3\rangle\langle 34\rangle41\rangle\langle q1\rangle}+
\frac{\langle q|x|3]^2}{\langle q4\rangle\langle 41\rangle12\rangle\langle q2\rangle}+
\frac{\langle q|x|4]^2}{\langle q1\rangle\langle 12\rangle23\rangle\langle q3\rangle}\\&+
\frac{(\langle q|x|3]\langle 3q\rangle+\langle q|x|4]\langle 4q\rangle)^2}{\langle q1\rangle\langle q2\rangle\langle q3\rangle\langle q4\rangle\langle 34\rangle\langle 12\rangle}\\&+
\frac{(\langle q|x|3]\langle 3q\rangle+\langle q|x|2]\langle 2q\rangle)}{\langle q1\rangle\langle q2\rangle\langle q3\rangle\langle q4\rangle\langle 41\rangle\langle 23\rangle}.
\end{split} 
\ee 
We want to demonstrate that this vanishes by momentum conservation and Schouten identities. Momentum conservation gives us the following 4 relations (not all independent)
\be 
\begin{split} 
\langle q|x|2]\langle 21\rangle + \langle q|x|3]\langle 31\rangle + \langle q|x|4]\langle 41\rangle = 0,\\
\langle q|x|1]\langle 12\rangle + \langle q|x|3]\langle 32\rangle + \langle q|x|4]\langle 42\rangle = 0,\\
\langle q|x|1]\langle 13\rangle + \langle q|x|2]\langle 23\rangle + \langle q|x|4]\langle 43\rangle = 0,\\
\langle q|x|1]\langle 14\rangle + \langle q|x|2]\langle 24\rangle + \langle q|x|3]\langle 34\rangle = 0.
\end{split} 
\ee 
These can be used to express two of the quantities $\langle q|x|i]$ in terms of the other two. For example, we
can decide to express $\langle q|x|1],\langle q|x|2]$ in terms of $\langle q|x|3],\langle q|x|4]$. We have
\be 
\langle q|x|1]\langle 12\rangle  = \langle q|x|3]\langle 23\rangle + \langle q|x|4]\langle 24\rangle,\nonumber\\\langle q|x|2]\langle 12\rangle  = \langle q|x|3]\langle 31\rangle + \langle q|x|4]\langle 41\rangle.
\ee 
Using this, we can collect coefficients in front of $\langle q|x|3]^2$, $\langle q|x|4]^2$ and $\langle q|x|3]\langle q|x|4]$. For example, the
coefficient in front of $\langle q|x|4]^2$
is
\be 
\begin{split} 
\frac{\langle 24\rangle^2}{\langle q2\rangle\langle 23\rangle\langle 34\rangle\langle 12\rangle^2\langle q4\rangle}+\frac{\langle 41\rangle}{\langle q3\rangle\langle 34\rangle\langle 12\rangle^2\langle q1\rangle}+\frac{1}{\langle q1\rangle\langle 12\rangle\langle 23\rangle\langle q3\rangle}\\
+\frac{\langle q4\rangle}{\langle q1\rangle\langle q2\rangle\langle q3\rangle\langle 12\rangle\langle 34\rangle}+\frac{\langle q2\rangle\langle 41\rangle}{\langle q1\rangle\langle q3\rangle\langle q4\rangle\langle 23\rangle\langle 12\rangle^2}.
\end{split} 
\ee 
This is further transformed using Schouten identities. Writing everything with the common denominator $\langle q1\rangle\langle q2\rangle\langle q3\rangle\langle q4\rangle\langle 23\rangle\langle 34\rangle\langle 12\rangle^2$ we get the following expression
\be 
\label{te}
\langle 24\rangle^2
\langle q1\rangle \langle q3\rangle + \langle 23\rangle \langle 41\rangle \langle q2\rangle \langle q4\rangle
+
\langle 12\rangle \langle 34\rangle \langle q2\rangle \langle q4\rangle\\
+
\langle 12\rangle\langle 23\rangle\langle q4\rangle^2+
\langle 34\rangle\langle 41\rangle\langle q2\rangle^2.
\ee
We can transform the first term using
\be 
\langle q1\rangle \langle 24\rangle=\langle q2\rangle\langle 14\rangle-\langle q4\rangle\langle 12\rangle, ~~\langle q3\rangle\langle 24\rangle=\langle q2\rangle\langle 34\rangle-\langle q4\rangle\langle 32\rangle. 
\ee 
This gives four terms that cancel the remaining terms in (\ref{te}). The coefficients in front of $\langle q|x|3]^2$ and
$\langle q|x|3]\langle q|x|4]$ are checked to vanish analogously.

\section{Five point case}
Let us now show that the 5-point amplitude is region momenta independent. To this end, we parametrise the region momenta in terms of one of them and the external momenta 
\be 
p_1=x,~~p_2=2+x,~~p_3=3+2+x,~~p_4=4+3+2+x,~~p_5=x-1.
\ee 
Let us start by analysing the terms linear in $x$. We exclude terms containing the region momenta $p_1$ because such terms are always quadratic in $x$. We extract the linear terms in $x$ from the remaining ones. For example, there is such term coming from
$\langle q|p_4\circ p_3|q\rangle^2$. Indeed, we have 
\be 
\langle q|p_4\circ p_3|q\rangle^2=\langle q|(4+3+2+x)\circ(3+2+x)|q\rangle^2=\langle q|(3+2+x)\circ 4|q\rangle^2\nonumber\\
=\langle q|(3+2)\circ 4|q\rangle^2+2\langle q|3+2\circ 4|q\rangle\langle q|x\circ 4|q\rangle+\langle q|x\circ 4|q\rangle^2.
\ee 
We can see that the second term on right hand side of the above equation is linear in $x$. Collecting such terms, we have 
\be 
\label{x1}
\mathcal{A}_5^x=2\langle q|3\circ x|q\rangle\Bigg(\frac{[32]}{\langle 3q\rangle\langle q4\rangle\langle 45\rangle\langle 51\rangle\langle 12\rangle}+\frac{[21]}{\langle q3\rangle\langle 5q\rangle\langle 45\rangle\langle 34\rangle\langle 12\rangle}+
\frac{[42]}{\langle q3\rangle\langle 5q\rangle\langle 34\rangle\langle 51\rangle\langle 12\rangle}\nonumber\\+
\frac{[32]}{\langle q5\rangle\langle 4q\rangle\langle 34\rangle\langle 51\rangle\langle 12\rangle}\Bigg)\nonumber
\ee
\be 
+
2\langle q|4\circ x|q\rangle\Bigg(\frac{[43]}{\langle 4q\rangle\langle q5\rangle\langle 23\rangle\langle 51\rangle\langle 12\rangle}+\frac{[42]\langle 2q\rangle}{\langle 3q\rangle\langle 4q\rangle\langle 5q\rangle\langle 51\rangle\langle 23\rangle\langle 12\rangle}+
\frac{[42]}{\langle q3\rangle\langle 5q\rangle\langle 34\rangle\langle 51\rangle\langle 12\rangle}\nonumber\\+
\frac{[32]}{\langle q5\rangle\langle 4q\rangle\langle 34\rangle\langle 51\rangle\langle 12\rangle}
-
\frac{[12]}{\langle 5q\rangle\langle q3\rangle\langle 34\rangle\langle 45\rangle\langle 12\rangle}
-
\frac{[41]}{\langle q5\rangle\langle 3q\rangle\langle 23\rangle\langle 45\rangle\langle 12\rangle}\nonumber\\
-
\frac{[51]}{\langle q4\rangle\langle 3q\rangle\langle 23\rangle\langle 45\rangle\langle 12\rangle}
\Bigg)\nonumber
\ee 
\be
\begin{split} 
&+
2\langle q|5\circ x|q\rangle\Bigg(\frac{[21]}{\langle q3\rangle\langle 5q\rangle\langle 45\rangle\langle 34\rangle\langle 12\rangle}+\frac{[51]}{\langle q4\rangle\langle 5q\rangle\langle 23\rangle\langle 34\rangle\langle 12\rangle}-
\frac{[41]}{\langle q5\rangle\langle 3q\rangle\langle 23\rangle\langle 45\rangle\langle 12\rangle}\\
&-
\frac{[51]}{\langle q4\rangle\langle 3q\rangle\langle 23\rangle\langle 45\rangle\langle 12\rangle}\Bigg).
\end{split} 
\ee 
Let us compute the bracketed terms for each of the coefficients $\langle q|3\circ x|q\rangle, \langle q|4\circ x|q\rangle, \langle q|5\circ x|q\rangle$. After bringing to a common denominator, we have for the coefficient of $\langle q|3\circ x|q\rangle$
\be 
\frac{[23]\langle34\rangle\langle q5\rangle+[12]\langle 51\rangle\langle q4\rangle+[24]\langle 45\rangle\langle 4q\rangle+[23]\langle 45\rangle\langle q3\rangle}{\langle 45\rangle\langle 51\rangle\langle 12\rangle\langle 34\rangle\langle q3\rangle\langle q4\rangle\langle q5\rangle}.
\ee 
Using Schouten identity, we can write the first and fourth terms in the numerator as 
\be 
\label{e1}
[23](\langle 34\rangle\langle q5\rangle+\langle 45\rangle\langle q3\rangle)=-[23]\langle 35\rangle\langle 4q\rangle.
\ee 
We next use momentum conservation to write the second and third terms in the numerator as 
\be 
\label{e2}
\langle q4\rangle([21]\langle 15\rangle+[24]\langle 45\rangle)=-\langle q4\rangle[23]\langle 35\rangle. 
\ee 
Clearly, the computed terms in (\ref{e1}) and (\ref{e2}) cancel each other. Thus, the coefficient of $\langle q|3 \circ x|q\rangle$ in (\ref{x1}) vanishes.Let us now compute the coefficients of $\langle q|4\circ x|q\rangle$. Putting the coefficients in (\ref{x1}) over a common denominator, we have
\be
\begin{split} 
\label{t}
&\frac{1}{\langle 45\rangle\langle 51\rangle\langle 12\rangle\langle 34\rangle\langle 23\rangle\langle 3q\rangle\langle 4q\rangle\langle 5q\rangle}\Big([43]\langle34\rangle\langle 45\rangle\langle 3q\rangle\\&+[42]\langle 34\rangle\langle 45\rangle\langle 2q\rangle+[42]\langle 45\rangle\langle 23\rangle\langle 4q\rangle+[32]\langle 45\rangle\langle 23\rangle\langle 3q\rangle+[12]\langle 51\rangle\langle 23\rangle\langle 4q\rangle\\&+[41]\langle 51\rangle\langle 34\rangle\langle 4q\rangle+[51]\langle 51\rangle\langle 34\rangle\langle 5q\rangle\Big).
\end{split} 
\ee 
First, let us apply momentum conservation on the fifth and sixth terms in the above. Thus we have  
\be 
\label{t1}
[12]\langle 23\rangle\langle 51\rangle\langle 4q\rangle+[14]\langle 43\rangle\langle 51\rangle\langle 4q\rangle=[15]\langle 35\rangle\langle51\rangle\langle 4q\rangle.
\ee 
Let us then combine the term on the right hand side in (\ref{t1}) with the last term in (\ref{t}) using Schouten identity. We have 
\be 
\begin{split} 
[15]\langle 35\rangle\langle51\rangle\langle 4q\rangle+[51]\langle 51\rangle\langle 34\rangle\langle 5q\rangle&=[15]\langle 51\rangle (\langle 35\rangle\langle 4q\rangle-\langle 34\rangle\langle 5q\rangle)\\
&=[15]\langle 51\rangle\langle 54\rangle\langle 3q\rangle.
\end{split} 
\ee 
We also apply Schouten identity to combine the second and third terms in (\ref{t})
\be 
\begin{split} 
[42]\langle 45\rangle\langle 34\rangle\langle 2q\rangle+[42]\langle 45\rangle\langle 23\rangle\langle 4q\rangle&=[42]\langle 45\rangle(\langle 34\rangle\langle 2q\rangle+\langle 23\rangle\langle 4q\rangle)\\
&=[42]\langle 45\rangle\langle 24\rangle\langle 3q\rangle. \end{split} 
\ee 
All in all, we finally reduced the numerator in (\ref{t}) to the following sum of terms
\be 
\label{sum1}
\Big(-[15]\langle 15\rangle +[32]\langle 32\rangle
+[43]\langle 43\rangle
+[42]\langle 42\rangle\Big)\langle 45\rangle\langle 3q\rangle.
\ee 
However, momentum conservation gives us
\be 
(1+5)^2=(2+3+4)^2
\implies [15]\langle 15\rangle=[32]\langle 32\rangle+[43]\langle 43\rangle+[42]\langle 42\rangle.
\ee 
Thus the terms in (\ref{sum1}) add up to zero. Therefore we find that the coefficient of $\langle q|4\circ x|q\rangle)$ vanishes. Next we compute the coefficient of $\langle q|5\circ x|q\rangle$. Once again, we put the relevant coefficient in (\ref{x1}) over a common denominator. We have 
\be 
\label{t2}
\frac{[21]\langle 23\rangle\langle q4\rangle+[51]\langle 45\rangle\langle q3\rangle+[41]\langle 43\rangle\langle q4\rangle+[51]\langle 34\rangle\langle q5\rangle}{\langle 12\rangle\langle 23\rangle\langle 34\rangle\langle 45\rangle\langle q3\rangle\langle q4\rangle\langle q5\rangle}.
\ee 
Consider the numerator in (\ref{t2}). Using Schouten identity for the second and fourth terms, we write it as
\be 
\label{t3} 
[51]\langle 45\rangle\langle q3\rangle+[51]\langle 34\rangle\langle q5\rangle=[51]\langle 53\rangle\langle 4q\rangle.
\ee 
We can then write the numerator in (\ref{t2}) as 
\be 
[21]\langle 23\rangle\langle q4\rangle +[41]\langle 43\rangle\langle q4\rangle-[51]\langle 53\rangle\langle q4\rangle\nonumber\\
=([1|2+4|3\rangle +[1|5|3\rangle)\langle q4\rangle
=0.
\ee 

Thus we have shown that all the linear terms in $x$ vanish, i.e 
\be 
\mathcal{A}^x_5=0
\ee 
Consider now the quadratic terms in $x$. The relevant ones are
\be
\begin{split} 
\mathcal{A}^{x^2}_5&=\langle q|1\circ x|q\rangle^2\Big(J(1)J(2,3,4,5)+J(5,1)J(2,3,4)+J(1,2)J(3,4,5)\Big)\\&+
\langle q|2\circ x|q\rangle^2 \Big(J(2,3)J(4,5,1)+J(1,2)J(3,4,5)+J(2)J(3,4,5,1)\Big)\\&+
\langle q|3\circ x|q\rangle^2\Big(J(3)J(4,5,1,2)+J(3,4)J(5,1,2)+J(2,3)J(4,5,1)\Big)\\&+\langle q|4\circ x|q\rangle^2\Big(J(4)J(5,1,2,3)+J(3,4)J(5,1,2)+J(4,5)J(1,2,3)\Big)\\
&+
\langle q|5\circ x|q\rangle^2\Big(J(5)J(1,2,3,4)+J(4,5)J(1,2,3)+J(5,1)J(2,3,4)\Big)
\\&+2\langle q|1\circ x|q\rangle\langle q|2\circ x|q\rangle J(1,2)J(3,4,5)+2\langle q|4\circ x|q\rangle\langle q|3\circ x|q\rangle J(3,4)J(5,1,2)\\&+2\langle q|4\circ x|q\rangle\langle q|5\circ x|q\rangle J(4,5)J(1,2,3)
+2\langle q|3\circ x|q\rangle\langle q|2\circ x|q\rangle J(2,3)J(4,5,1)\\&+2\langle q|1\circ x|q\rangle\langle q|5\circ x|q\rangle J(5,1)J(2,3,4).
\end{split} 
\ee 
There are two kinds of terms. One is of the form $\langle q|i\circ x|q\rangle^2, i=1,2,..5$ and another of the form $\langle q|i\circ x|q\rangle\langle q|j\circ x|q\rangle$. 
We want to demonstrate $\mathcal{A}^{x^2}_5$ vanishes using momentum conservation and Schouten identity. First, we have the following set of relations on using momentum conservation identities 
\be 
\langle q|x|2]\langle 21\rangle+\langle q|x|3]\langle 31\rangle+\langle q|x|4]\langle 41\rangle+\langle q|x|5]\langle 51\rangle=0,\nonumber\\
\langle q|x|1]\langle 12\rangle+\langle q|x|3]\langle 32\rangle+\langle q|x|4]\langle 42\rangle+\langle q|x|5]\langle 52\rangle=0,\nonumber\\
\langle q|x|2]\langle 23\rangle+\langle q|x|1]\langle 13\rangle+\langle q|x|4]\langle 43\rangle+\langle q|x|5]\langle 53\rangle=0,\nonumber\\
\langle q|x|2]\langle 24\rangle+\langle q|x|3]\langle 34\rangle+\langle q|x|1]\langle 14\rangle+\langle q|x|5]\langle 54\rangle=0,\nonumber\\
\langle q|x|2]\langle 25\rangle+\langle q|x|3]\langle 35\rangle+\langle q|x|4]\langle 45\rangle+\langle q|x|1]\langle 15\rangle=0.
\ee 
Using this, we can express any two of the quantities $\langle q|x|i]$ in terms of the other three. Thus we express 
\be 
\langle q|x|1]\langle 12\rangle=\langle q|x|3]\langle 23\rangle+\langle q|x|4]\langle 24\rangle+\langle q|x|5]\langle 25\rangle,\nonumber\\
\langle q|x|2]\langle 21\rangle=\langle q|x|3]\langle 13\rangle+\langle q|x|4]\langle 14\rangle+\langle q|x|5]\langle 15\rangle.
\ee 
We next collect coefficients of terms like $\langle q|3\circ x|q\rangle^2, \langle q|4\circ x|q\rangle^2, \langle q|4\circ x|q\rangle\langle|q|5\circ x|q\rangle$, etc. For instance, the coefficient of $\langle q|4\circ x|q\rangle\langle|q|5\circ x|q\rangle$ is given by 
\be 
\begin{split} 
\label{x11}
&-\Bigg(\frac{\langle 1q\rangle\langle 2q\rangle\langle 15\rangle\langle 24\rangle}{\langle 4q\rangle\langle 5q\rangle\langle12\rangle^2}+\frac{\langle 1q\rangle\langle 2q\rangle\langle 25\rangle\langle 14\rangle}{\langle 4q\rangle\langle 5q\rangle\langle12\rangle^2}\Bigg)J(1,2)J(3,4,5)\\&+\frac{\langle 24\rangle\langle 25\rangle\langle 1q\rangle^2}{\langle 12\rangle^2\langle 4q\rangle\langle 5q\rangle}\Bigg(J(1)J(2,3,4,5)+J(5,1)J(2,3,4)+J(1,2)J(3,4,5)\Bigg)\\&+\frac{\langle 14\rangle\langle 15\rangle\langle 2q\rangle^2}{\langle 21\rangle^2\langle 4q\rangle\langle 5q\rangle}\Bigg(J(2,3)J(4,5,1)+J(1,2)J(3,4,5)+J(2)J(3,4,5,1)\Bigg)\\&+J(4,5)J(1,2,3)+\frac{\langle 24\rangle\langle 1q\rangle}{\langle 4q\rangle\langle 12\rangle} J(5,1)J(2,3,4).
\end{split}
\ee
Let us simplify the above terms. Note that the numerator of the coefficients of $J(1,2)J(3,4,5)$ in the second and third bracketed terms can be expanded using the  Schouten identity. We have 
\be 
\langle 24\rangle\langle 25\rangle\langle 1q\rangle^2=\langle 1q\rangle\langle 2q\rangle\langle 15\rangle\langle24\rangle-\langle 1q\rangle\langle5q\rangle\langle 24\rangle\langle 12\rangle,\nonumber\\
\langle 14\rangle\langle15\rangle\langle 2q\rangle^2=\langle 2q\rangle\langle 5q\rangle\langle 12\rangle\langle 14\rangle+\langle 1q\rangle\langle 2q\rangle\langle 25\rangle\langle 14\rangle. 
\ee 
Some of the terms in the right hand side now cancel the terms of the first bracketed expression in (\ref{x1}). In a similar way we combine the coefficients of $J(5,1)J(2,3,4)$ using Schouten identity.
Then we are left with the following terms 
\be 
\begin{split} 
\label{x2}
&\frac{\langle 24\rangle\langle 25\rangle\langle 1q\rangle^2}{\langle 12\rangle^2\langle 4q\rangle\langle 5q\rangle}\Bigg(J(1)J(2,3,4,5)\Bigg)\\&+\frac{\langle 14\rangle\langle 15\rangle\langle 2q\rangle^2}{\langle 21\rangle^2\langle 4q\rangle\langle 5q\rangle}\Bigg(J(2,3)J(4,5,1)+J(2)J(3,4,5,1)\Bigg)+J(4,5)J(1,2,3)\\&+J(5,1)J(2,3,4)+ J(1,2)J(3,4,5).
\end{split}
\ee
Putting everything over the common denominator $\langle 1q\rangle^2\langle 2q\rangle^2\langle 4q\rangle^2\langle 5q\rangle^2\langle 3q\rangle\langle 12\rangle^2\langle 23\rangle\langle34\rangle\langle45\rangle\langle51\rangle$ we get the following terms in the numerator 
\be 
\label{num1}
\langle 24\rangle\langle 25\rangle\langle 51\rangle\langle 1q\rangle^2\langle 2q\rangle\langle 3q\rangle\langle 4q\rangle
+
\langle 14\rangle\langle 1\rangle\langle 23\rangle\langle 1q\rangle\langle 4q\rangle\langle 5q\rangle\langle 2q\rangle^2
\nonumber\\+
\langle 14\rangle\langle 15\rangle\langle 34\rangle\langle 2q\rangle^3\langle 1q\rangle\langle 5q\rangle
+
\langle 12\rangle\langle 34\rangle\langle 51\rangle\langle 1q\rangle\langle 2q\rangle^2\langle 4q\rangle\langle 5q\rangle
\nonumber\\+
\langle 24\rangle\langle 45\rangle\langle 15\rangle\langle 1q\rangle^2\langle 2q\rangle^2\langle 3q\rangle
+
\langle 12\rangle\langle 23\rangle\langle 51\rangle\langle 1q\rangle\langle 2q\rangle\langle 4q\rangle^2\langle 5q\rangle.
\ee 
We next show that all these terms add up to zero by repeated use of Schouten identities. First, consider the term $\langle 14\rangle\langle 15\rangle\langle 34\rangle\langle 2q\rangle^3\langle 1q\rangle\langle 5q\rangle$. Using Schouten identity for the pair $\langle 14\rangle\langle 2q\rangle$, we can expand it as 
\be 
\langle 14\rangle\langle 15\rangle\langle 34\rangle\langle 2q\rangle^3\langle 1q\rangle\langle 5q\rangle=\langle 12\rangle\langle 15\rangle\langle 34\rangle\langle 2q\rangle^2\langle 1q\rangle\langle 5q\rangle\langle 4q\rangle\nonumber\\-\langle 42\rangle\langle 15\rangle\langle 34\rangle\langle 2q\rangle^2\langle 1q\rangle^2\langle 5q\rangle.
\ee 
We again transform the second term on the right hand side to get
\be 
\label{num2}
\langle 14\rangle\langle 15\rangle\langle 34\rangle\langle 2q\rangle^3\langle 1q\rangle\langle 5q\rangle=\langle 12\rangle\langle 15\rangle\langle 34\rangle\langle 2q\rangle^2\langle 1q\rangle\langle 5q\rangle\langle 4q\rangle\nonumber\\-\langle 42\rangle\langle 15\rangle\langle 35\rangle\langle 2q\rangle^2\langle 1q\rangle^2\langle 4q\rangle\nonumber\\+
\langle 42\rangle\langle 15\rangle\langle 45\rangle\langle 2q\rangle^2\langle 1q\rangle^2\langle 3q\rangle.
\ee 
The first and the third terms on the right hand side of (\ref{num2}) cancel the fourth and the fifth terms in (\ref{num1}) respectively. We are then left with the following terms 
\be 
\label{num3}
\langle 24\rangle\langle 25\rangle\langle 51\rangle\langle 1q\rangle^2\langle 2q\rangle\langle 3q\rangle\langle 4q\rangle
+
\langle 14\rangle\langle 15\rangle\langle 23\rangle\langle 1q\rangle\langle 4q\rangle\langle 5q\rangle\langle 2q\rangle^2
\nonumber\\
-
\langle 42\rangle\langle 15\rangle\langle 35\rangle\langle 2q\rangle^2\langle 1q\rangle^2\langle 4q\rangle
+
\langle 12\rangle\langle 23\rangle\langle 51\rangle\langle 1q\rangle\langle 2q\rangle\langle 4q\rangle^2\langle 5q\rangle.
\ee 
Let us now expand the term $\langle 12\rangle\langle 23\rangle\langle 51\rangle\langle 1q\rangle\langle 2q\rangle\langle 4q\rangle^2\langle 5q\rangle$ using Schouten identity for the pair $\langle 12\rangle\langle 4q\rangle$. We have 
\be 
\label{num4}
\langle 12\rangle\langle 23\rangle\langle 51\rangle\langle 1q\rangle\langle 2q\rangle\langle 4q\rangle^2\langle 5q\rangle=\langle 14\rangle\langle 23\rangle\langle 51\rangle\langle 1q\rangle\langle 2q\rangle^2\langle 4q\rangle\langle 5q\rangle\nonumber\\
-\langle 24\rangle\langle 23\rangle\langle 51\rangle\langle 1q\rangle^2\langle 2q\rangle\langle 4q\rangle\langle 5q\rangle.
\ee 
The first term on the right hand side in (\ref{num4}) cancels the second term in (\ref{num3}). Finally, we have the remaining terms 
\be 
\langle 24\rangle\langle 25\rangle\langle 51\rangle\langle 1q\rangle^2\langle 2q\rangle\langle 3q\rangle\langle 4q\rangle
-
\langle 42\rangle\langle 15\rangle\langle 35\rangle\langle 2q\rangle^2\langle 1q\rangle^2\langle 4q\rangle
\nonumber\\-\langle 24\rangle\langle 23\rangle\langle 51\rangle\langle 1q\rangle^2\langle 2q\rangle\langle 4q\rangle\langle 5q\rangle.
\ee 
Taking a factor of $\langle 24\rangle\langle 51\rangle\langle 1q\rangle^2\langle 4q\rangle\langle 2q\rangle$, we can write the above as 
\be 
\langle 24\rangle\langle 51\rangle\langle 1q\rangle^2\langle 4q\rangle\langle 2q\rangle\Big(\langle 25\rangle\langle 3q\rangle-\langle 23\rangle\langle 5q\rangle-\langle 35\rangle\langle 2q\rangle\Big).
\ee 
The bracketed terms then vanish due to the Schouten identity. Thus we have shown that the coefficient of $\langle q|4\circ x|q\rangle\langle q|5\circ x|q\rangle$ vanish completely. Analogously, it is checked that all the coefficients of the other quadratic terms vanish similarly. Thus the linear and quadratic terms in $x$ for the five point amplitude all vanish, establishing that it is region momenta independent.
\chapter{BRST closure} 
In this part, we verify that the BRST transformations outlined in (\ref{brstn}) are nilpotent. We already described this for the trivial set of transformations in the main text. Here, we detail out the rest of the non-trivial sets. 

\section{Diffeomorphism ghost} 
Let us start with the second transformation in (\ref{brstn}), which is that of the diffeomorphism ghost. We have 
\be 
\begin{split} 
s^2c_{\mu}&=sc^{\lambda}\nabla^T_{\lambda}c_{\mu}-c^{\lambda}\nabla^T_{\lambda}sc^{\mu}\\
&=sc^{\lambda}\nabla_{\lambda}c_{\mu}-c^{\lambda}\nabla_{\lambda}sc^{\mu}\\
&=c^{\gamma}\nabla_{\gamma}c^{\lambda}\nabla_{\lambda}c^{\mu}-c^{\lambda}\nabla_{\lambda}c^{\nu}\nabla_{\nu}c^{\mu}-c^{\lambda}c^{\nu}\nabla_{\lambda}\nabla_{\nu}c^{\mu}\\
&=-c^{\lambda}c^{\nu}\nabla_{\lambda}\nabla_{\nu}c^{\mu}=-\frac{1}{2}(c^{\lambda}c^{\nu}-c^{\nu}c^{\lambda})\nabla_{\lambda}\nabla_{\nu}c^{\mu}\\
&=-\frac{1}{2}c^{\lambda}c^{\nu}[\nabla_{\lambda},\nabla_{\nu}]c^{\mu}\\&=-\frac{1}{2}c^{\lambda}c^{\nu}R^{\mu}_{~\alpha\nu\lambda}c^{\alpha}\\
&=-\frac{1}{2}c^{\alpha}c^{\lambda}c^{\nu}R^{\mu}_{~\alpha\nu\lambda}\\
&=-\frac{1}{6}c^{\alpha}c^{\lambda}c^{\nu}(R^{\mu}_{~\alpha\nu\lambda}+R^{\mu}_{~\nu\lambda\alpha}+R^{\mu}_{~\lambda\alpha\nu})=0.
\end{split}
\ee 
where we have used the sign carefully to take care of the anti-commutativity of the ghosts. In the fourth line of the above, we used this property to write the single term as a combination of two terms, which then becomes the commutator of the covariant derivatives. Finally, we express this commutator in terms of the Riemann tensor and used the symmetry properties of it to deduce that the expression vanishes due to the first Bianchi identity.
\section{Tetrad}
For the transformation of the tetrad, we have 
\be 
\label{tetrad}
\begin{split}
s^2h^i_{\mu}&=sc^{\nu}\nabla_{\nu}h^{i}_{\mu}(x)-c^{\nu}\nabla_{\nu}sh^{i}_{\mu}(x)+sh^{i}_{\nu}(x)\nabla_{\mu}c^{\nu}+h^{i}_{\nu}(x)\nabla_{\mu}sc^{\nu}\\~\\
&=c_{\gamma}\nabla^{\gamma}c^{\delta}\nabla_{\delta} h^i_{\mu}-c^{\nu}\nabla_{\nu}c^{\delta}\nabla_{\delta}h^i_{\mu}-c^{\nu}c^{\delta}\nabla_{\delta}\nabla_{\nu}h^i_{\mu}\\&-c^{\gamma}\nabla_{\gamma}h^i_{\nu}\nabla_{\mu}c^{\nu}-c^{\gamma}h^i_{\nu}\nabla_{\gamma}\nabla_{\mu}c^{\nu}+c^{\gamma}\nabla_{\gamma}h^i_{\nu}\nabla_{\mu}c^{\nu}+h^i_{\gamma}\nabla_{\nu}c^{\gamma}\nabla_{\mu}c^{\nu}\\&+h^i_{\nu}\nabla_{\mu}c^{\gamma}\nabla_{\gamma}c^{\nu}+h^i_{\nu}c^{\gamma}\nabla_{\mu}\nabla_{\gamma}c^{\nu}\\
&=h^i_{\nu}c^{\gamma}\nabla_{\mu}\nabla_{\gamma}c^{\nu}-c^{\gamma}h^i_{\nu}\nabla_{\gamma}\nabla_{\mu}c^{\nu}-c^{\gamma}c^{\delta}\nabla_{\delta}\nabla_{\gamma}h^i_{\mu}\\
&=h^i_{\nu}c^{\gamma}[\nabla_{\mu},\nabla_{\gamma}]c^{\nu}+\frac{1}{2}c^{\delta}c^{\gamma}[\nabla_{\delta},\nabla_{\gamma}]h^i_{\mu}\\
&=h^i_{\nu}c^{\gamma} R^{\nu}_{~\alpha\gamma\mu}c^{\alpha}+\frac{1}{2}c^{\delta}c^{\gamma}R^{\beta}_{~\mu\gamma\delta}h^i_{\beta}
=h^i_{\nu}c^{\gamma}c^{\alpha} R^{\nu}_{~\alpha\gamma\mu}+\frac{1}{2}c^{\delta}c^{\gamma}R^{\beta}_{~\mu\gamma\delta}h^i_{\beta}\\
&=\frac{1}{2}h^i_{\nu}(c^{\gamma}c^{\alpha}-c^{\alpha}c^{\gamma})R^{\nu}_{~\alpha\gamma\mu}+\frac{1}{2}c^{\delta}c^{\gamma}R^{\beta}_{~\mu\gamma\delta}h^i_{\beta}\\
&=\frac{1}{2}h^i_{\nu}c^{\gamma}c^{\alpha} R^{\nu}_{~\alpha\gamma\mu}+\frac{1}{2}h^i_{\nu}c^{\gamma}c^{\alpha} R^{\nu}_{~\gamma\mu\alpha}+\frac{1}{2}c^{\delta}c^{\gamma}R^{\beta}_{~\mu\gamma\delta}h^i_{\beta}\\
&=\frac{1}{2}h^i_{\nu}c^{\gamma}c^{\alpha} (R^{\nu}_{~\alpha\gamma\mu}+R^{\nu}_{~\gamma\mu\alpha}+R^{\nu}_{~\mu\alpha\gamma})=0.
\end{split}
\ee 
In the second line of the above, many terms get cancelled and we are left with only three terms. These terms are rearranged in the form of the commutator of covariant derivatives which are further expressed again in terms of the Riemann tensor. We then used the anti-commutativity of the ghosts to expand one of the terms. The three terms now have the same form and thus we used the first Bianchi identity for the Riemann tensor in the last expression to yield a vanishing result. 
\section{Lorentz ghost} 
For the transformation of the Lorentz ghost field, we find 
\be 
\label{lg}
\begin{split}
s^2b^{ij}&=-\frac{1}{2}s[b,b]^{ij}+\frac{1}{2}s(c^{\mu}c^{\nu}F^{ij}_{~~\mu\nu})\\
&=-sb^{i}_{~k}b^{kj}+b^{i}_{~k}sb^{kj}+\frac{1}{2}sc^{\mu}c^{\nu}F^{ij}_{~~\mu\nu}\\
&-\frac{1}{2}c^{\mu}sc^{\nu}F^{ij}_{~~\mu\nu}+\frac{1}{2}c^{\mu}c^{\nu}sF^{ij}_{~~\mu\nu}\\
&=b^{i}_{~m}b^m_{~k}b^{kj}-\frac{1}{2}c^{\mu}c^{\nu}F^{i}_{~k\mu\nu}b^{kj}-b^i_{~k}b^{k}_{~m}b^{mj}\\
&+\frac{1}{2}b^i_{~k}c^{\mu}c^{\nu}F^{kj}_{~~\mu\nu}+\frac{1}{2}c^{\lambda}\nabla^T_{\lambda}c^{\mu}c^{\nu}F^{ij}_{~~\mu\nu}
-\frac{1}{2}c^{\mu}c^{\lambda}\nabla^T_{\lambda}c^{\nu}F^{ij}_{~~\mu\nu}\\
&+\frac{1}{2}c^{\mu}c^{\nu}c^{\lambda}\nabla^T_{\lambda}F^{ij}_{~~\mu\nu}+\frac{1}{2}c^{\mu}c^{\nu}F^{ij}_{~~\mu\lambda}\nabla^T_{\nu}c^{\lambda}+\frac{1}{2}c^{\mu}c^{\nu}F^{ij}_{~~\lambda\nu}\nabla^T_{\mu}c^{\lambda}+\frac{1}{2}c^{\mu}c^{\nu}[F_{\mu\nu},b]^{ij}\\
&=\frac{1}{2}c^{\mu}c^{\nu}c^{\lambda}\nabla^T_{\lambda}F^{ij}_{~~\mu\nu}
=\frac{1}{6}c^{\mu}c^{\nu}c^{\lambda}(\nabla^T_{\lambda}F^{ij}_{~~\mu\nu}+\nabla^T_{\mu}F^{ij}_{~~\nu\lambda}+\nabla^T_{\nu}F^{ij}_{~~\lambda\mu})\\
&=0.
\end{split}
\ee
where in the last line, we used the Bianchi identity and the fact that the Lorentz ghost fields anti commute and hence can be appropriately permuted to give rise to three such terms.

\section{Connection}
Let us now check the closure for the connection field. We have \\
\be
\label{connection}
\begin{split}
s^2\omega^{ij}_{~\mu}&=\nabla^T_{\mu}sb^{ij}+sc^{\nu}F^{ij}_{~~\nu\mu}-c^{\nu}sF^{ij}_{~~\nu\mu}\\
&=-\nabla^T_{\mu}(b^i_{~k}b^{kj})+\frac{1}{2}\nabla^T_{\mu}(c^{\mu}c^{\nu}F^{ij}_{~~\mu\nu})+c^k\nabla^T_{k}c^{\nu}F^{ij}_{~~\nu\mu}-c^{\nu}[F_{\nu\mu},b]^{ij}\\
&-c^{\nu}c^{\lambda}\nabla^T_{\lambda}F^{ij}_{~~\nu\mu}-c^{\nu}F^{ij}_{~~\nu\lambda}\nabla^T_{\mu}c^{\lambda}-c^{\nu}F^{ij}_{~~\lambda\mu}\nabla^T_{\nu}c^{\lambda}\\
&=-\nabla^T_{\mu}(b^i_{~k}b^{kj})-\frac{1}{2}c^{\nu}\nabla^T_{\mu}c^{\mu}F^{ij}_{~~\mu\nu}+\frac{1}{2}c^{\mu}\nabla^T_{\mu}c^{\nu}F^{ij}_{~~\mu\nu}\\&+\frac{1}{2}c^{\mu}c^{\nu}\nabla^T_{\mu}F^{ij}_{~~\mu\nu}+c^k\nabla^T_{k}c^{\nu}F^{ij}_{~~\nu\mu}-c^{\nu}c^{\lambda}\nabla^T_{\lambda}F^{ij}_{~~\nu\mu}\\&-c^{\nu}F^{ij}_{~~\nu\lambda}\nabla^T_{\mu}c^{\lambda}-c^{\nu}F^{ij}_{~~\lambda\mu}\nabla^T_{\nu}c^{\lambda}-c^{\nu}[F_{\nu\mu},b]^{ij}\\
&=-\nabla^T_{\mu}(b^i_{~k}b^{kj})-c^{\nu}[F_{\nu\mu},b]^{ij}\\
&-\frac{1}{2}c^{\nu}c^{\lambda}(\nabla^T_{\lambda}F^{ij}_{~~\nu\mu}+\nabla^T_{\nu}F^{ij}_{~~\mu\lambda}+\nabla^T_{\mu}F^{ij}_{~~\lambda\nu})\\
&=0.
\end{split}
\ee 
where in the above, we used the Bianchi identity for the curvature, interchanged dummy indices and used the anti-commutative nature of the ghost fields.
This completes the check of the nilpotency of the BRST operator.
\chapter{Self energy computation in gravity}
\section{Shift Computation}
\subsection{Linear part}
The linear part of the shift is given by 
\be 
-i\lim_{x\to\infty}\int\frac{d\Omega}{(2\pi)^4}s'_{\mu}x^{\mu}\frac{(\langle q|x+s'|p]\langle q|x+k|p])^2}{x^2}\Bigg(1-\frac{2x.(s'+s)}{x^2}\Bigg).
\ee 
The non-zero contribution can only come from the quadratic and quartic in $x$ terms of this shift, after taking an average over all directions. Note, the $x^6$ term do not contribute since it vanishes after the possible contractions. The quadratic term is 
\be 
2k'_{\mu}x^{\mu}\langle q|x|p]\Big(\langle q|s|p]\langle q|k'|p]^2+\langle q|s|p]^2\langle q|s'|p]\Big).
\ee 
This gives the shift 
\be 
-\frac{2i}{32\pi^2}\langle q|s'|p]^2\langle q|s|p]\Big(\langle q|s|p]+\langle q|s'|p]\Big).
\ee 
\\The quartic part of the shift is given by 
\be 
2i\lim_{x\to\infty}\int\frac{d\Omega}{(2\pi)^4}s'_{\mu}x^{\mu}(s'+s)_{\nu}x^{\nu}\frac{\langle q|x|p]\langle q|x|p]\Big(\langle q|s|p]^2+\langle q|s'|p]^2+4\langle q|s|p]\langle q|s'|p]\Big)}{x^4}.
\ee 
The integral is computed using 
\be
\label{quart}
\int\frac{d\Omega}{(2\pi)^4}\frac{x_{\mu}x_{\nu}x_{\rho}x_{\sigma}}{x^4}=\frac{1}{32.6\pi^2}(\eta_{\mu\nu}\eta_{\rho\sigma}+\eta_{\mu\rho}\eta_{\nu\sigma}+\eta_{\mu\sigma}\eta_{\rho\nu}).
\ee 
 The $x^6$ contractions vanish while the $x^4$ contractions give \be
\begin{split} 
\frac{i}{32.3\pi^2}\langle q|s'|p]\langle q|s'+s|p]\times\Big(\langle q|s|p]^2+\langle q|s'|p]^2+4\langle q|s|p]\langle q|s'|p]\Big).
\end{split}
\ee 
\subsection{Quadratic part}
For the quadratic part of the shift, the integral is given by 
\be 
\label{quadratic}
\frac{i}{2}\lim_{x\to\infty}\int\frac{d\Omega}{(2\pi)^4}s'_{\mu}s'_{\nu}x^{\mu}x^2\frac{\partial}{\partial x_{\nu}}\frac{(\langle q|x+s'|p]\langle q|x+s|p])^2}{(x+s')^2(x+s)^2}.
\ee 
When the derivative hits the denominator, it produces a factor proportional to 
\be 
\frac{i}{2}(-4)\lim_{x\to\infty}\int\frac{d\Omega}{(2\pi)^4}k'_{\mu}k'_{\nu}x^{\mu}x^{\nu}\frac{\langle q|x|p]^2\Big(\langle q|s|p]^2+\langle q|s'|p]^2+4\langle q|s|p]\langle q|s'|p]\Big)}{x^4}.
\ee 
The quartic part of the numerator is the only one which contributes. Using (\ref{quart}) this gives
\be 
-\frac{i}{32.3\pi^2}\langle q|s'|p]^2\Big(\langle q|s|p]^2+\langle q|s'|p]^2+4\langle q|s|p]\langle q|s'|p]\Big).
\ee 
When the derivative in (\ref{quadratic}) hits the numerator, in the large $x$ limit, we get 
\be 
i\lim_{x\to\infty}\int\frac{d\Omega}{(2\pi)^4}s'_{\mu}x^{\mu}\langle q|s'|p]\frac{\Big(\langle q|x+s'|p]\langle q|x+s|p]^2+\langle q|x+s|p]\langle q|x+s'|p]^2\Big)}{x^2}
\ee 
Only the quadratic in $x$ terms in the numerator contribute, while the others vanish. Using the relevant contraction, this gives 
\be 
\frac{i}{32\pi^2}\langle q|s'|p]^2\Big(\langle q|s|p]^2+\langle q|s'|p]^2+4\langle q|s|p]\langle q|s'|p]\Big).
\ee
\subsection{Cubic part}
For the cubic part of the shift, the integral is given by 
\be 
\label{cubic}
\frac{i}{6}\lim_{x\to\infty}\int\frac{d\Omega}{(2\pi)^4}s'_{\mu}s'_{\alpha}s'_{\nu}x^{\mu}x^2\frac{\partial}{\partial x_{\alpha}}\frac{\partial}{\partial x_{\nu}}\frac{(\langle q|x+s'|p]\langle q|x+s|p])^2}{(x+s')^2(x+s)^2}.
\ee 
In the quadratic part of the shift, we evaluated the action of the derivative on the function in two parts, one in which the derivative acts on the numerator and another in which it acts on the denominator. Here, we will take these parts separately and act the second derivative on each part. Let us first consider the part which comes as a result of the action of the derivative on the denominator of the function. It is given by  
\be 
\label{der}
\frac{i}{6}(-4)\lim_{x\to\infty}\int\frac{d\Omega}{(2\pi)^4}s'_{\mu}s'_{\nu}s'_{\alpha}x^{\mu}x^{\nu}\frac{\partial}{\partial x_{\alpha}}\frac{(\langle q|x+s'|p]\langle q|x+s|p])^2}{x^4}.
\ee 
When the derivative hits the denominator, in the large $x$-limit, it gives 
\be 
\label{cc}
\frac{16i}{6}\lim_{x\to\infty}\int\frac{d\Omega}{(2\pi)^4}s'_{\mu}s'_{\nu}s'_{\alpha}x^{\mu}x^{\nu}x^{\alpha}\frac{(\langle q|x+s'|p]\langle q|x+s|p])^2}{x^6}.
\ee 
Only the $x^6$ part in the numerator contributes. Using
\be 
\lim_{x\to\infty}\int\frac{d\Omega}{(2\pi)^4}\frac{x_{\mu} x_{\nu}x_{\alpha}x_{\beta}x_{\gamma}x_{\delta}}{x^6}=\frac{1}{32.48\pi^2}\Big(\eta_{\mu\nu}\eta_{\alpha\beta}\eta_{\gamma\delta}+\textrm{perm} \Big)
\ee 
the only non vanishing contractions are six in total, giving
\be 
\frac{2i}{32.3\pi^2}\langle q|s'|p]^3\Big(\langle q|s|p]+\langle q|s'|p]\Big).
\ee 
When the derivative in (\ref{der}) hits the numerator, in the large $x$-limit, it gives 
\be 
\label{cc2}
\frac{-4i}{3}\lim_{x\to\infty}\int\frac{d\Omega}{(2\pi)^4}s'_{\mu}s'_{\nu}x^{\mu}x^{\nu}\frac{\Big(\langle q|x+s'|p]\langle q|x+s|p]^2+\langle q|x+s|p]\langle q|x+s'|p]^2\Big)}{x^4}
\ee 
Using the relevant contraction, this gives 
\be 
\frac{-2i}{32\pi^2}\langle q|s'|p]^3\Big(\langle q|s|p]+\langle q|s'|p]\Big).
\ee 
Next consider the part which comes as a result of the action of the first derivative on the numerator of the function in (\ref{cubic}). Only the $x^4$ parts in the numerator contributes, giving 
\be 
\frac{-4i}{32\pi^2}\langle q|s'|p]^3\Big(\langle q|s|p]+\langle q|s'|p]\Big).
\ee 
When the derivative hits the numerator,
only the $x^2$ terms in the numerator contribute, giving 
\be 
\begin{split} 
\frac{2i}{6}\frac{1}{32\pi^2}\Big(\langle q|s'|p]^3\langle q|s|p]+2\langle q|s'|p]^3(\langle q|s|p]+\langle q|s'|p])\\+\langle q|s'|p]^4+\langle q|s'|p]^3(\langle q|s|p]+\langle q|s'|p])\Big).
\end{split}
\ee 
Simplifying a bit, we get 
\be 
\frac{4i}{32.3\pi^2}\langle q|s'|p]^3\Big(\langle q|s|p]+\langle q|s'|p]\Big).
\ee 
\subsection{Quartic part} 
For the quartic part of the shift, the integral is given by 
\be 
\label{quartic}
\frac{i}{24}\lim_{x\to\infty}\int\frac{d\Omega}{(2\pi)^4}s'_{\beta}s'_{\mu}s'_{\alpha}s'_{\nu}x^{\mu}x^2\frac{\partial}{\partial x_{\beta}}\frac{\partial}{\partial x_{\alpha}}\frac{\partial}{\partial x_{\nu}}\frac{(\langle q|x+s'|p]\langle q|x+s|p])^2}{(x+s')^2(x+s)^2}.
\ee 
We use the same evaluation technique for the shift as before. For the cubic part of the shift, we evaluated the action of the second derivative on the resulting functions from the quadratic part in two parts, one in which the derivative acts on the numerator and another in which it acts on the denominator. Here, we will take the corresponding functions from the cubic part and act the third derivative on each part. Let us first consider the part which comes as a result of the action of the derivative on the denominator of the function in (\ref{cc}). It is given by 
\be 
-\frac{16.6i}{24}\lim_{x\to\infty}\int\frac{d\Omega}{(2\pi)^4}s'_{\beta}s'_{\mu}s'_{\nu}s'_{\alpha}x^{\mu}x^{\nu}x^{\alpha}x^{\beta}\frac{(\langle q|x+s'|p]\langle q|x+s|p])^2}{x^8}.
\ee 
Only the $x^8$ factor in the numerator contributes, with sixty non-vanishing contractions, giving 
\be 
-\frac{i}{32.120\pi^2}\langle q|s'|p]^4\times 60\\
=-\frac{i}{32.2\pi^2}\langle q|s'|p]^4.
\ee 
Next, when the third derivative in (\ref{quartic}) hits the numerator of (\ref{cc}), we have 
\be 
\frac{32.6i}{24}\lim_{x\to\infty}\int\frac{d\Omega}{(2\pi)^4}s'_{\mu}s'_{\nu}s'_{\alpha}x^{\mu}x^{\nu}x^{\alpha}\langle q|s'|p]\frac{\Big(\langle q|x+s'|p]\langle q|x+s|p]^2}{x^6}\nonumber\\+\frac{\langle q|x+s|p]\langle q|x+s'|p]^2\Big)}{x^6}
\ee 
Only the $x^6$ terms in the numerator contribute with six non-vanishing contractions, giving 
\be 
\begin{split} 
\frac{32.12i}{192.24.8\pi^2}\langle q|s'|p]^4\times 6\nonumber\\
=\frac{2i}{32\pi^2}\langle q|s'|p]^4.
\end{split}
\ee 
We now consider the action of the third derivative on the next function, in (\ref{cc2}). When the derivative hits the denominator of the function, it gives
\be 
\label{quartic3}
\frac{-32i}{24}\lim_{x\to\infty}\int\frac{d\Omega}{(2\pi)^4}s'_{\beta}s'_{\mu}k'_{\nu}x^{\mu}x^{\nu}x^{\beta}\frac{\langle q|s'|p]\Big(\langle q|x+s'|p]\langle q|x+s|p]^2}{x^6}\nonumber\\+\frac{\langle q|x+s|p]\langle q|x+s'|p]^2\Big)}{x^6}.
\ee 
Terms with only $x^6$ factor in the numerator contribute, giving \be 
\frac{-i}{32.3\pi^2}\langle q|s'|p]^4.
\ee 
When the derivative hits the numerator of the function, it gives in the large $x$-limit
\be 
\label{quartic4}
\frac{-8i}{24}\lim_{x\to\infty}\int\frac{d\Omega}{(2\pi)^4}s'_{\mu}s'_{\nu}x^{\mu}x^{\nu}\frac{\langle q|s'|p]^2\Big(\langle q|x+s|p]^2+4\langle q|x+s|p]\langle q|x+s'|p]}{x^4}\nonumber\\+\frac{\langle q|x+s'|p]^2\Big)}{x^4}.
\ee 
The $x^4$ terms in the numerator contributes, giving 
\be 
\frac{-i}{32\pi^2}\langle q|s'|p]^4.
\ee 
Next consider the function in (\ref{cc2}). When the third derivative hits the denominator in the large $x$-limit, it gives
\be 
\label{cubic22}
\frac{32i}{24}\lim_{x\to\infty}\int\frac{d\Omega}{(2\pi)^4}s'_{\beta}s'_{\mu}s'_{\alpha}x^{\mu}x^{\alpha}x^{\beta}\langle q|s'|p]\frac{\Big(\langle q|x+s'|p]\langle q|x+s|p]^2}{x^6}\nonumber\\+\frac{\langle q|x+s|p]\langle q|x+s'|p]^2\Big)}{x^6}.
\ee 
The $x^6$ terms in the numerator contributes here, giving 
\be 
\frac{i}{32.18\pi^2}\langle q|s'|p]^4.
\ee 
When the derivative hits the numerator of the function in (\ref{cc2}), it gives 
\be 
\label{cubic23}
\frac{-8i}{24}\lim_{x\to\infty}\int\frac{d\Omega}{(2\pi)^4}s'_{\mu}s'_{\alpha}x^{\mu}x^{\alpha}\langle q|s'|p]^2\frac{\Big(\langle q|x+s|p]^2+\langle q|x+s'|p]^2}{x^4}\nonumber\\+\frac{4\langle q|x+s|p]\langle q|x+s'|p]\Big)}{x^4}.
\ee 
Only the $x^4$ terms in the numerator survive, giving 
\be 
\frac{-i}{32\pi^2}\langle q|s'|p]^4.
\ee 
Next consider integral in (\ref{quartic}). When the third derivative acts on the numerator of this function, it gives 
\be 
\label{cubic32}
\frac{2i}{24}\lim_{x\to\infty}\int\frac{d\Omega}{(2\pi)^4}s'_{\mu}s'_{\beta}x^{\mu}\langle q|s'|p]\frac{\frac{\partial}{\partial x_{\beta}}\Big(\langle q|x+s|p]^2+\langle q|x+s'|p]^2}{x^2}\nonumber\\+\frac{4\langle q|x+s|p]\langle q|x+s'|p]\Big)}{x^2}.
\ee 
Evaluating the derivative and simplifying, we have
\be 
\label{cubic33}
\frac{2.6i}{24}\lim_{x\to\infty}\int\frac{d\Omega}{(2\pi)^4}s'_{\mu}s'_{\beta}x^{\mu}\langle q|s'|p]^3\frac{\Big(\langle q|x+s|p]+\langle q|x+s'|p]\Big)}{x^2};
\ee 
This gives 
\be 
\frac{i}{32.2\pi^2}\langle q|s'|p]^4.
\ee 
\subsection{Result} 
Adding all the contributions, we have for the SDGR self energy bubble, upto numerical factors
\be 
\label{shiftgr2}
\Pi^{--}_{SDGR}\sim\Bigg(\frac{\langle q|s|p]\langle q|s'|p]}{\langle qp\rangle^2}\Bigg)^2. 
\ee
\end{appendices}

\end{fmffile}
\end{document}